\documentclass[useAMS,usenatbib,usegraphicx,twocolumn]{mn2e}
  
\usepackage{float}
\usepackage{aas_macros}
\usepackage{threeparttable}
\usepackage{grffile}
\usepackage{amsmath, amssymb}
\usepackage{multirow}
\usepackage{wallpaper}
\usepackage{mathbbol}
\usepackage{rotating}
\usepackage{blindtext}
\usepackage[T1]{fontenc}

\usepackage{tocloft}
\newcommand{\fat}{\textbf}
\newcommand{\ita}{\textit}

\newcommand{\beq}{\begin{equation}}
\newcommand{\eeq}{\end{equation}}  
\newcommand{\RNum}[1]{\uppercase\expandafter{\romannumeral #1\relax}}

  \title[Deuterium fractionation in cloud cores]{Deuterium fractionation and H$_2$D$^+$ evolution in turbulent and magnetized cloud cores}
  
  \author[B. K\"ortgen]{Bastian K\"ortgen$^{1}$\thanks{bkoertgen@hs.uni-hamburg.de}, Stefano Bovino$^{1}$\thanks{stefano.bovino@uni-hamburg.de},
        Dominik~R.~G.~Schleicher$^{2}$,  \newauthor{Andrea Giannetti$^{3,4}$, and Robi Banerjee$^{1}$}\\
  $^{1}$ Hamburger Sternwarte, Universit\"at Hamburg, Gojenbergsweg 112, D-21029 Hamburg, Germany \\
  $^{2}$ Departamento de Astronom\'{i}a, Facultad Ciencias F\'{i}sicas y Matem\'{a}ticas, Universidad de Concepci\'{o}n, \\
  \  \ Av. Esteban Iturra s/n Barrio Universitario, Casilla 160--C, Concepci\'{o}n, Chile\\
  $^{3}$ INAF-Istituto di Radioastronomia, Via P. Gobetti, 101, I-40129 Bologna, Italy\\
  $^{4}$ Max-Panck Institut f\"ur RadioAstronomie, Auf dem H\"ugel 69
D-53121 Bonn, Germany\\
  }

\date{Released 2016}

\pagerange{\pageref{firstpage}--\pageref{lastpage}} \pubyear{2016}

\begin{document}

\label{firstpage}
\maketitle

\begin{abstract}
High--mass stars are expected to form from dense prestellar cores. Their precise formation conditions are widely discussed, including their virial condition, 
which results in slow collapse for super--virial cores with strong support by turbulence or magnetic fields, or fast collapse for sub--virial sources. To disentangle their formation processes, measurements of the deuterium fractions are frequently employed to approximately estimate the ages of these cores and to obtain constraints on their dynamical evolution. We here present 3D magneto--hydrodynamical simulations including for the first time an accurate non--equilibrium chemical network with 21 gas--phase species plus dust grains and 213 reactions.  With this network we model the deuteration process in fully depleted prestellar cores in great detail and determine its 
response to variations in the initial conditions. We explore the dependence on the initial gas column density, the turbulent Mach number, the mass--to--magnetic flux ratio and the distribution of the magnetic field, as well as the initial ortho--to--para ratio of H$_2$. We find excellent agreement with recent observations of 
deuterium fractions in quiescent sources. Our results show that deuteration is rather efficient, even when assuming a conservative ortho--to--para ratio of 3 and highly sub--virial initial conditions, leading to large deuterium fractions already within roughly a free--fall time. We discuss the implications of our results and give an outlook to relevant future investigations.
\end{abstract}
\begin{keywords}
\end{keywords}

\section{Introduction}
Stars form in dense and cold regions inside molecular clouds. Over the years, a basic understanding of the formation of \mbox{low--mass} stars has been established \citep{Bonnell08, Hennebelle2011, Bate12, Riaz13, Krumholz16}, while the formation of high--mass stars remains more controversial. 
In particular, one may distinguish the turbulent core accretion model by \citet{McKee2003}, suggesting the support of massive cores by supersonic turbulence and/or magnetic fields against gravity, as well as the competitive accretion scenario by \citet{Bonnell01}. 
High--mass stars are fundamental to the evolution of galaxies, as they impact the interstellar medium via their strong radiation and the injection of heavy elements into it, and affect the star formation rate by regulating the formation of dense, molecular gas out of the diffuse atomic environment.\\
However, studying the places where high--mass stars form is a challenging task as they reside in distant and rare Giant Molecular Clouds (GMCs), form preferentially in 
clusters, and have shorter evolutionary timescales compared to low--mass stars. In a series of studies \citet{Girichidis2012} showed that the initial conditions of collapsing cloud cores largely determine the subsequent evolution of individual fragments as well as the 
resulting star formation properties. They emphasize the crucial role of supersonic turbulence that can lead to a reduction of the star formation 
efficiency or the complete halt of the star formation process due to the so called fragmentation induced starvation \citep[see also][]{Peters11}. 
For highly evolved filamentary structures, recent observations have shown that magnetic, turbulent and gravitational energy are comparable \citep{Leurini2011,Busquet16,Stutz16}.
Based on such results, \citet{Seifried2015} investigated the role of turbulence and magnetic fields for star formation in interstellar filaments. They find a crucial 
influence of both turbulent fluctuations and the orientation of the magnetic field on the fragmentation of those filaments. The authors point out that 
magnetic fields can only support filaments against radial contraction when they are aligned with the filaments' major axis. In any other case, the filaments 
heavily fragment and eventually form stars with the rate, efficiency and location of the star formation process being primarily determined by the initial conditions.\\
In addition, the feedback produced by a young massive star rapidly reshapes the region where it forms making the environment dynamically even more complex \citep[see e.g.][]{Peters11,Ochsendorf16}. The latter process has been 
 extensively studied via numerical simulations including both the self--consistent formation of jets and outflows \citep{Banerjee2006,Banerjee2007,Seifried2011,Seifried2012b} as well as ionising radiation \citep{Kuiper2010,Kuiper2011,Peters11,Rosen2016}. Those 
 studies underline the destructive power of young high--mass stars, the resulting difficulties of gas accretion onto the central protostellar object and the 
 increase of complexity of the morphology of the surrounding environment. \\
To make further progress in this area, it is particularly important to understand both the lifetimes of the cores, specifically whether collapse is fast or slow, as well as the local dynamical parameters. By using deuteration as a chemical clock, \citet{Kong2015} derived core lifetimes of several free--fall times $t_\mathrm{ff}$, where
\beq
 t_\mathrm{ff}=\sqrt{\frac{3\pi}{32\,G\left<\varrho\right>}},
 \eeq
 $G$ is Newton's constant and $\left<\varrho\right>$ denotes the average gas density. The findings by \citet{Kong2015} are 
 indicative of a slow collapse, even though they find a strong dependence on the assumed chemical conditions. In contrast, other groups such as \citet{Hernandez2011} or \citet{Peretto2013} suggested collapse in about one free--fall time \citep[see also][]{Fontani2006,Fontani2011,Giannetti2014,Lackington2016}. Recently \citet{Battersby2017} reported a timescale in between 0.6-4.0 $\times t_\mathrm{ff}$ obtained from a sample of observational data which then confirm the short duration of the prestellar phase.\\ 
Infrared Dark Clouds (IRDCs) are the ideal candidates to host high--mass starless cores (HMSCs), i.e. the densest (column densities of $\sim$10$^{23}$--10$^{25}$ cm$^{-2}$) and coldest ($T < 25$~K) regions where the star formation process should occur. A recent detailed investigation by \citet{Ohashi16} of cores in the IRDC G14.225-0.506 points towards sub--virial cores that are not supported against collapse, unless magnetic fields are very strong, of the order of $4-20$~mG. In addition, observations by \citet{Busquet16} of the hub--N and hub--S region in this IRDC report field strengths of about $1$~mG implying sub--alfv\'{e}nic conditions, and also an extrapolation of the relation found by \citet{Crutcher10} leads to a similar value for 
 field strengths in such regions. 
 HMSCs are further characterized by high depletion, i.e. most of the metals and in particular C--bearing species are frozen--out onto dust grains \citep{Chen2011, Hernandez2011,Giannetti2014}. Due to these particular chemical conditions previous studies suggested that deuterium fractionation, i.e.  the ratio of deuterated to non-deuterated species, can be  considered as an appropriate chemical clock to assess the evolutionary stage of these cores \citep{Caselli2002,Fontani2011,Pagani2011,Kong2015}. In fact, deuterium fractionation is started by the following (slightly exothermic) reaction
\begin{equation}\label{eq:key}
	\mathrm{H_3^+ + HD \rightleftharpoons H_2D^+ + H_2 +  232\,\, K}
\end{equation}
where we have omitted the different nuclear spin states. In cold regions ($T \leq 20$ K) such as in starless or prestellar cores, the backward reaction is inhibited and under high depletion conditions also the following reactions  
\begin{align}
\mathrm{H_3^+ + CO} &\rightarrow \mathrm{HCO^+ + H_2}\\
\mathrm{H_2D^+ + CO} &\rightarrow \mathrm{DCO^+ + H_2}
\end{align}

\noindent do not take place, thus favouring the deuterium fractionation process.\\
Due to the difficulties in observing species without a permanent dipole moment, as for instance H$_3^+$, the current studies on deuterium fractionation in starless cores have been focused on different tracers, such as N$_2$H$^+$ and its isotopologue N$_2$D$^+$.  
Several studies reported a [D/H] ratio well above the standard cosmic value of $\sim$10$^{-5}$, with values spanning a rather large range from 0.001 to 0.1, both in low and high mass cores \citep[][]{Fontani2006,Caselli2008,Chen2011,Kong2016ApJ}. 
Despite this, the H$_3^+$ isotopologue H$_2$D$^+$ is considered an unambiguous and better tracer of the densest gas \citep{Walmsley2004} and it is also one of the main charge carriers in those regions. Hence, knowing how this species evolves is crucial not only because it is a tracer which survives longer and can probe dense central regions compared to other species, but also to assess the ionization fraction of the gas and thus to provide estimates of the ambipolar diffusion timescale \citep{Caselli2002P&SS}.\\
Ortho--H$_2$D$^+$ (henceforth o--H$_2$D$^+$) has been found to be very abundant in cold clouds \citep{Pagani1992_1} and has been observed both in low-- and high--mass cores \citep{Caselli2003,Harju2006,Vastel2006,Caselli2008,Friesen2014}, with column densities of \mbox{$N$(o--H$_2$D$^+$)$\sim$10$^{12}$--10$^{14}$cm$^{-2}$}.  
\citet{Caselli2003} reported values of 4.8$\times$10$^{13}$cm$^{-2}$ within a radius of 2800 au, while \citet{Vastel2006} observed high \mbox{$N$(o--H$_2$D$^+$)$\sim$ 1.8$\times$10$^{13}$cm$^{-2}$} at radii of about 5000 au in low--mass cores. Extended \mbox{o--H$_2$D$^+$} was also observed by \citet{Pillai2012} within a radius of 7000 au in the high--mass star--forming region Cygnus--X, and by \citet{Harju2006} in a massive prestellar core in Orion B. 
This is in agreement with the fact that high abundances of \mbox{o--H$_2$D$^+$} should be observed in the densest regions which are already highly depleted, i.e. within radii of about 6000 au as suggested by previous studies \citep{Caselli1999, Tafalla2002}.\\
From an observational point of view o--H$_2$D$^+$ is the only species that can be easily observed with its transition at 372~GHz, while \mbox{p--H$_2$D$^+$}  needs higher frequencies only available with SOFIA (THz) \citep[see e.g.][]{Brunken2014}. Even considering the observational challenge, H$_2$D$^+$ is an important tool to fully understand the molecular deuteration process and to estimate timescales related to the early evolution of (massive) prestellar cores. In addition,  the H$_2$D$^+$ ortho--to--para ratio (OPR)  linearly correlates with the H$_2$ OPR, and can then be used to probe this important physical parameter. \\
The deuterium fraction $D_{\mathrm{frac}}\equiv\left[\mathrm{H}_2\mathrm{D}^+/\mathrm{H}_3^+\right]$ is in fact strongly affected by the initial H$_2$ OPR: the lower the latter the higher the $D_{\mathrm{frac}}$. 
While it is well known that the H$_2$ OPR decreases with time and becomes very small reaching equilibrium values $\ll 1$, unfortunately its initial value is unknown. Observations in molecular clouds provide an estimate of OPR$\leq$0.1 \citep{Troscompt2009}. Besides the initial H$_2$ OPR, 
deuteration is affected by a series of physical uncertainties \citep{Kong2015}. In general the degree of deuterium fractionation depends on many 
parameters such as i) the already mentioned initial H$_2$ OPR, ii) the depletion factor, iii) the cosmic--ray flux, iv) the gas temperature, v) the gas density, and vi) the chemical time.  
The cosmic--ray flux $\zeta$ affects the formation of deuterated species, as higher $\zeta$ boosts the formation of H$_3^+$ which represents, via Eq. \ref{eq:key}, the first step to start deuterium fractionation. The gas temperature is instead related to the backward reaction in Eq. \ref{eq:key} which becomes accessible at higher temperatures, then competing with the forward reaction. It is also strongly connected to the freeze--out process which is enhanced by low temperatures and high densities \citep{Tafalla2002,Bergin2007}. Finally, depending on the initial H$_2$ OPR and cosmic--ray flux, the time needed to reach a
 high level of deuteration changes. In general, this process takes longer when starting with high H$_2$ OPR and low $\zeta$. \\
Most of the previous works make use of simplified one--zone and 1D models \citep[e.g.][]{Vastel2006,Sipila2010,Pagani2013}. However, a first set of 3D simulations to explore the deuteration in massive cores has recently been presented by \citet{Goodson2016}. While their simulations are still based on an approximated chemical network and a limited dynamical range, they claim that the observed deuterium fractions can be reproduced after 3--4 free--fall times. However, they caution that this timescale can be reduced with increasing \ita{chemical age} of the core, that is, when the core is dense enough to promote deuterium chemistry but still has not build a centrally concentrated density profile, typical of the prestellar phase.\\
 Similar to this study, we here explore the deuteration in prestellar cores in a large sample of numerical simulations employing, for the first time, an accurate chemical network, and by varying the core mass, the turbulent Mach number, the mass--to--magnetic flux ratio, the initial surface density of the cores, as well as the chemical initial conditions (namely the H$_2$ OPR). In addition, we further 
 aim to expand over the dynamical range using the adaptive mesh refinement (AMR) approach and employing a sink particle scheme to ensure an accurate numerical treatment when the Jeans resolution is reached.\\
The structure of our paper is thus as follows. In section~\ref{chemistry}, we present our chemical model. In section~\ref{method}, we explain the numerical methodology adopted in our simulations. The results are presented and described in section~\ref{results}, and a final discussion is given in section~\ref{summary}.
\section{Chemistry}\label{chemistry}
To model the evolution of H$_2$D$^+$ and $D_\mathrm{frac}$ we employ the network provided by \cite{Walmsley2004}. This model assumes full depletion, i.e. heavy elements condensed out on the grain surface, which is supposed to occur in the densest and coldest regions of molecular clouds. CO freeze--out on icy mantles for instance has been estimated to be effective at a density of $\sim$ few $10^4$ cm$^{-3}$ (e.g. \citealt{Caselli1999,Bergin2001,Bacmann2002,Tafalla2002,Pagani2005,Giannetti2014}).
Under  the ``full depletion" assumption, the only important gas--phase reactions involve primordial species: H, He, and D. In addition to the gas--phase reactions the model also includes the formation of H$_2$, and HD on grains as reported in \citet{LePetit2002}. The adsorption rate coefficient $k_\mathrm{ads,i}$  is expressed as

\begin{equation}
	k_\mathrm{ads,i} = S n_\mathrm{gr} \pi \langle a \rangle ^2 v_\mathrm{i}\,.
\end{equation} 

\noindent where $S$ is the sticking probability here assumed to be one, $\pi\langle a\rangle^2$ is the averaged grain cross--section ($\sigma_\mathrm{gr}$), $n_\mathrm{gr}$ the grain number density, and $v_\mathrm{i}$ the thermal velocity of the ith atom, in this case H or D\footnote{Note that $v_\mathrm{D}$ = $v_\mathrm{H}/\sqrt{2}$.}. We are not considering the formation of D$_2$ on grains as this is mainly regulated by gas--phase reactions \citep{Sipila2010}.
We assumed that the ortho and para forms of H$_2$ are produced on grains according to their statistical population ratios 3:1 (i.e. an H$_2$ OPR of 3). Note that this is a very conservative assumption as observations suggested values of $\leq$0.1. We explore the effect of reducing the H$_2$ OPR in section~\ref{initialop}.\\
For the grains we assume an average size of $\langle a \rangle = 0.1\,\mu\mathrm{m}$, a dust grain density $\rho_0$ of 2 g cm$^{-3}$, and a dust to gas mass ratio $\mathcal{D}=0.013$ \citep{Walmsley2004}. 
With these values we obtain the initial number density of the grains, which we assume being all neutrals, as

\begin{equation}
	n_\mathrm{gr} = \frac{\mathcal{D} \mu  m_\mathrm{H}}{M_\mathrm{dust}}
\end{equation}

\noindent with $M_\mathrm{dust} = \frac{4}{3} \pi \rho_0 \langle a\rangle^3$, $m_\mathrm{H}$ being the hydrogen mass, and $\mu$ being the mean molecular weight.
With the above parameters we obtain an initial fractional grain abundance of $n_\mathrm{gr}/n_\mathrm{H} = 3.5 \times 10^{-12}$, and a grain surface area per hydrogen nucleus $n_\mathrm{gr} \sigma_\mathrm{gr} /n_\mathrm{H} = 1.1\times 10^{-21}$ cm$^2$. \\
Electron attachment and recombination of positive ions on grains are also included in the chemical network, and are taken from \citet{Flower2003}. As already discussed by \citet{Sipila2010}, the rate coefficients reported in the former work were calculated assuming an MRN distribution \citep{Mathis1977} with $a_\mathrm{min} = 0.01$ $\mu$m, and $a_\mathrm{max} = 0.3$ $\mu$m, which correspond to an effective grain radius of $a_\mathrm{eff} = 0.02$ $\mu$m. The values reported in Appendix A of \citet{Walmsley2004} should then be rescaled by a factor $(\langle a_\mathrm{eff} \rangle / 0.02\,\mu\mathrm{m})^2$. In our case $\langle a_\mathrm{eff} \rangle = 0.1$ $\mu$m, and the rescaling factor is 25. These rates also depend on the Coulomb factor, $\tilde{J}$, which takes into account the electric interaction between dust grains and gas phase neutrals and/or ions, and has been calculated following \citet{Draine1987}.\\
The final chemical network includes 21 gas--phase species plus charged/neutral grains, as reported in Table \ref{table:species}, with a total number of 213 reactions\footnote{We note that, as we are considering isothermal conditions ($T=\,$const.), adding more reactions/species is not going to introduce a large computational overhead, so this network can be easily extended in the future to include for instance N--bearing species.}. A complete list of reactions is reported in Appendix A of \citet{Walmsley2004}.

\begin{table}
\centering
\caption{List of the species included in the chemical network. \emph{gr} refers to dust grains.}
	\begin{tabular}{llll}
		\hline	
		\hline	 
		 H & D & H$^+$ & e$^-$\\
		He & He$^+$ & D$^+$ & HD$^+$ \\
		p-H$_2$ & o-H$_2$ & p-H$_2^+$ & o-H$_2^+$ \\
		p-H$_3^+$ & o-H$_3^+$ & p-H$_2$D$^+$ & o-H$_2$D$^+$ \\
		D$_2$ & D$_2^+$ & D$_2$H$^+$ & D$_3^+$ \\
       gr & gr$^+$ & gr$^-$\\
	\hline
	\hline
	\end{tabular}
	\label{table:species}
\end{table}

\subsection{Chemical network benchmark}
\begin{figure}
	\includegraphics[scale=0.35]{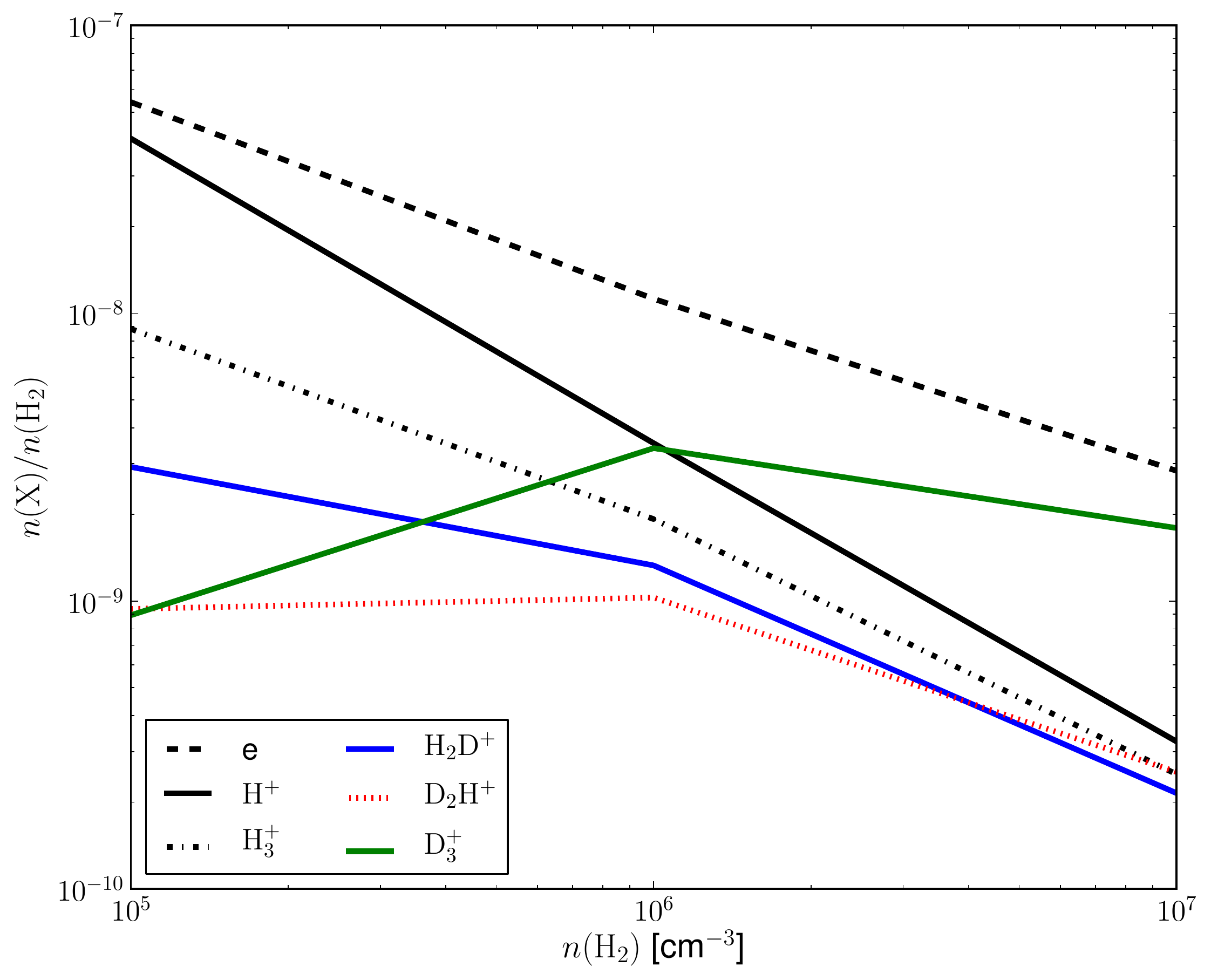}
	\caption{Equilibrium abundances for different species as a function of the gas density. The results are in agreement with the ones reported by \citet{Walmsley2004}, which represents our benchmark.}\label{fig:figure1}
\end{figure}

The full set of rate equations is solved by employing the astrochemistry package \textsc{krome} \citep{Grassi2014}, which has already been used in a variety of numerical simulations for modelling galaxy formation and evolution \citep{Bovino16b}, molecular clouds \citep{Grassi17}, ISM collapsing filaments \citep{Seifried16}, formation of primordial supermassive black holes \citep{Latif2014,Latif16b}, and the formation of low--metallicity stars \citep{Bovino14, Bovino16a}. Complex microphysics (e.g. dust processes), and the capability to be coupled with radiative transport algorithms, together with the high optimization to speed--up the solution of large networks, makes the package suitable to study deuterium fractionation in molecular cloud cores. The non--equilibrium chemistry is fully coupled with the hydrodynamical equations through the public \textsc{flash} patch released with the package. As we aim to follow an isothermal collapse, cooling and heating processes are not included.
To test the reliability of the chemical network we have re--done the same calculations as in \citet{Walmsley2004}, starting from the same initial conditions, i.e. \mbox{$\langle a \rangle = 0.1$~$\mu$m}, gas temperature $T = 10$ K, and cosmic--ray flux \mbox{$\zeta = 3 \times 10^{-17}$ s$^{-1}$}, evolving the species until the equilibrium is reached. The results are presented in Fig.~\ref{fig:figure1} and show perfect agreement with the ones reported in Fig. 2 of \citet{Walmsley2004}.

\section{Numerical Model}\label{method}
\subsection{Method}
We use the FLASH code (v4.2.2) to solve the magneto--hydrodynamic (MHD) equations as well as the Poisson equation for self--gravity \citep{Fryxell00,Dubey08}. The (magneto--)hydrodynamical system is solved with a HLL5R--solver, which 
preserves positivity of density and internal energy \citep{Bouchut10, Waagan11}, while we use a tree--solver for the self--gravity \citep[optimised for GPU][]{Lukat16}. We further employ outflow boundary conditions for the 
hydrodynamics and isolated ones for the self--gravity. \\
\subsection{Initial conditions}
The spherical core is initialized using a Bonnor--Ebert (BE) density profile \citep{Ebert55, Bonnor56}. The BE radius for the cores is $R_\mathrm{BE}\sim 7-8$, indicating that the sphere is slightly unstable as $R_\mathrm{BE,crit}\sim6.45$. The ambient density is one percent of the density at the cut--off radius of the core, $R_\mathrm{c}$ and constant throughout. \\
The gas is isothermal at $T=15\,\mathrm{K}$ and turbulent. The turbulent Mach number ranges from $\mathcal{M}_\mathrm{rms}=0.5-12$, which is rather at the low regime, but still in agreement with observations \citep[][]{Heiles2003,Heiles2005,Padoan11,Federrath16b}. The turbulent velocity field is 
set up with an energy spectrum si\-milar to Burgers type turbulence
\beq
E\left(k\right)\propto\left\{
\begin{array}{c}
 k^{10}\quad k<k_\mathrm{int},\\
 k^{-2}\quad k>k_\mathrm{int}
 \end{array}
 \right.
 \eeq
 where the integral wave number is chosen to be $k_\mathrm{int}=6$, giving $\lambda_\mathrm{int}=L_\mathrm{box}/k_\mathrm{int}=0.097\,\mathrm{pc}\sim R_\mathrm{c}$.\\
 The core is magnetized. Although the field strength in cores can be inferred from Zeeman observations or dust polarisation measurements \citep[with bias due to the method used,][]{Crutcher10}, the radial profile is difficult to reveal because of the need for 
 high spatial resolution. We hence implement the magnetic field according to 
 \beq
 B_\mathrm{z}\left(R_\perp\right)\propto \left(\frac{R_\perp}{R_\mathrm{c}}\right)^{-\kappa},
 \eeq
 where $R_\perp$ is the perpendicular distance to the symmetry axis of the core (z--axis), $\kappa$ is an input parameter, which we take to be $\kappa=0.5$ or $\kappa=2$ and $B_\mathrm{z}$ is the component of the field parallel to the symmetry axis. This 
 setup ensures the solenoidal constraint of the magnetic field. The field is then normalized to yield a specific mass--to--magnetic flux ratio in the core.\\
 Table \ref{tabIC} gives an overview of the performed simulations and the initial conditions, including e.g. the core's free--fall time, virial ratio, and mass--to--flux ratio.\\
The simulation volume is set up with a side length of $L_\mathrm{box}=0.6\,\mathrm{pc}$ and we use 9 levels of refinement\footnote{Note that the values for the refinement presented here take into account the default initial grid refinement of two levels in the FLASH code.}. The root grid has a resolution of $\Delta x_\mathrm{root}=0.009\,\mathrm{pc}=1887\,\mathrm{au}$, equivalent to 6 levels of refinement. We allow for additional three levels of adaptive refinement during runtime -- a compromise of 
numerical resolution and feasible simulation run times due to very small timesteps -- giving $\Delta x_\mathrm{max}=235\,\mathrm{au}$. The grid is refined once the local Jeans length is resolved with less than 16 grid cells. On the highest level of refinement, 
gas accumulates and the local Jeans length will be unresolved with time\footnote{We emphasize that a radial density profile has different free--fall times at different radii. It is thus inappropriate to simulate up to the free--fall time of the average gas density as the innermost regions have $\varrho_\mathrm{inner}>\left<\varrho\right>$ and consequently $t_\mathrm{ff}\left(\varrho_\mathrm{inner}\right)<t_\mathrm{ff}\left(\left<\varrho\right>\right)$.}. Hence, we include sink particles \citep{Federrath10}, which are formed when gravitationally bound gas exceeds the threshold density of $n_\mathrm{sink}=3\times10^7\,\mathrm{cm}^{-3}$ besides 
further tests of gravitational boundedness. At this density, the local Jeans length is still resolved with four grid cells and the Truelove--criterion is not violated \citep{Truelove97}. We point out that the sink particle is allowed to accrete gas within a volume with radius of $R_\mathrm{accr}=470\,
\mathrm{AU}=2\times\Delta x$, which inevitably affects the evolution of the gas and thus the chemical species in the innermost part of the core. The cosmic--ray 
flux is set to $\zeta=3\times10^{-17}\,\mathrm{s}^{-1}$. \\
\begin{table*}
	\begin{center}
		\caption{List of the performed simulations and the main initial dynamical parameters. Marked bold are the simulations in which a sink particle has formed. In all simulations, the maximum resolution is $\Delta x=235\,\mathrm{au}$.}
		\begin{tabular}{p{2.5cm}ccccccp{1.cm}ccc}
		\hline
		\hline
		Run 	 &Surface 	&Core 	&Core  &Av. Field &Mass--to--	&Mach 	&Virial&\# Jeans  &Free--fall 	&B--field	\\
			&density	&radius	 &mass &strength	&flux ratio$^a$ 		&number	&parameter&masses$^b$	&time$^c$	& slope \\
			&$\left(\mathrm{g/cm}^2\right)$ &$\left(\mathrm{pc}\right)$ &$\left(\mathrm{M}_\odot\right)$ &$\left(\mu\mathrm{G}\right)$&$\mu/\mu_\mathrm{crit}$ &$\mathcal{M}_\mathrm{turb}$ &$\alpha_\mathrm{vir}$& &$\left(\mathrm{kyr}\right)$ &$\kappa$ \\
		\hline
		\fat{Lmu10M1}		&0.14	&0.17	&60	&27&10	&1	&0.16 &24	&149		&0.5	\\
		\fat{Lmu10M2}		&0.14	&0.17	&60	&27&10	&2	&0.64 &24	&149		&0.5	\\
		\fat{Lmu10M2OPR0.1}$^d$&0.14	&0.17&60&27	&10	&2	&0.64 &24	&149		&0.5	\\
		\fat{Lmu10M2OPR1}$^d$&0.14	&0.17	&60&27	&10	&2	&0.64 &24	&149		&0.5	\\
		\fat{Lmu10M2S2}	&0.14	&0.17	&60	&27&10	&2	&0.64 &24	&149		&2	\\
		Lmu10M4			&0.14	&0.17	&60&27	&10	&4	&2.56 &24	&149		&0.5	\\
		Lmu10M6			&0.14	&0.17	&60&27	&10	&6	&5.76 &24	&149		&0.5	\\
		Lmu10M12		&0.14	&0.17	&60&27	&10	&12	&23.04 &24	&149		&0.5	\\
		\fat{Lmu5M2}		&0.14	&0.17	&60	&54&5	&2	&0.64 &24	&149		&0.5	\\
		Lmu5M4			&0.14	&0.17	&60	&54&5	&4	&2.56 &24	&149		&0.5	\\
		\fat{Lmu2.5M0.5}	&0.14	&0.17	&60	&108&2.5	&0.5	&0.04 &24	&149		&0.5	\\
		Lmu2.5M2		&0.14	&0.17	&60	&108&2.5	&2	&0.64 &24	&149		&0.5	\\
		Lmu2.5M6		&0.14	&0.17	&60	&108&2.5	&6	&5.76 &24	&149		&0.5	\\
		\hline
		\fat{Mmu10M2}		&0.24	&0.08		&27	&49 &10	&2	&0.71 &20	&72		&0.5	\\
		Mmu5M2			&0.24	&0.08		&27	&98 &5	&2	&0.71 &20	&72		&0.5	\\
		\hline
		\fat{Hmu10M0.5}	&0.39	&0.1		&60	&76&10	&0.5	&0.03 &60	&67		&0.5	\\
		\fat{Hmu10M2}		&0.39	&0.1		&60	&76&10	&2	&0.48 &60	&67		&0.5	\\
		\fat{Hmu10M2S2}	&0.39	&0.1		&60	&76&10	&2	&0.48 &60	&67		&2	\\
		\fat{Hmu5M2}		&0.39	&0.1		&60	&152&5	&2	&0.48 &60	&67		&0.5	\\
		Hmu2.5M2		&0.39	&0.1		&60	&304&2.5	&2	&0.48 &60	&67		&0.5	\\
		Hmu2.5M4		&0.39	&0.1		&60	&304&2.5	&4	&1.92 &60	&67		&0.5	\\
		\hline
		\hline
		\end{tabular}\\
		\label{tabIC}
		\begin{flushleft}
		\tiny{$^a$ Calculated using the average magnetic field in the core.}\\
		\tiny{$^b$ We only take into account thermal support here.}\\
		\tiny{$^c$ Calculated using the average density in the core.}\\
		\tiny{$^d$ In this simulation we employ an initial H$_2$ OPR of 0.1 or 1, respectively.}\\
		\end{flushleft}
	\end{center}
	
\end{table*}

\section{Results}\label{results}
In the following, we will present the results of this study. We will start with the description of the overall evolution, and subsequently explore the impact of different parameters on deuteration and o--H$_2$D$^+$ evolution. A set of representative simulations will then be analyzed in greater detail. We finally attempt to qualitatively compare the deuterium fraction obtained in our simulations to observational results.

\subsection{Overview of the evolution}
\begin{figure*}
	\begin{center}
		\begin{tabular}{cccc}
		Total column density &Average D$_\mathrm{frac}^{\mathrm{H_2D^+}}$	&o--H$_2$D$^+$ column density \\
		\includegraphics[width=0.30\textwidth]{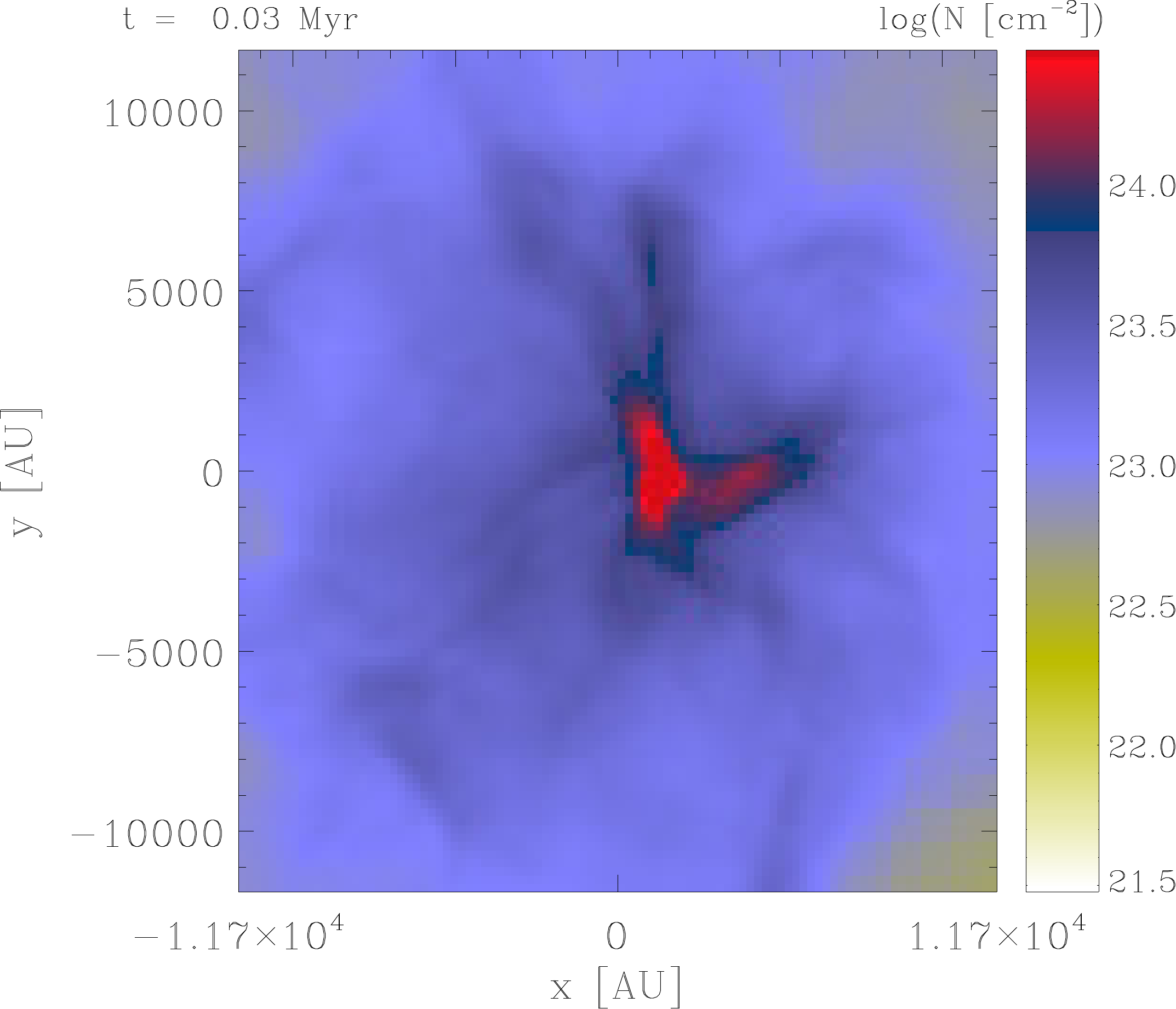}&\includegraphics[width=0.30\textwidth]{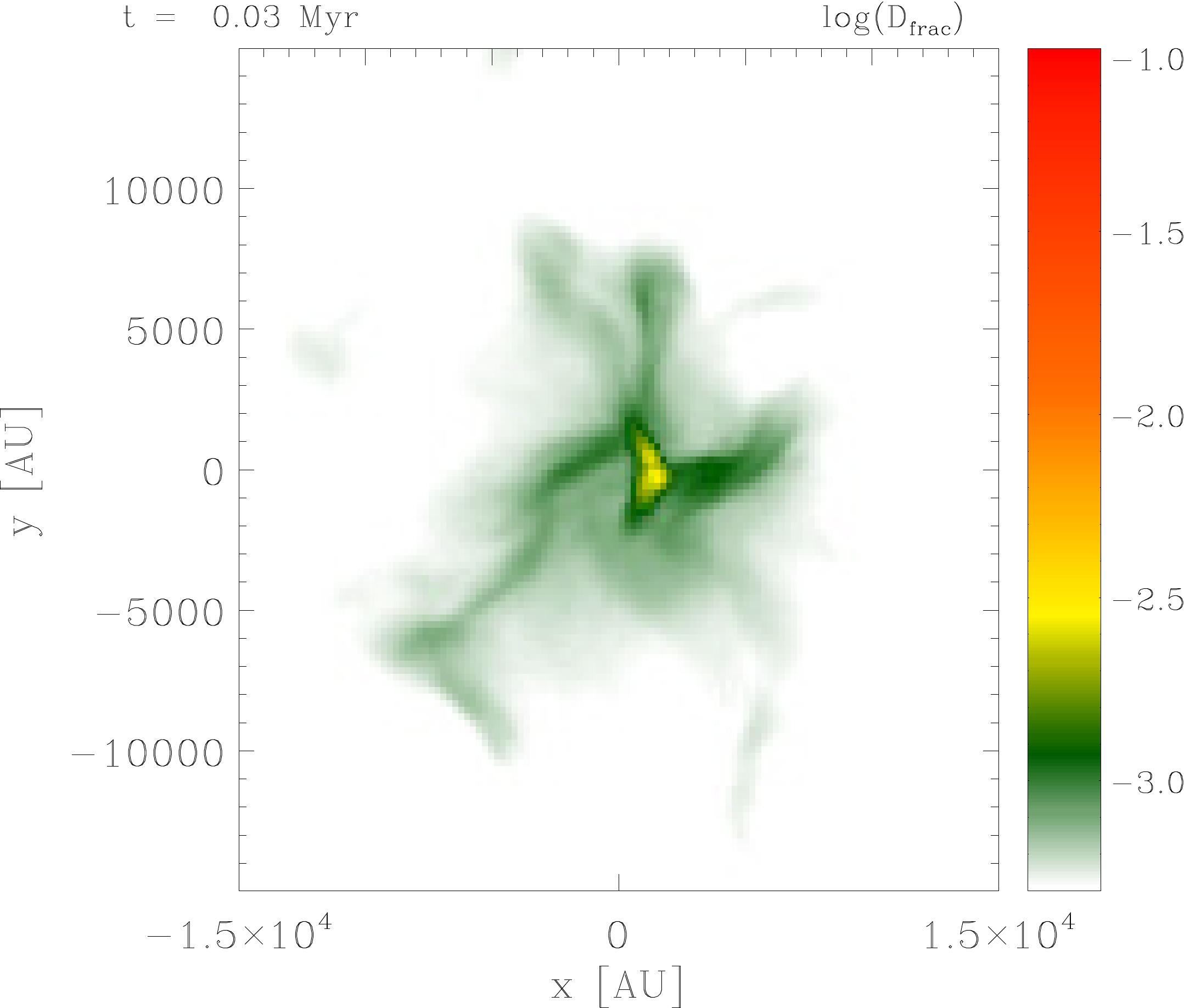}&\includegraphics[width=0.30\textwidth]{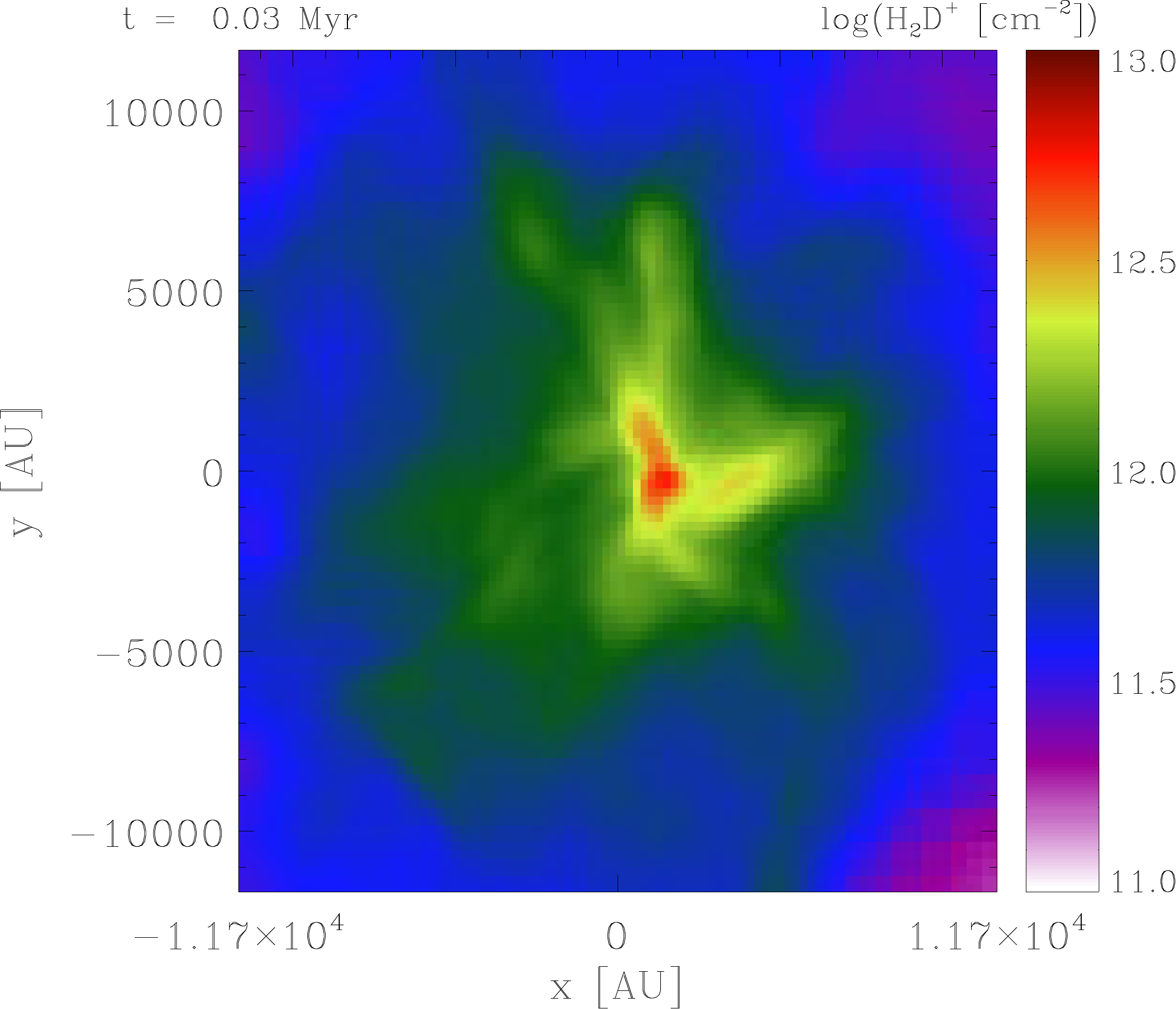}\\
		\includegraphics[width=0.30\textwidth]{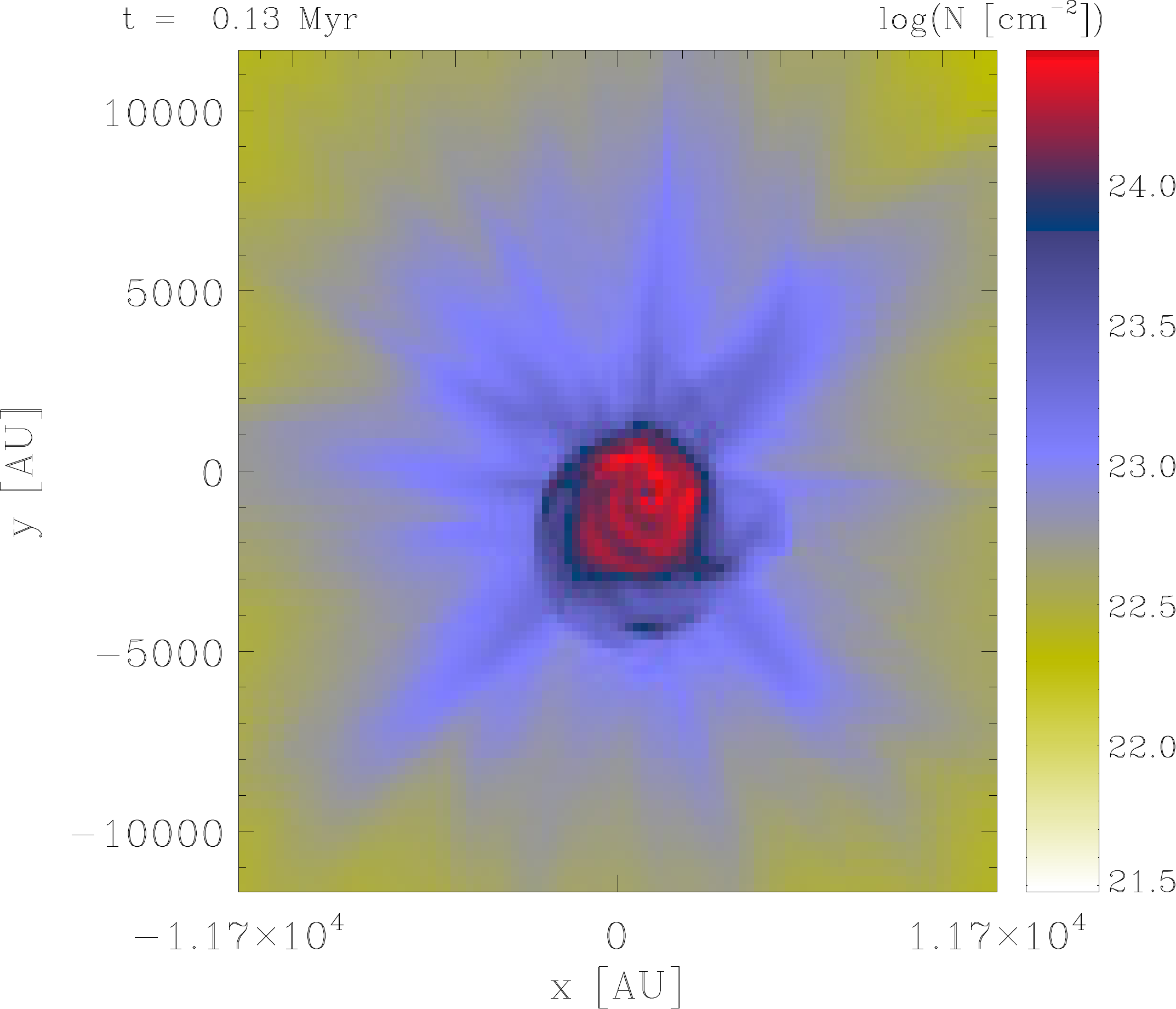}&\includegraphics[width=0.30\textwidth]{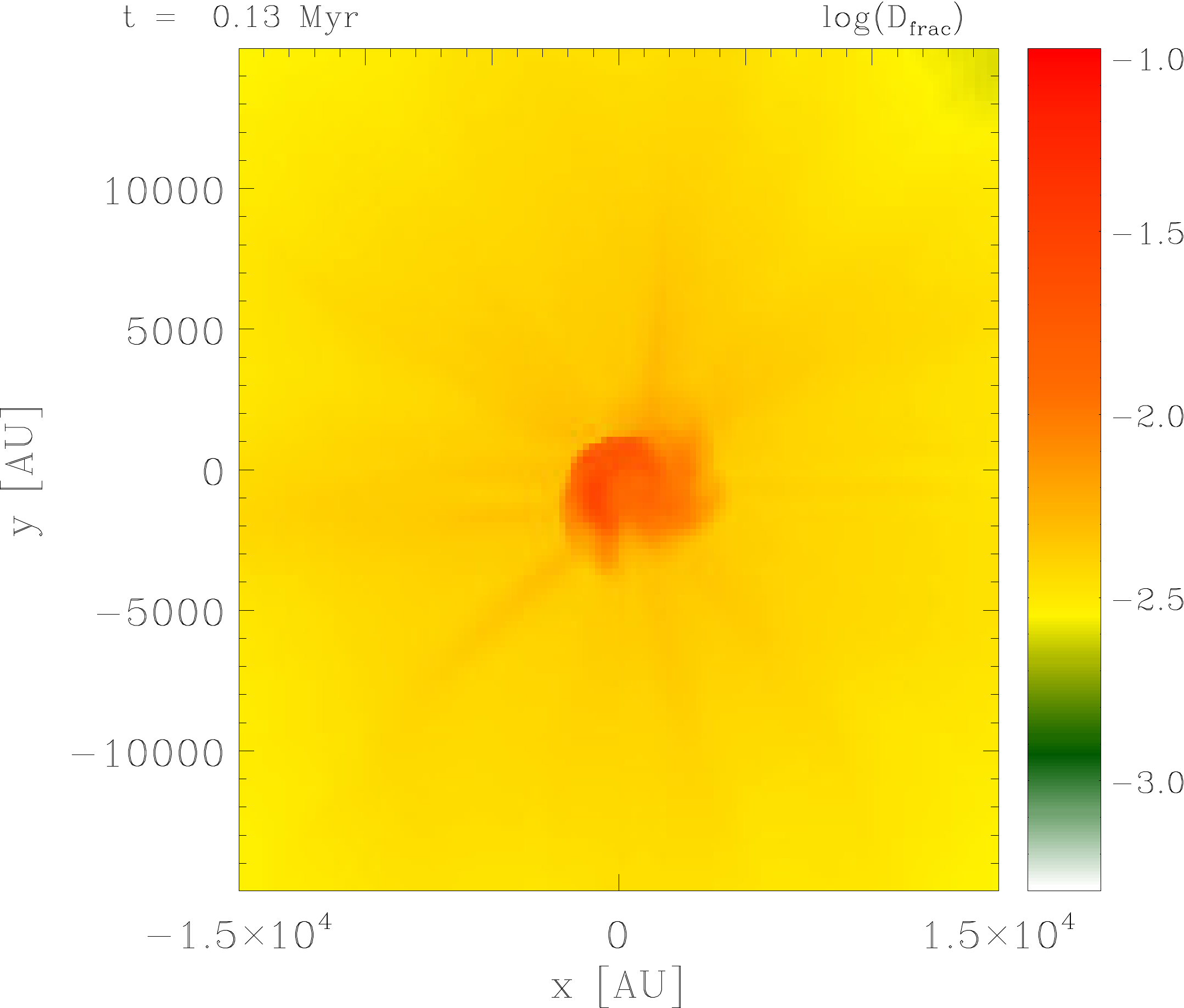}&\includegraphics[width=0.30\textwidth]{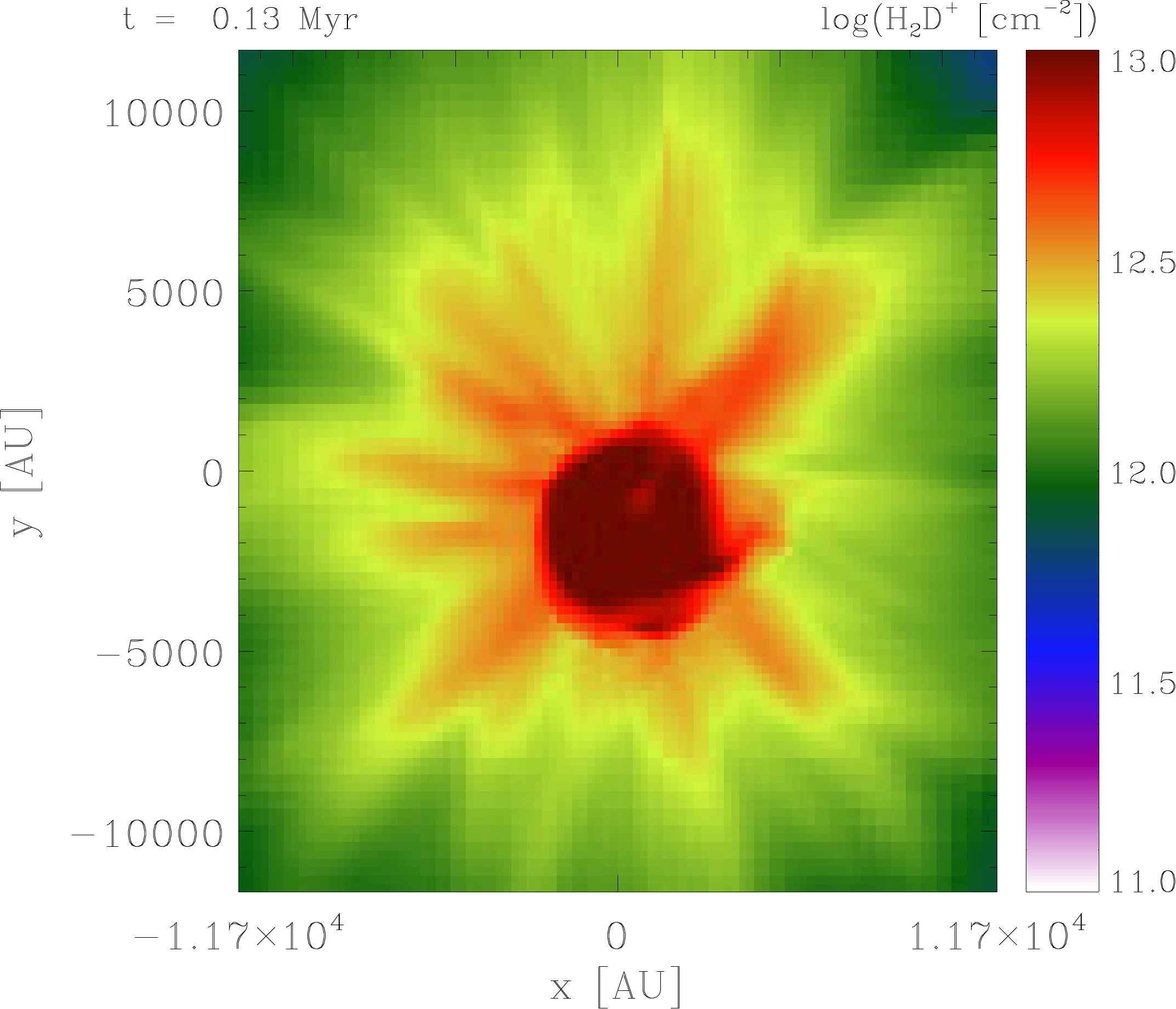}\\
		\includegraphics[width=0.30\textwidth]{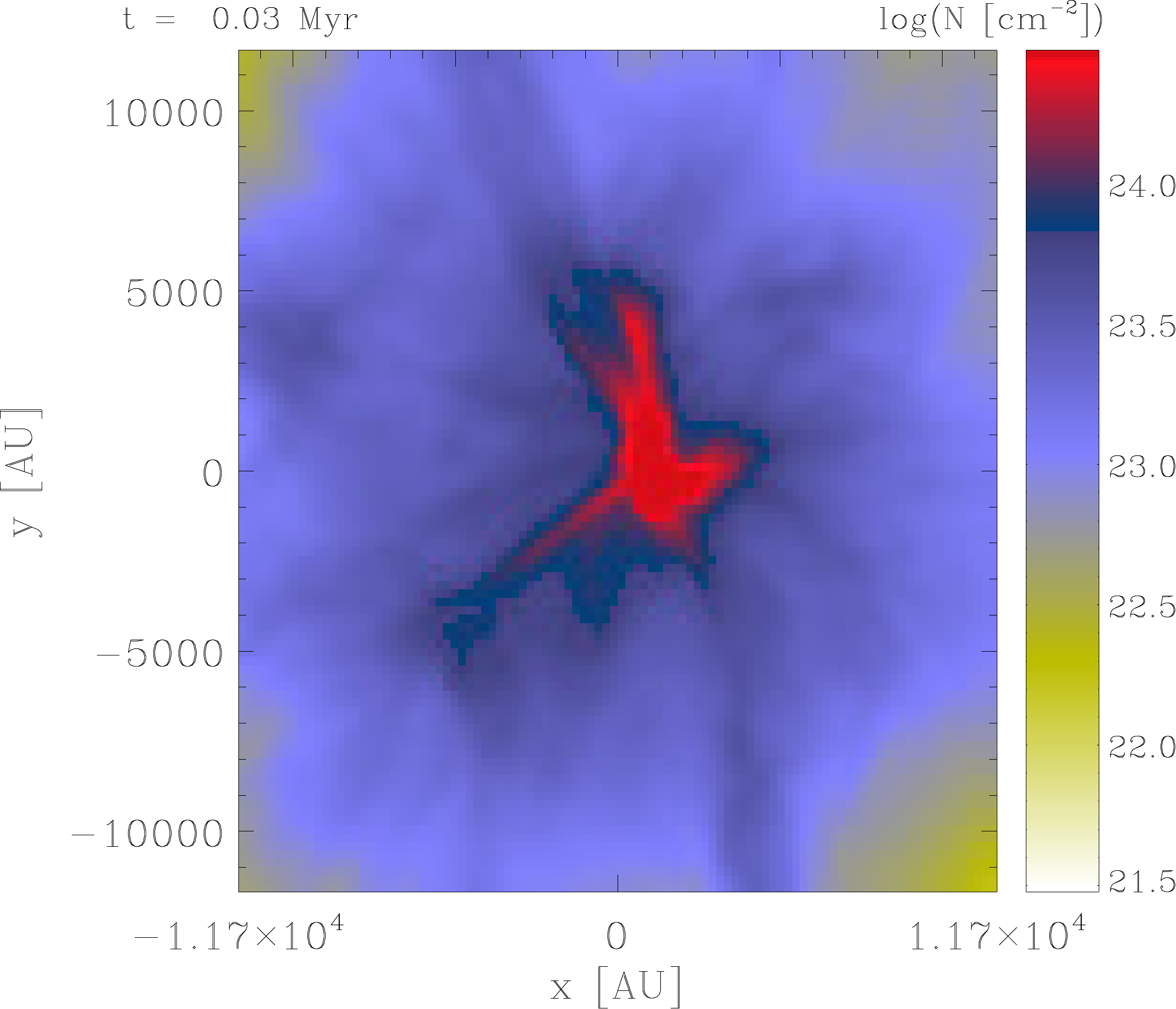}&\includegraphics[width=0.30\textwidth]{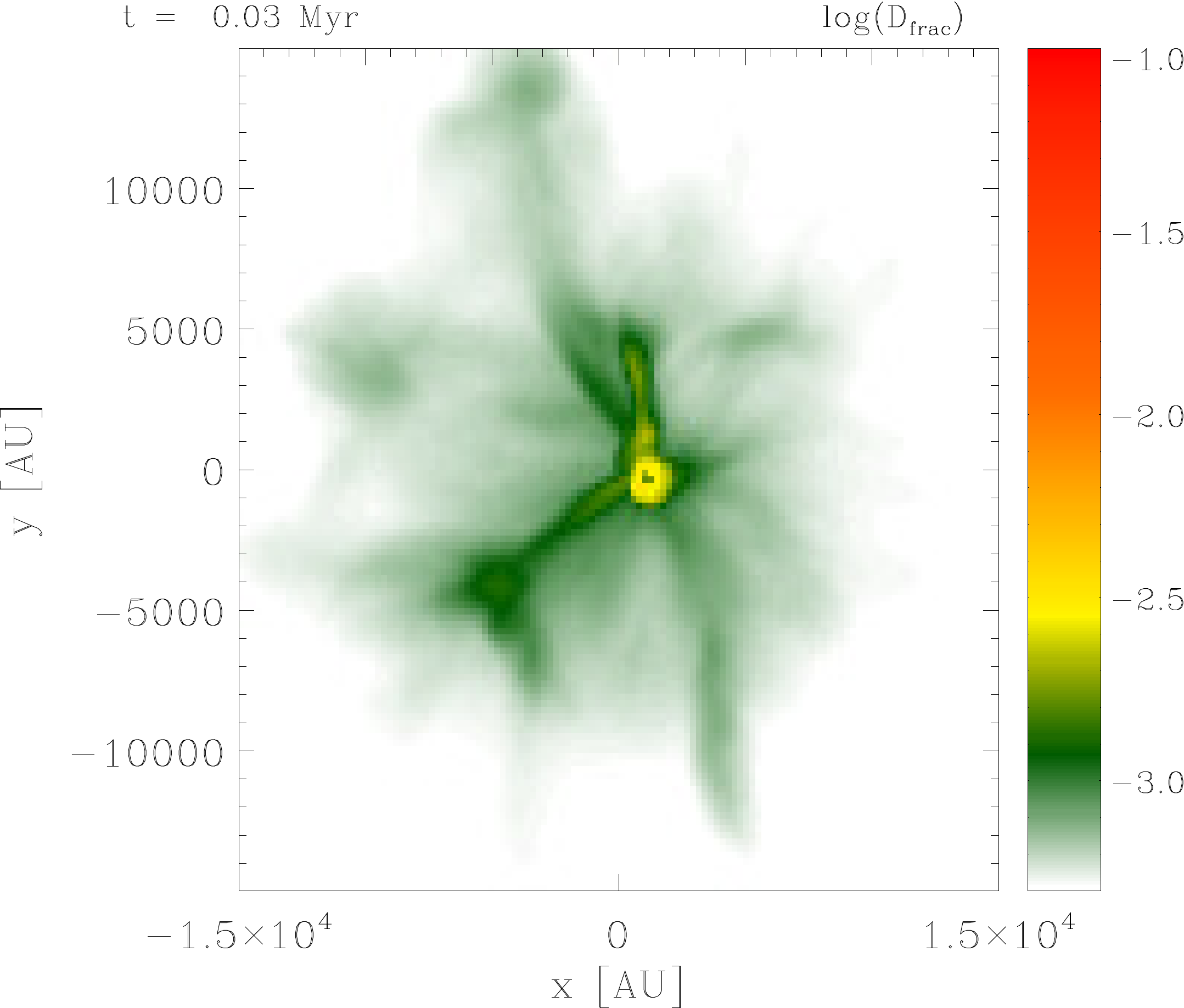}&\includegraphics[width=0.30\textwidth]{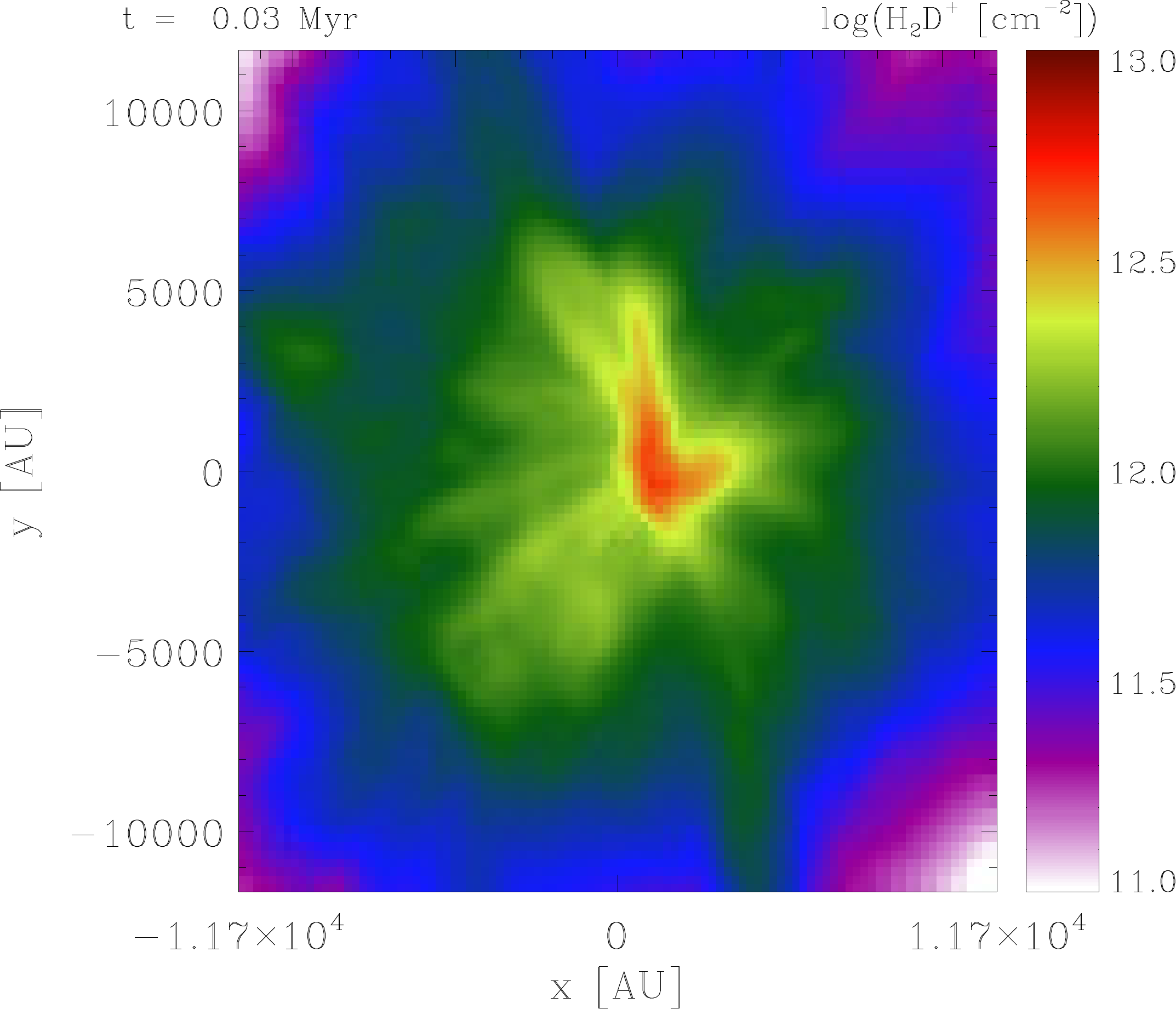}\\
		\includegraphics[width=0.30\textwidth]{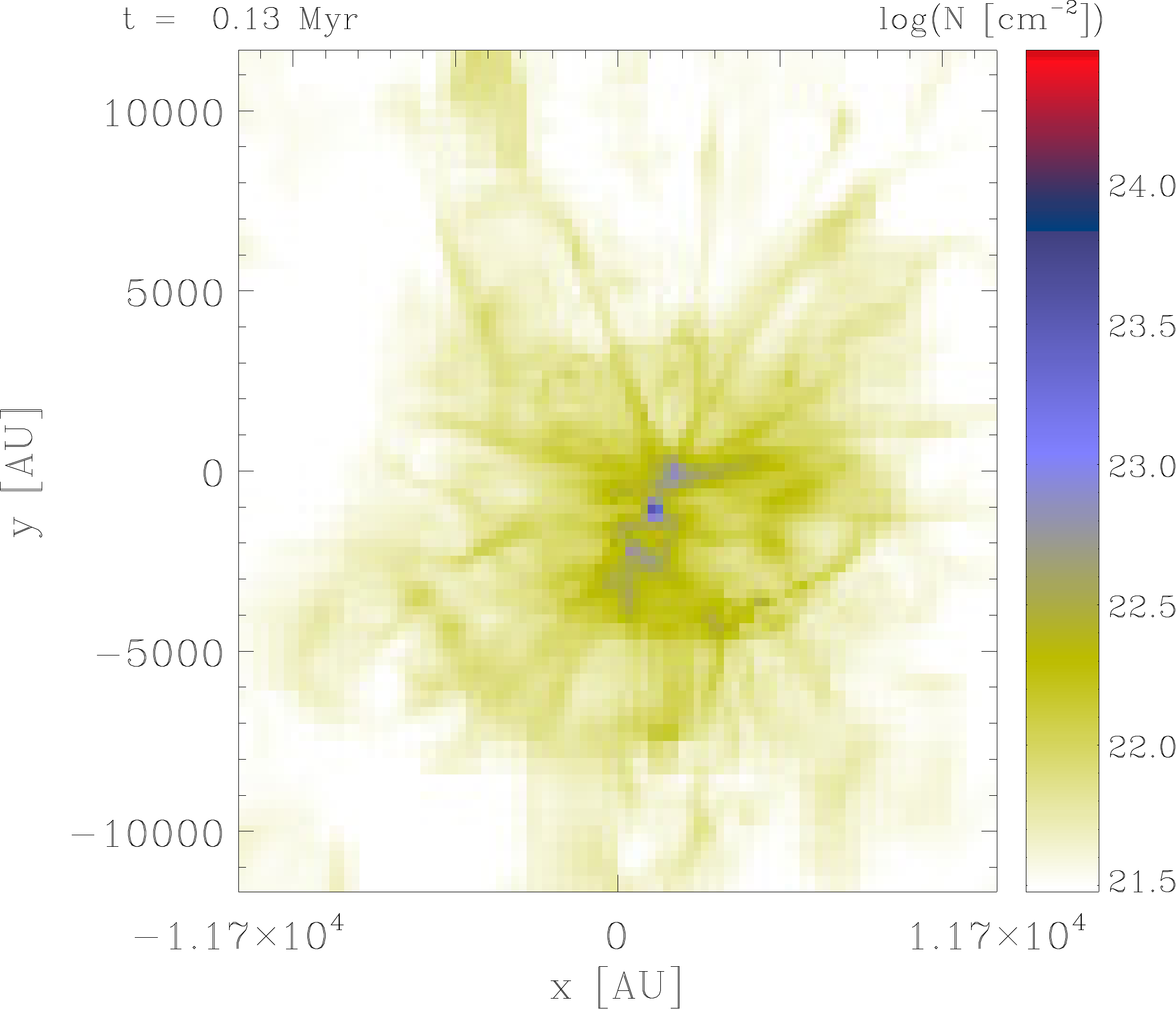}&\includegraphics[width=0.30\textwidth]{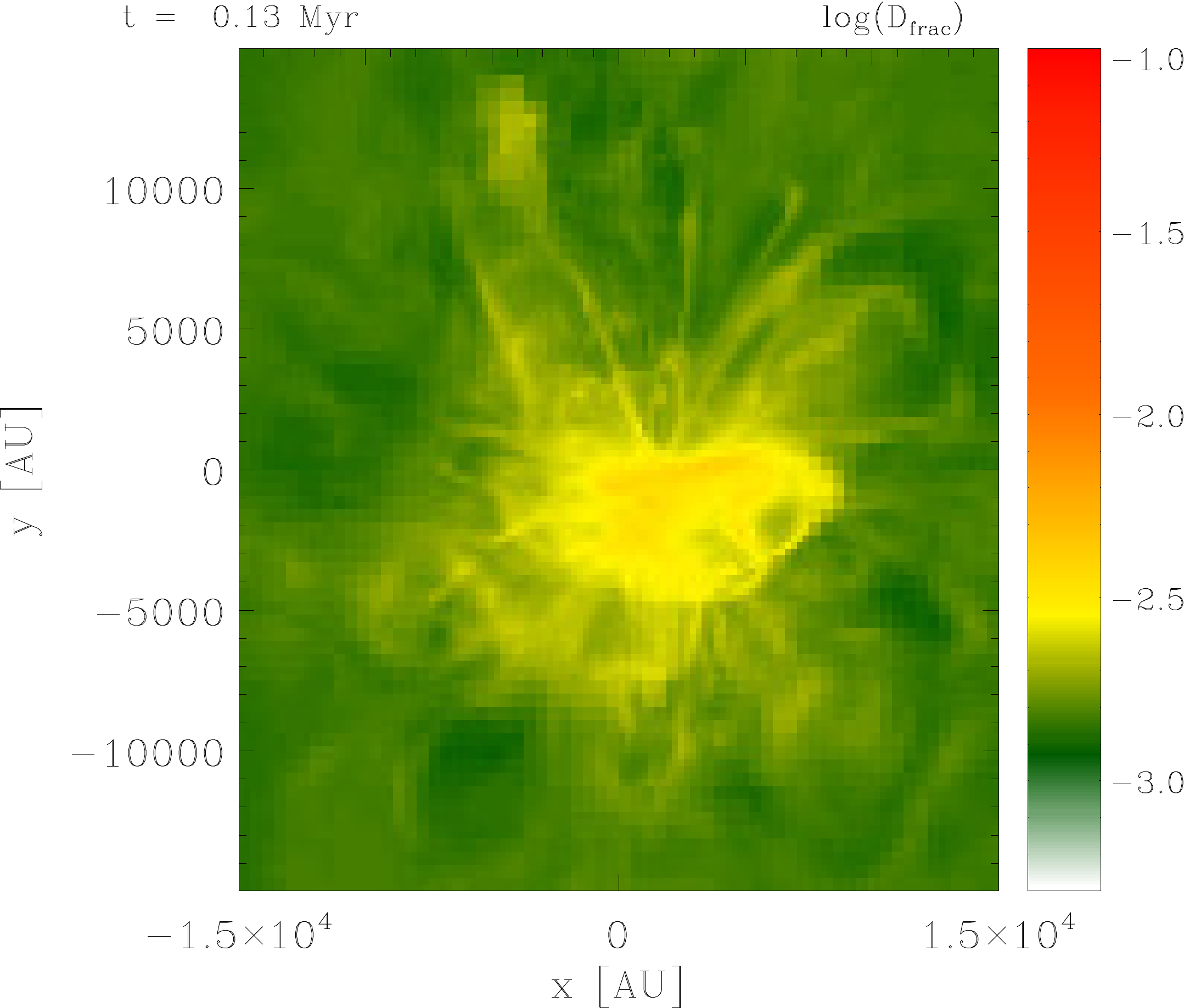}&\includegraphics[width=0.30\textwidth]{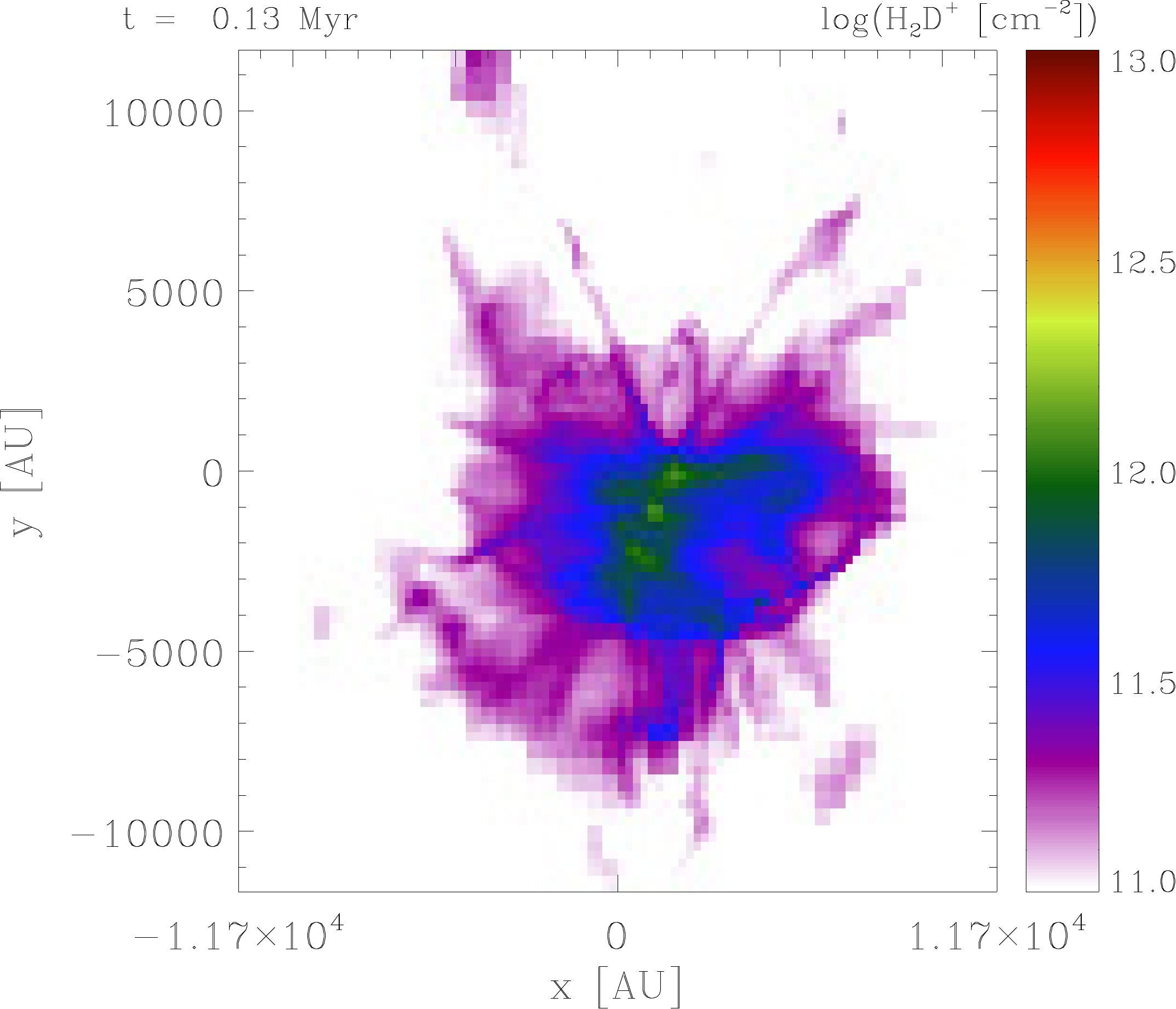}\\
		\end{tabular}
	\end{center}
	\caption{\ita{Rows 1 \& 2}: Data for run Lmu10M2 at $t=30\,\mathrm{kyr}$ (row 1) and at $t=130\,\mathrm{kyr}$ (row 2). \ita{Rows 3 \& 4}: Data for run Hmu10M2 at $t=30\,\mathrm{kyr}$ (row 3) and at $t=130\,\mathrm{kyr}$ (row 4). \ita{Left to right}: 
	Total gas column density integrated along a sightline of 0.2\,pc length in z--direction, that is, parallel to the initial magnetic field; column density weighted deuteration ratio in the core and o--H$_2$D$^+$ column density in the core. In case of 
	Hmu10M2 the gas is absorbed by the sink particle at late times (sink particle not shown), whereas in run Lmu10M2 a disc--like feature forms near the center in which the deuteration is enhanced. At later times, deuteration is greatly enhanced throughout the core. Further note that, although the core density is enhanced at early times in run Hmu10M2, the collapse timescale is shorter than the timescale for the build--up of sufficiently high deuteration.}
	\label{figCore1}
\end{figure*}

\begin{figure*}
	\begin{center}
		\begin{tabular}{cccc}
		Total column density &Average D$_\mathrm{frac}^{\mathrm{H_2D^+}}$	&o--H$_2$D$^+$ column density \\
		\includegraphics[width=0.28\textwidth]{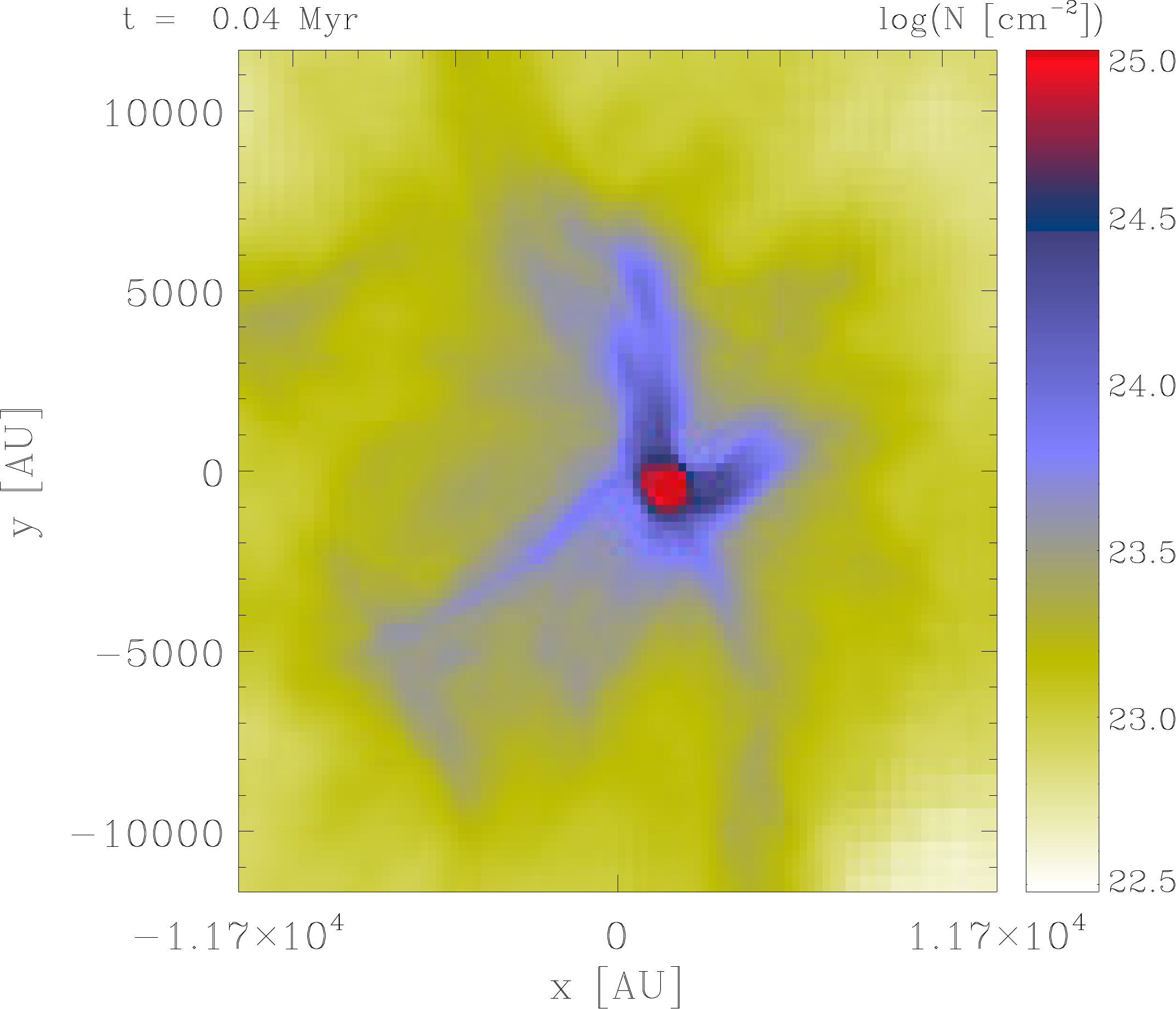}&\includegraphics[width=0.28\textwidth]{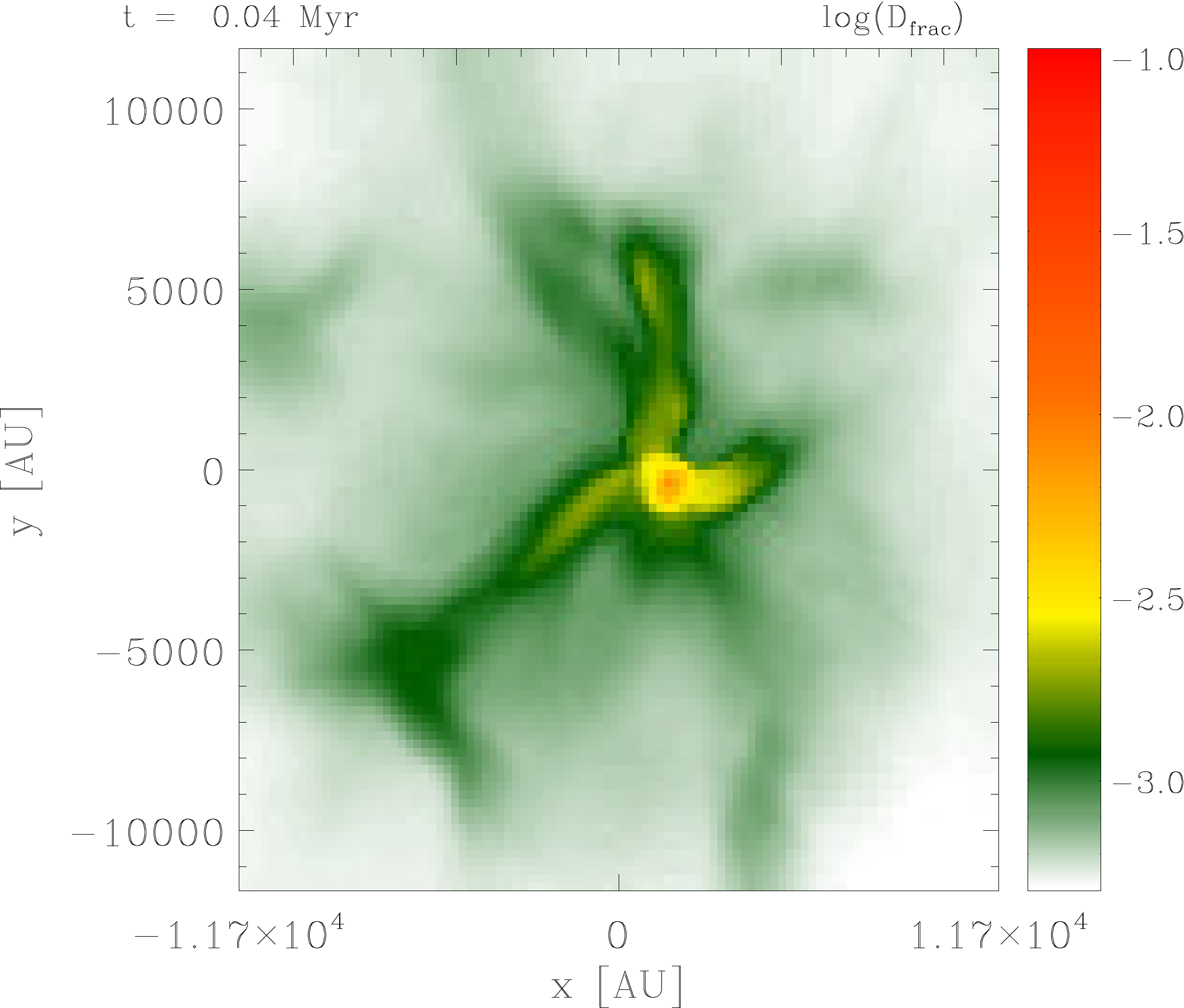}&\includegraphics[width=0.28\textwidth]{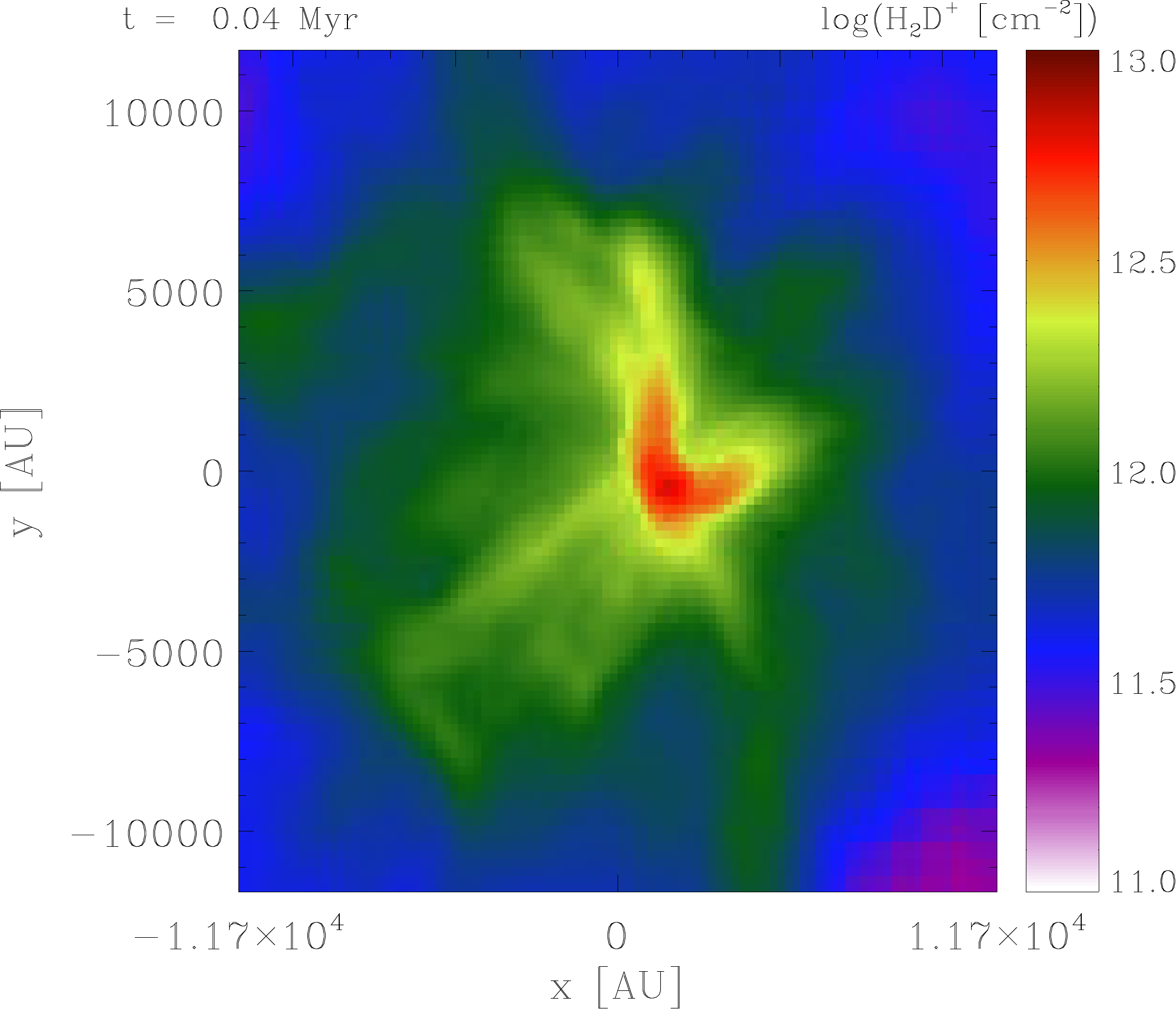}\\
		\includegraphics[width=0.28\textwidth]{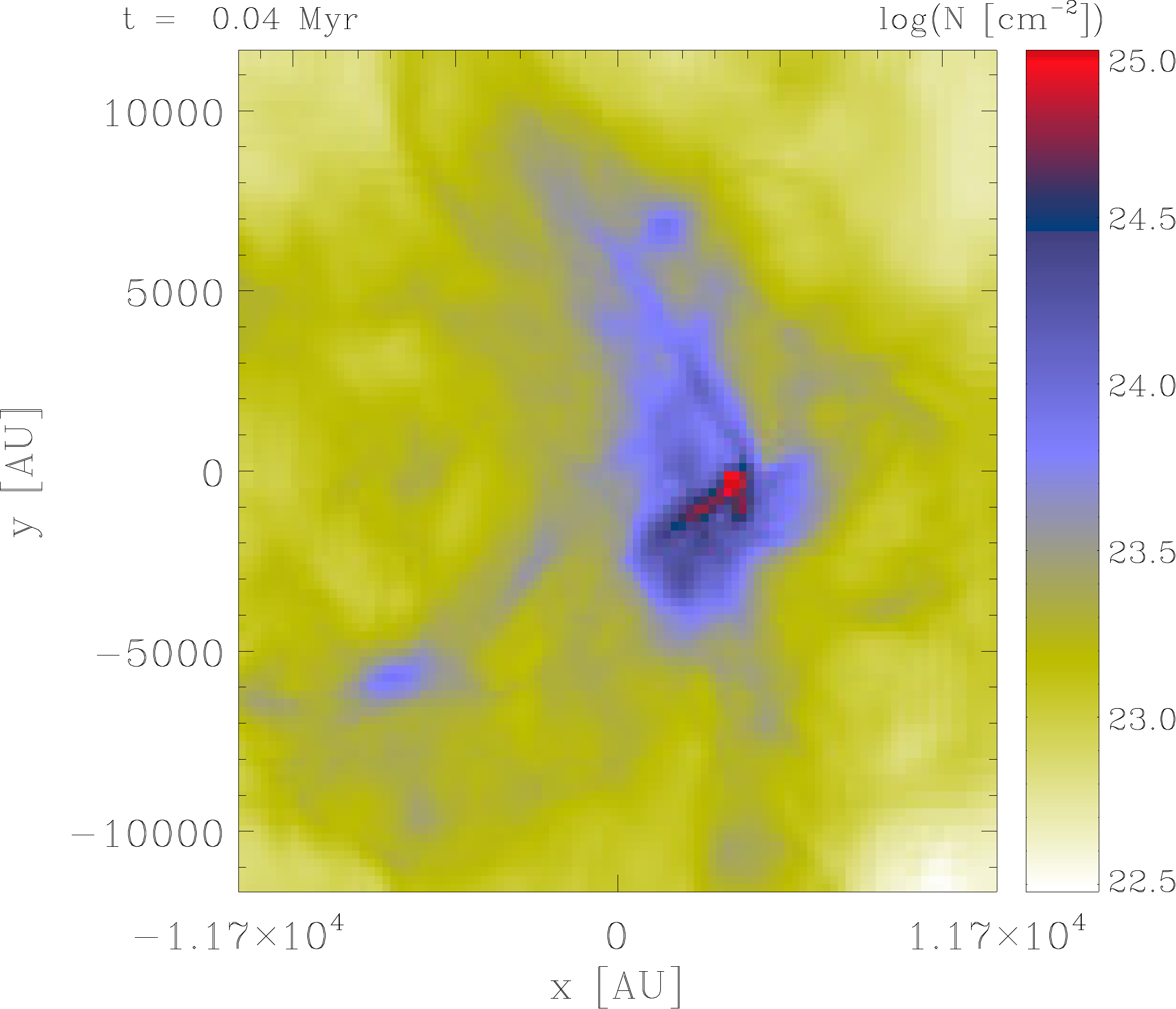}&\includegraphics[width=0.28\textwidth]{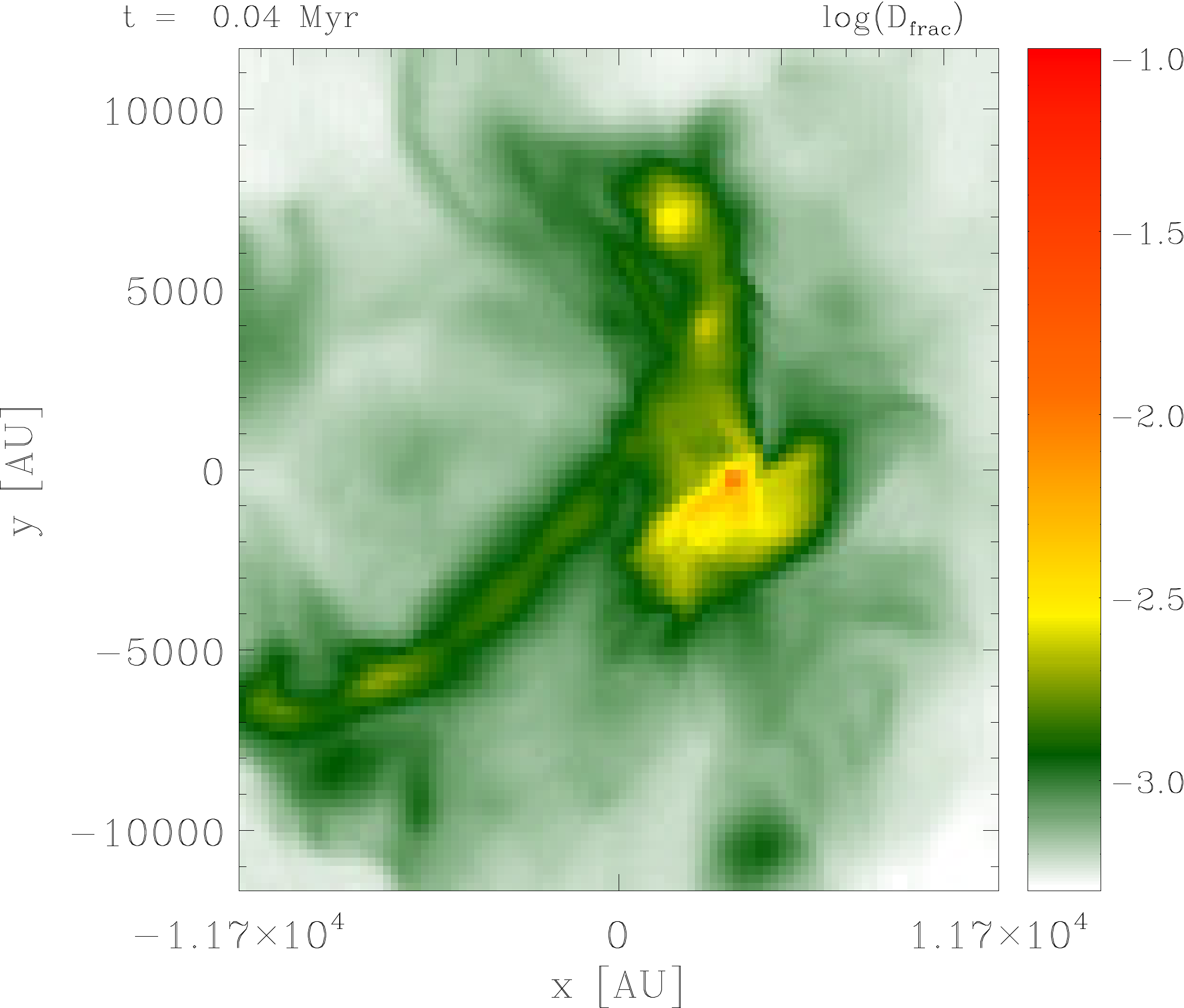}&\includegraphics[width=0.28\textwidth]{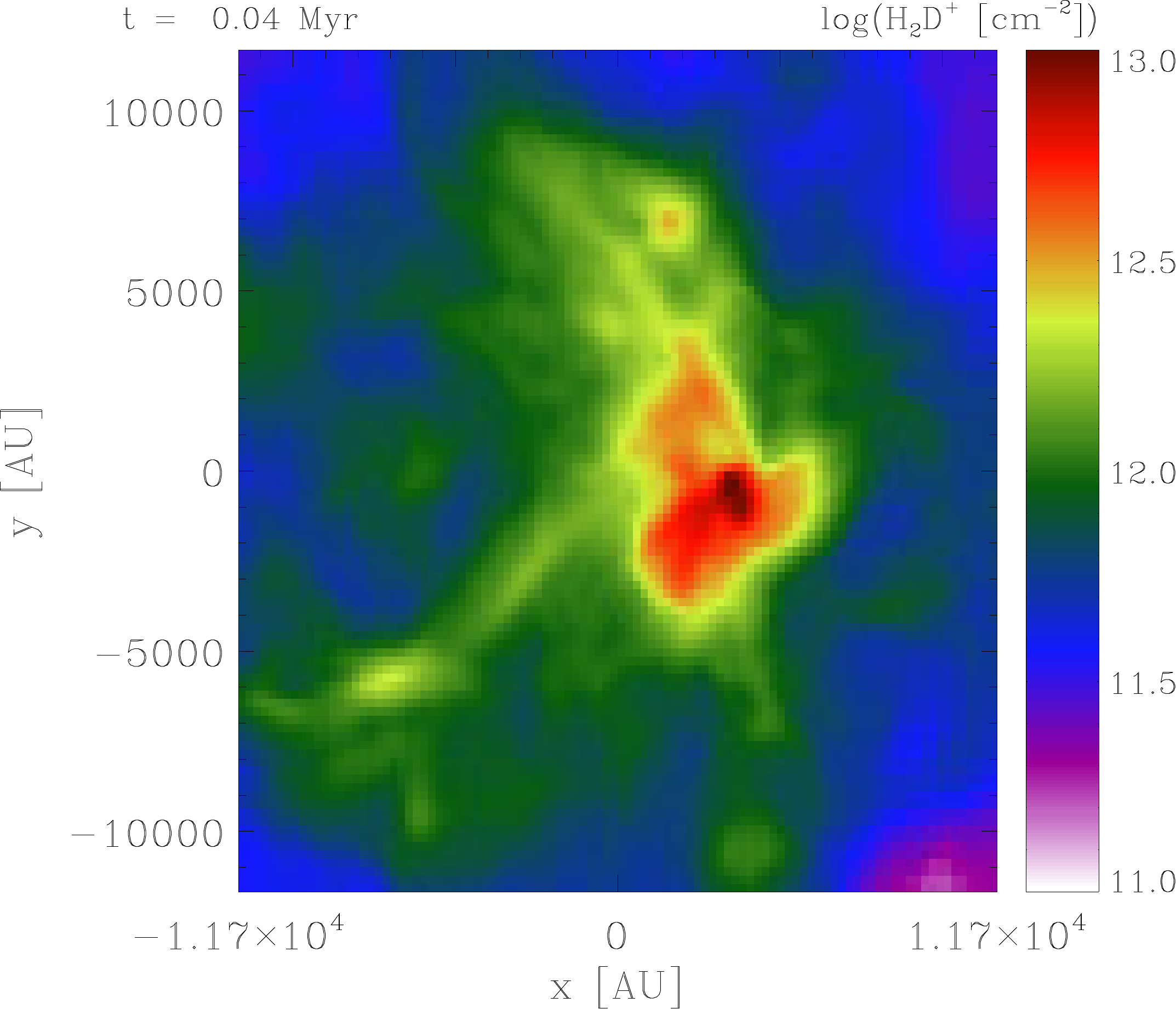}\\
		\includegraphics[width=0.28\textwidth]{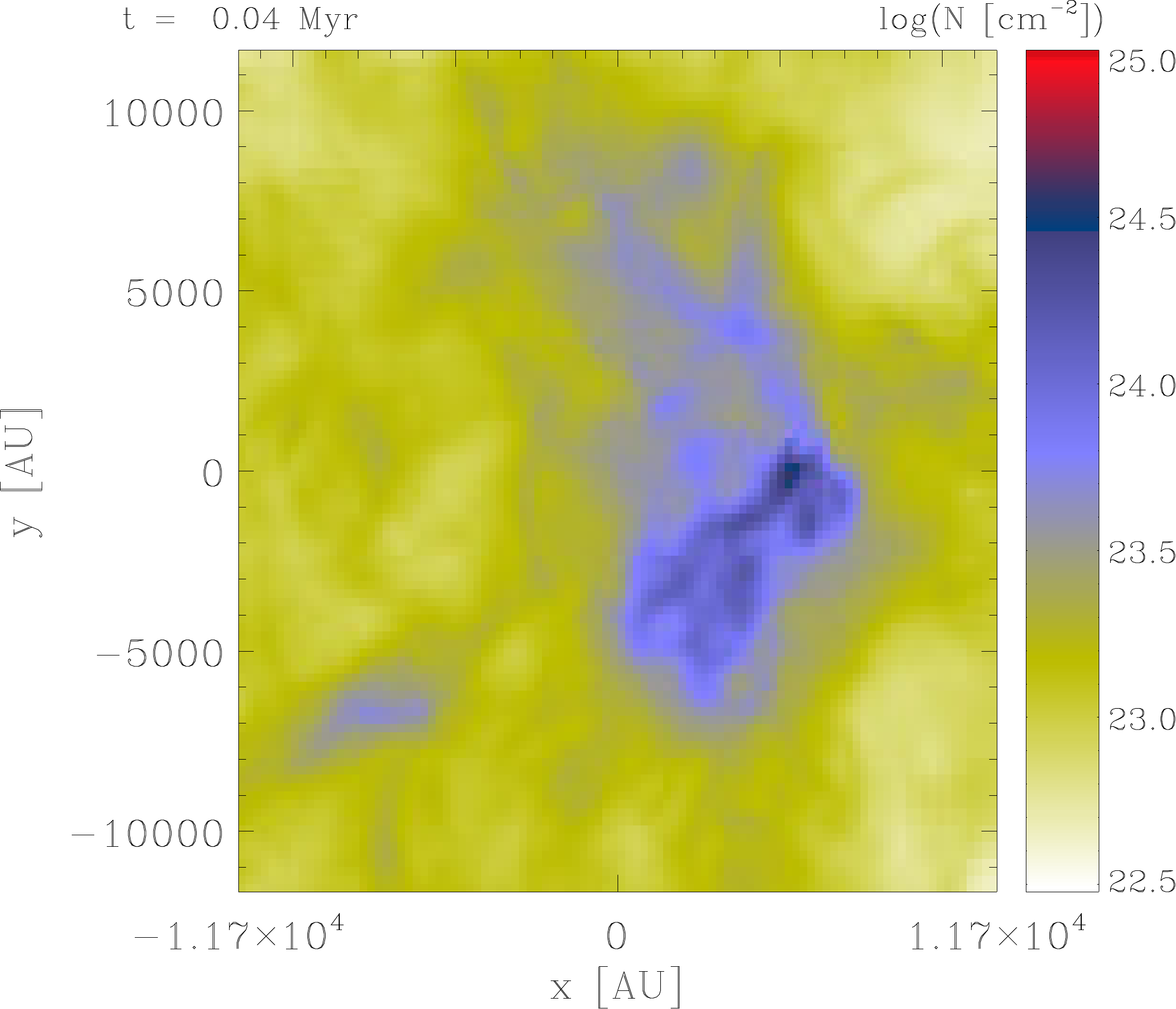}&\includegraphics[width=0.28\textwidth]{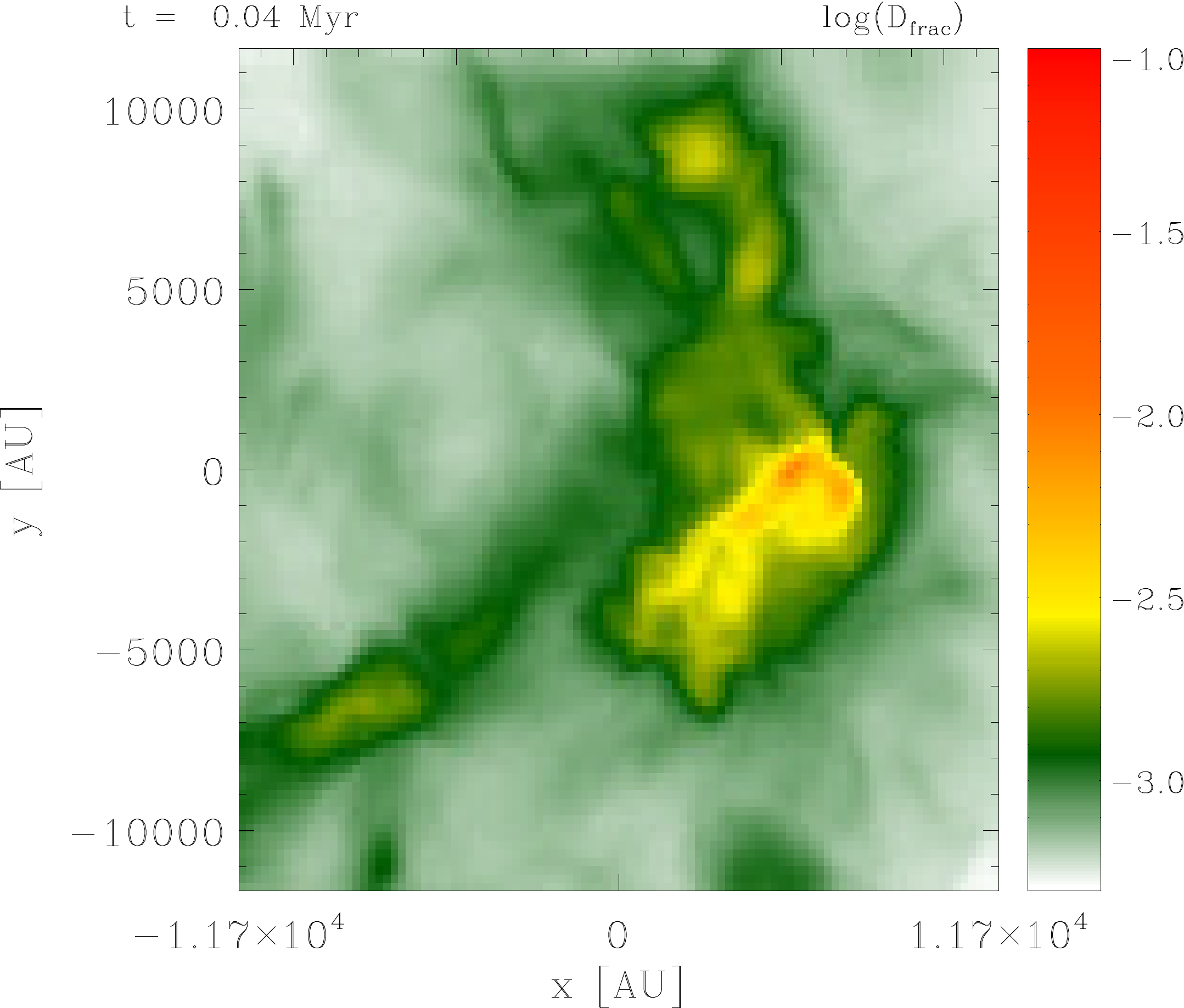}&\includegraphics[width=0.28\textwidth]{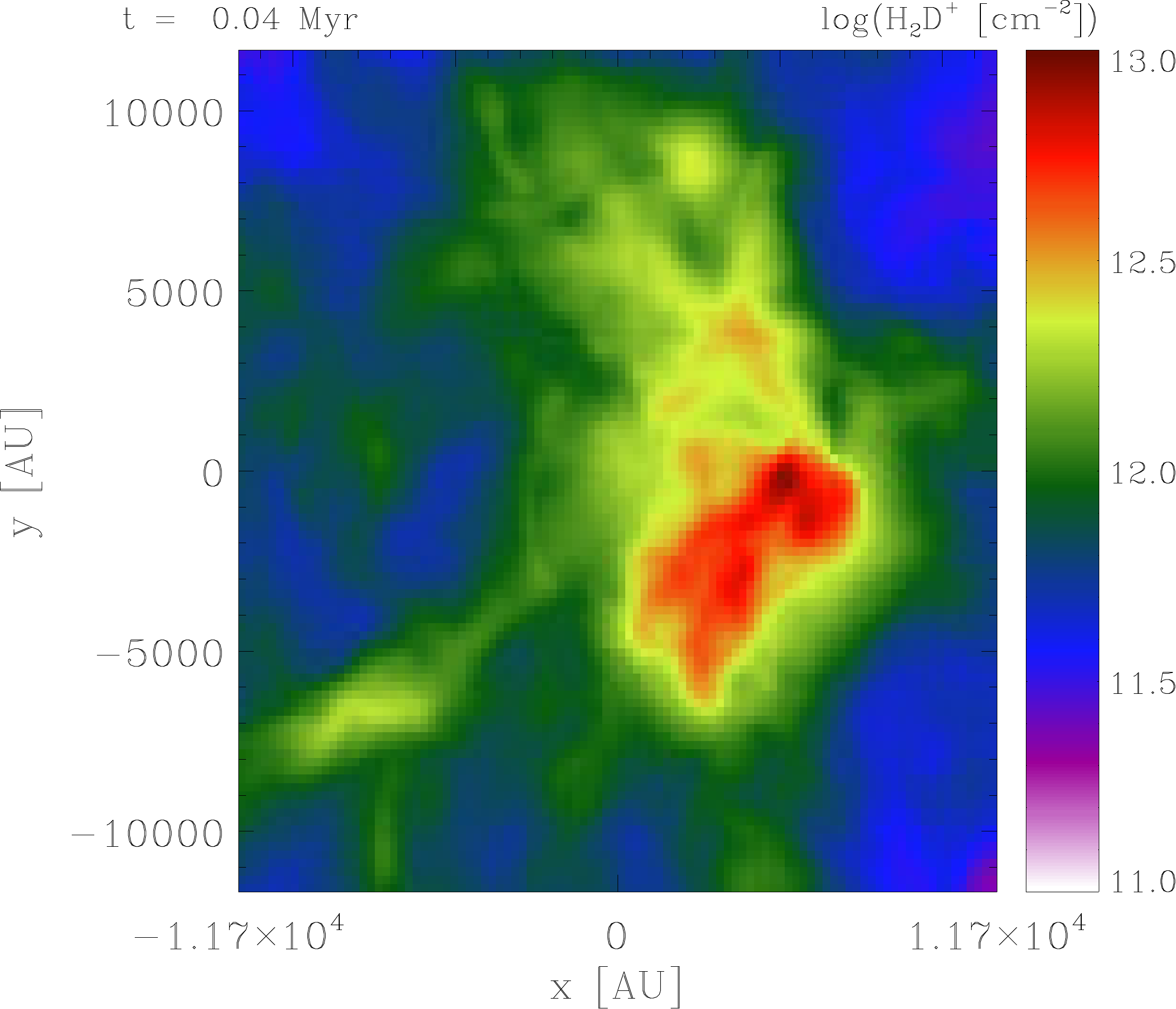}\\
		
		\end{tabular}
	\end{center}
	\caption{Total gas column density, deuterium fraction and o--H$_2$D$^+$ column density for the innermost region of the cores at $t=40\,\mathrm{kyr}$. From top to bottom the turbulent Mach number increases from $\mathcal{M}=2$ to  $\mathcal{M}=4$ and $\mathcal{M}=6$. For stronger turbulent fluctuations, there appears a more widespread region of enhanced deuteration. Note that, although the column density is less peaked for the core with $\mathcal{M}=6$ fluctuations, 
	the deuterium fraction seems to be of similar magnitude.}
	\label{figCore2}
\end{figure*}

As our reference runs, we consider Lmu10M2 and Hmu10M2, which both have a core mass of $60$~M$_\odot$, a mass--to--flux ratio $\mu/\mu_{\rm crit}=10$ and a Mach number $\mathcal{M}=2$. For Lmu10M2, the core radius is $0.17$~pc and the gas surface density is $0.14$~g~cm$^{-2}$, implying a virial parameter of 0.64, a free--fall time of 149~kyrs as well as 24 Jeans masses within the core. For Hmu10M2, the core radius is 0.1~pc and the gas surface density is $0.39$~g~cm$^{-2}$, implying a virial parameter of 0.48, a free--fall time of 67~kyrs and 60 Jeans masses within the core. Both cores are thus sub--virial, magnetically super--critical and highly Jeans unstable and can be expected to collapse, with the second one collapsing faster than the first. \\
Figure \ref{figCore1} compares the innermost 20,000 au after 30~kyrs and 130~kyrs, respectively, during the evolution of the collapsing core for Lmu10M2 (top two rows) and Hmu10M2 (bottom rows). We choose absolute timescales here in order to compare the state of the core at the same chemical time. The early stages show a rather disordered core structure as a result of turbulence stirring up the gas. Although the surface density of the cores differs by a factor of 3, the cores look very similar with a high--density filamentary structure in its center and striations near the outskirts, indicating an initially similar evolution.  Note the larger region of increased column density in the high surface density core, which appears since the core has an initially higher average density. These 
high--density regions are also readily observed in the deuterium fraction (middle column) where the latter is increased in such regions. Again, the spread of the 
region with enhanced deuterium fraction is explained by an initially higher density in the core. \\
The last column depicts the o--H$_2$D$^+$ column density. Here it is seen that this species traces the dense gas very well and shows column densities of \mbox{$N$(o--H$_2$D$^+$)=10$^{12}$--10$^{13}$ cm$^{-2}$} near the core center, which is already comparable to observations \citep[][]{Caselli2003,Vastel2006}. Note that at $t=30\,\mathrm{kyr}$ a sink particle has already formed in the core of run Hmu10M2, while it is not yet present in Lmu10M2, hence the two cores are in a different dynamical stage.\\
At later times the two cores evolved quite differently, and a sink particle is also present in the core of run Lmu10M2. The cores now show further filaments as well as striations. However, while the core in Lmu10M2 is observed to form a disc--like feature\footnote{We emphasize that the compressive modes in the velocity field 
decay faster than their solenoidal counterparts, which favors the formation of vortices on small scales.} in its center with high density gas collapsing radially onto it, the core in run Hmu10M2 is almost entirely accreted by the sink particle, i.e. it has not been stabilized by a sufficient amount of rotation, turbulence or magnetic pressure. In spite of the different dynamical evolution, both cores show widespread deuteration. In the case of Lmu10M2, the whole core is highly deuterated with further enhanced deuteration in the disc--like region. 
The core in run Hmu10M2 only shows widespread deuteration in the innermost 10,000 au as the density in this regions is slightly enhanced and not yet accreted onto the sink particle. Note, however, that the core in run Lmu10M2 shows almost up to an order of magnitude higher deuteration throughout the core. Further note that the core with initially higher surface density (Hmu10M2), if it was traced by o--H$_2$D$^+$, is much less extended than its lower--surface density counterpart. This result suggests that higher deuterium fractions are reached if part of the dense gas can be stabilized within the core, without going into immediate collapse, and it is certainly plausible that both the shorter free--fall time as well as the lower virial ratio of Hmu10M2 were contributing to the more rapid accretion onto the center.\\
In Fig. \ref{figCore2} we compare three cores at $t=40\,\mathrm{kyr}$ with an initial surface density of $0.14$~g~cm$^{-2}$, but different Mach numbers ranging from $\mathcal{M}=2$ (top row) to 
$\mathcal{M}=6$ (bottom row). As in the previous run Lmu10M2, these cores have an initial mass of $60$~M$_\odot$, a core radius of $0.17$~pc and a mass--to--flux ratio $\mu/\mu_{\rm crit}=10$. The virial parameter varies from 0.64 for $\mathcal{M}=2$ to 2.56 for $\mathcal{M}=4$ and 5.76 for $\mathcal{M}=6$. The latter two runs are initially strongly supported against collapse via supersonic turbulence, thus delaying the collapse to high density gas. While the turbulence does decay on roughly a crossing time, it still leaves an imprint on the subsequent evolution, as the dispersion of the collapsing gas introduces a significantly extended region of enhanced deuteration as well as of a greater o--H$_2$D$^+$ column density. It also appears that the o--H$_2$D$^+$ column density is enhanced in the central region of the core.\\
In summary, figures \ref{figCore1} and \ref{figCore2} indicate that the morphology and magnitude of the deuteration ratio and the chemical species depend 
strongly on time and the virial parameter, which is regulated here via the magnitude of the turbulent fluctuations. The latter two aspects will be discussed in greater detail below. 


\subsection{Effect of the initial turbulence}
\begin{figure*}
	\begin{center}
		\begin{tabular}{cccc}
		&D$_\mathrm{frac}^{\mathrm{H_2D^+}}$	&o--H$_2$D$^+$&p--H$_2$D$^+$\\
		\rotatebox[origin=r]{90}{$t=22\,\mathrm{kyr}=0.15\,t_\mathrm{ff}$\qquad}&\includegraphics[width=0.26\textwidth,angle=-90]{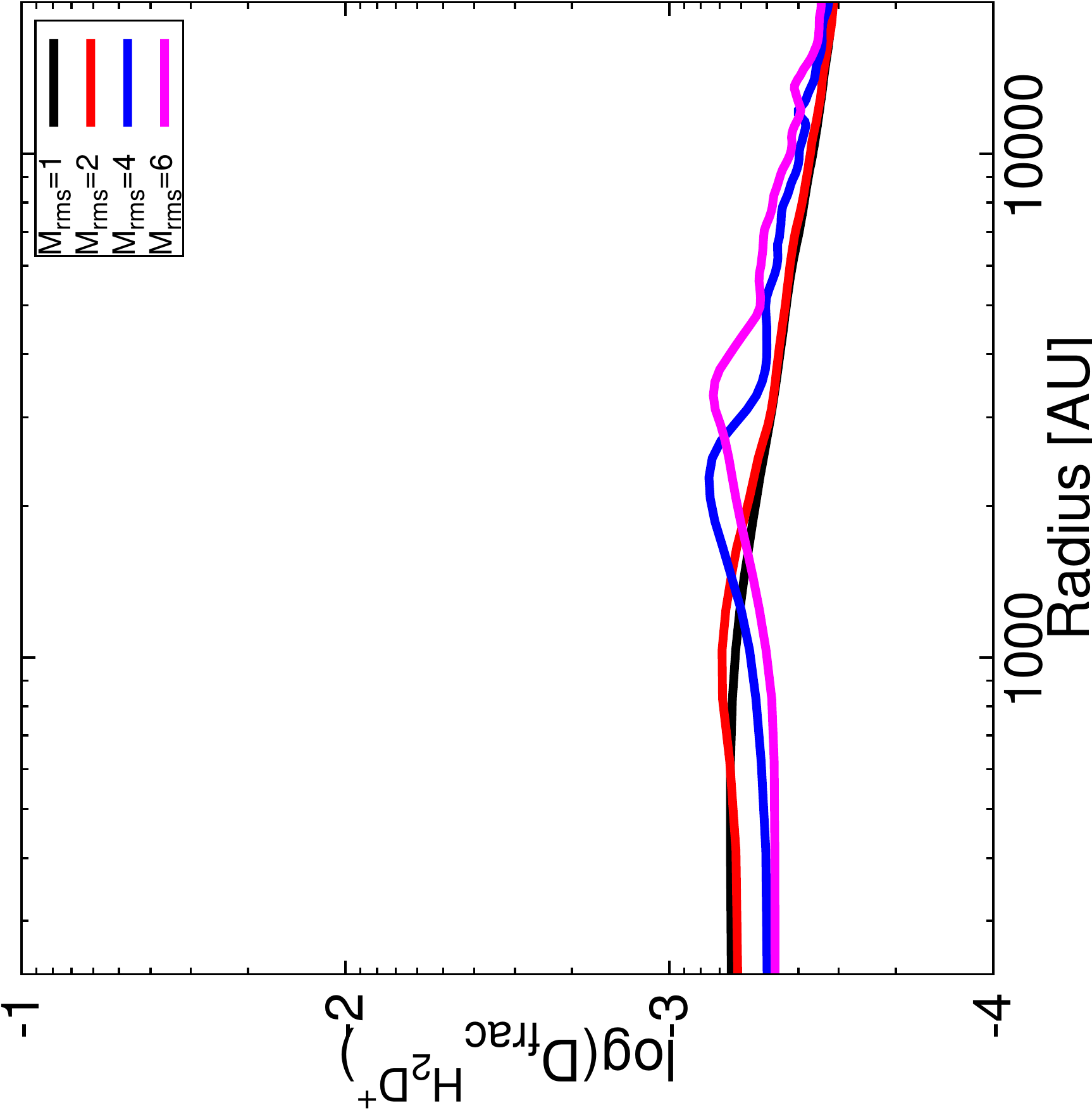}&\includegraphics[width=0.26\textwidth,angle=-90]{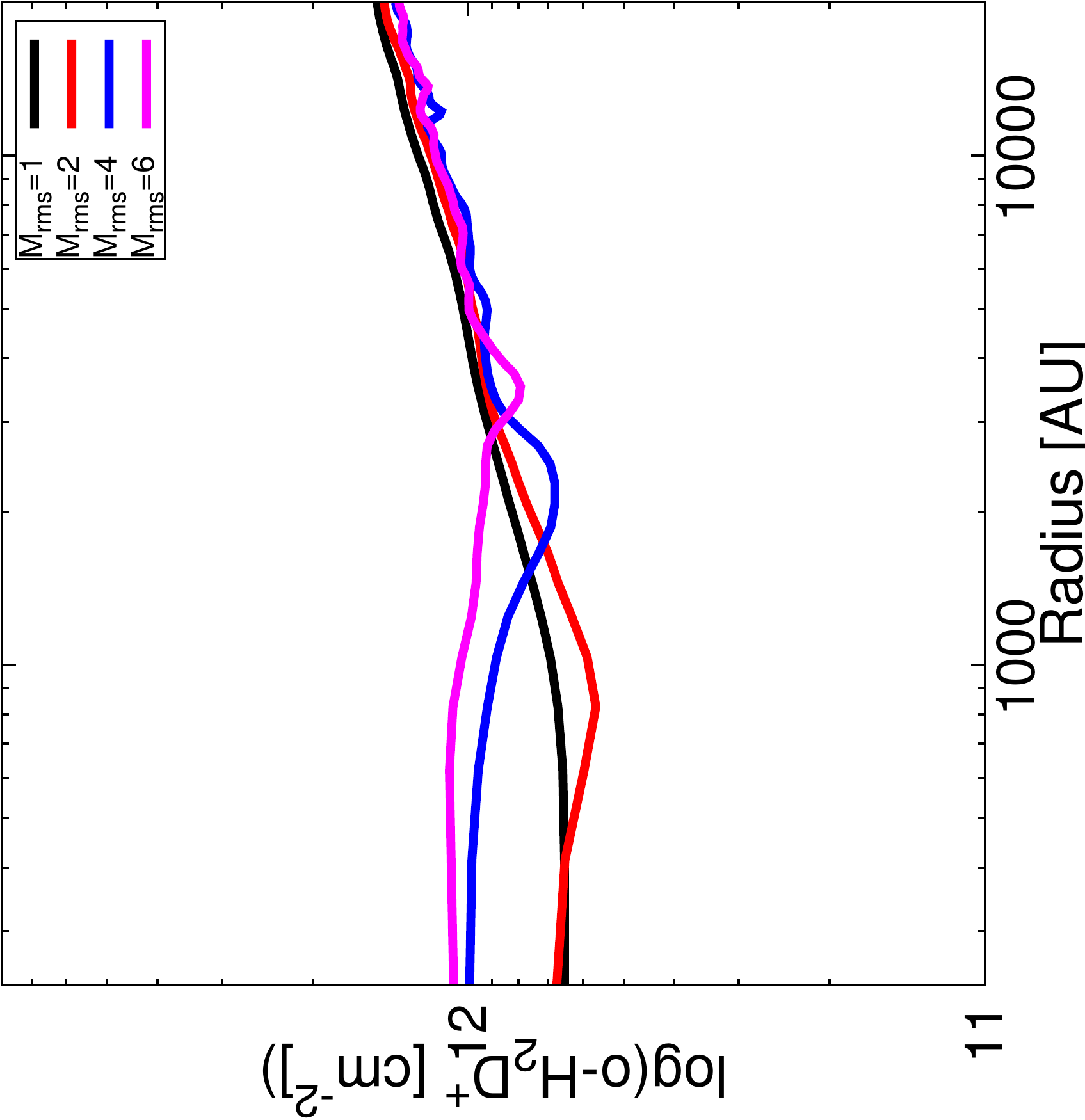}&\includegraphics[width=0.26\textwidth,angle=-90]{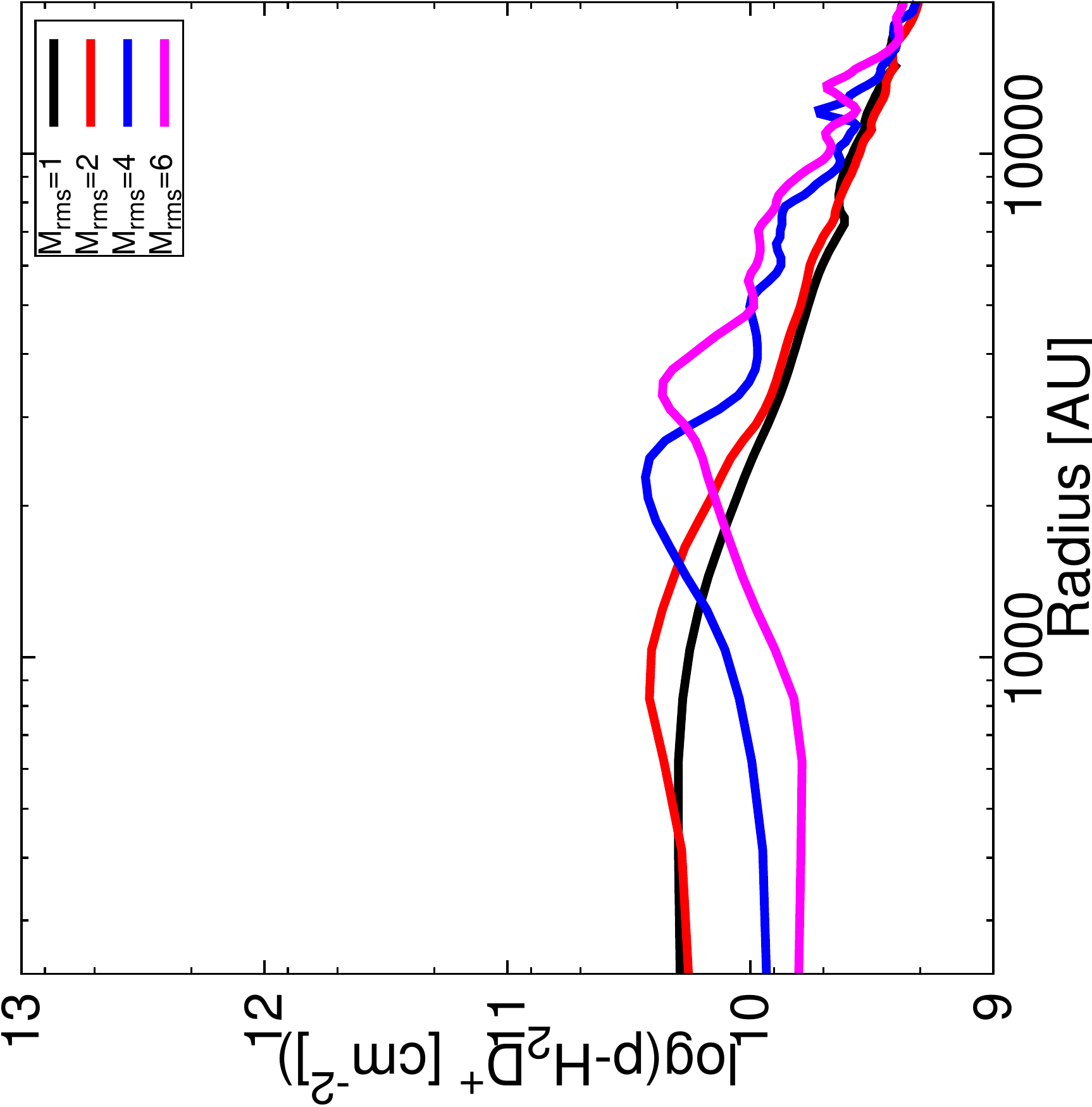}\\
		\rotatebox[origin=r]{90}{$t=32\,\mathrm{kyr}=0.21\,t_\mathrm{ff}$\qquad}&\includegraphics[width=0.26\textwidth,angle=-90]{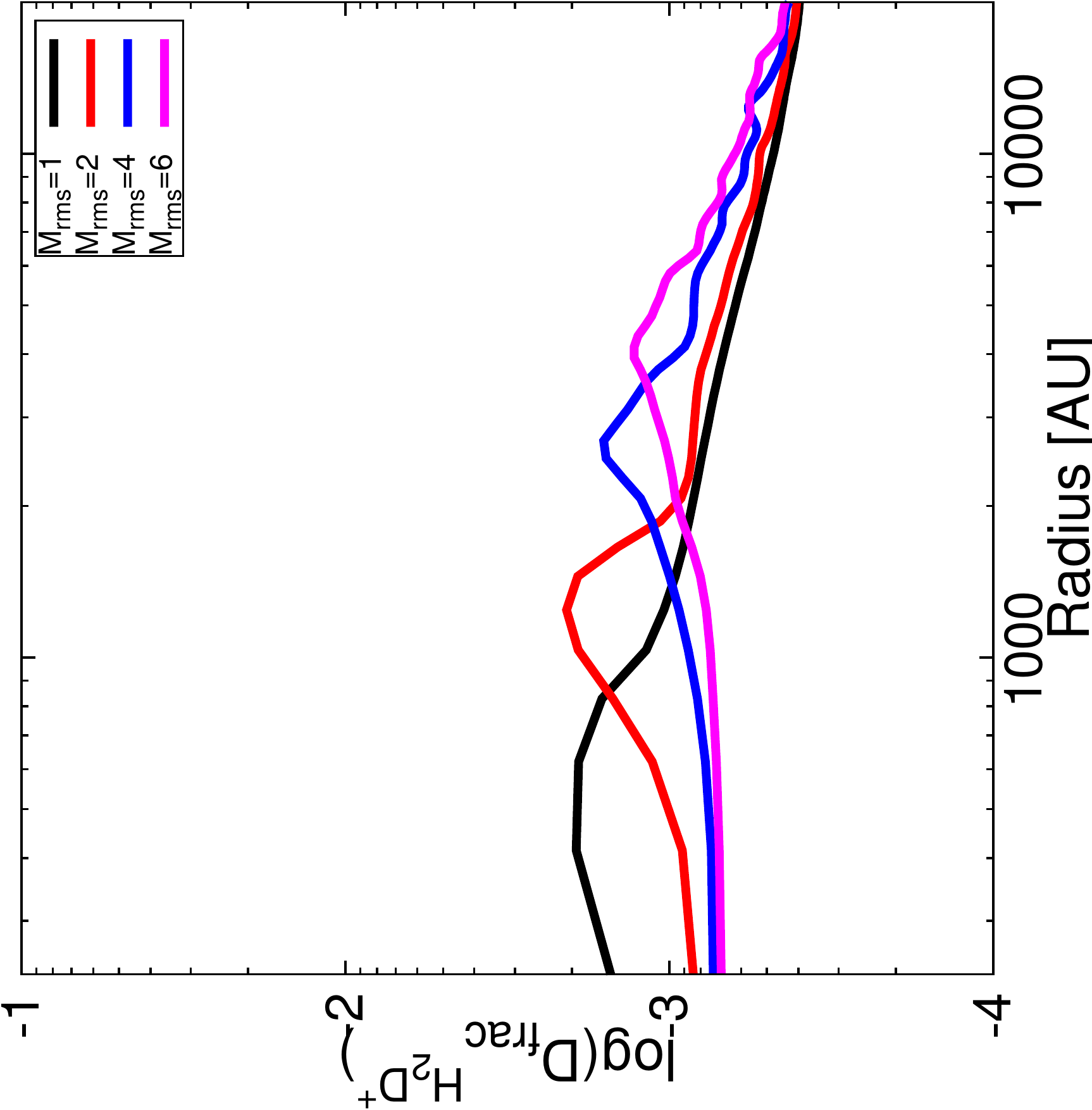}&\includegraphics[width=0.26\textwidth,angle=-90]{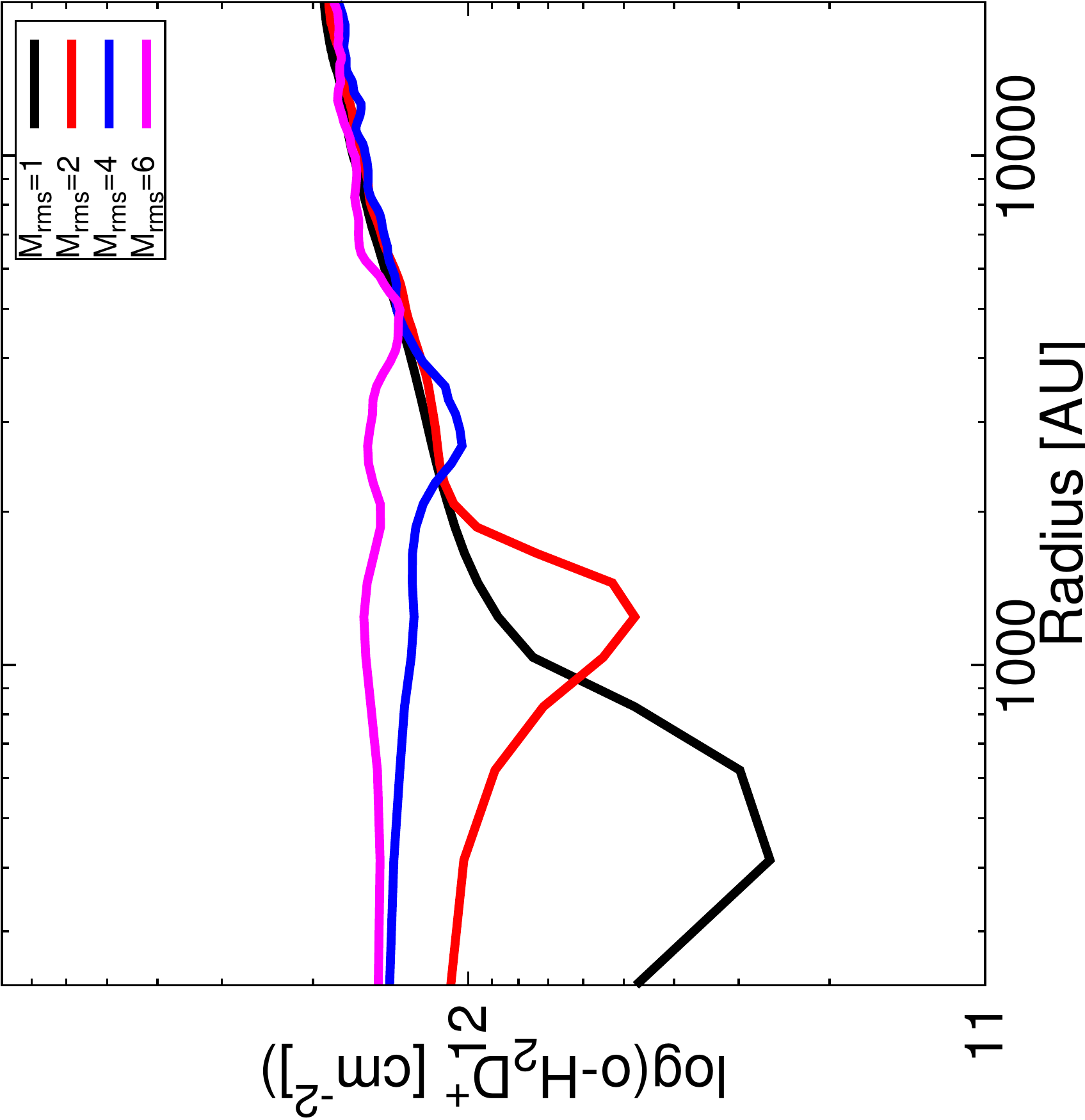}&\includegraphics[width=0.26\textwidth,angle=-90]{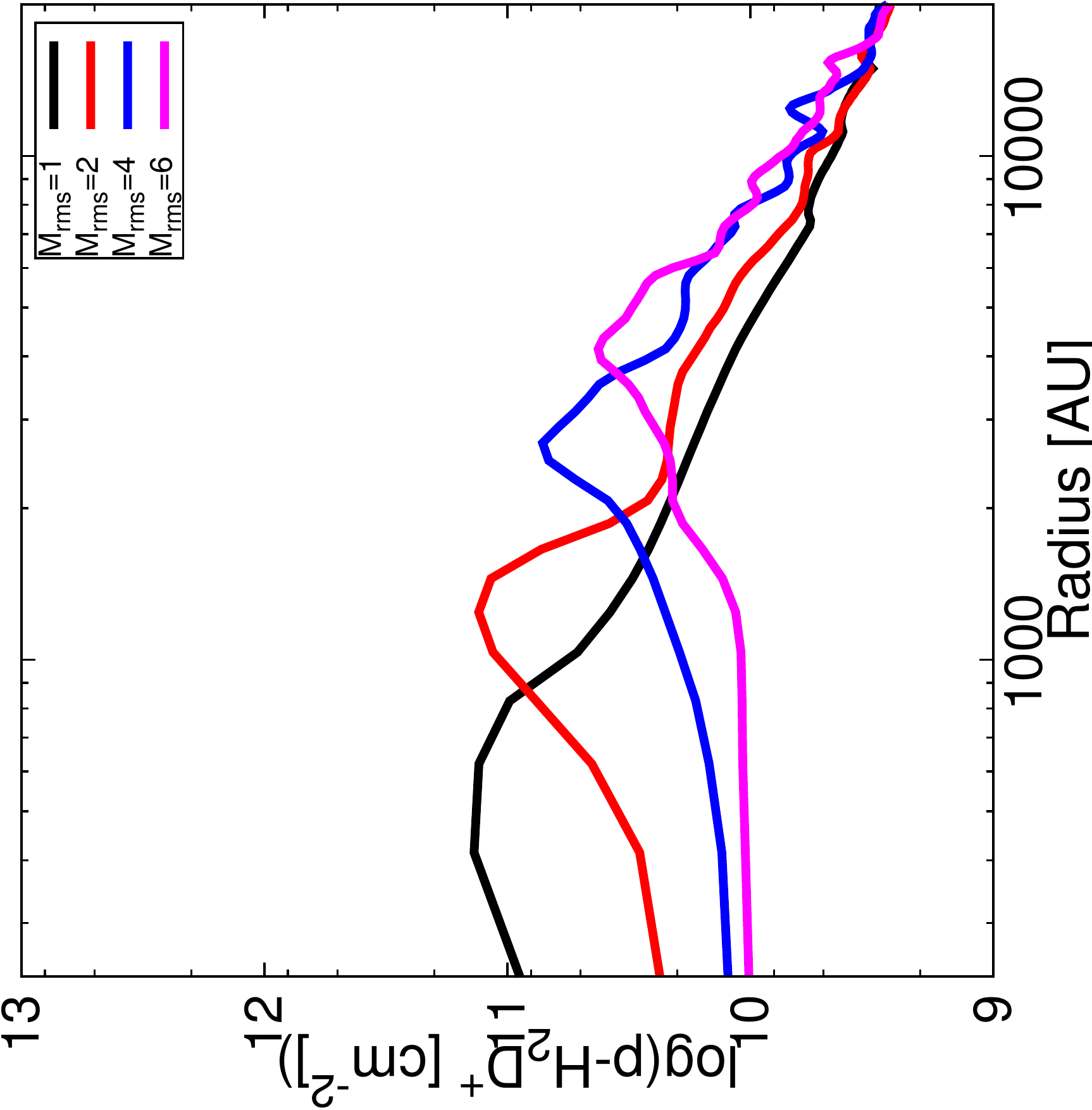}\\
		\rotatebox[origin=r]{90}{$t=42\,\mathrm{kyr}=0.28\,t_\mathrm{ff}$\qquad}&\includegraphics[width=0.26\textwidth,angle=-90]{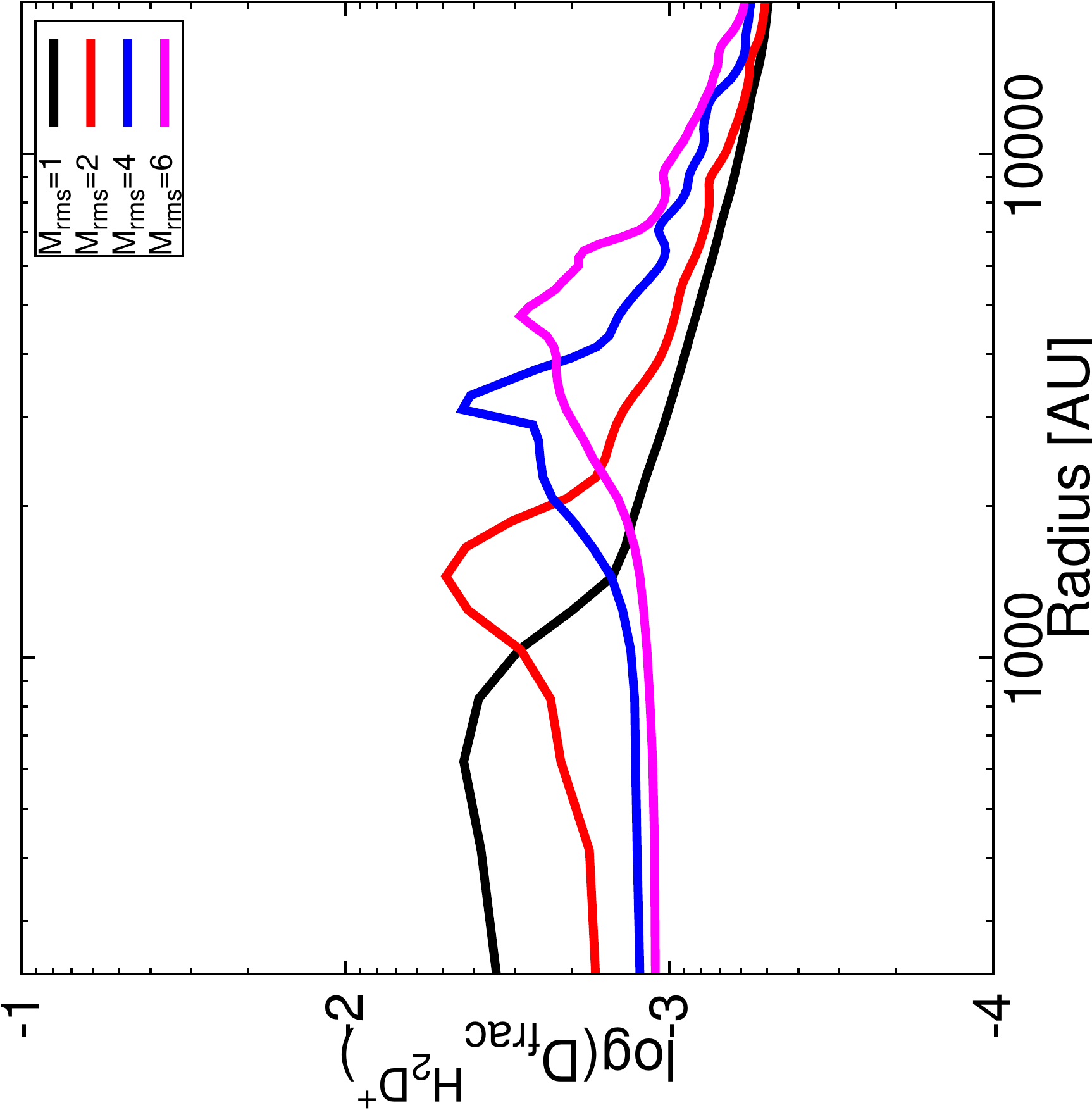}&\includegraphics[width=0.26\textwidth,angle=-90]{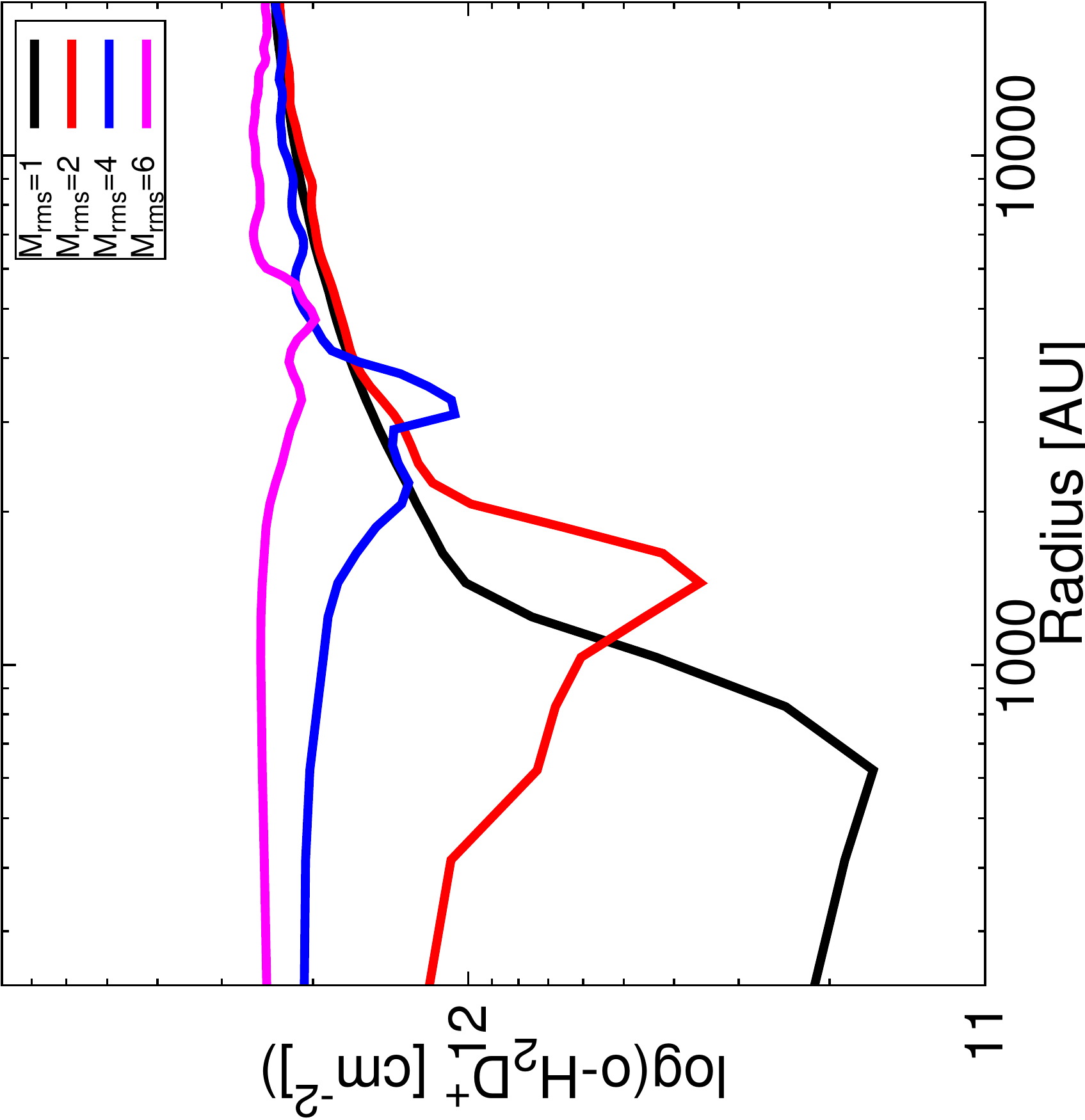}&\includegraphics[width=0.26\textwidth,angle=-90]{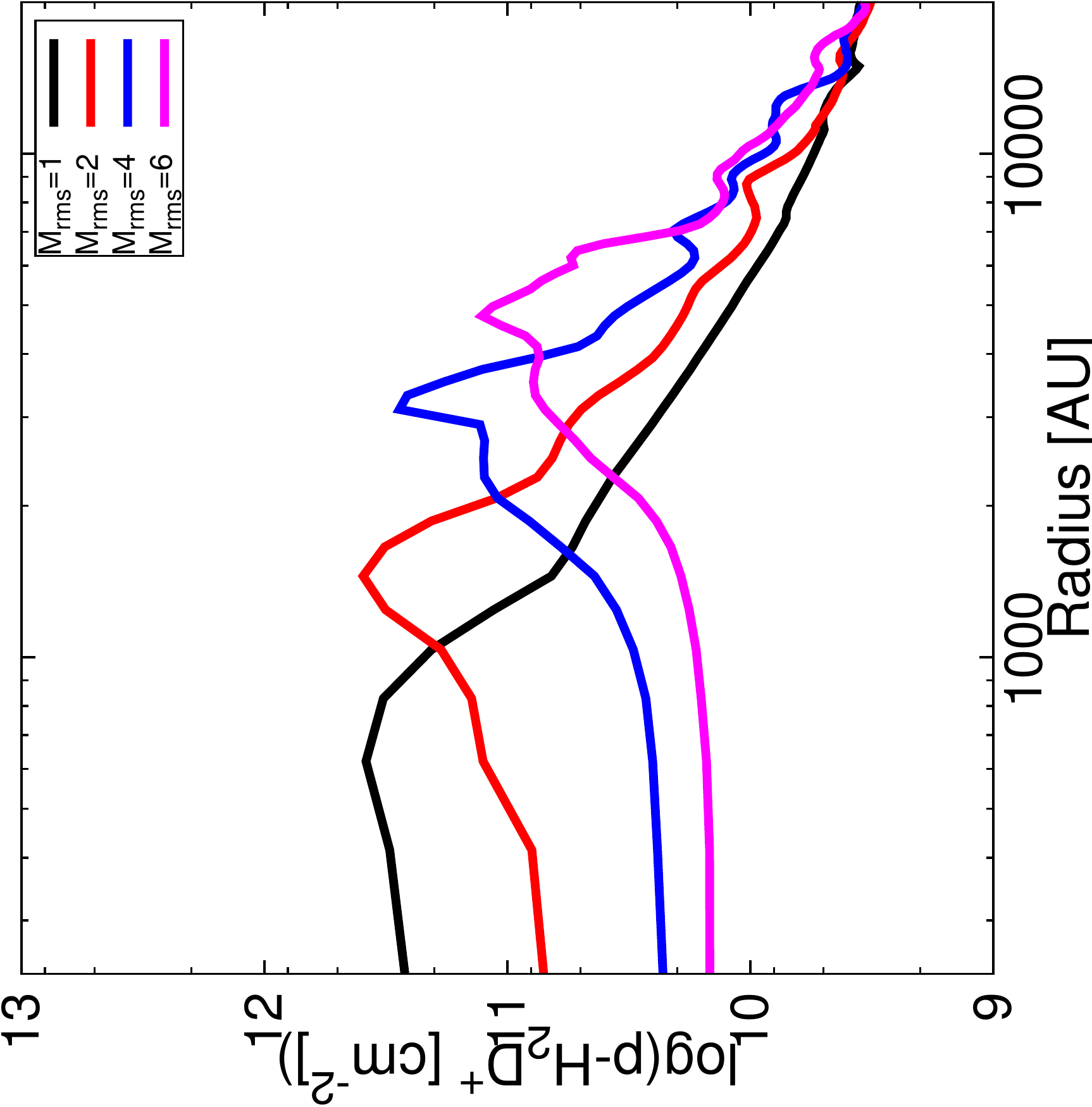}\\
		\rotatebox[origin=r]{90}{$t=66\,\mathrm{kyr}=0.44\,t_\mathrm{ff}$\qquad}&\includegraphics[width=0.26\textwidth,angle=-90]{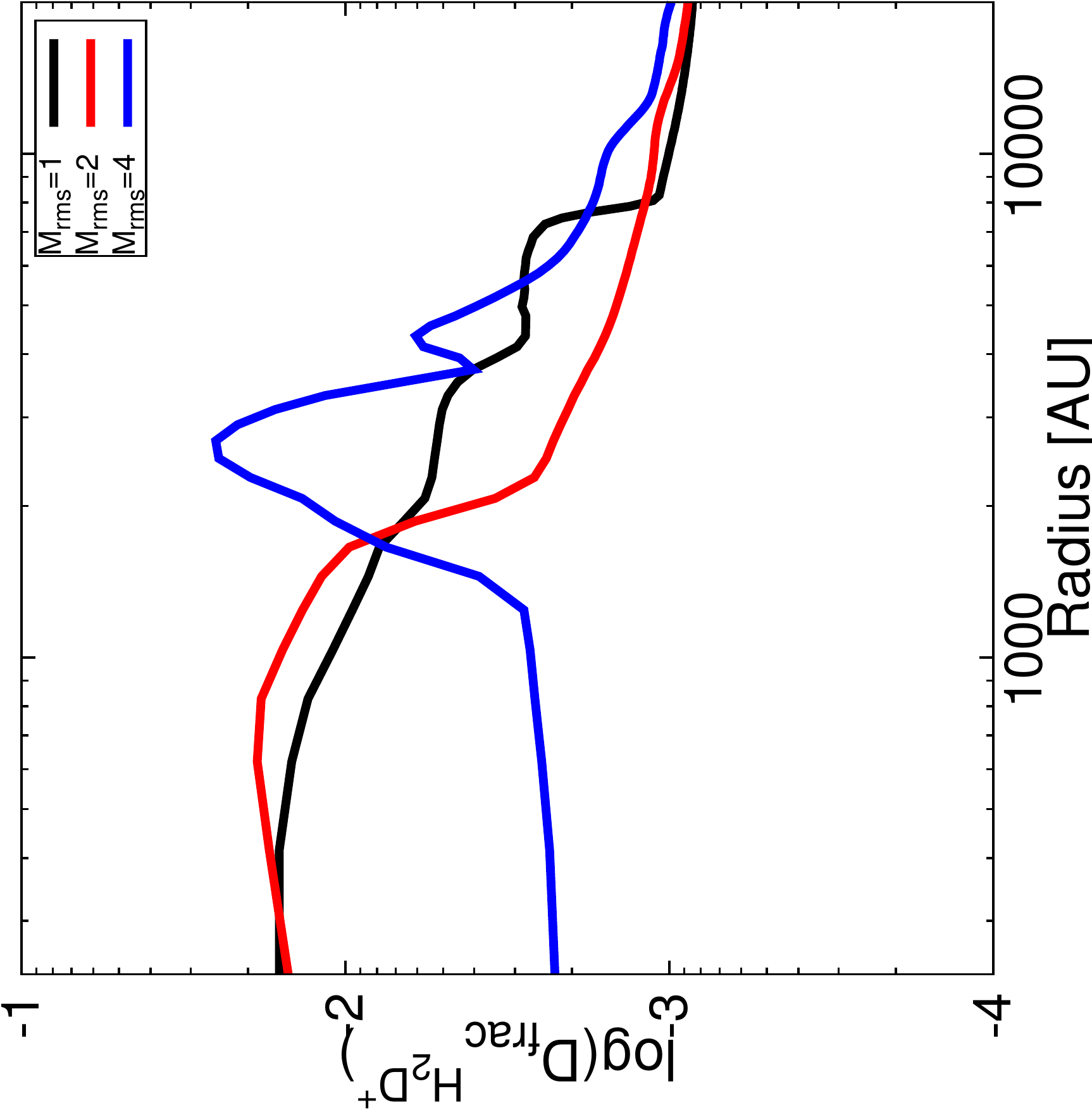}&\includegraphics[width=0.26\textwidth,angle=-90]{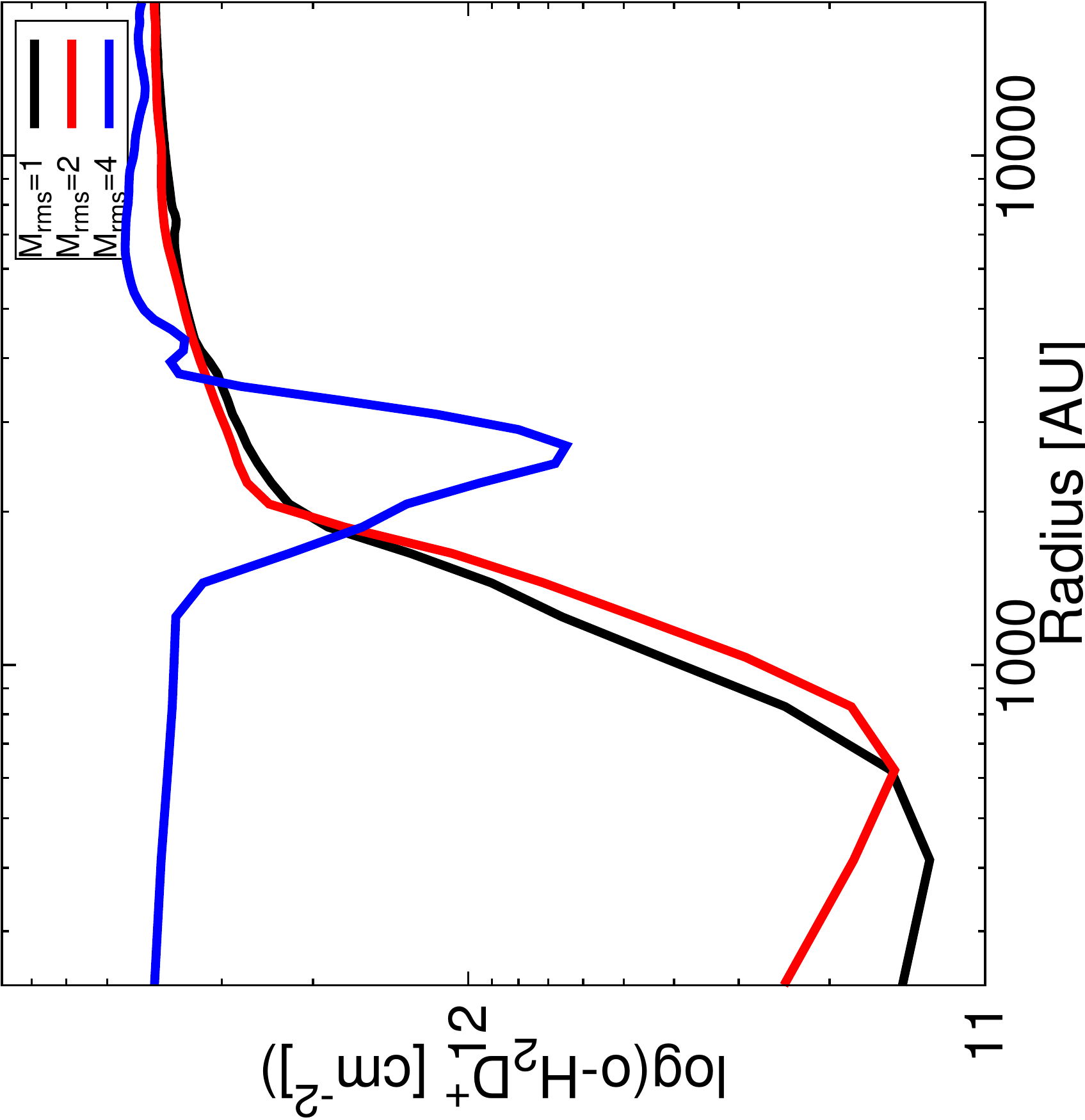}&\includegraphics[width=0.26\textwidth,angle=-90]{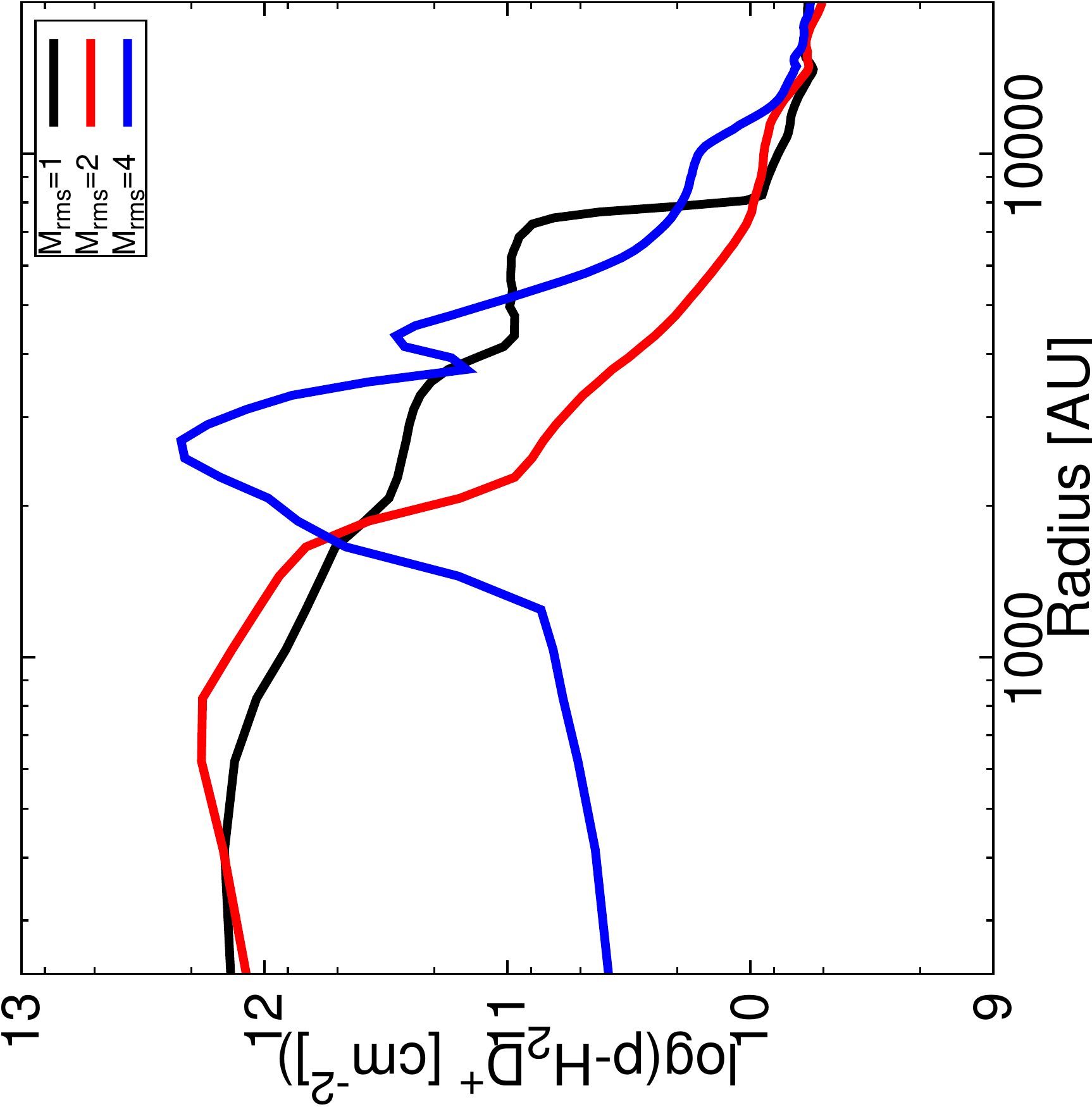}\\
		\end{tabular}
	\end{center}
	\caption{Data for the cores with lowest surface density ($\Sigma=0.14\,\mathrm{g\,cm}^{-2}$) and $\mu/\mu_\mathrm{crit}=10$: Column density weighted radial profiles of the deuterium fraction (left) and the spin states of H$_2$D$^+$ (middle,right) for different times and different initial turbulent Mach numbers. The radius is estimated with respect to the center of mass of the core. Top to bottom shows evolution in time over the course of approximately 40\,kyr. Relative time with respect to the initial free--fall time of the average density is also stated. The effect of turbulence is clearly seen as e.g. the peak D$_\mathrm{frac}^{H_2D^+}$ is shifted outwards and the overall magnitude is decreased. Note that more evolved cores (bottom row) show a shift in the deuteration peak towards the core centre. Also the radial profiles take similar forms.}\label{Mach}
\end{figure*}

The initial turbulence, parameterized here via the Mach number, has a central impact on the overall evolution. We consider here Mach numbers of 1, 2, 4 and 6, with otherwise the same properties as Lmu10M2. We do not show results for run Lmu10M12 as it has not evolved to more than $t\gtrsim30\,\mathrm{kyrs}$ due to too 
small timesteps. However, the observed trend, which will be discussed below, is confirmed. Due to the different Mach numbers, the virial parameters of these runs vary considerably, from 0.16, 0.64 and 2.56 up to 5.76 \citep[see also][]{Girichidis2011,Girichidis2012}. As a result, the last two cores will be initially stabilized by the supersonic turbulence. We however do note that the latter decays within roughly a crossing time, so that also these cores eventually go into collapse. The collapse is however delayed and there is more time for chemical evolution before the density increases substantially.\\
For a systematic comparison of these runs, we calculate column--density weighted radial profiles for the deuterium fraction as measured via H$_2$D$^+$, the column density of \mbox{o--H$_2$D$^+$} as well as of \mbox{p--H$_2$D$^+$} after 22~kyrs, 32~kyrs, 42~kyrs and 66~kyrs for all four simulations. The radial profiles are evaluated with respect to the center of mass of the core. This is a natural choice as it is at best comparable with observations, where the core center is often chosen to be the location of highest surface density, which coincides with the center of mass \citep{Butler12}. The results of this averaging procedure are given in Fig.~\ref{Mach}. \\
While the runs show a time--dependence, with the average deuterium fraction as well as the column densities of \mbox{o--H$_2$D$^+$} and \mbox{p--H$_2$D$^+$} increasing with time, a distinct feature is introduced via the turbulent Mach number, which can be clearly recognized in the radial profiles. In particular, for each Mach number, there is a characteristic and initially small peak, which occurs on larger scales for larger values of the Mach number, due to the increasing amount of stabilization via supersonic turbulence\footnote{We have re--run parts of the simulation with a different turbulent seed field, which shows similar behavior. The location of the peak 
slightly varies, but the trend is the same.}. The strength of the initial turbulence  seems to imply a characteristic relation between the center of the core, defined via the gas column density maximum, and the extent of the inner region currently supported by the turbulent pressure, therefore giving rise to a characteristic length scale. A close inspection of our deeper investigation in section~\ref{detailedcomp} reveals that the density maxi\-mum within that region is displaced from the maximum column density, and therefore the peak of the deuterium fraction is offset from the point with the highest column densities.
A similar behavior is shown for the column density of \mbox{o--H$_2$D$^+$}, where a characteristic dip in the column density is present depending on the Mach number, and the dip again occurs on larger scales for higher Mach numbers. This phenomenon is also visible from the column density of \mbox{p--H$_2$D$^+$}. It is thus evident that turbulence and the unresolved structure within the beam will have a strong impact on these quantities in observations.

\subsection{Effect of the initial surface density}
\begin{figure*}
	\begin{center}
		\begin{tabular}{ccc}
		D$_\mathrm{frac}^{\mathrm{H_2D^+}}$	&o--H$_2$D$^+$&p--H$_2$D$^+$\\
		\includegraphics[width=0.26\textwidth,angle=-90]{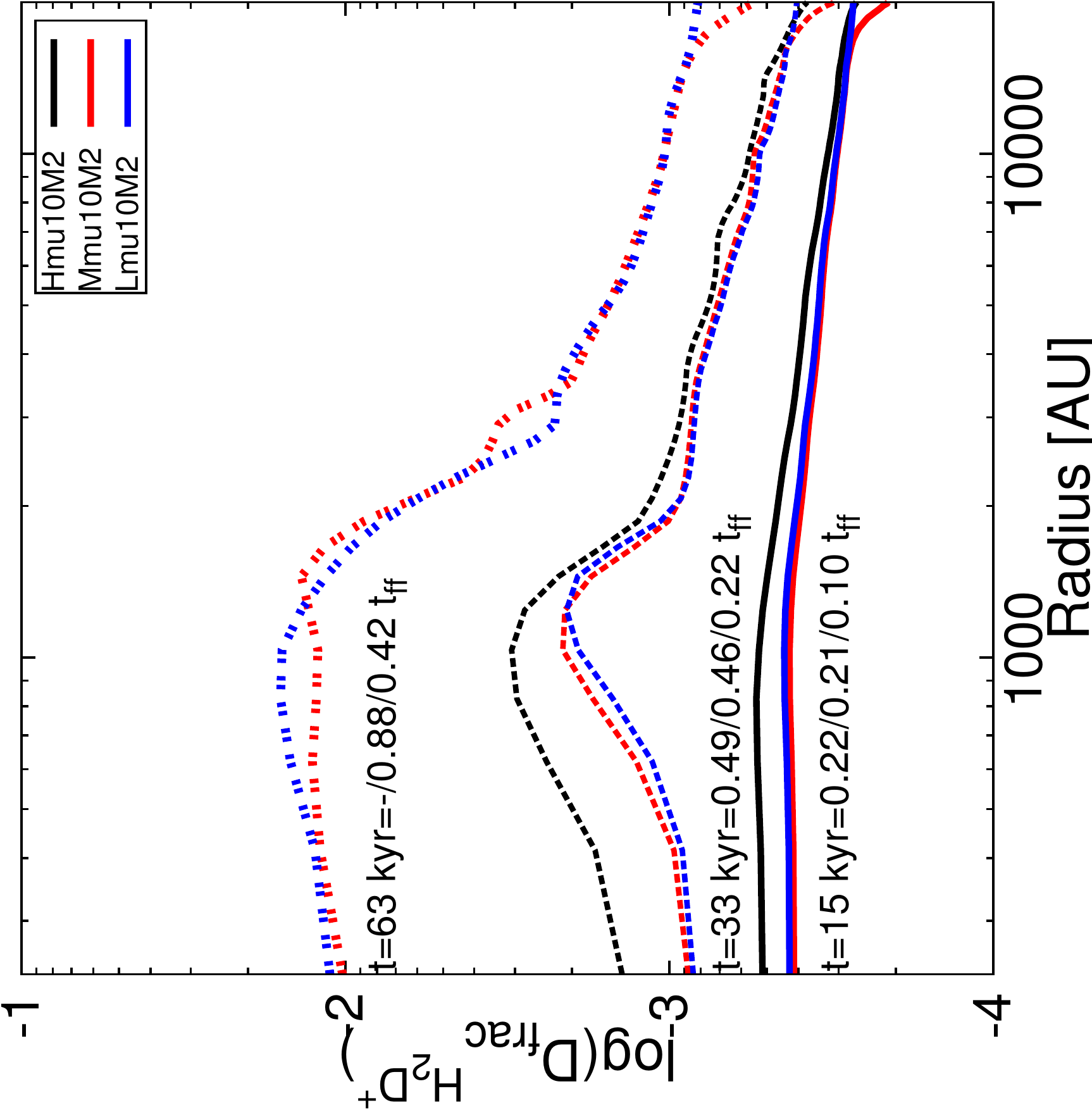}&\includegraphics[width=0.26\textwidth,angle=-90]{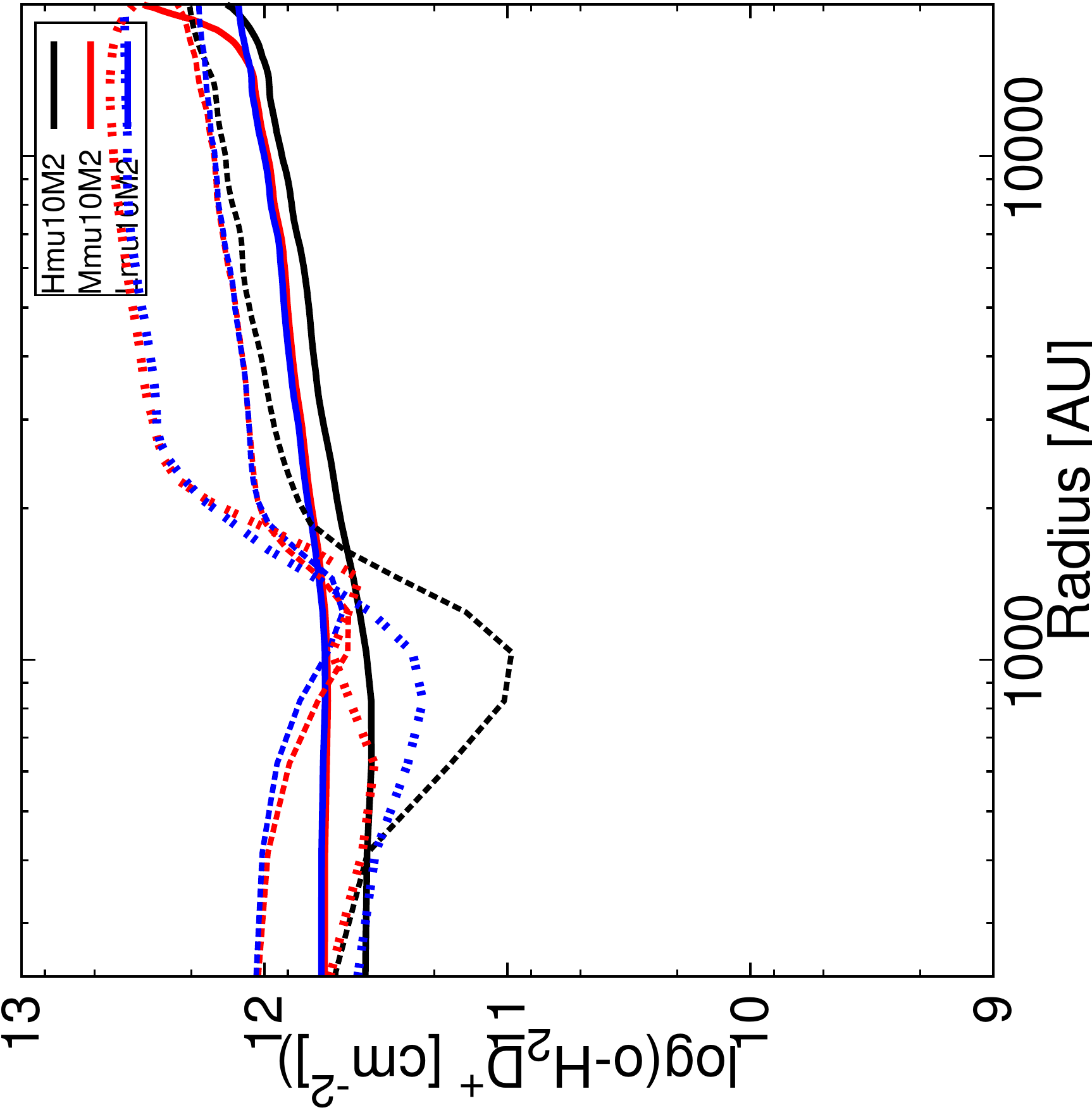}&\includegraphics[width=0.26\textwidth,angle=-90]{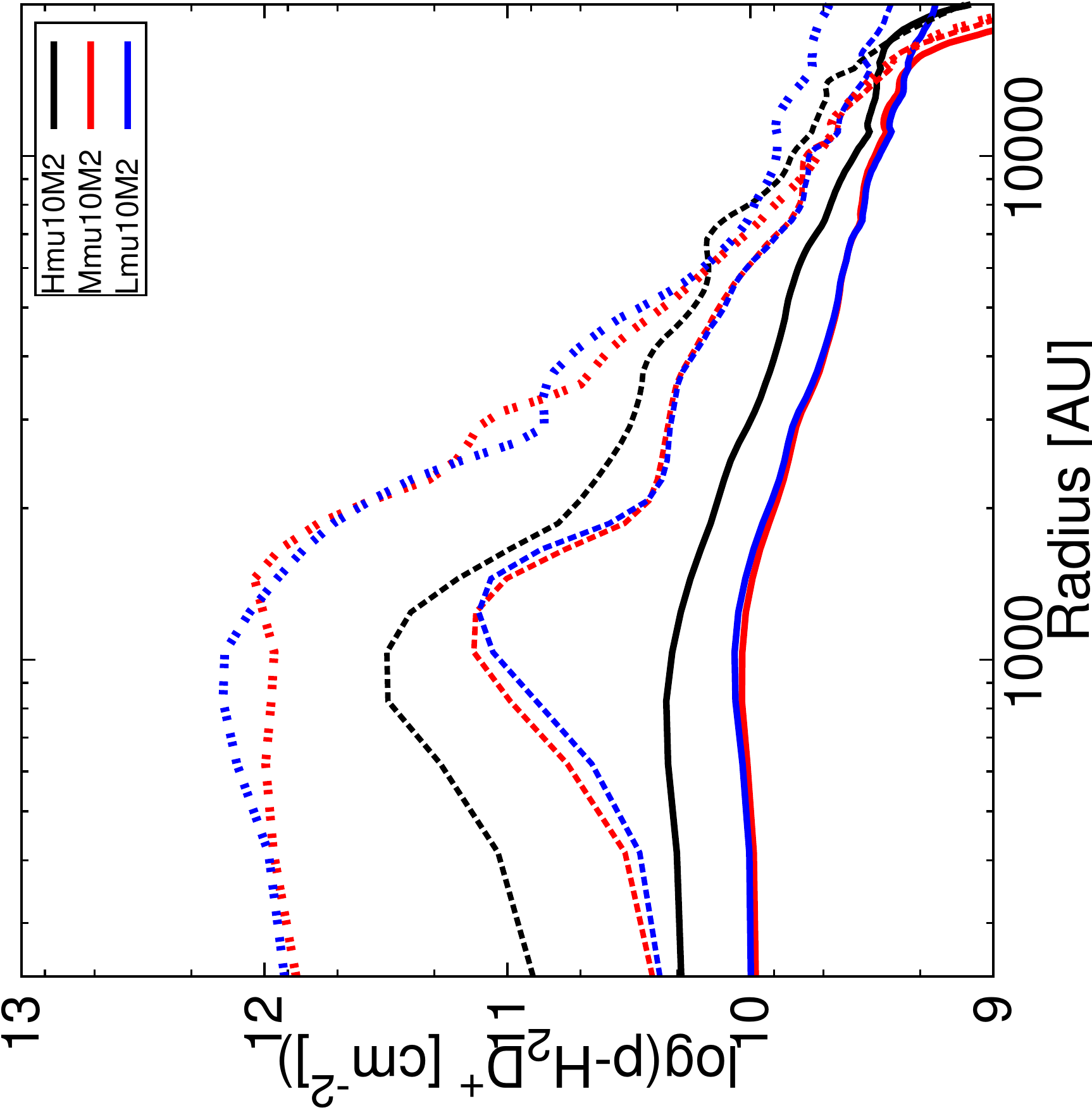}\\
		\end{tabular}
	\end{center}
	\caption{Column density weighted radial profiles of the deuterium fraction (left) and the spin states of H$_2$D$^+$ (middle,right) for different times and initial core surface densities. Time is given in absolute times as well as relative to the free--fall time of the respective run. $t_1/t_2/t_3$ refers to times of Hmu/Mmu/Lmu. Radius is estimated with respect to the center of mass of the core. There 
	appears some small difference when the initial surface density is high. However, at lower surface densities this difference vanishes and it is only restricted to the inner part of the core. At the outer parts, the radial profiles converge. With time the 
	profiles do not change significantly. Please note that at $t=63\,\mathrm{kyr}$ we avoid to show a profile for Hmu10M2 as this is influenced by the sink particle.}\label{sigmaprofile}
\end{figure*}

We now explore the effect of the initial surface density. For this purpose, we investigate three reference cases, Lmu10M2 as above with a gas surface density of $0.14$~g~cm$^{-2}$, Mmu10M2 with a gas surface density of $0.24$~g~cm$^{-2}$ and Hmu10M2 with a gas surface density of $0.39$~g~cm$^{-2}$. While Lmu10M2 and Hmu10M2 have the same mass but different core radii, Mmu10M2 has a similar core radius as Hmu10M2, but a lower mass of $27$~M$_\odot$. The virial parameter for these runs corresponds to 0.71, 0.64 and 0.48, respectively, and the free--fall time corresponds to 149, 72 and 67~kyrs. \\
For a systematic comparison of these runs, we calculate column--density weighted radial profiles for the deuterium fraction as measured via H$_2$D$^+$, the column density of \mbox{o--H$_2$D$^+$} as well as of \mbox{p--H$_2$D$^+$} after 15~kyrs, 33~kyrs and 63~kyrs for all three simulations and show the results in Fig. \ref{sigmaprofile}. It is almost directly visible within the figure that the overall deuterium fraction at each radius depends mostly on time. Both at 15~kyrs as well as at 63~kyrs, the lines are close to identical, with thus very little dependence on the surface density of the gas, while at 33~kyrs, at least within the central 1,000~au there is a variation of about a factor of 2. While the deuterium fraction is increasing with time at every distance from the core center, this increase is particularly pronounced between the snapshots at 33 and 63~kyrs, presumably reflecting the larger differences in time. 
In all plots, we still see the characteristic peak for the deuteration or dip for the column densities of o--H$_2$D$^+$, which we found to depend on the turbulent Mach number in the previous subsection, or more precisely, on the overall amount of virial support. Here it appears that the simulations Lmu10M2 and Mmu10M2 are showing this peak at a quite similar position, in spite of having different virial parameter. However, while in all previous simulations the free--fall time was the same and therefore the virial ratio was the main governing parameter of the collapse, here both the free--fall time and the virial parameter are different in all simulations. Hmu10M2 has both the lowest virial parameter and the lowest free--fall time, and its characteristic peak is thus closer to the interior. The difference in the free--fall times also explains that the peak or dip appearing in the respective quantities in Mmu10M2 occurs somewhat closer to the center of the core than for Lmu10M2.
Considering the column density of o--H$_2$D$^+$, on the other hand, the variation as a function of time, radius or column density is substantially reduced, and all lines range between column densities of at least $10^{11}$~cm$^{-2}$ and at most $5\times10^{12}$~cm$^{-2}$, with an average around $5\times10^{11}$~cm$^{-2}$. While on large scales it is recognizable that the o--H$_2$D$^+$ column density increases with time, a characteristic dip occurs at around 1,000~au, where the more evolved runs show lower o--H$_2$D$^+$ column densities. This should not be confused with a decrease of the abundance of o--H$_2$D$^+$, but rather shows the effect of the collapse, due to which a smaller amount of mass is available on these scales. The latter reflects the difference in the dynamical time, which is still large enough at larger radii to avoid a substantial mass decrease. Towards the center of the core, these differences are decreasing again, with only a non--systematic variation by about a factor of 3.
For p--H$_2$D$^+$, the column densities again show a clear dependence on time, as here the increase of the chemical abundance with time overcompensates for the decreasing mass during the chemical evolution. It is clearly recognizable that they in fact increase at all radii as a function of time, though showing a certain drop towards the center, presumably due to the lower mass within the central region. The dependence on the gas surface density for this quantity appears very weak, and the evolution seems to depend almost entirely on time. We point out that these findings agree with previous 
results by \citet{Goodson2016}.

\subsection{Effect of the initial magnetic field strength}
\begin{figure*}
	\begin{center}
		\begin{tabular}{cccc}
		&D$_\mathrm{frac}^{\mathrm{H_2D^+}}$	&o--H$_2$D$^+$&p--H$_2$D$^+$\\
		\rotatebox[origin=r]{90}{$\Sigma=0.14\,\mathrm{g\, cm^{-2}}$\qquad\qquad}&\includegraphics[width=0.26\textwidth,angle=-90]{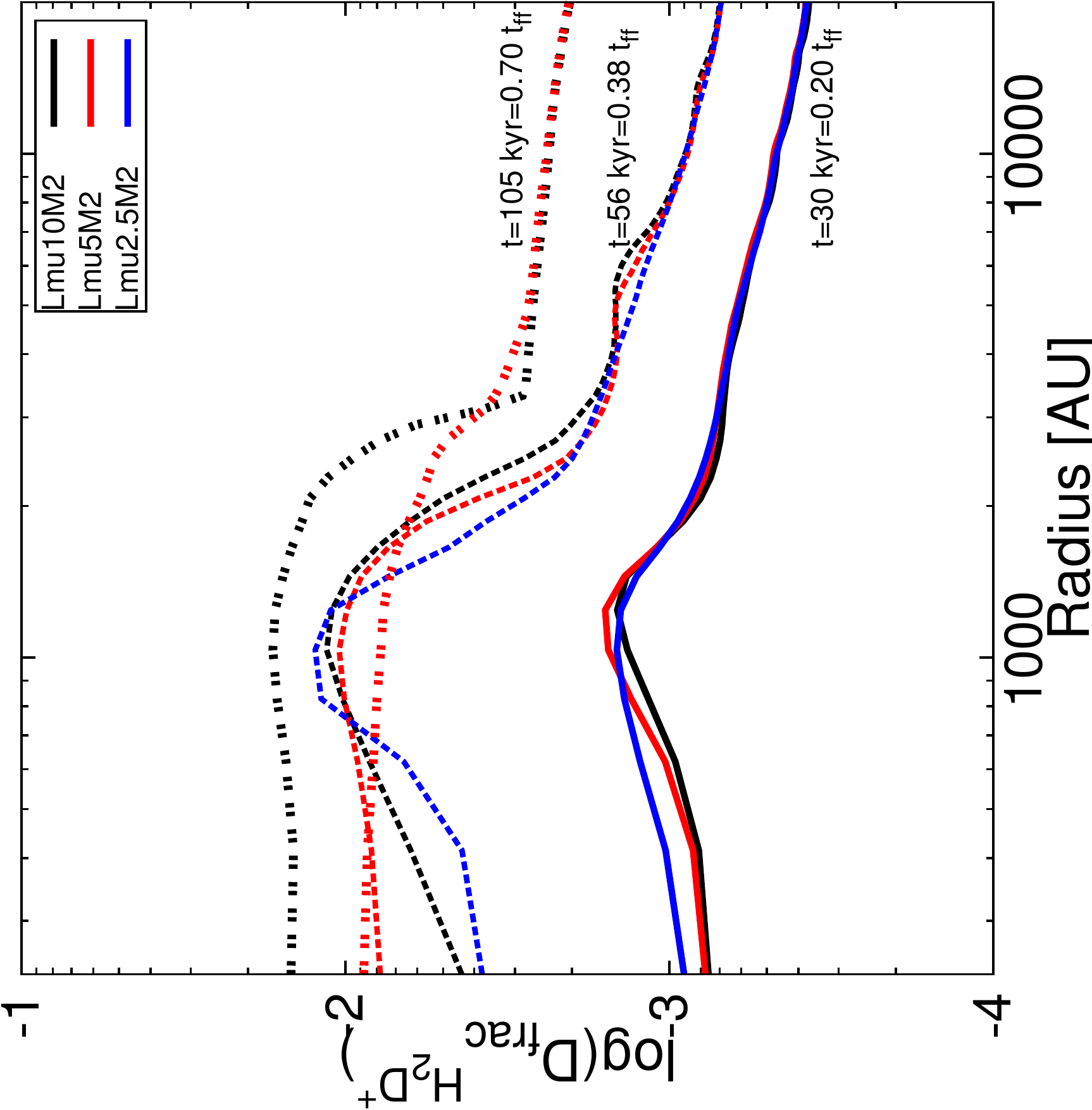}&\includegraphics[width=0.26\textwidth,angle=-90]{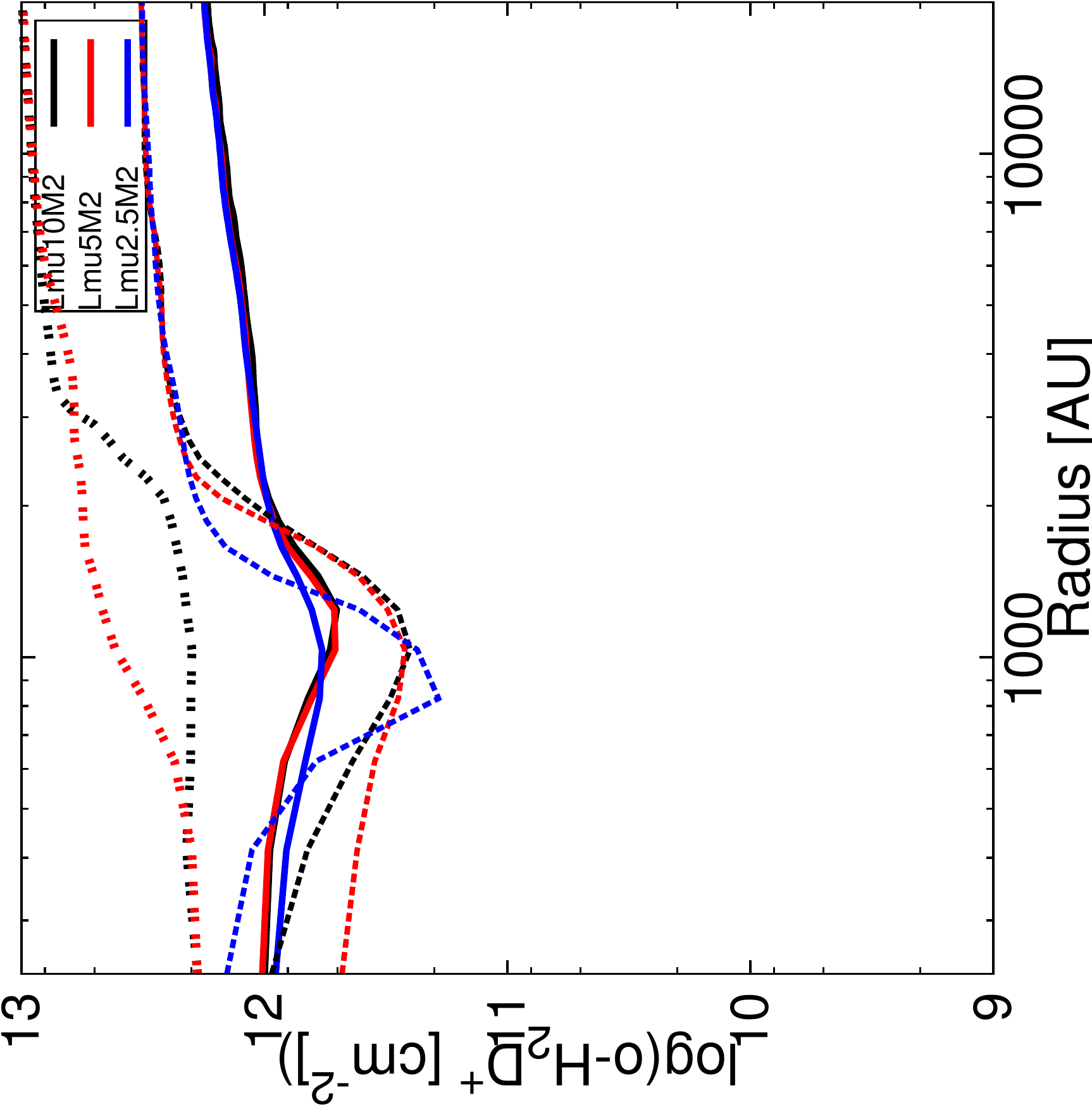}&\includegraphics[width=0.26\textwidth,angle=-90]{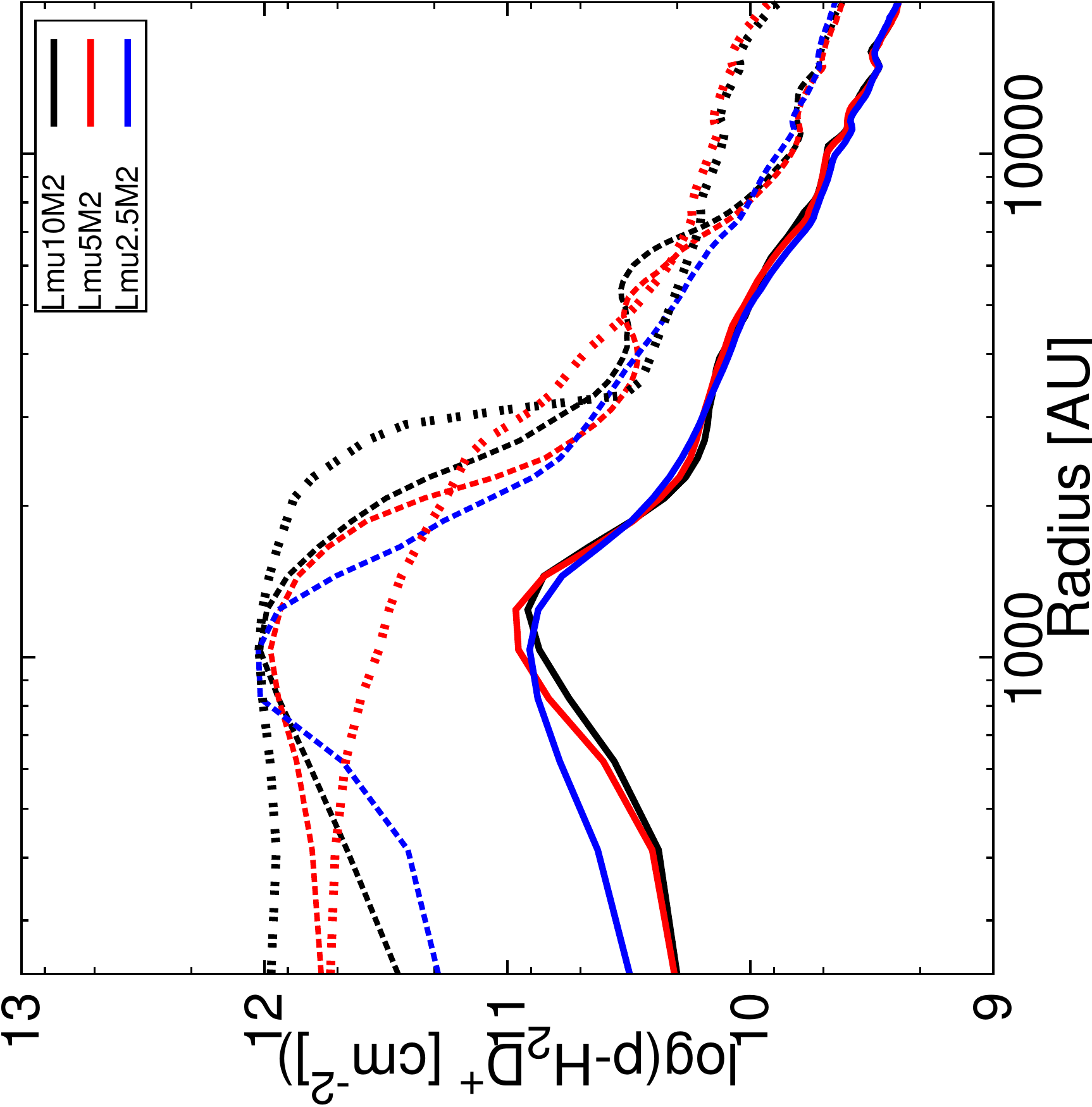}\\
		\rotatebox[origin=r]{90}{$\Sigma=0.24\,\mathrm{g\, cm^{-2}}$\qquad\qquad}&\includegraphics[width=0.26\textwidth,angle=-90]{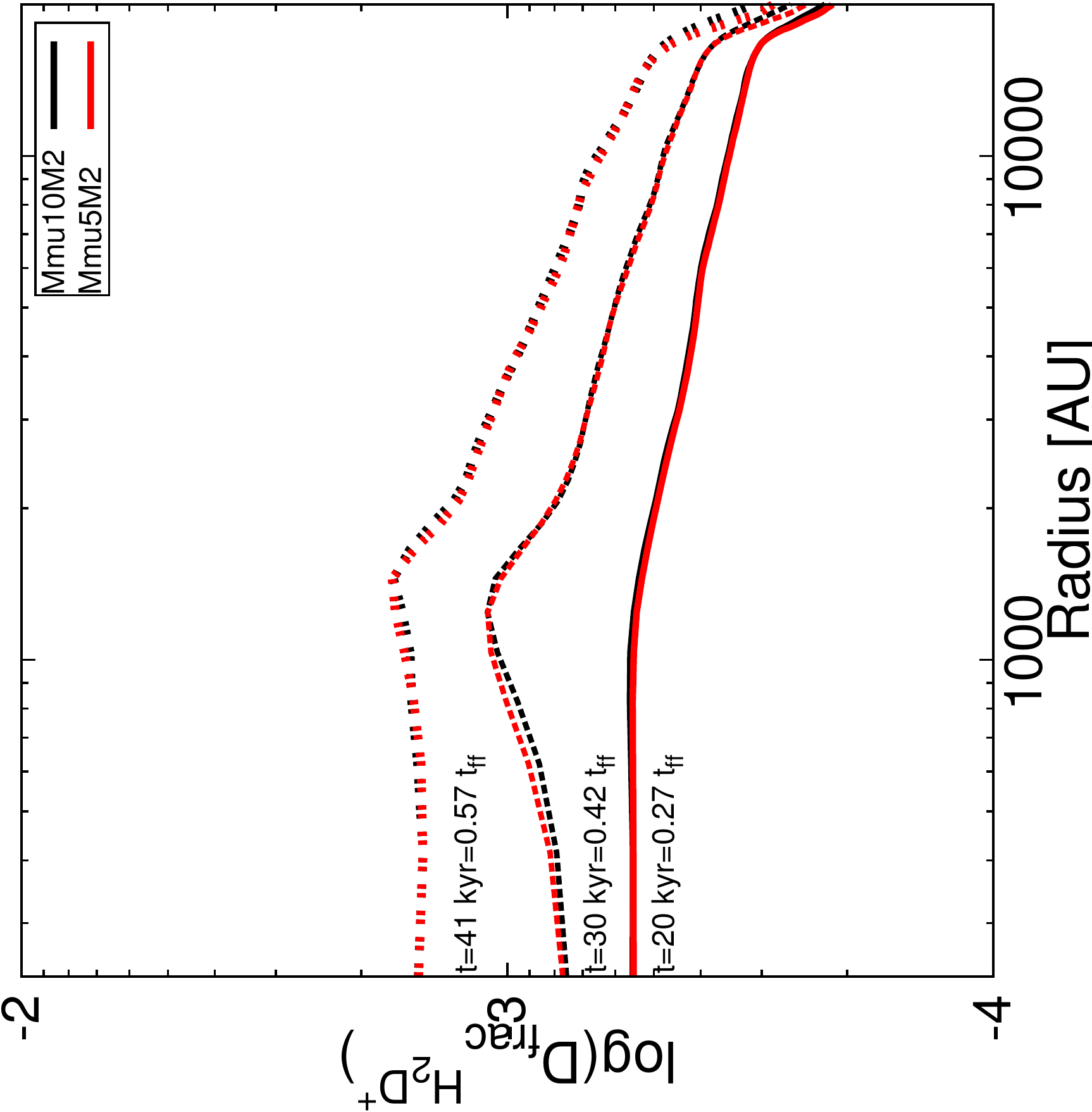}&\includegraphics[width=0.26\textwidth,angle=-90]{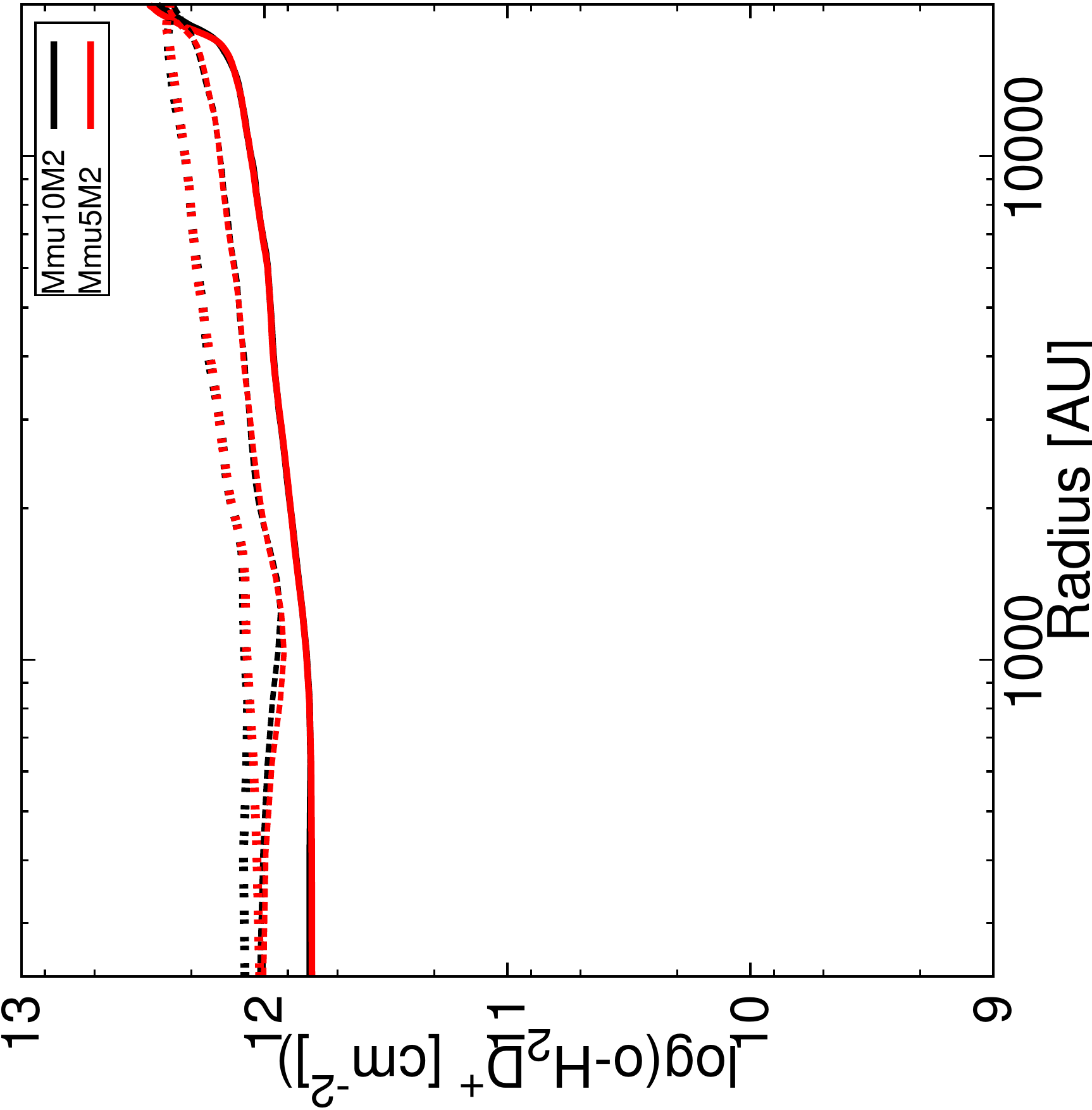}&\includegraphics[width=0.26\textwidth,angle=-90]{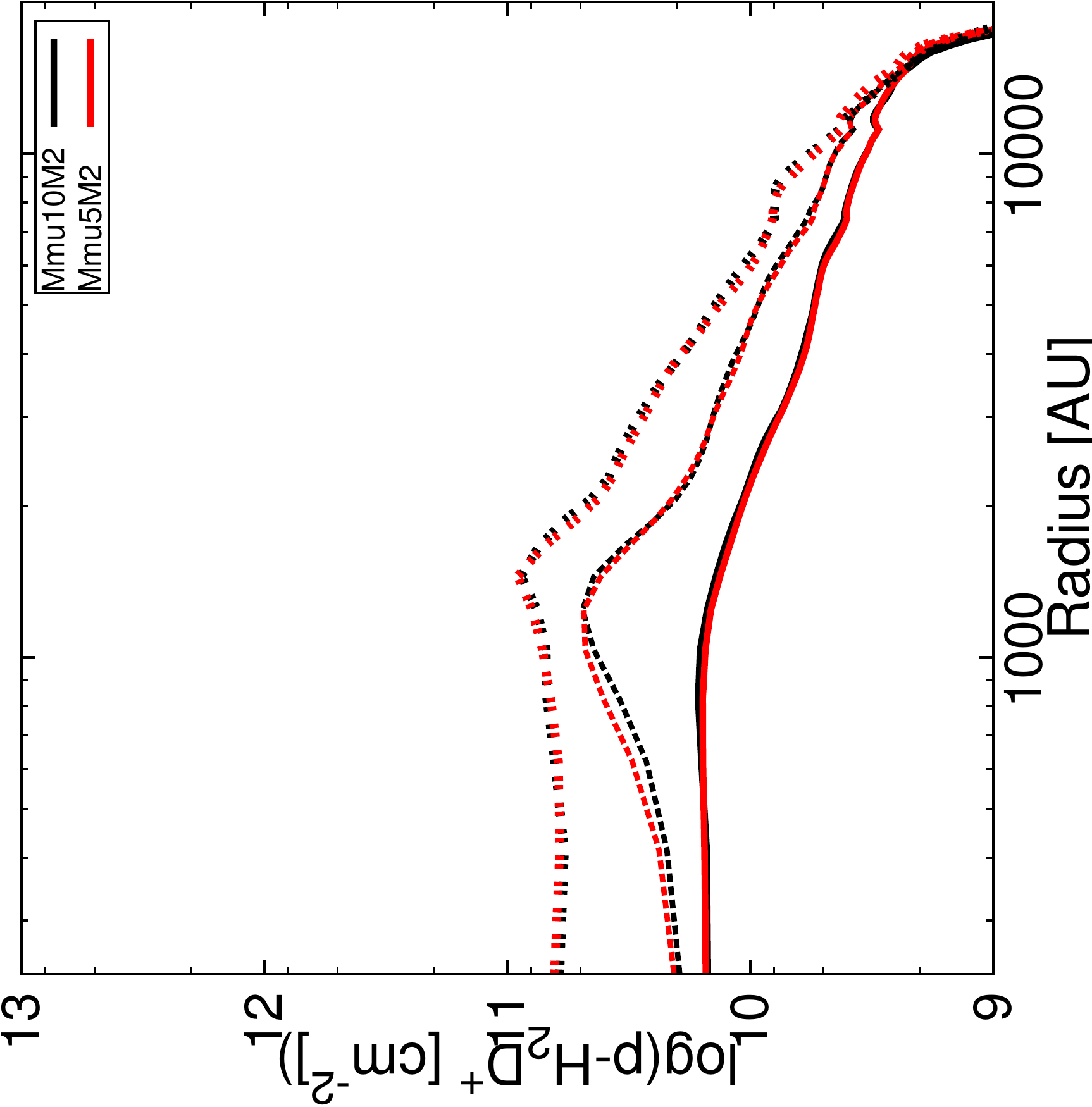}\\
		\rotatebox[origin=r]{90}{$\Sigma=0.39\,\mathrm{g\, cm^{-2}}$\qquad\qquad}&\includegraphics[width=0.26\textwidth,angle=-90]{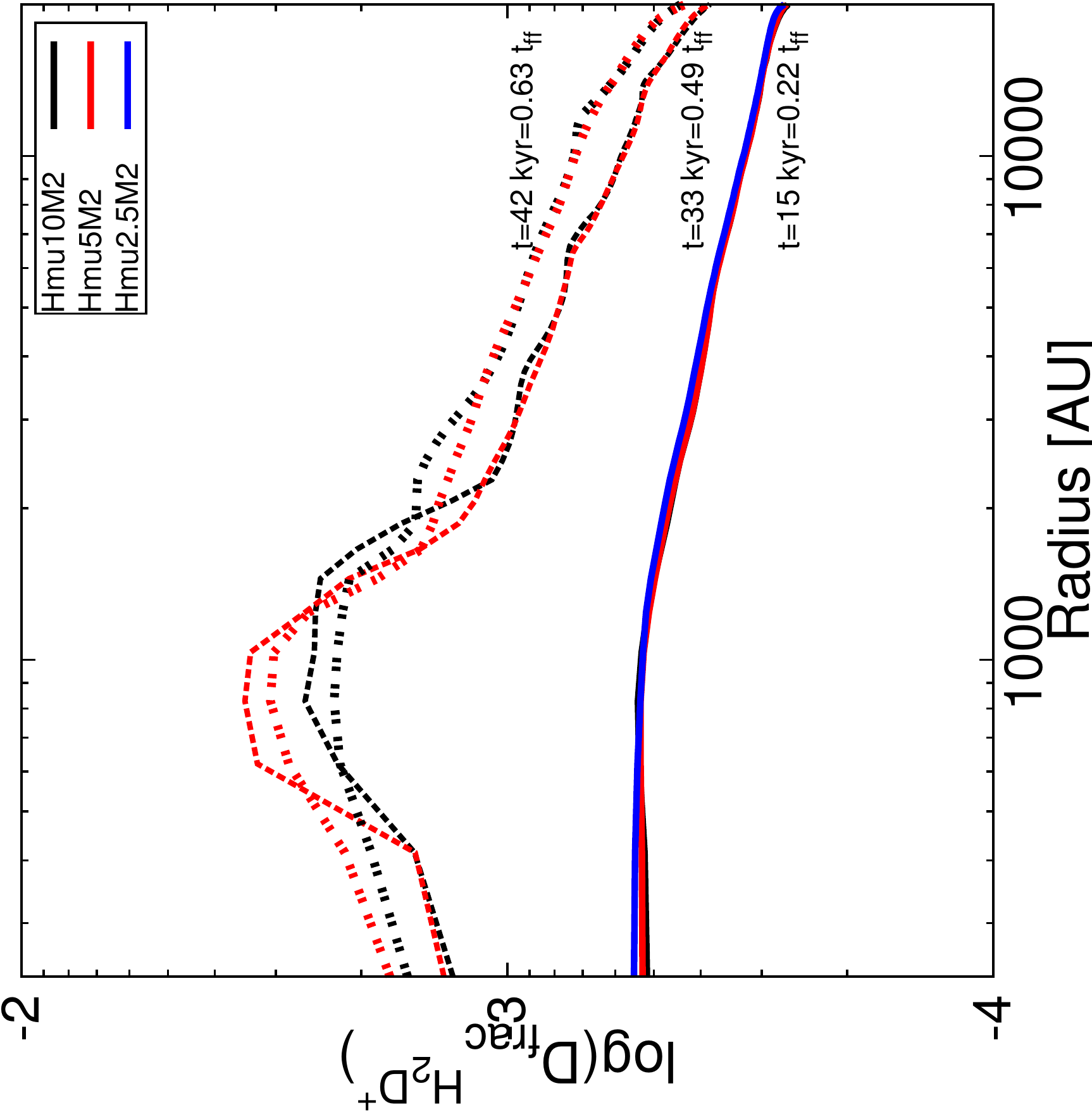}&\includegraphics[width=0.26\textwidth,angle=-90]{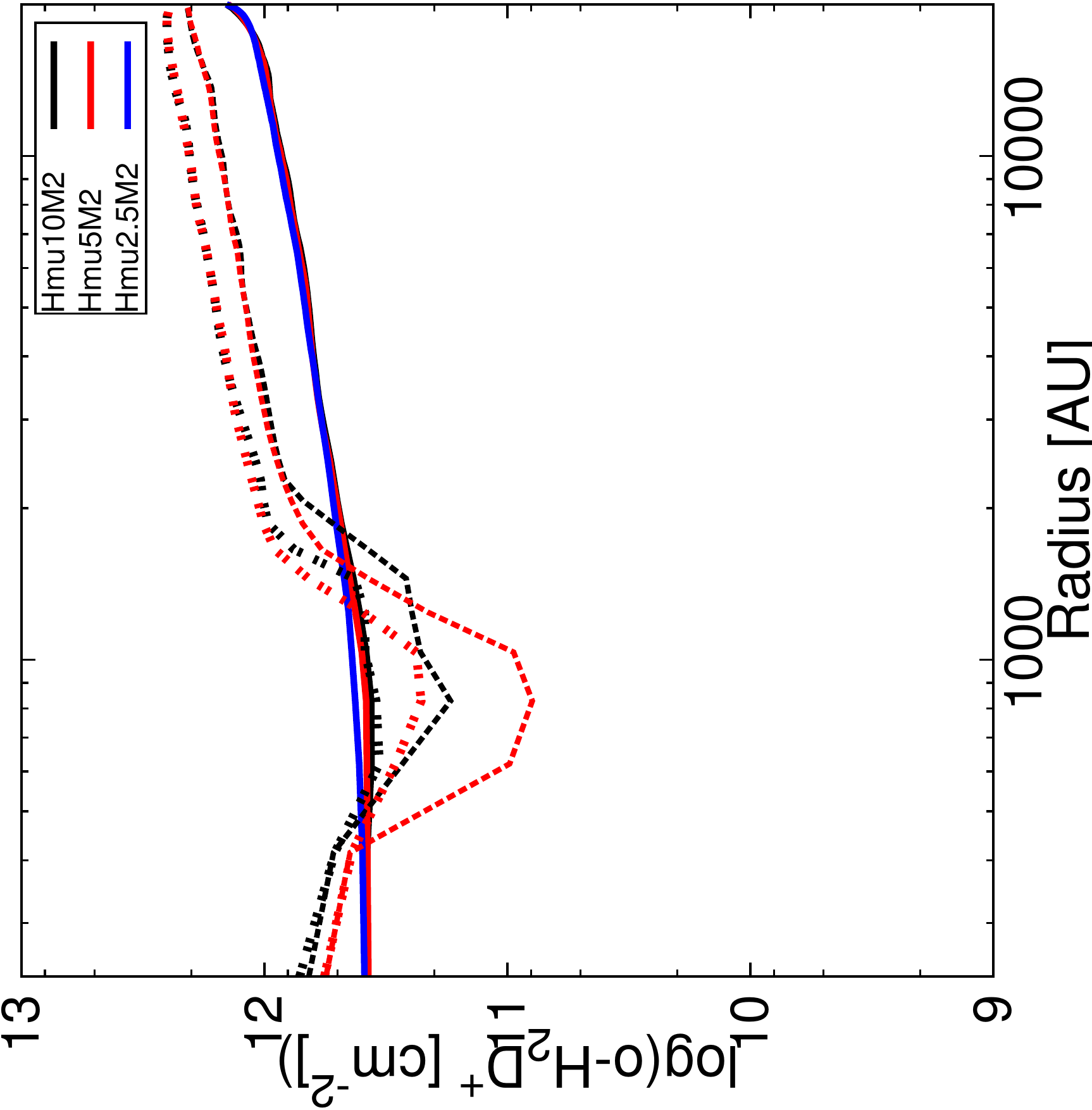}&\includegraphics[width=0.26\textwidth,angle=-90]{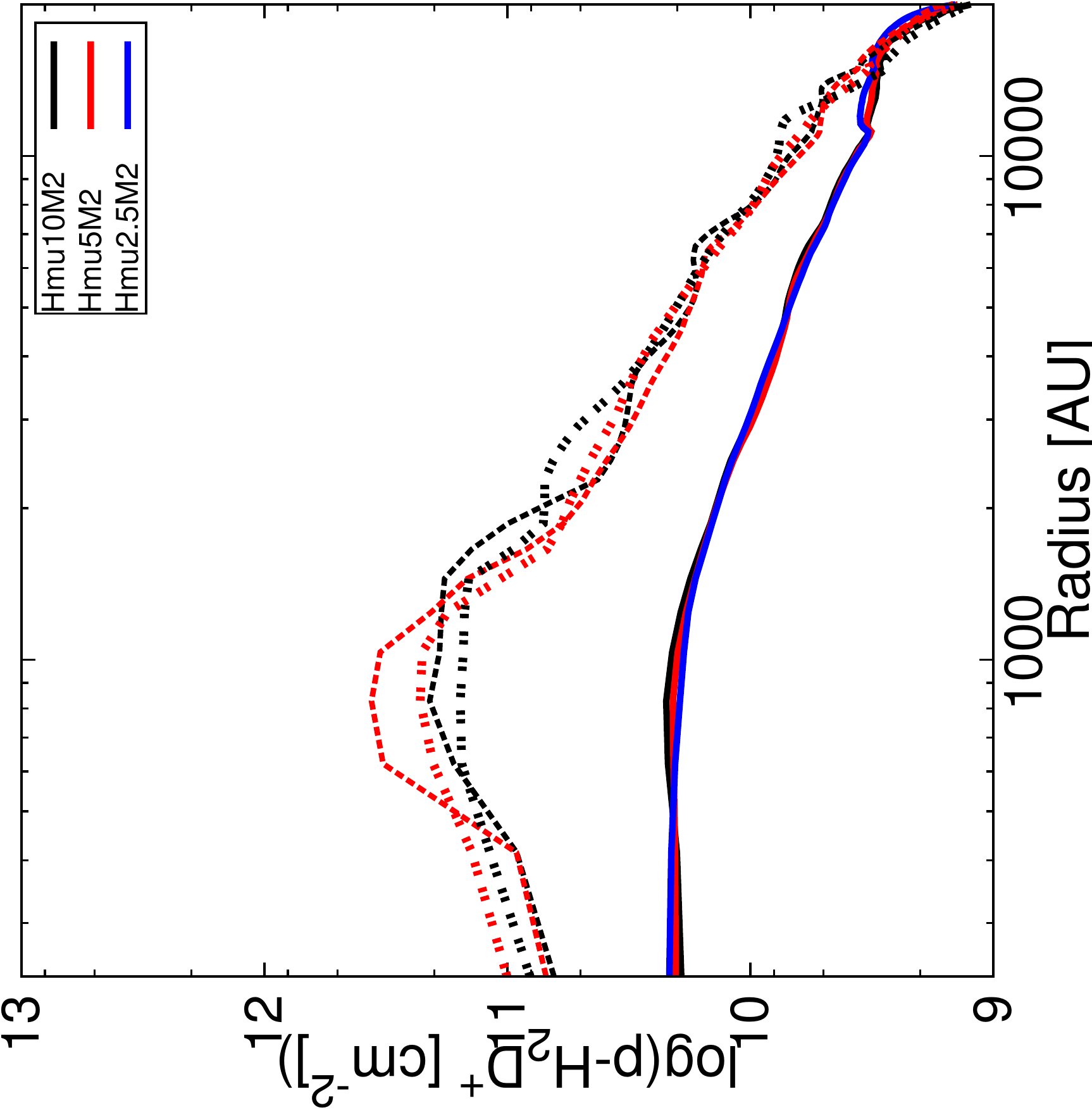}
		\end{tabular}
	\end{center}
	\caption{Column density weighted radial profiles of the deuterium fraction (left) and the spin states of H$_2$D$^+$ (middle,right) for different times and different initial mass--to--flux ratios. Radius is estimated with respect to the center of mass of the core. Note that changes in the radial distribution do not occur for times $t<50\,\mathrm{kyr}$. However, for evolved states the difference in the deuterium fraction is small. We further note that in the low and high surface density cases the runs with $\mu/\mu_\mathrm{crit}=\left(2.5,5\right)$ could not be evolved much farther due to very small 
	timesteps. The lower deuterium fraction for run Lmu5M2 compared to Lmu10M2 is the result of the suppression of accumulation of gas by the magnetic field.}\label{Bprofile}
\end{figure*}

Another central parameter to explore is the magnetic field strength in the core, which we parameterize here through the mass--to--flux ratio $\mu/\mu_{\rm crit}$. For this purpose, we consider a suite of simulations in which we vary $\mu/\mu_{\rm crit}$ as well as the gas surface density. Specifically, for each of the runs discussed in the previous subsection, we adopt values for $\mu/\mu_{\rm crit}$ of 10, 5 and 2.5, respectively (i.e. increasing the magnetic field strength by roughly a factor of 4). To investigate the results, we again create column density weighted radial profiles of the deuterium fraction as well as the column densities of \mbox{o--H$_2$D$^+$} and \mbox{p--H$_2$D$^+$}. 
As we are here predominantly interested in comparing the effect of the magnetic field in runs with otherwise the same properties, the radial profiles are shown at timescales that roughly reflect the expected evolutionary times of the core, with 30~kyrs, 56~kyrs and 105~kyrs for the low surface density runs ($0.14$~g~cm$^{-2}$), 20~kyrs, 30~kyrs and 40~kyrs for the intermediate surface density runs ($0.24$~g~cm$^{-2}$), and 15~kyrs, 33~kyrs and 42~kyrs for the high surface density runs ($0.39$~g~cm$^{-2}$). The corresponding radial profiles are given in Fig.~\ref{Bprofile}.\\
It is clearly visible again that the deuterium fraction depends predominantly on time, while other variables appear as secondary parameters. This is particularly prominent in the intermediate surface density runs, where the radial profiles shown at the same time are close to identical for the deuterium fraction as well as the column densities of \mbox{o--H$_2$D$^+$} and \mbox{p--H$_2$D$^+$}, where the increase particularly in the center appears roughly proportional to the amount of time 
passed from the beginning of the simulation. On large scales, the deuterium fraction increases with decreasing radius towards 1,000~au, where then some flattening occurs, sometimes accompanied by a weak drop in the deuterium fraction. This behavior may in particular reflect the density distribution within the core, as material tends to accumulate around 1,000~au in some simulations. \\
The dependence on time is also clearly visible in the high surface density runs, even though the outputs at 33 and 42~kyrs are here more similar, presumably due to the smaller differences in time. While the output at 15~kyrs shows almost no dependence on  $\mu/\mu_{\rm crit}$, there is a difference at later times, with a peak in the deuterium fraction that decreases with the mass--to--flux ratio, and a dip in the \mbox{o--H$_2$D$^+$} column density, corresponding to the characteristic feature that we discussed before, reflecting the enhanced effective virial parameter in this simulation. The displaced peak in deuteration and the dip in \mbox{o--H$_2$D$^+$} are likely an effect of inhomogeneities within the density distribution, which introduce dense gas with an enhanced deuterium fraction, while the average column densities are still low. It is conceivable that the stronger magnetic field here contributes both to the transport of angular momentum and damping of the turbulence, thus driving the mass efficiently towards the center and also reducing the virial parameter. For the column density of \mbox{p--H$_2$D$^+$}, on the other hand, we note that all lines basically overlap at 15~kyrs. The column density is also increasing with time, even though the increase becomes modest between 32 and 42~kyrs, so that these lines essentially overlap within the scatter. The increase in the chemical abundances appears to be approximately compensated by the decrease in the overall mass.\\
For the low surface density runs, finally, a clear dependence on time can be recognized in the deuterium fraction, where the mass--to--flux ratio appears to play little role at 30 and 56~kyrs, while at 106~kyrs, the run with the stronger field has a somewhat reduced deuterium fraction in the center by about a factor of 2. For the same snapshot, the column density of \mbox{o--H$_2$D$^+$} is slightly higher around 1,000~au, implying that there is still somewhat more mass, and in particular for these lower--density cores, the combined impact of a stronger magnetic field with already existing turbulent support may help to delay the collapse. For the \mbox{p--H$_2$D$^+$}, the dependence on time can also be recognized, even though the simulation with the stronger magnetic field at 106~kyrs overlaps with the radial profiles at 56~kyrs. As a result, we can in particular deduce that the relation between time and the deuterium fraction is not monothonic, so there can be degeneracies in the interpretation. It appears as quite natural that these differences become more pronounced at late times, i.e. after the initial collapse, as well as for cores with higher virial parameters, which are closer to being stable and where any additional contribution makes a larger difference.

\subsubsection{Possible effect of ambipolar diffusion}
Modelling AD in numerical simulations, where the cores are subcritical or mildly supercritical, is not trivial as a series of complex processes, as for example the evaluation of the grain charge, and the momentum transfer processes should be taken into account, together with the usual modifications of the MHD equations \citep[see e.g.][]{Pinto20081,Pinto20082}. Both simplified and more complex approaches \citep{Kortgen2015,Marchand2016} have been proposed, though implying strong assumptions.
Following \citet{MacLow04} the AD timescale is
\beq
\begin{split}
\tau_\mathrm{AD}=25\,\mathrm{Myr}\times\left(\frac{3\,\mu\mathrm{G}}{B}\right)^2\times\left(\frac{n_\mathrm{n}}{100\,\mathrm{cm}^{-3}}\right)^2 \\
\times\left(\frac{R}{1\,\mathrm{pc}}\right)^2\times\left(\frac{x_\mathrm{ion}}{10^{-6}}\right)
\end{split}
\eeq 
with $B$ being the magnetic field strength in $\mu\mathrm{G}$, $n_\mathrm{n}$ being the neutral gas density, $R$ the core radius, and $x_\mathrm{ion}$ being the 
ionization fraction of the gas. For highly magnetized, small--sized cores with a sufficiently small ionization fraction, the AD timescale can be shorter than the core's free--fall time (the magnitude of the AD time also implicitly depends on e.g. the cosmic--ray rate). As the cores are turbulent, individual regions within the core can be prone to AD, while others are dominated by free--fall \citep{Heitsch04}.  For the 
typical high--density parts in our simulations, AD may indeed be the relevant physical process 
governing the evolution of the core. However, AD also changes the thermodynamics of the gas through additional heating \citep[e.g.][]{Schleicher09}, which impacts 
the chemistry in certain regions. To which extent AD will modify the dynamical and chemical evolution of the cores will be investigated in a future study.

\subsection{Effects of variation in the slope of the magnetic field scaling}
\begin{figure*}
	\begin{center}
		\begin{tabular}{ccc}
		D$_\mathrm{frac}^{\mathrm{H_2D^+}}$	&o--H$_2$D$^+$&p--H$_2$D$^+$\\
		\includegraphics[width=0.26\textwidth,angle=-90]{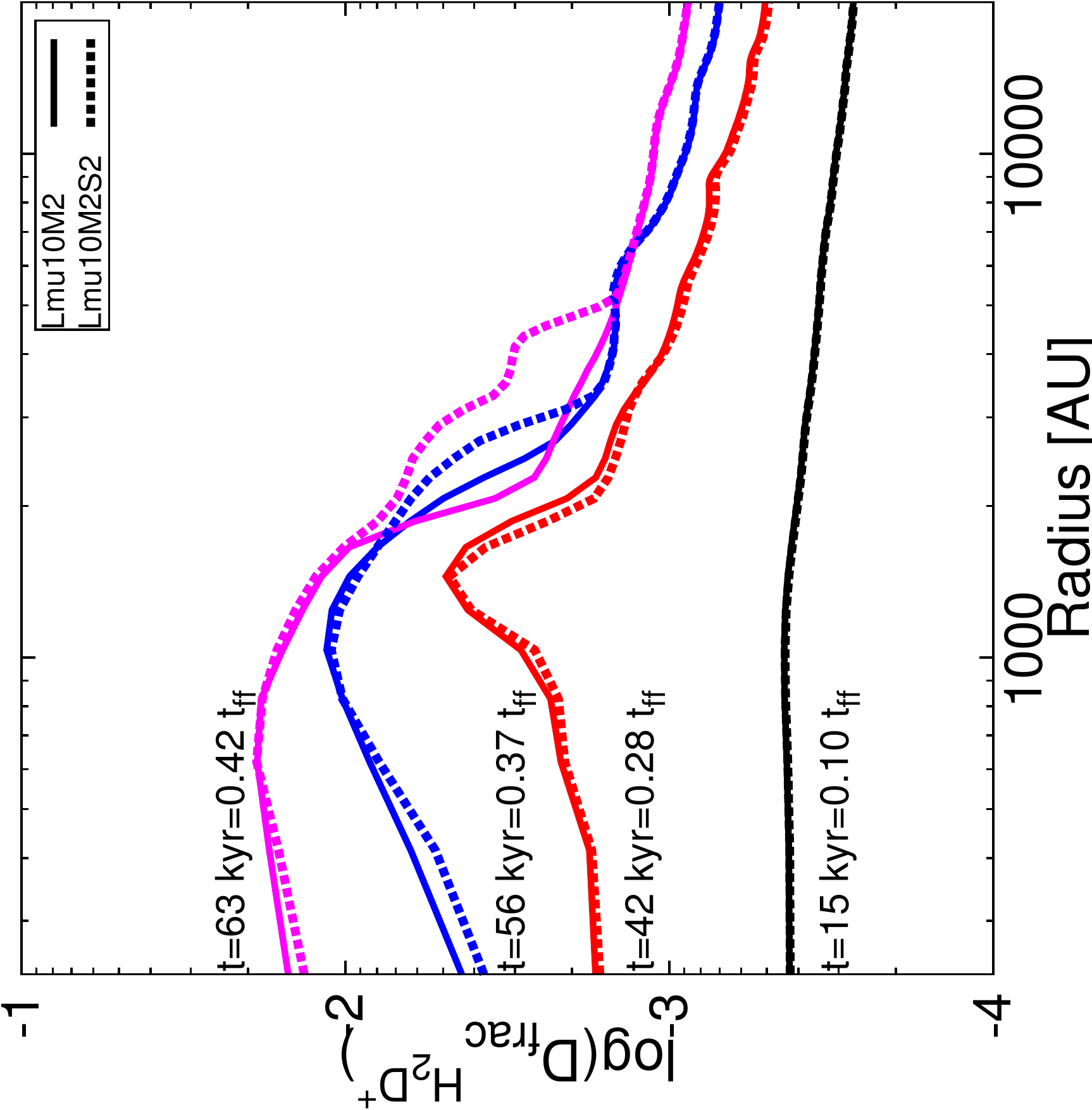}&\includegraphics[width=0.26\textwidth,angle=-90]{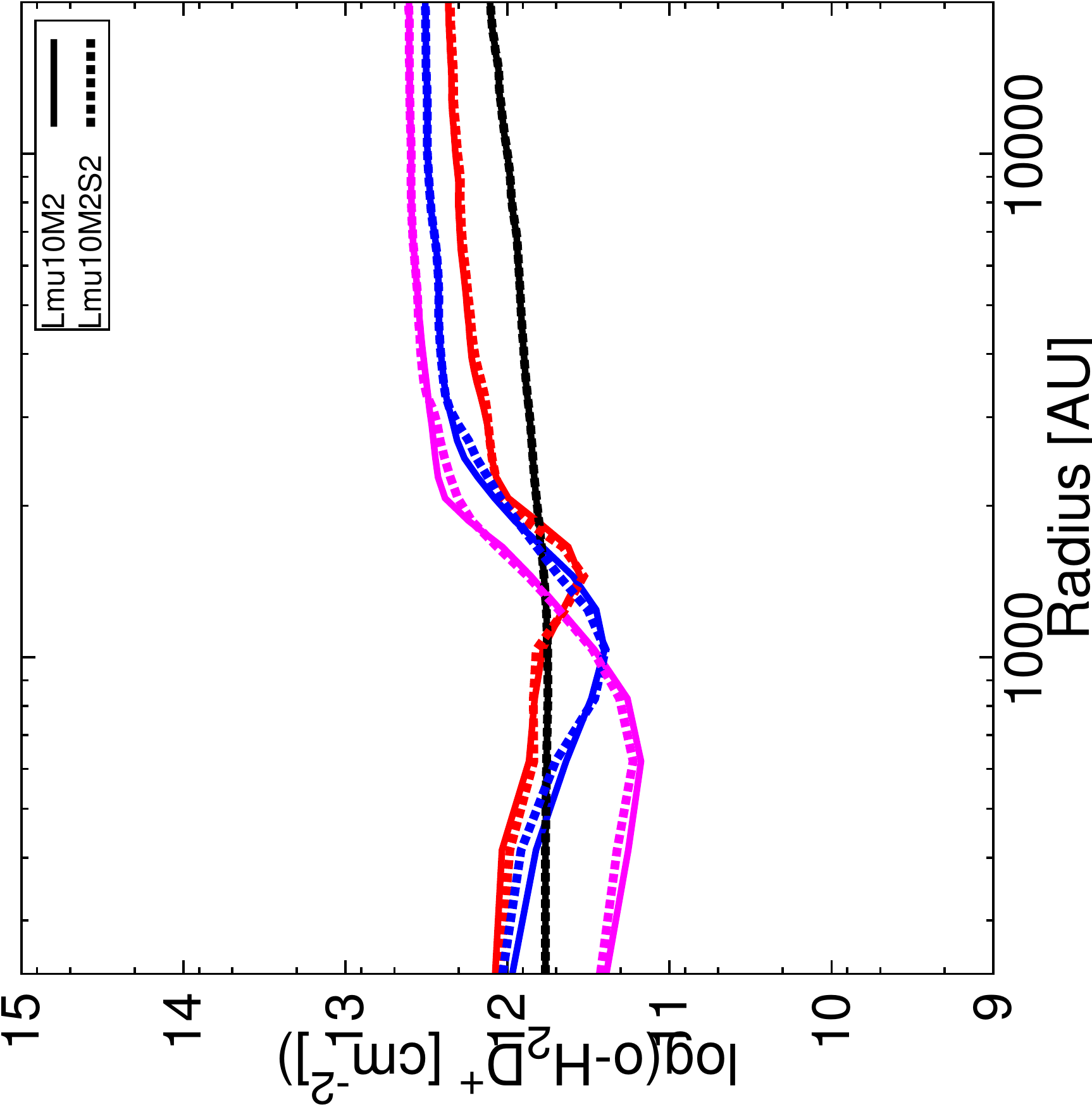}&\includegraphics[width=0.26\textwidth,angle=-90]{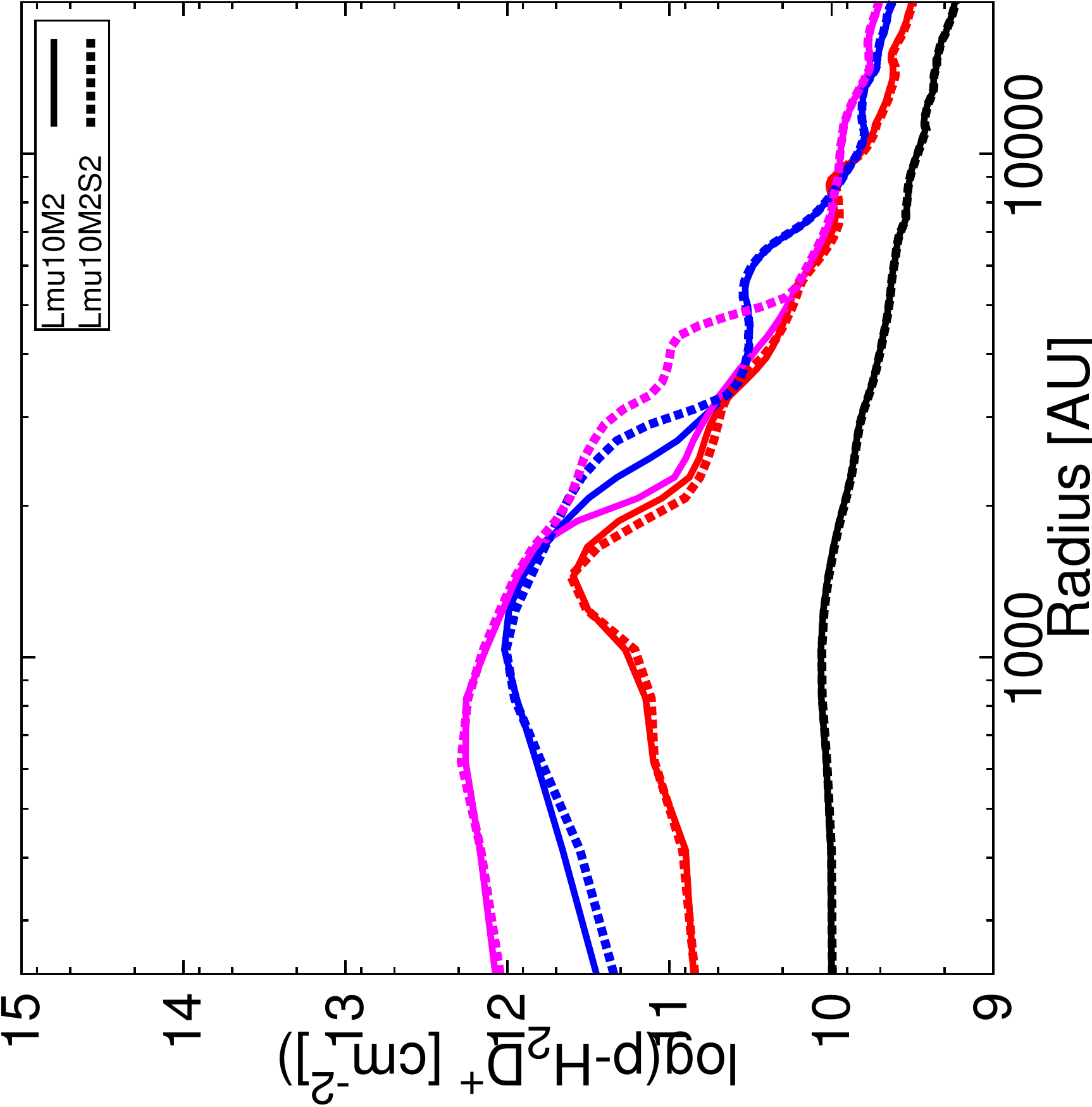}\\
		\end{tabular}
	\end{center}
	\caption{Varying the slope: Evolution of radial profiles of the deuteration ratio, as well as of the spin states of H$_2$D$^+$ for runs Lmu10M2 and Lmu10M2S2. Evolution is only shown for times before sink particle formation due to the bias induced by the accreting sink. It is 
	evident that the slope of the magnetic field does not impact the evolution of the chemical species too much. This is due to the core being highly supercritical. The magnetic field does not contribute significantly to the support against gravity. Differences in 
	the deuterium fraction are $\lesssim 5$.}\label{slope}
\end{figure*}

In addition to the mass--to--flux ratio of the cores, which determines the average magnetic field strength for a given core size and mass, also the radial distribution of the magnetic field can be relevant for the dynamics. In the simulations we considered so far, the adopted slope was $\kappa=0.5$, while in the following, we will for comparison investigate the impact of $\kappa=2$. Apart from the difference in this slope, the parameters in both simulations are the same as for Lmu10M2. In Fig.~\ref{slope}, we again plot column density averaged radial profiles of the deuterium fraction, and the \mbox{o--H$_2$D$^+$} and \mbox{p--H$_2$D$^+$} column density, at 15, 42, 56 and 63~kyrs, respectively. 
In spite of the large difference in the slope, it is clearly recognizable that the latter has a minor impact on the dynamical evolution, while instead a clear dependence on time can be established. In particular the deuterium fraction as well as the \mbox{p--H$_2$D$^+$} column density clearly increase with time, with a particularly pronounced increase within the central 1,000~au, where the density is higher. The variability of the \mbox{o--H$_2$D$^+$} column density, on the other hand, is weaker both in time and space, with only a small increase towards the center due to the overall lower mass. Even in this case, it can however be clearly recognized that the dependence on the slope is rather weak, which indicates that the mass--to--flux ratio is the governing quantity.

\subsection{Varying the initial ortho--to--para ratio of H$_2$}\label{initialop}
\begin{figure*}
	\begin{center}
		\begin{tabular}{ccc}
			\includegraphics[scale=0.26,angle=-90]{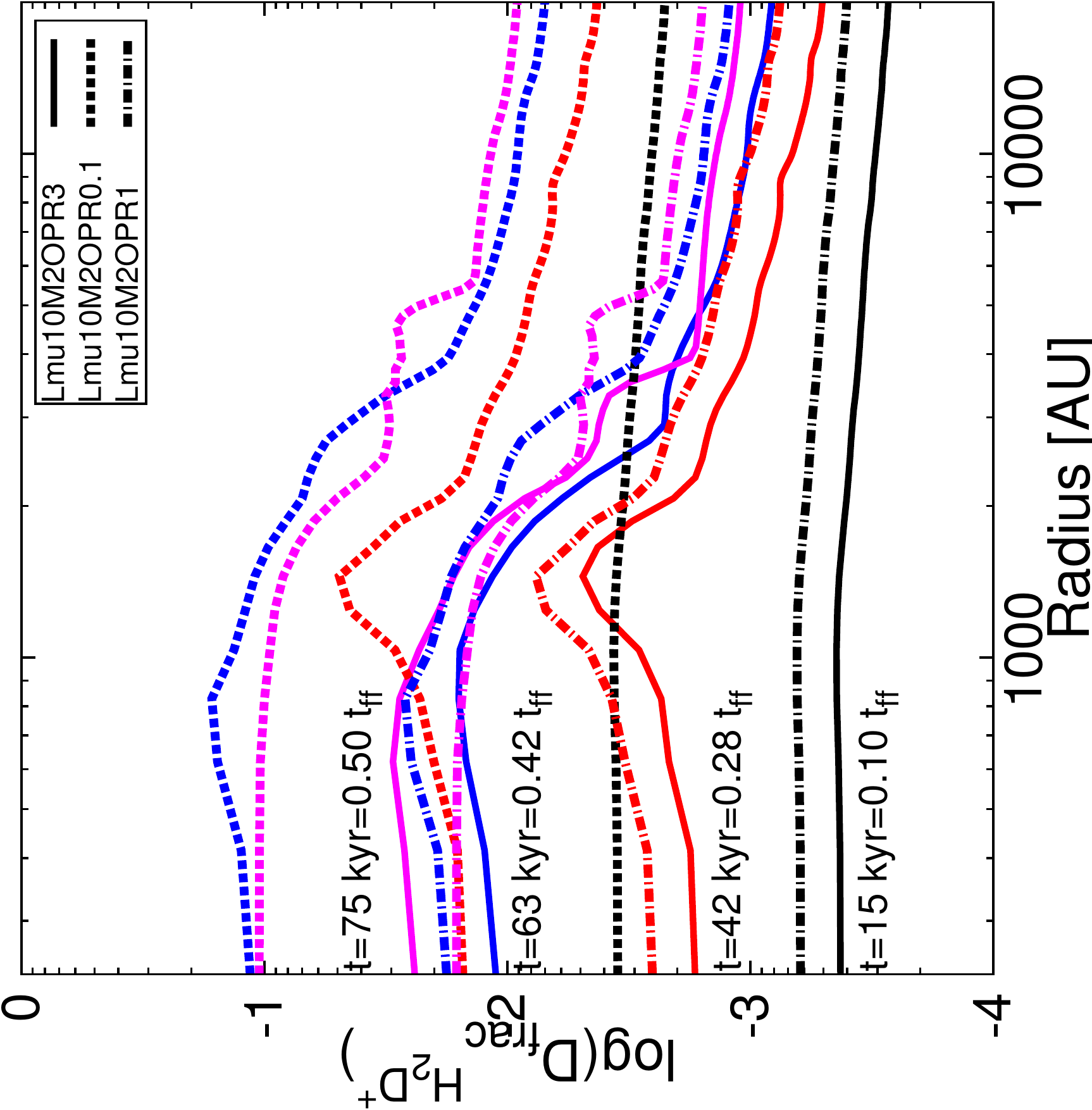}&\includegraphics[scale=0.26,angle=-90]{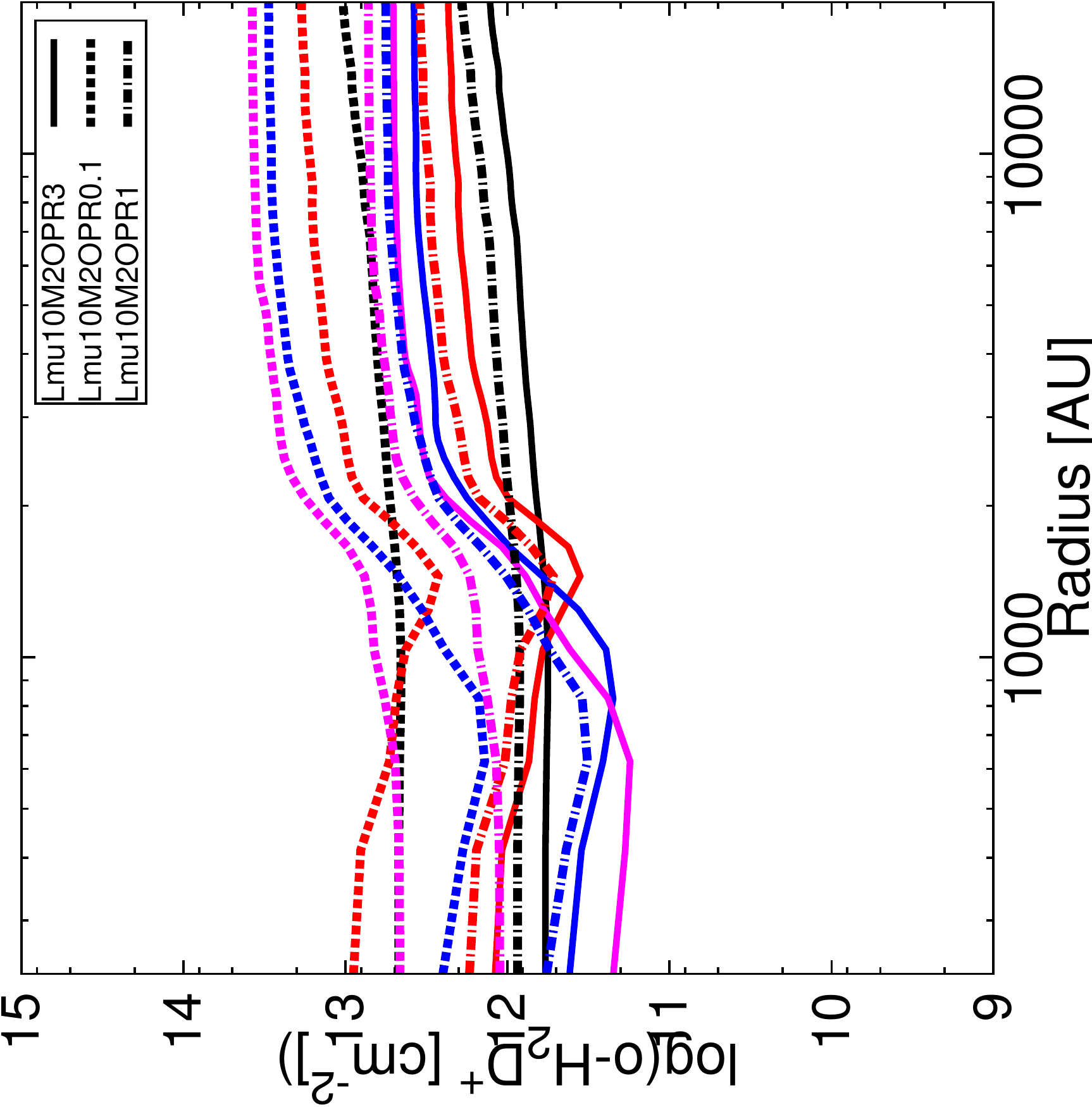}&\includegraphics[scale=0.26,angle=-90]{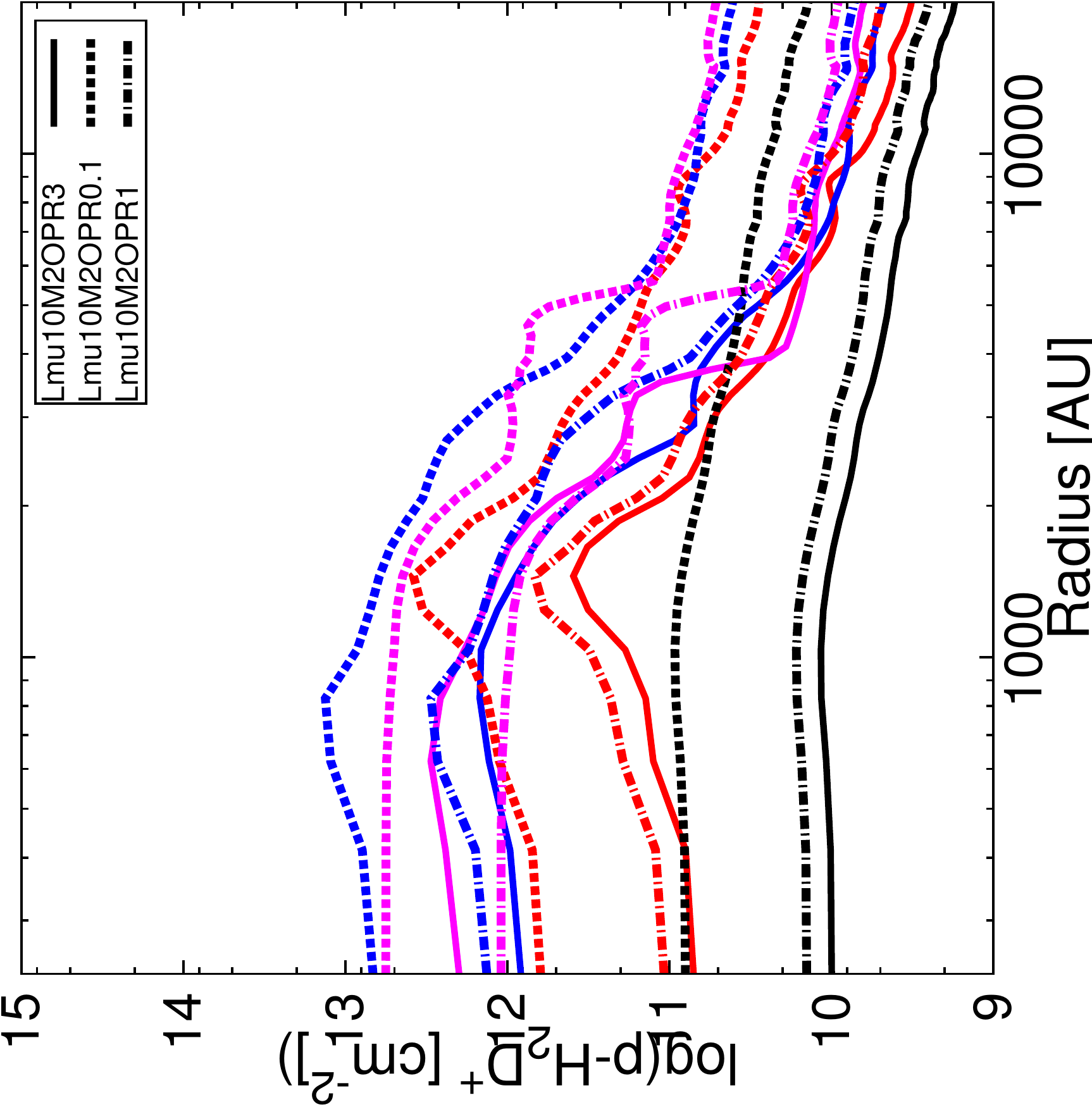}\\
		\end{tabular}
	\end{center}
	\caption{Comparison of radial profiles for runs with initial OPR=3 (solid), OPR=1 (dash--dotted) and OPR=0.1 (dashed). Note that a sink particle has formed between $t=63\,\mathrm{kyr}$ and $t=75\,\mathrm{kyr}$.}\label{ortho}
\end{figure*}

In addition to the dynamical quantities, on which we were focusing so far, the chemical initial conditions are quite relevant to determine how fast deuteration is going to occur \citep[see e.g.][]{Kong2015}. While we assume full depletion of the metals throughout the paper, another important parameter is the initial OPR of H$_2$, which we explore in the following. \\
As already shown by previous works \citep[e.g.][]{Kong2015}, the effect of reducing the H$_2$ OPR is to decrease the time needed to reach high $D_\mathrm{frac}$. While in the simulations we discussed so far, we adopted a generic OPR of 3, i.e. the least favorable case, we explore here the impact of OPR of 1 and 0.1, thus including values inferred by observations \citep{Troscompt2009}. The parameters in the simulation are otherwise the same as for Lmu10M2. The column density weighted radial profiles of the deuterium fraction and the \mbox{o--H$_2$D$^+$} and \mbox{p--H$_2$D$^+$} column densities are given in Fig.~\ref{ortho} at 15, 42, 63 and 75~kyrs. 
Also here, the evolution depends mostly on time, even though some distinct impact of the initial OPR can be recognized. While an OPR of 3 and 1 yields relatively similar results, with the second yielding a slightly enhanced deuterium fraction, the deuterium fraction is increased by almost an order of magnitude for an OPR of 0.1. The latter has a significant impact and can lead to a considerable speed--up of the deuteration process. 
Once considering this uncertainty, the interpretation of the deuterium fraction thus becomes degenerate between time and the OPR. It will therefore be important to determine the initial ratio with higher accuracy in the future with, for instance, measurements of the H$_2$D$^+$ OPR which linearly correlates with the H$_2$ OPR or by performing simulations which can follow the formation of molecular clouds and the embedded cores \citep[see for instance][]{Kortgen2015} consistently with the chemistry.\\
A similar behavior can also be recognized for the column density of \mbox{o--H$_2$D$^+$}, which, as noted before, has a generally weaker dependence on position and time, but which still can be strongly boosted by about an order of magnitude for a low OPR. This quantity, due to its weak dependence on time, predominantly depends on the OPR, as was also shown by previous models \citep{Brunken2014} where a linear correlation between H$_2$--OPR and H$_2$D$^+$--OPR has been reported. For the column density of \mbox{p--H$_2$D$^+$}, on the other hand, a clear dependence on time can again be recognized, though a considerable boost is possible for very low OPRs. Also for this quantity, there is a degeneracy with the time evolution.

\subsection{Comparison of selected runs}\label{detailedcomp}
\begin{figure*}
	\begin{center}
		\begin{tabular}{cccc}
		&Average density &D$_\mathrm{frac}^{\mathrm{H_2D^+}}$ &OPR--H$_2$D$^+$\\
		\rotatebox[origin=r]{90}{Lmu10M2\qquad\qquad} &\includegraphics[width=0.22\textwidth,angle=-90]{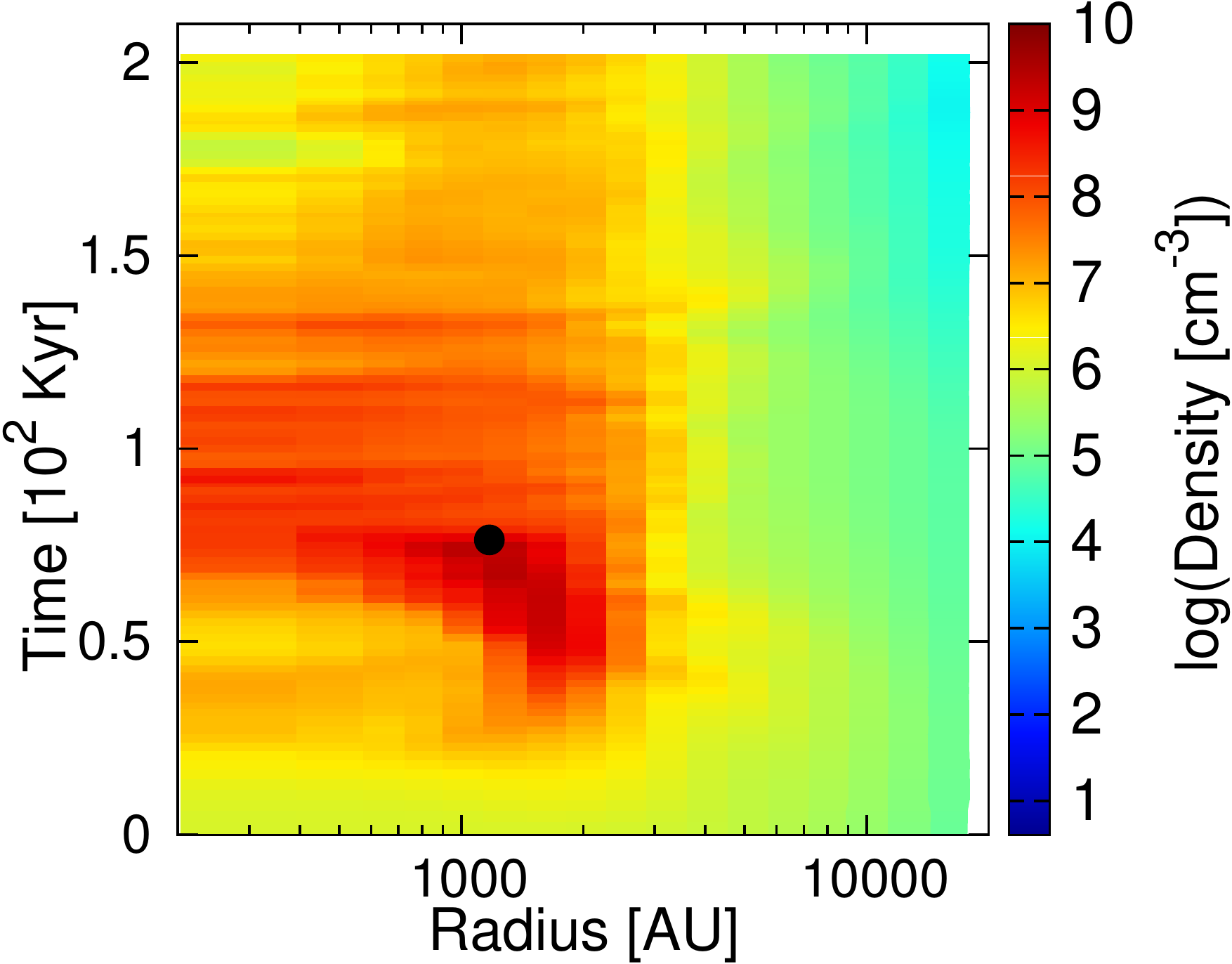} & \includegraphics[width=0.22\textwidth,angle=-90]{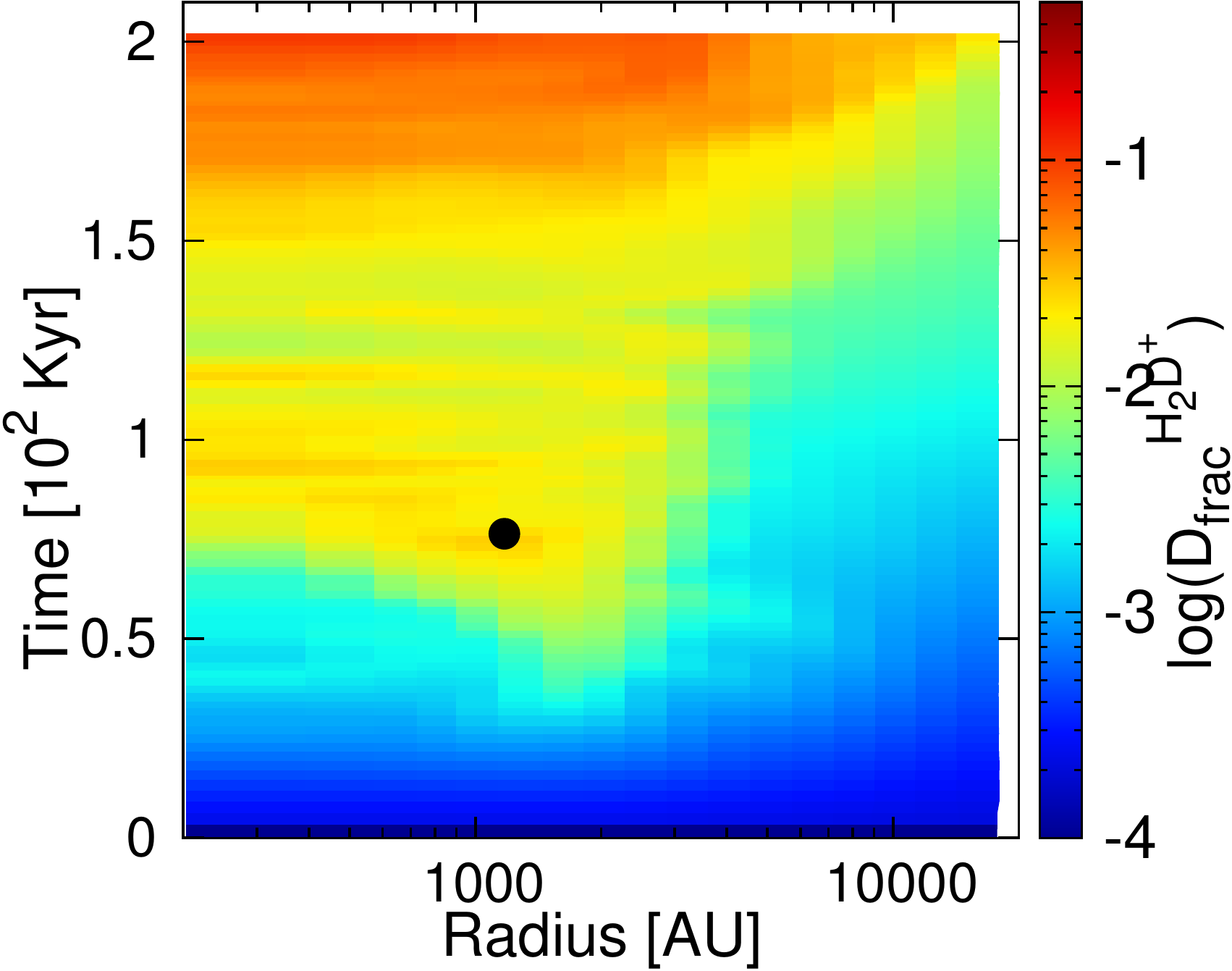}  &\includegraphics[width=0.22\textwidth,angle=-90]{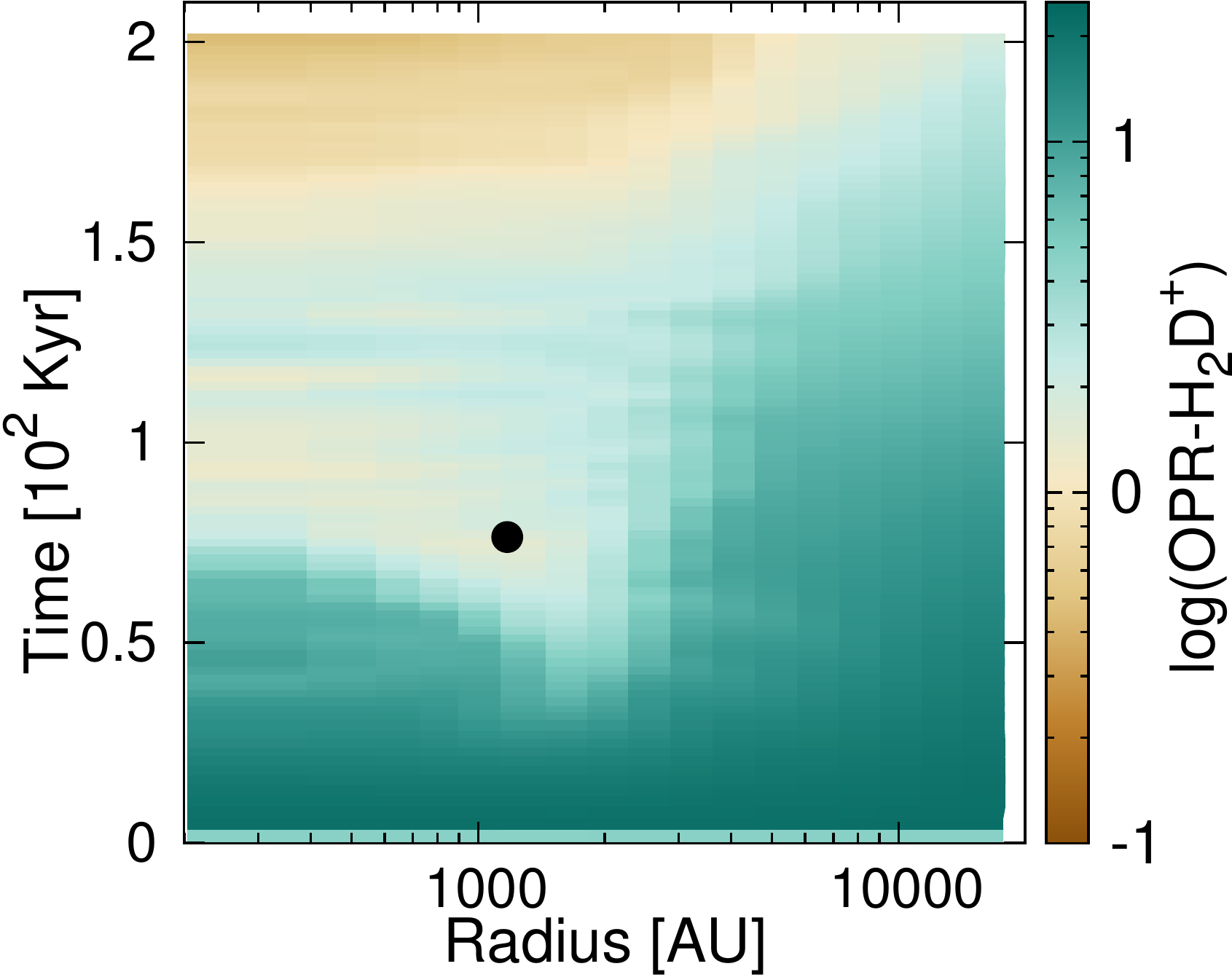} \\

		\rotatebox[origin=r]{90}{Lmu10M2S2\qquad\qquad} &\includegraphics[width=0.22\textwidth,angle=-90]{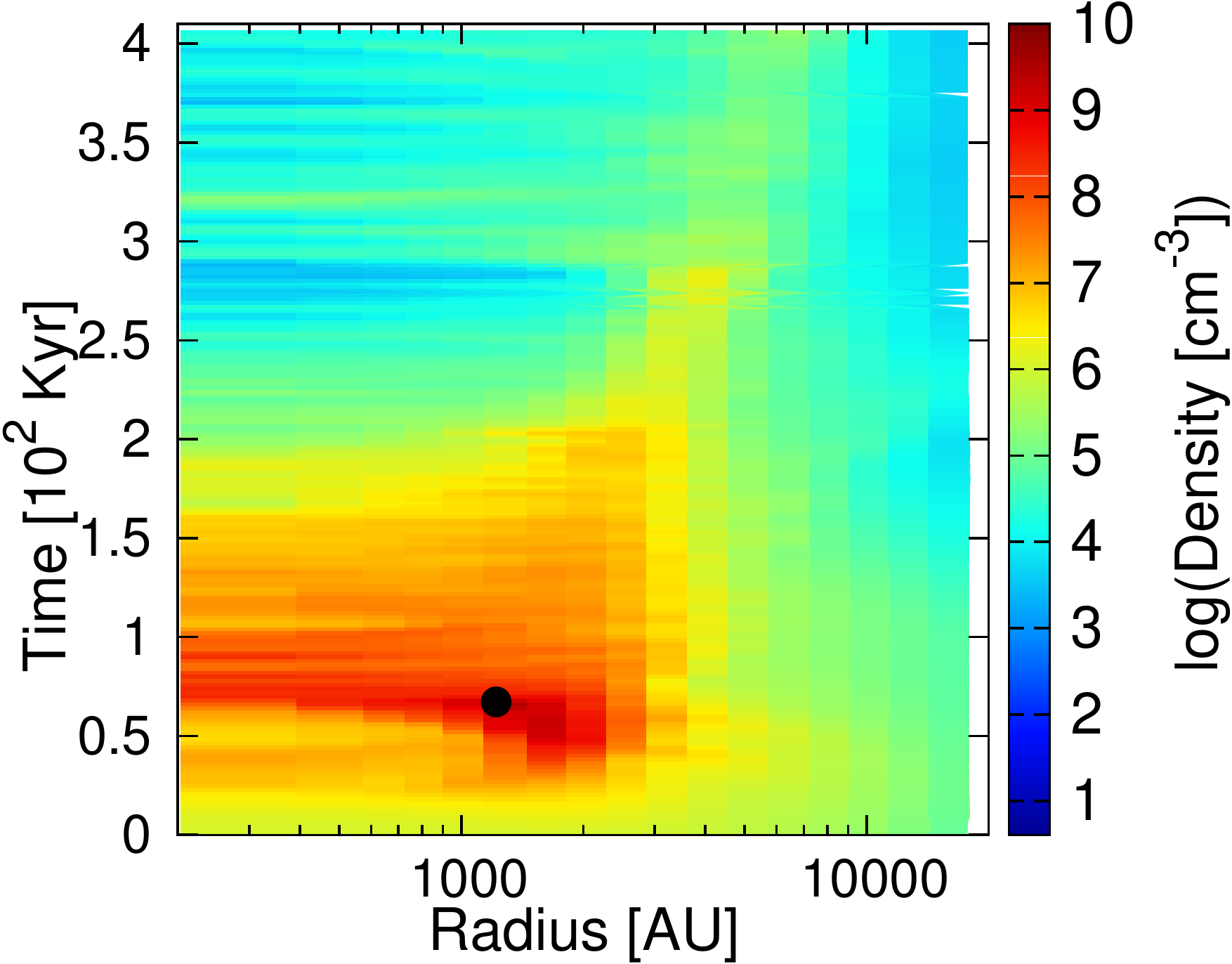} &
		\includegraphics[width=0.22\textwidth,angle=-90]{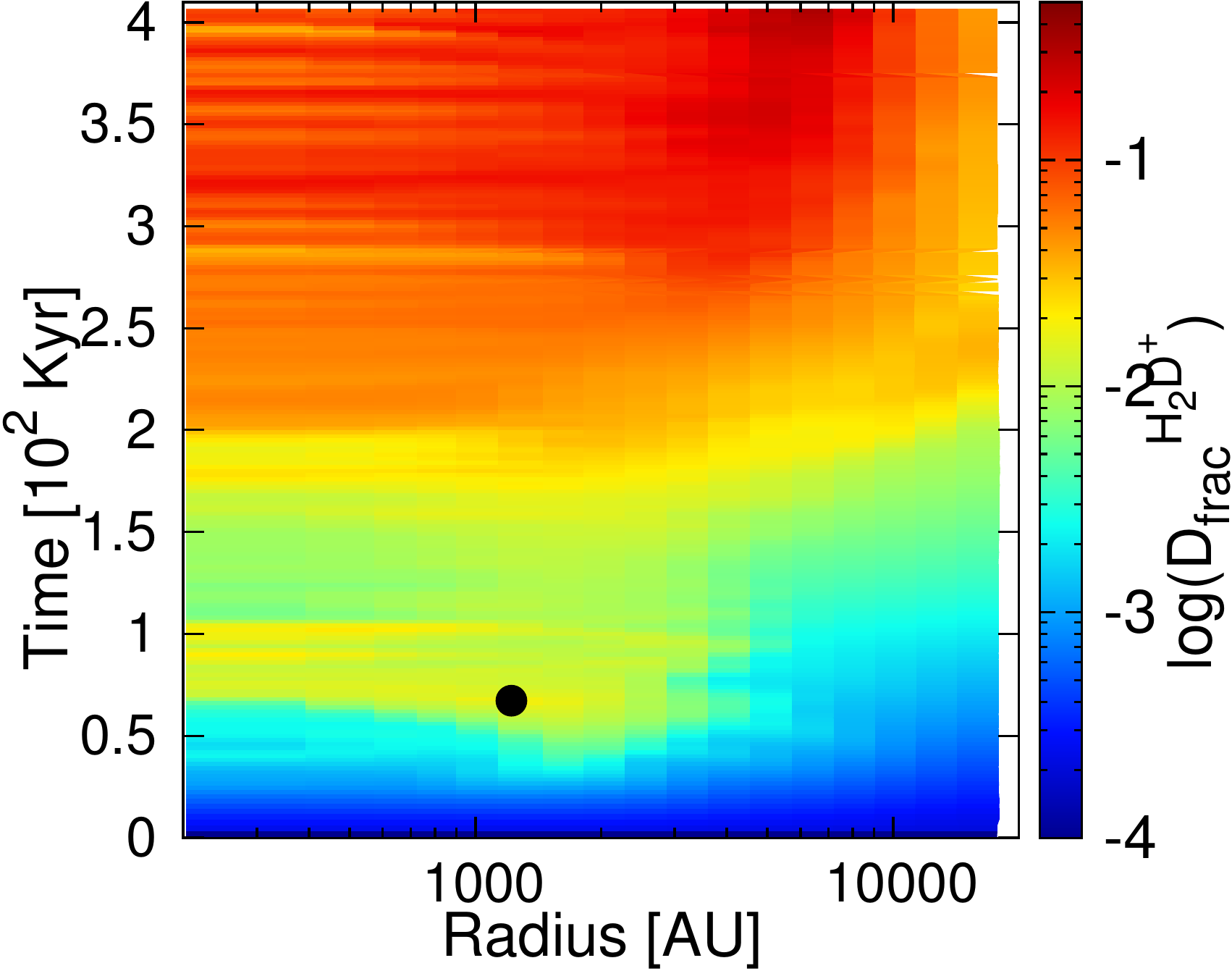} &\includegraphics[width=0.22\textwidth,angle=-90]{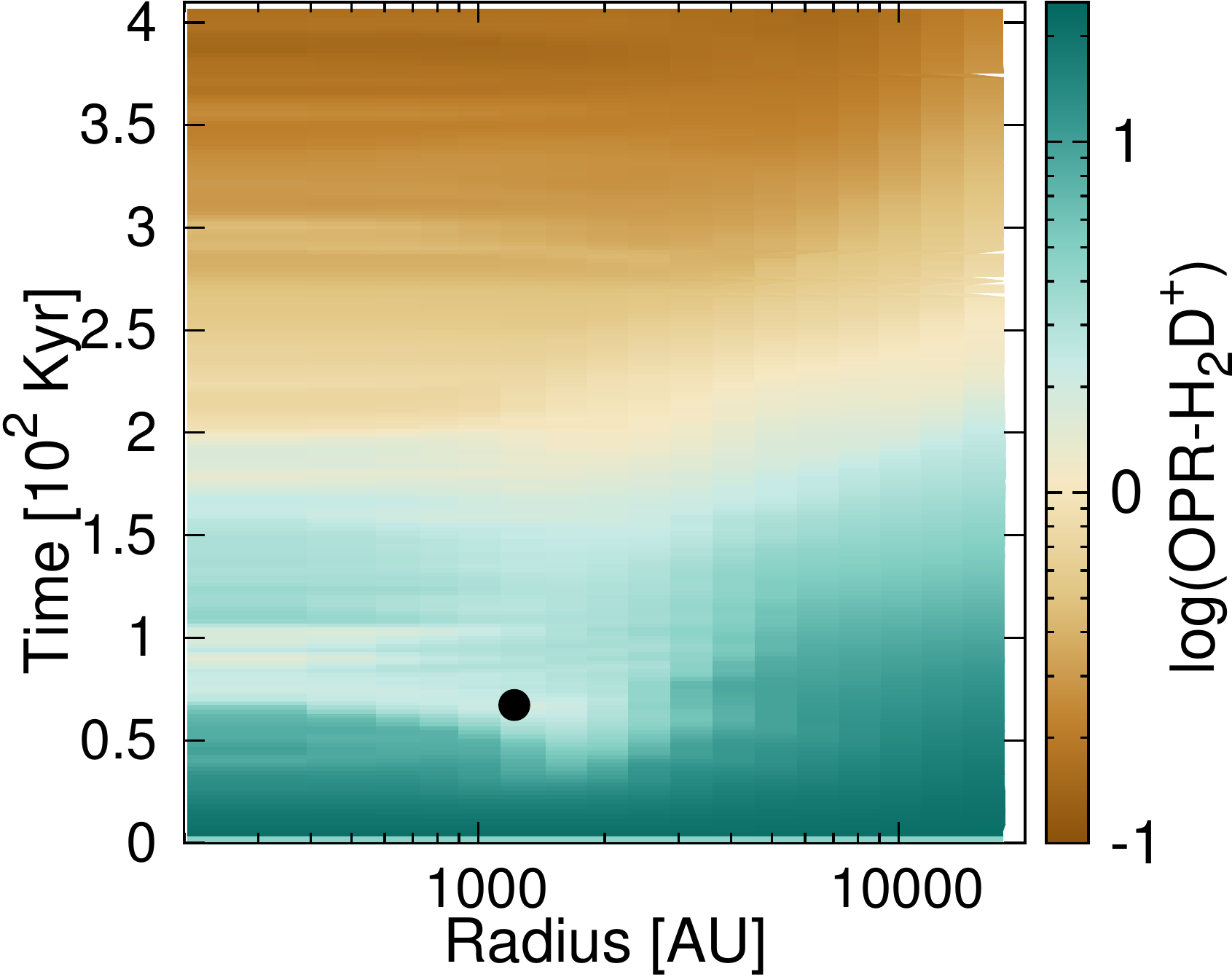} \\
		\rotatebox[origin=r]{90}{Lmu10M2OPR0.1\qquad} &\includegraphics[width=0.22\textwidth,angle=-90]{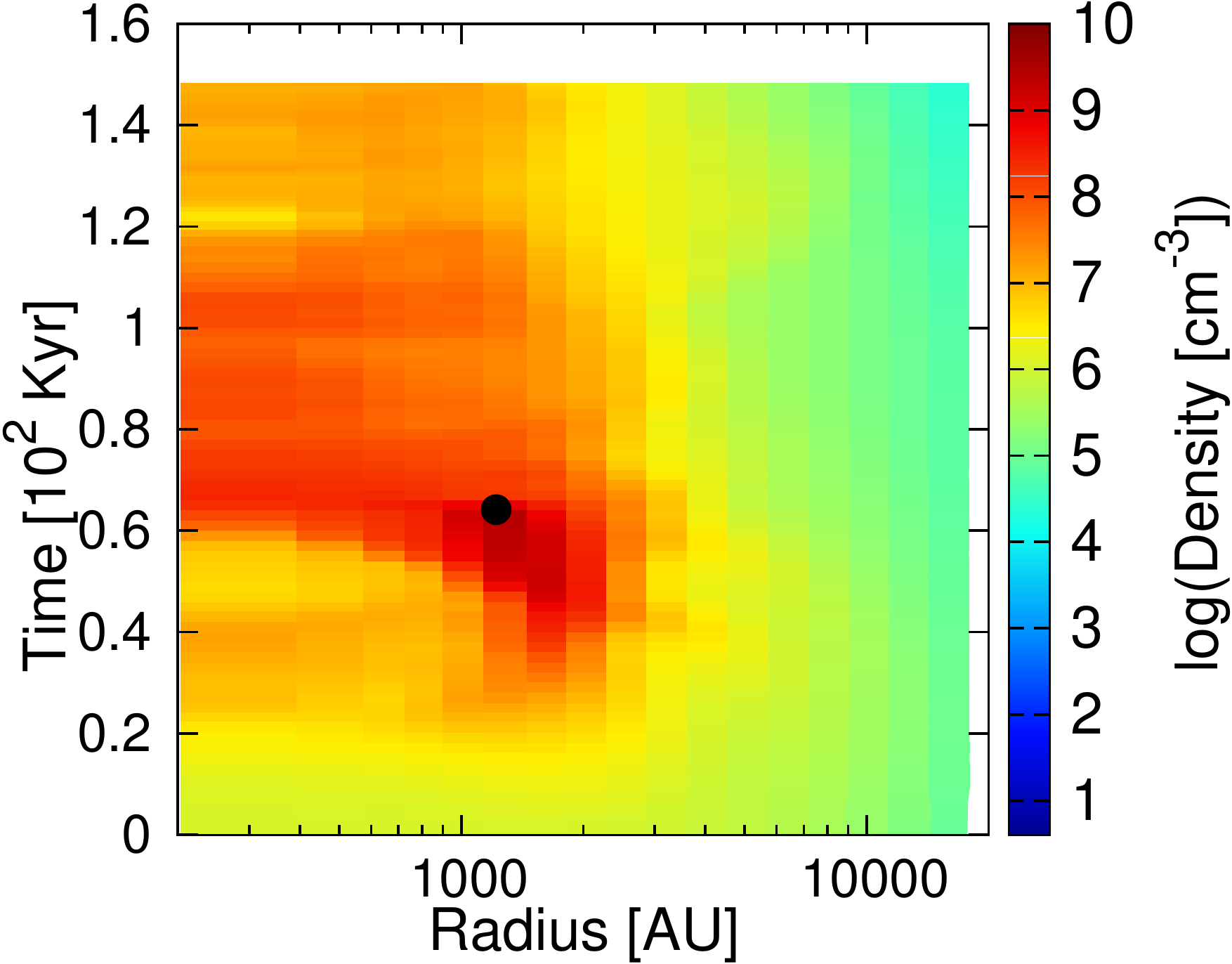} & \includegraphics[width=0.22\textwidth,angle=-90]{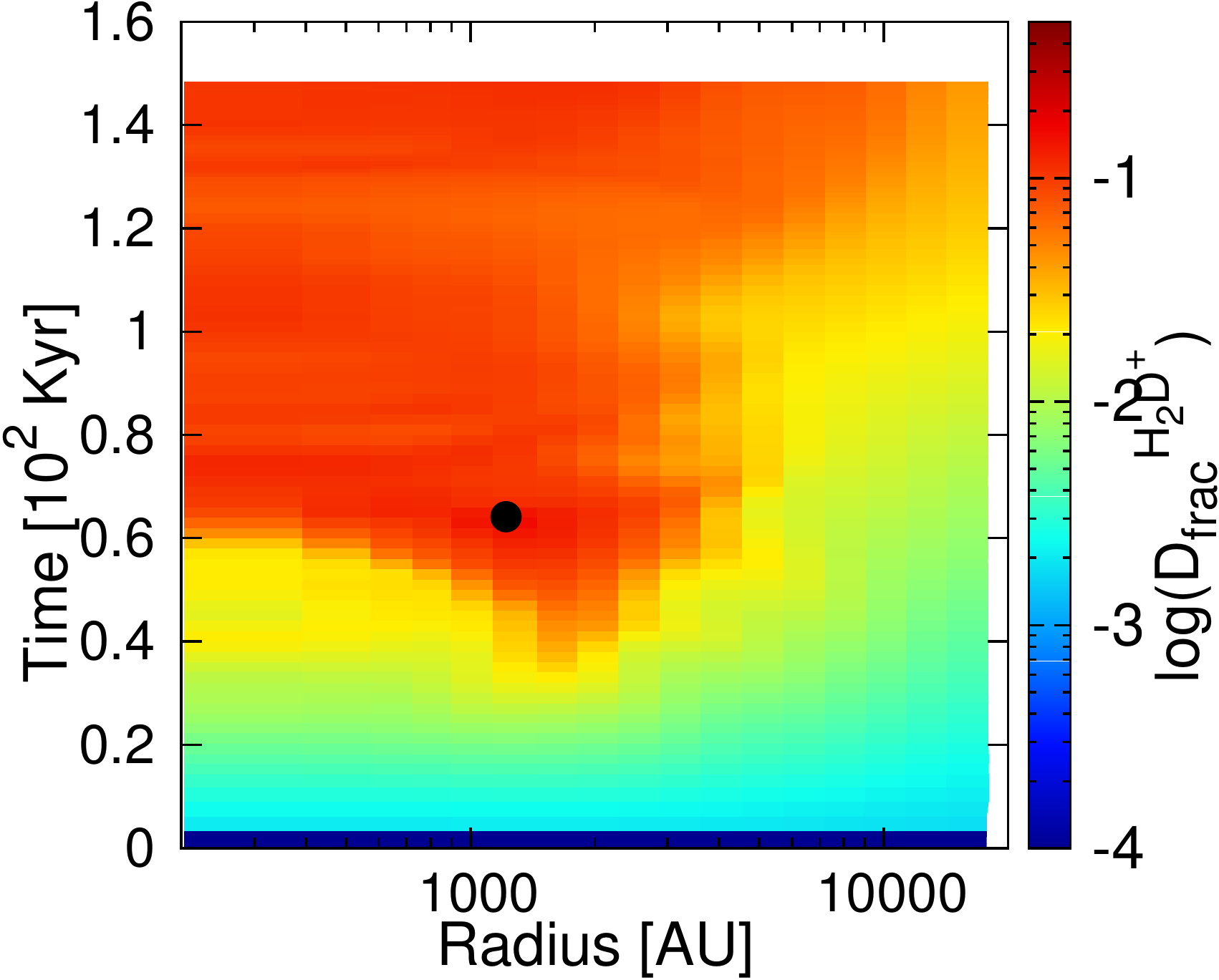}  &\includegraphics[width=0.22\textwidth,angle=-90]{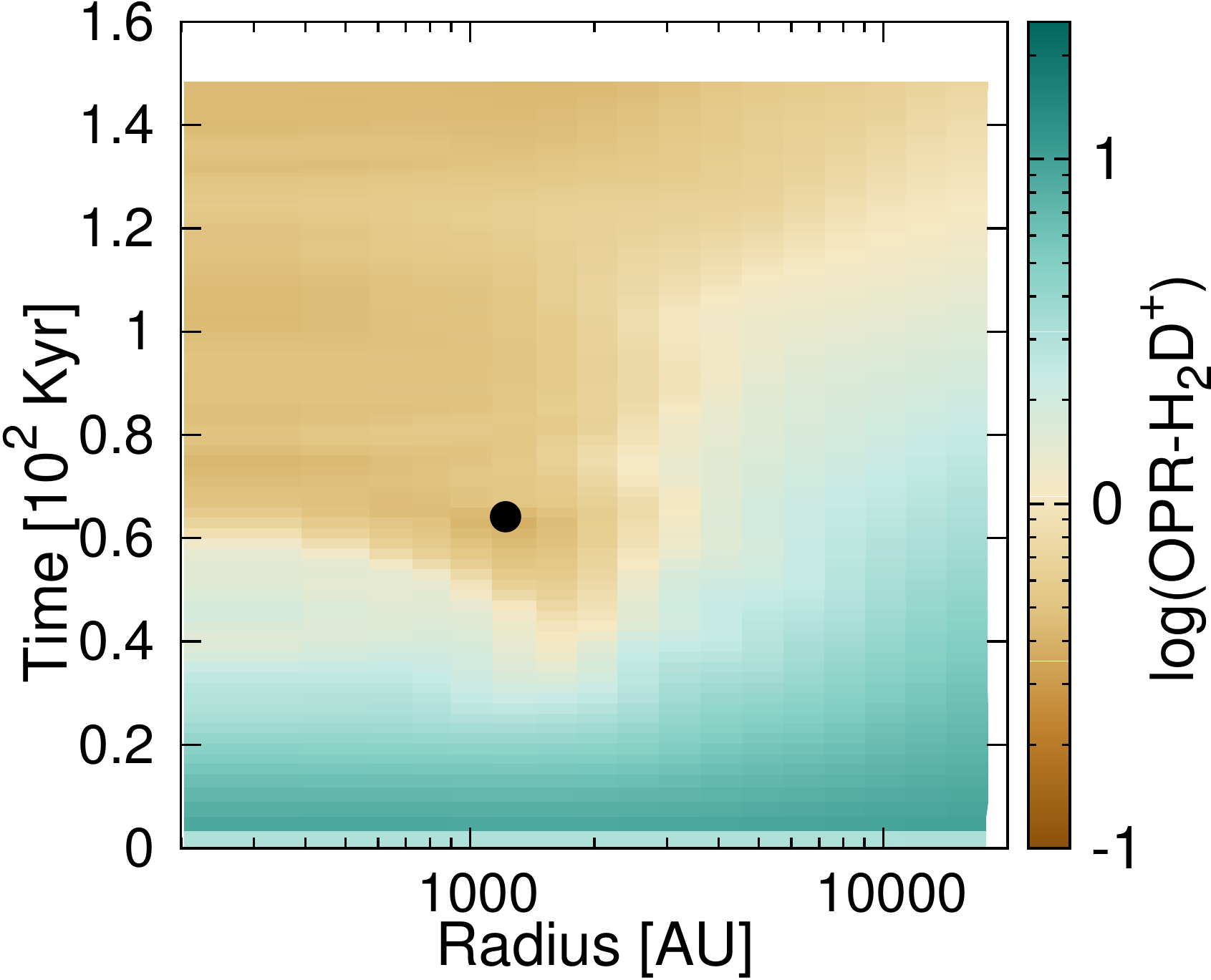} \\
				
		\rotatebox[origin=r]{90}{Lmu2.5M0.5\qquad\qquad} &\includegraphics[width=0.22\textwidth,angle=-90]{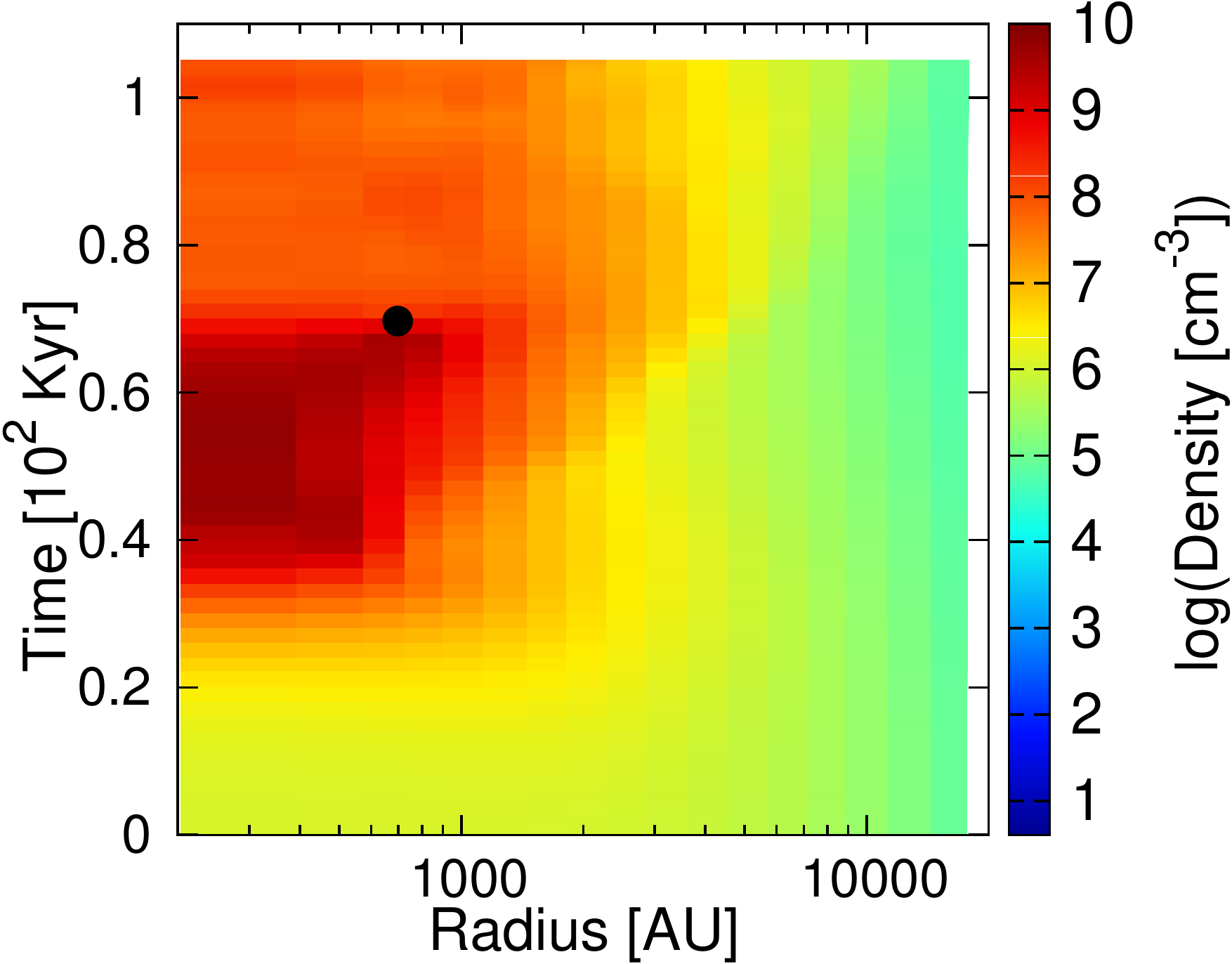} &
		\includegraphics[width=0.22\textwidth,angle=-90]{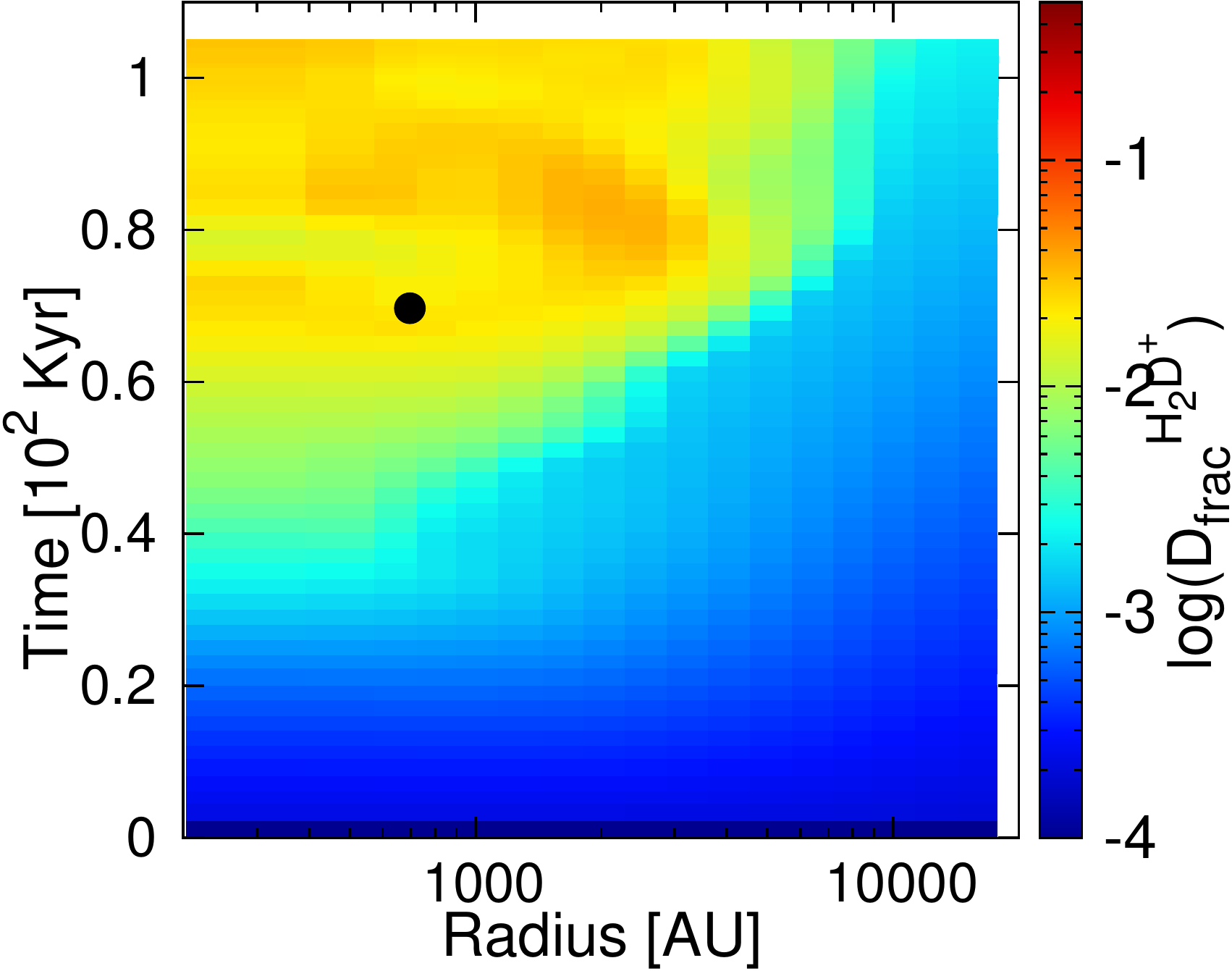}&\includegraphics[width=0.22\textwidth,angle=-90]{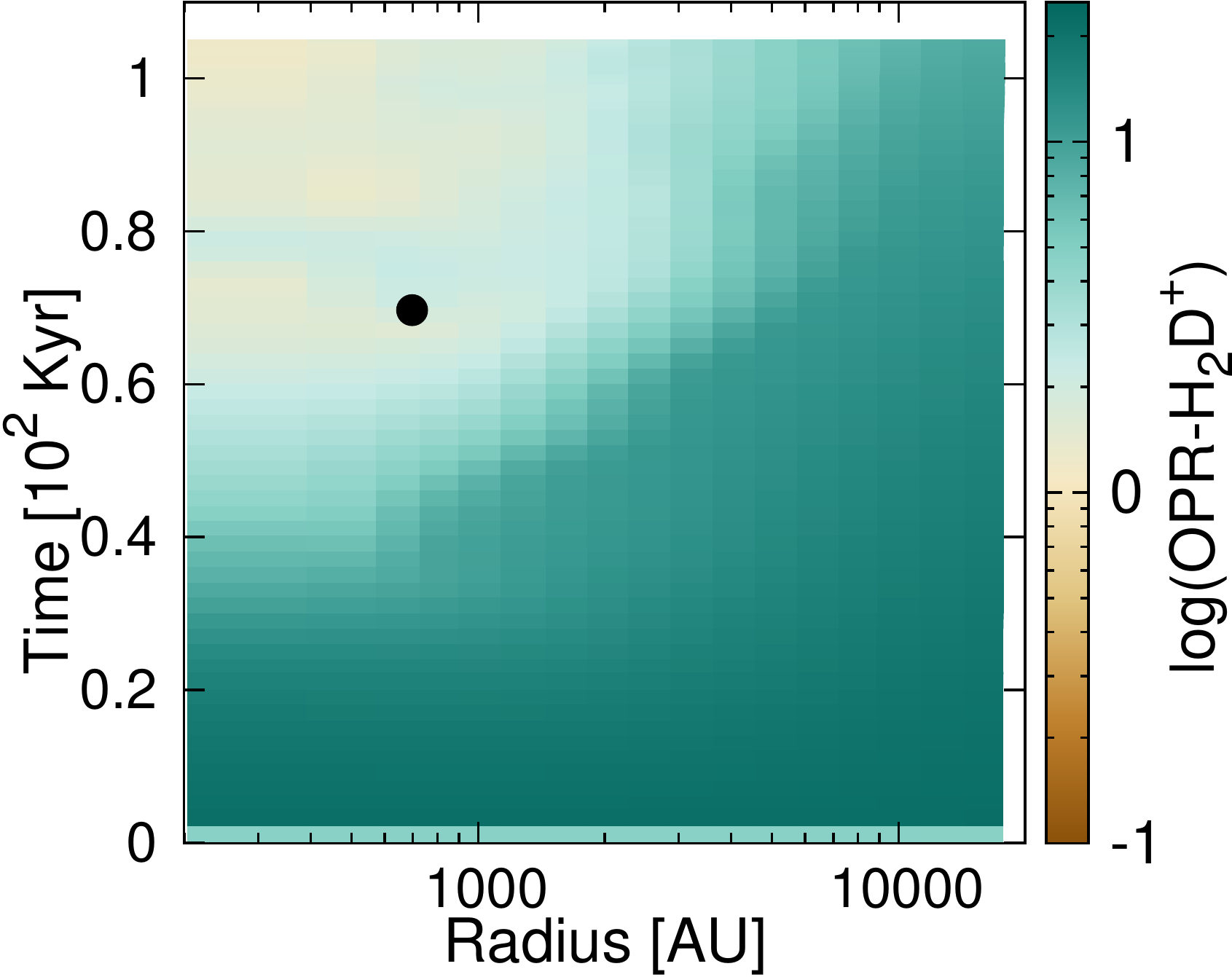} \\		\rotatebox[origin=r]{90}{Hmu10M2\qquad\qquad} &\includegraphics[width=0.22\textwidth,angle=-90]{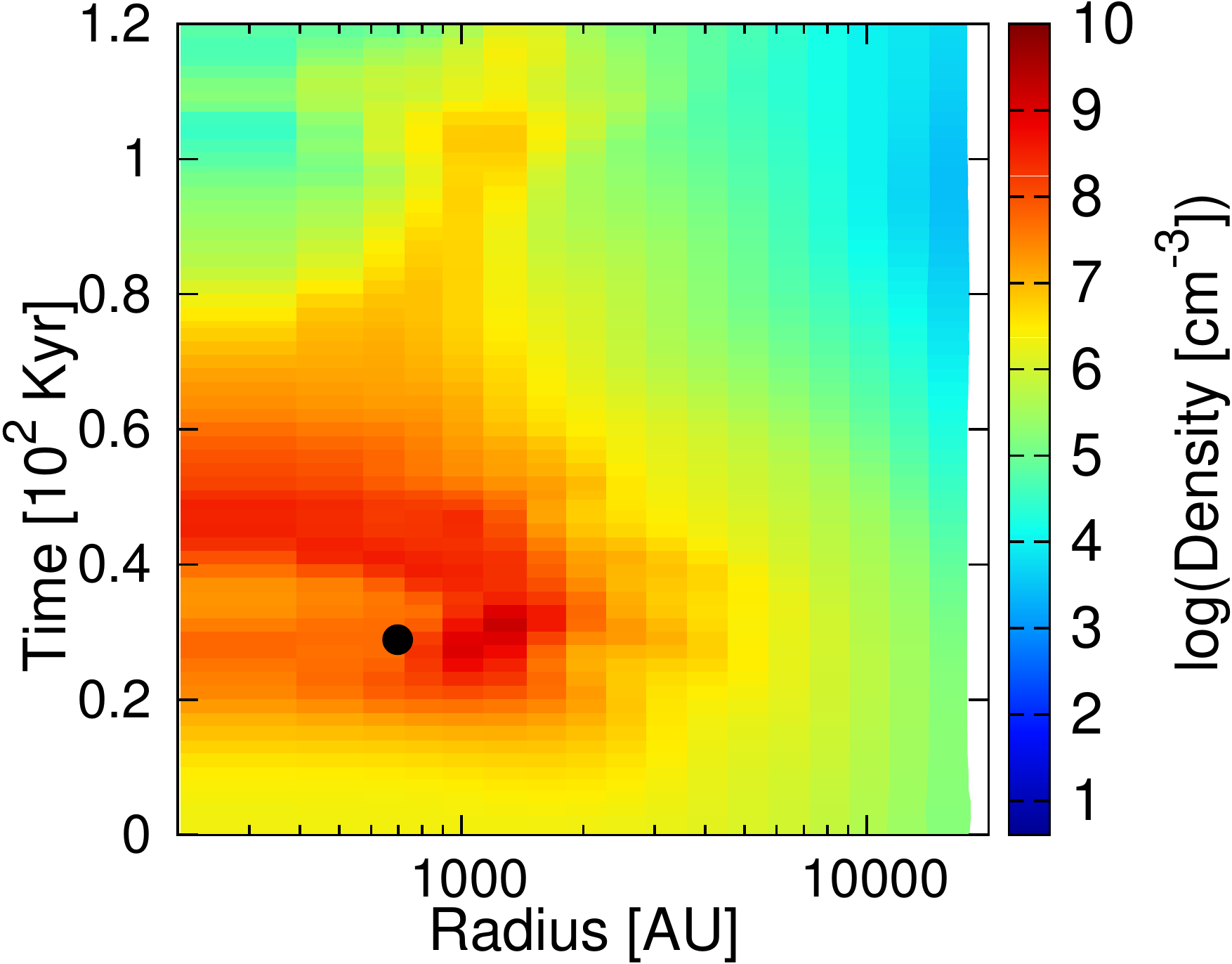} &		\includegraphics[width=0.22\textwidth,angle=-90]{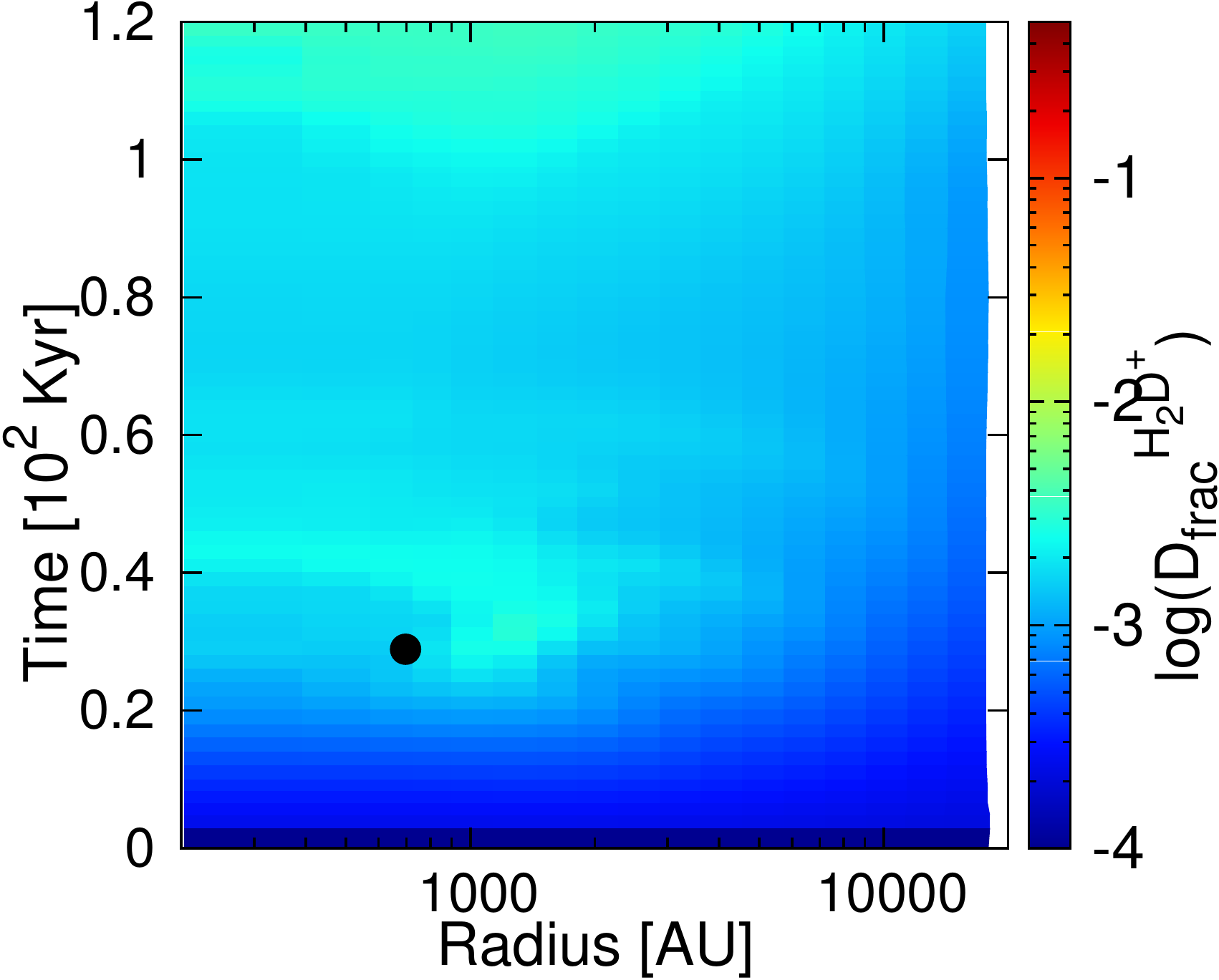}&\includegraphics[width=0.22\textwidth,angle=-90]{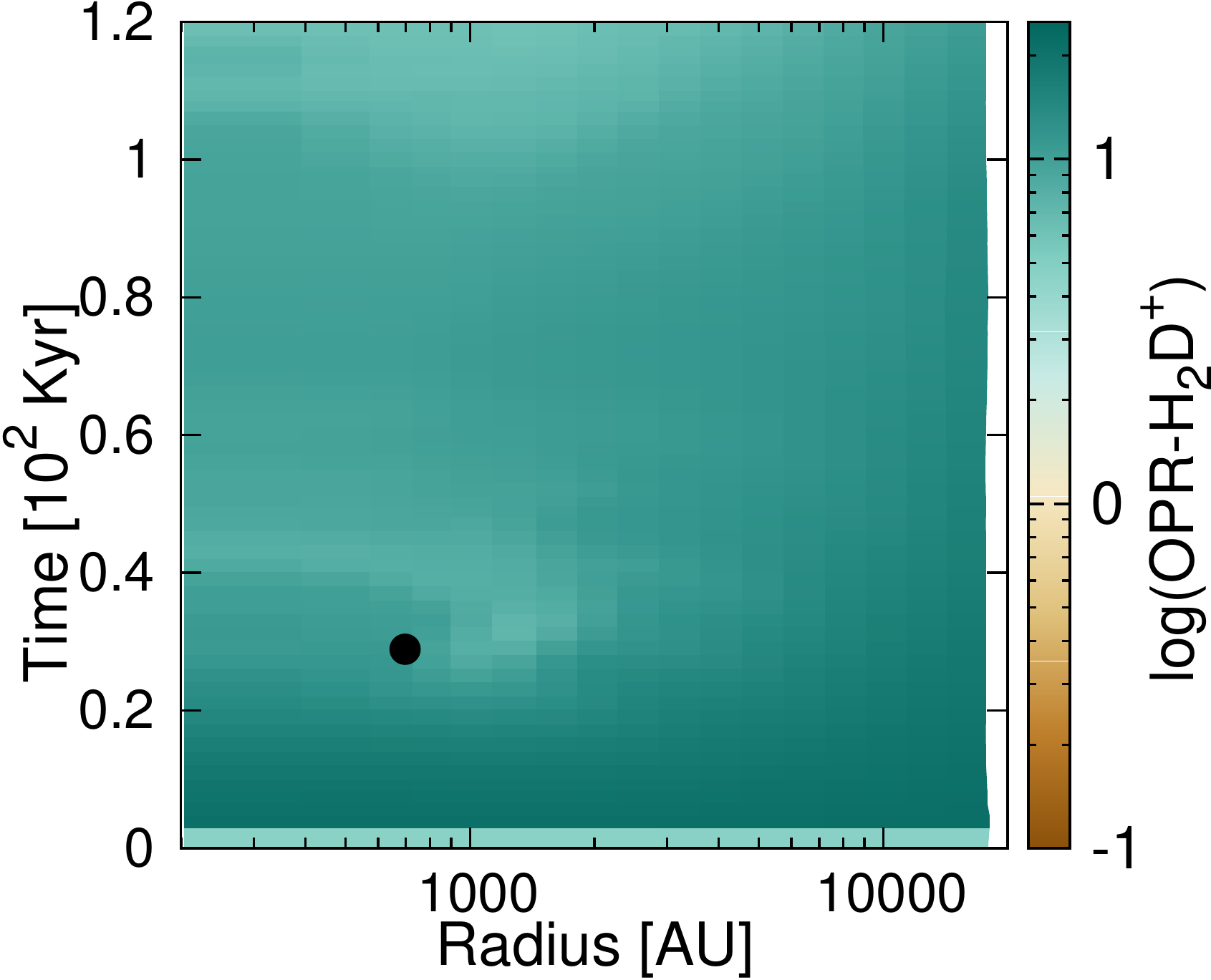}\\
		
		\end{tabular}
	\end{center}
	\caption{Radial profiles of the average gas density, the deuterium fraction, and the H$_2$D$^+$ OPR as function of time for different runs. The black dot represents the time when and position where the sink particle forms. Note that the H$_2$D$^+$ OPR is close to 
	equilibrium at times $t>200\,\mathrm{kyr}$.}
	\label{figBasti1}
\end{figure*}

After exploring the basic dependence on different parameters, we now pursue a more detailed comparison for a number of selected runs. For this purpose, we compare one of our reference runs Lmu10M2 with Lmu10M2S2, where the slope of the magnetic field is varied using $\kappa=2$, the simulation Lmu10M2OPR0.1 employing OPR=0.1, the simulation Lmu2.5M0.5 with both a lower mass--to--flux ratio and a lower turbulent Mach number, as well as the run Hmu10M2 in Fig.~\ref{figBasti1}. We note that these are the simulations that have been evolved for the longest time, which is particularly difficult for cases with low mass--to--flux ratios. The black dot denotes the time and position where a sink particle forms.
For these simulations, radially averaged gas density, deuterium fraction as well as the OPR of H$_2$D$^+$ is shown as a function of radius and time. It is clearly visible in all simulations that the deuterium fraction as well as the OPR of H$_2$D$^+$ depend on density as well as on time. Considering Lmu10M2, we see how density builds up towards the center. 
After sink formation, it is evident how the density in the center starts decreasing, as the gas is accreted onto the sink particle.
The deuterium fraction is initially quite low at all radii, and first starts to increase in the region where the dense gas starts to accumulate, and subsequently the region of high deuteration surrounds the sink particle. As a result of gas accretion onto the sink, the deuteration first slightly decreases with time, however on the long term, the additional time compensates for the decreasing density, leading to high deuterium fractions up to around $10\%$.  A very similar behavior is seen also for the OPR of H$_2$D$^+$. What is also noticeable for both quantities is that at late times, i.e. after 100,000~yrs ($\sim0.7\times t_{ff}$), high deuterium fractions (a few $\times10^{-3}$) are reached even at large radii with the lowest densities while the H$_2$D$^+$ OPR is about to approach its equilibrium value ($\sim$0.028 at 15~K).\\
The run Lmu10M2S2 shows a qualitatively similar behavior, except that the density starts to increase somewhat earlier, and also the sink particle forms already after about 50~kyrs. While this density increase is partly reflected in deuteration and the H$_2$D$^+$ OPR, the increase is less significant, as a much shorter amount of time was available for the deuteration. After 100,000~yrs (i.e. roughly 0.7$\times t_{ff}$), the deuterium fraction however has increased significantly with time and is widespread over a larger region.\\ 
The simulation Lmu10M2OPR0.1 with a low initial H$_2$ OPR of 0.1 shows the same density evolution as Lmu10M2, however the deuterium fraction is enhanced significantly, so that large deuterium fractions around 10\% are reached already after about 60~kyrs over a range of roughly 3,000~au. The latter provides the ideal conditions for a rapid deuteration scenario.\\
In case of Lmu2.5M0.5, the density evolution changes and the dense gas is more concentrated towards the center, due to the lower amount of turbulent support, which is not fully compensated by the increased magnetic field strength. Subsequently, the gas is accreted onto the sink as in the other runs, and the density starts decreasing. Deuterium fraction and H$_2$D$^+$ OPR show similar features as described before. As a result of the density distribution, the higher deuterium fractions are initially more concentrated towards the center. \\
The run Hmu10M2 is overall similar to Lmu10M2. The gas appears to be accreted more rapidly onto the sink particle, as a result of the shorter free--fall time. While the run starts showing an increasing deuterium fraction around 120,000~yrs, it is not yet evolved enough to reach very high deuterium fractions. We note, however, that the observed values of $D_\mathrm{frac}$ lie in between 10$^{-3}$ and 0.1 (see also Fig. \ref{obs}), and hence also this core appears to be in agreement with observations. It is also important to note that we reach observed values of $D_\mathrm{frac}$ in all our runs despite the high H$_2$ OPR and the assumption on the dynamical support by magnetic fields and turbulence. This picture of course could change if the timescale needed to reach a high level of freeze--out is very long.

\begin{figure*}
	\begin{center}
		\begin{tabular}{cccc}
		&o--H$_2$D$^+$ &p--H$_2$D$^+$ &D$_2$H$^+$\\
		\rotatebox[origin=r]{90}{Lmu10M2\qquad\qquad} &\includegraphics[width=0.22\textwidth,angle=-90]{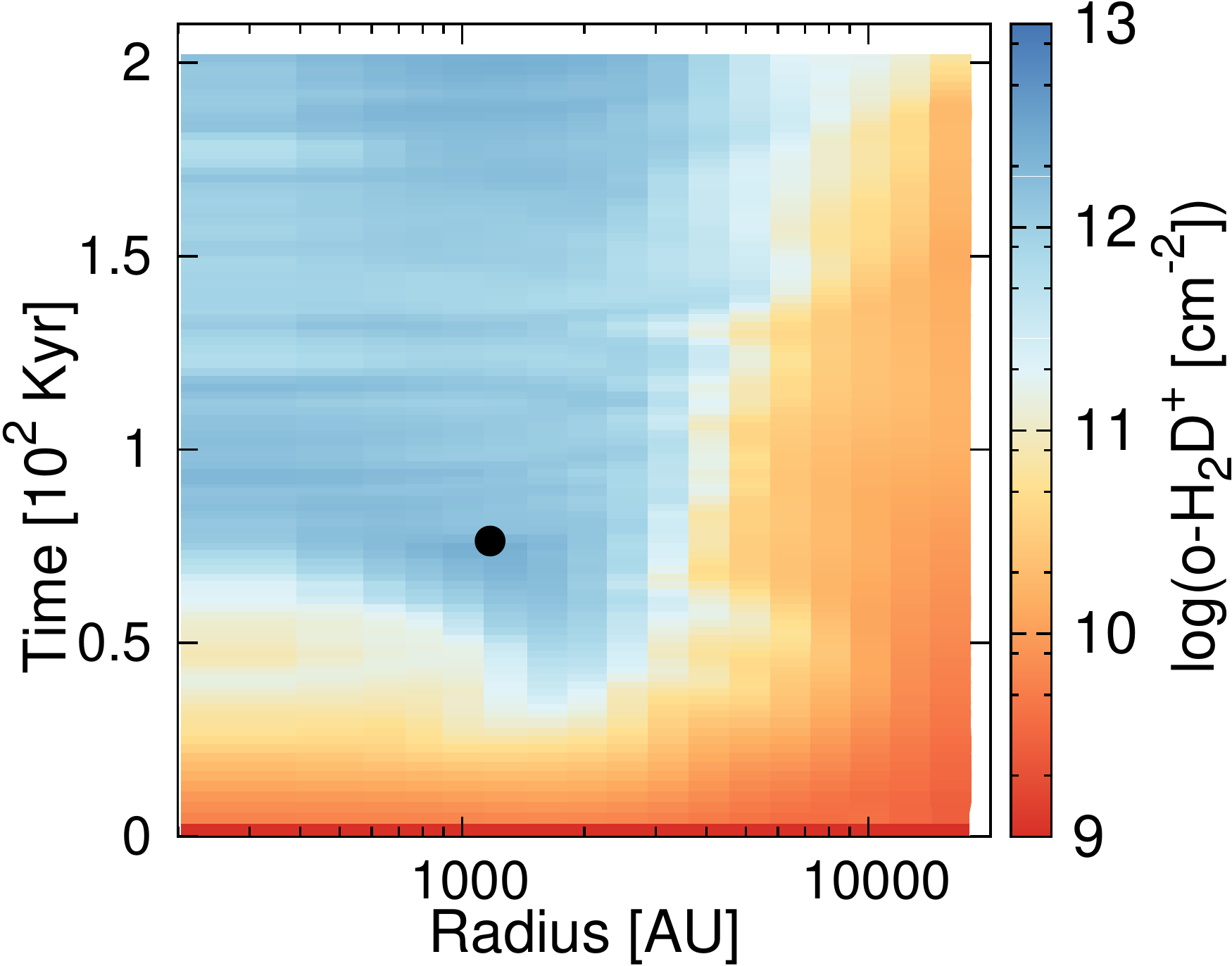} &\includegraphics[width=0.22\textwidth,angle=-90]{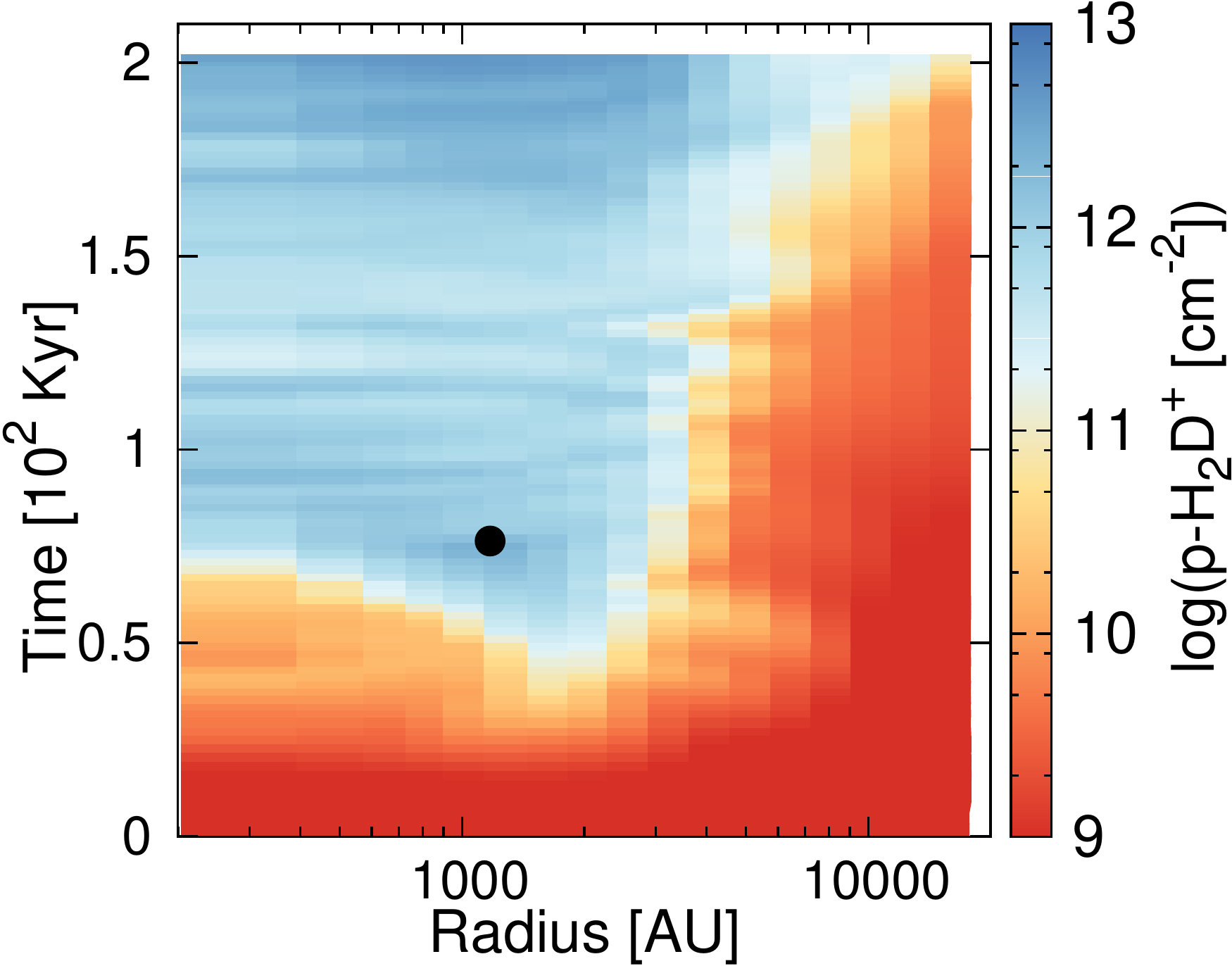} &\includegraphics[width=0.22\textwidth,angle=-90]{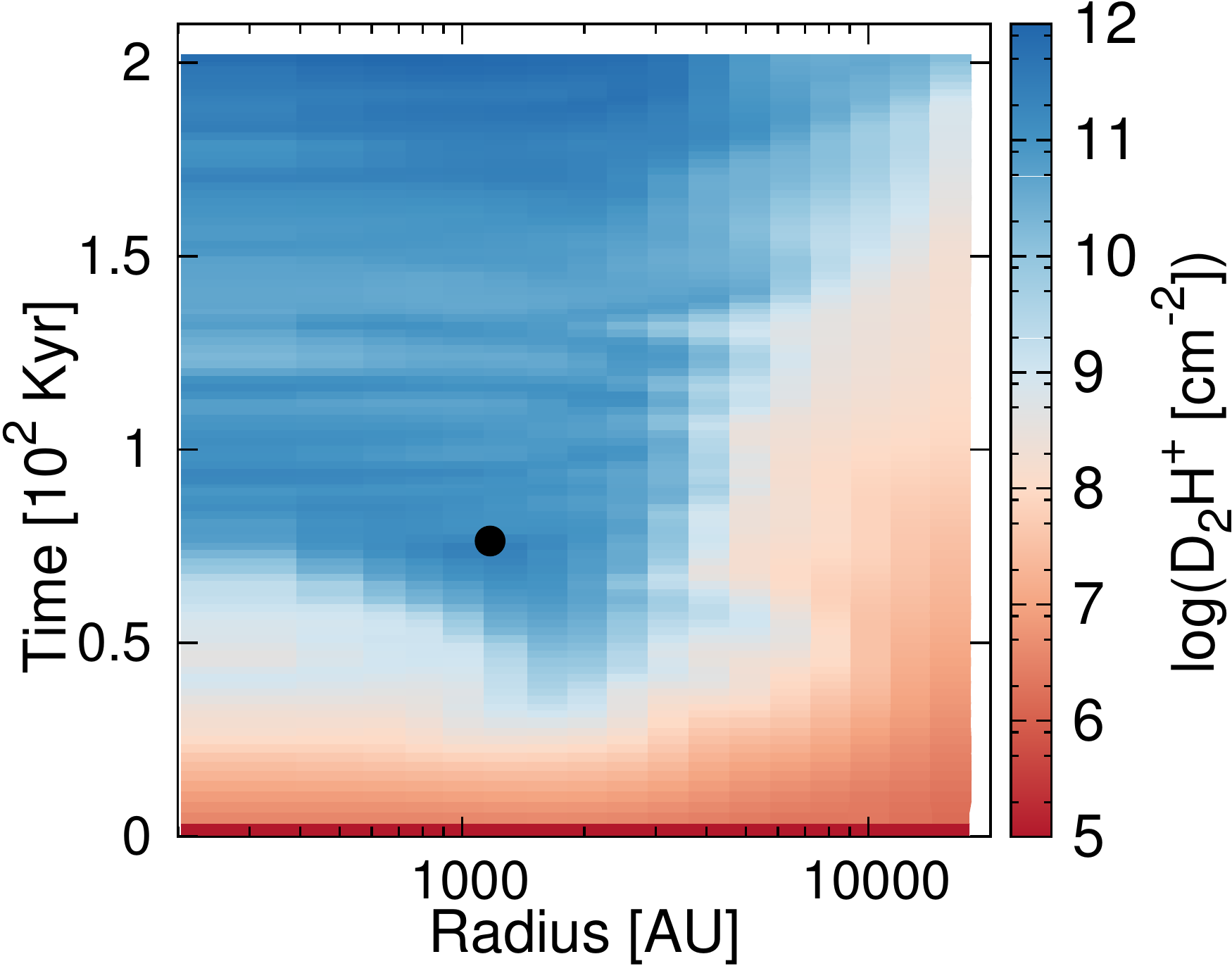} \\
		\rotatebox[origin=r]{90}{Lmu10M2S2\qquad\qquad} &\includegraphics[width=0.22\textwidth,angle=-90]{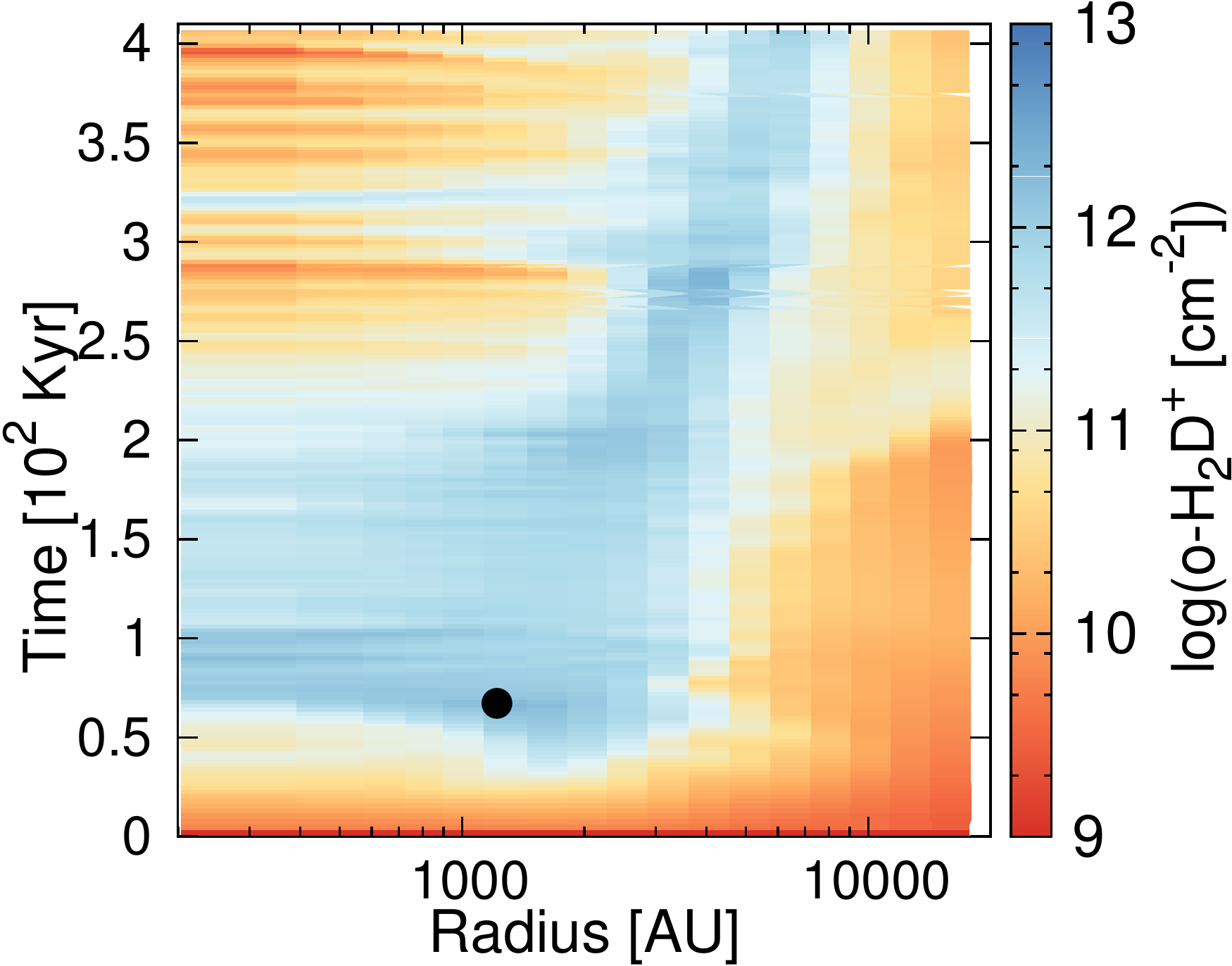} &\includegraphics[width=0.22\textwidth,angle=-90]{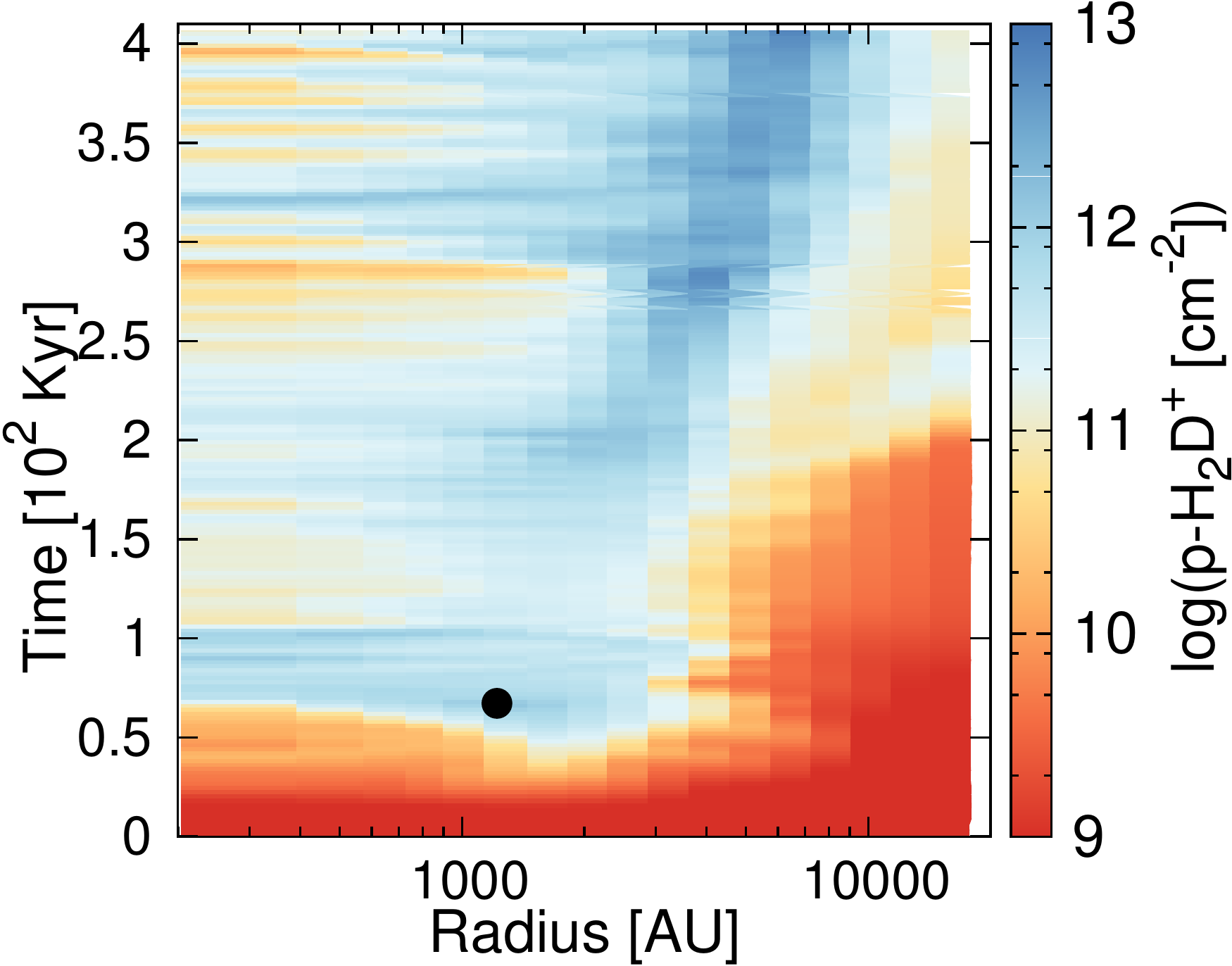} &\includegraphics[width=0.22\textwidth,angle=-90]{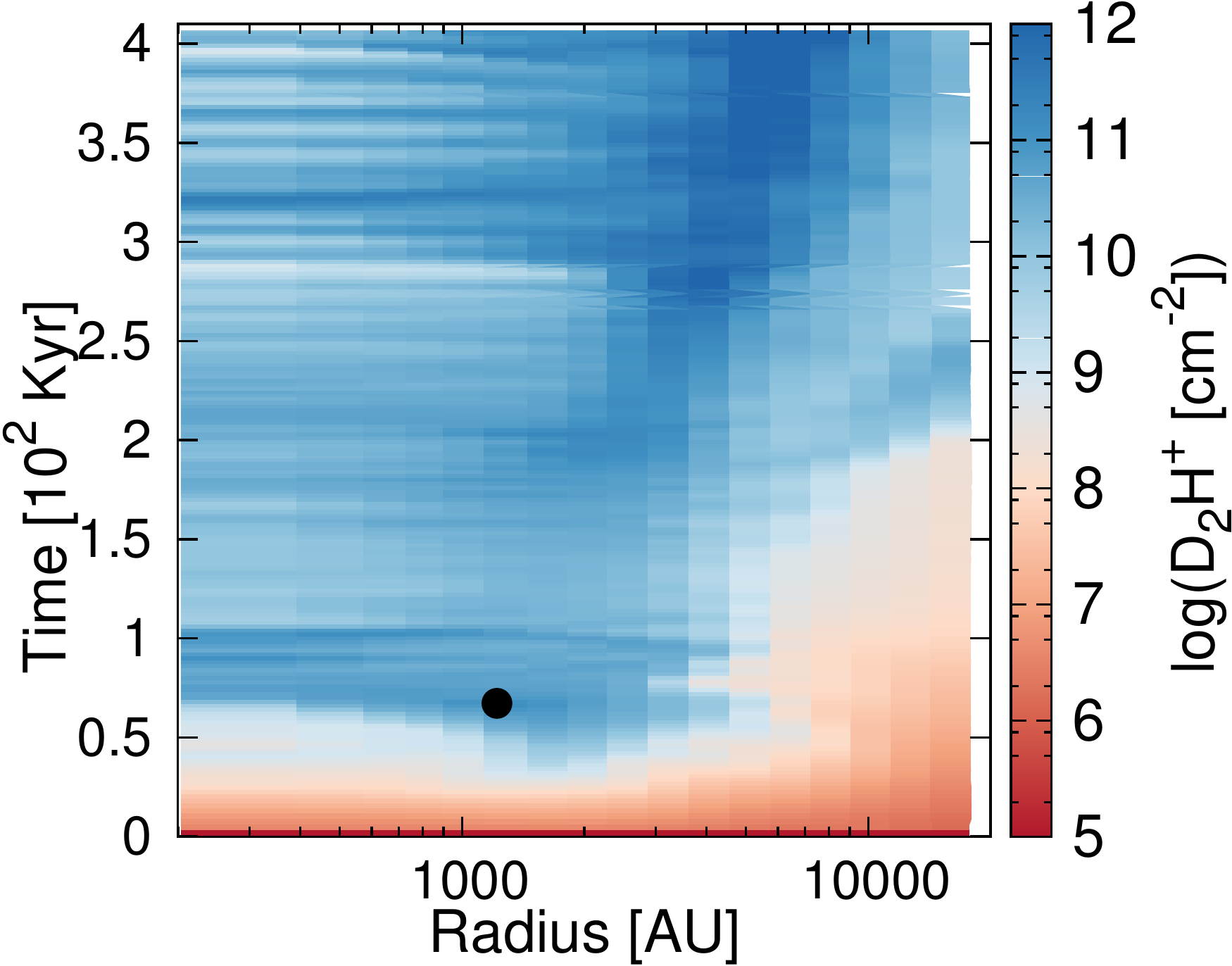} \\
		\rotatebox[origin=r]{90}{Lmu10M2OPR0.1\qquad} &\includegraphics[width=0.22\textwidth,angle=-90]{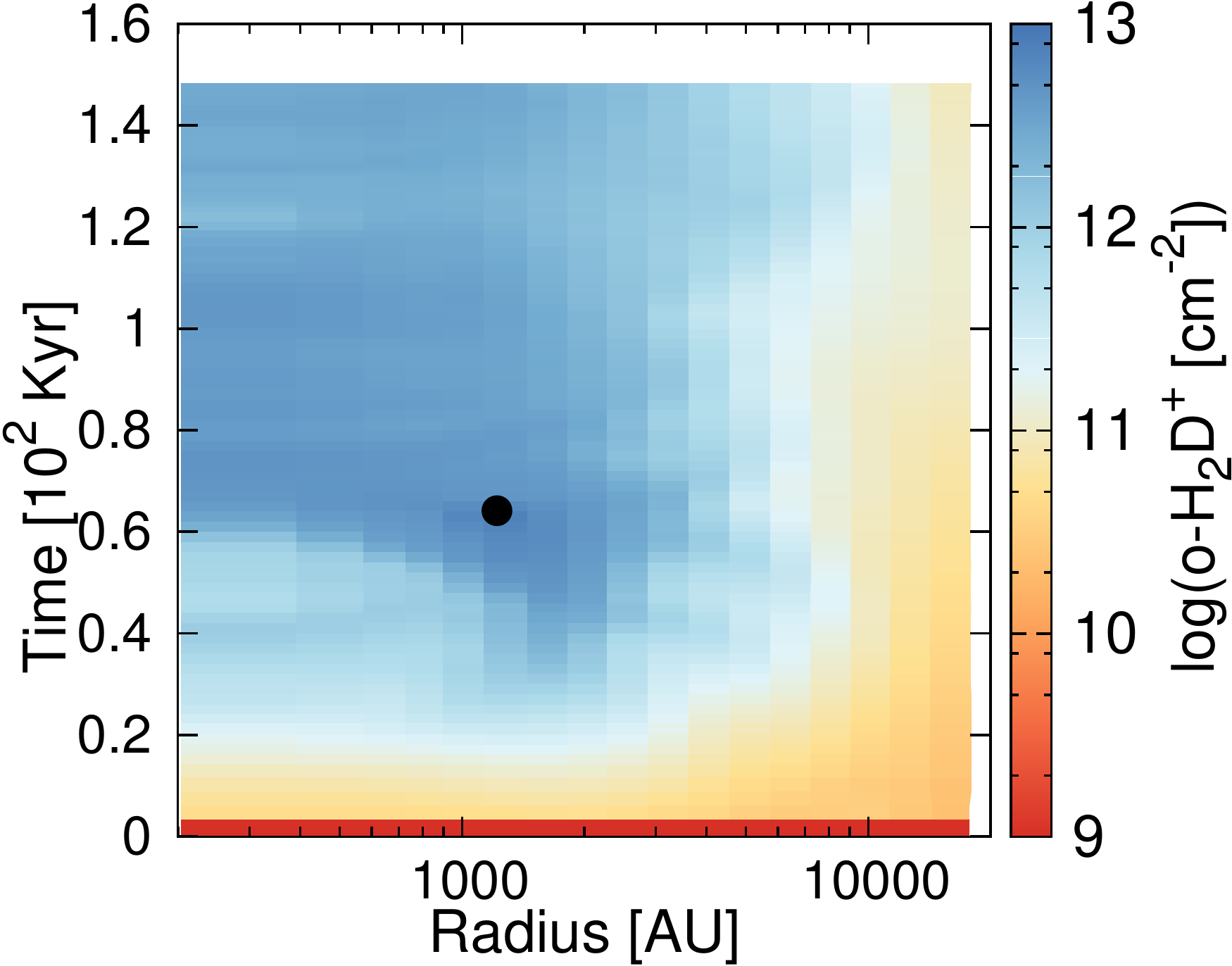} &\includegraphics[width=0.22\textwidth,angle=-90]{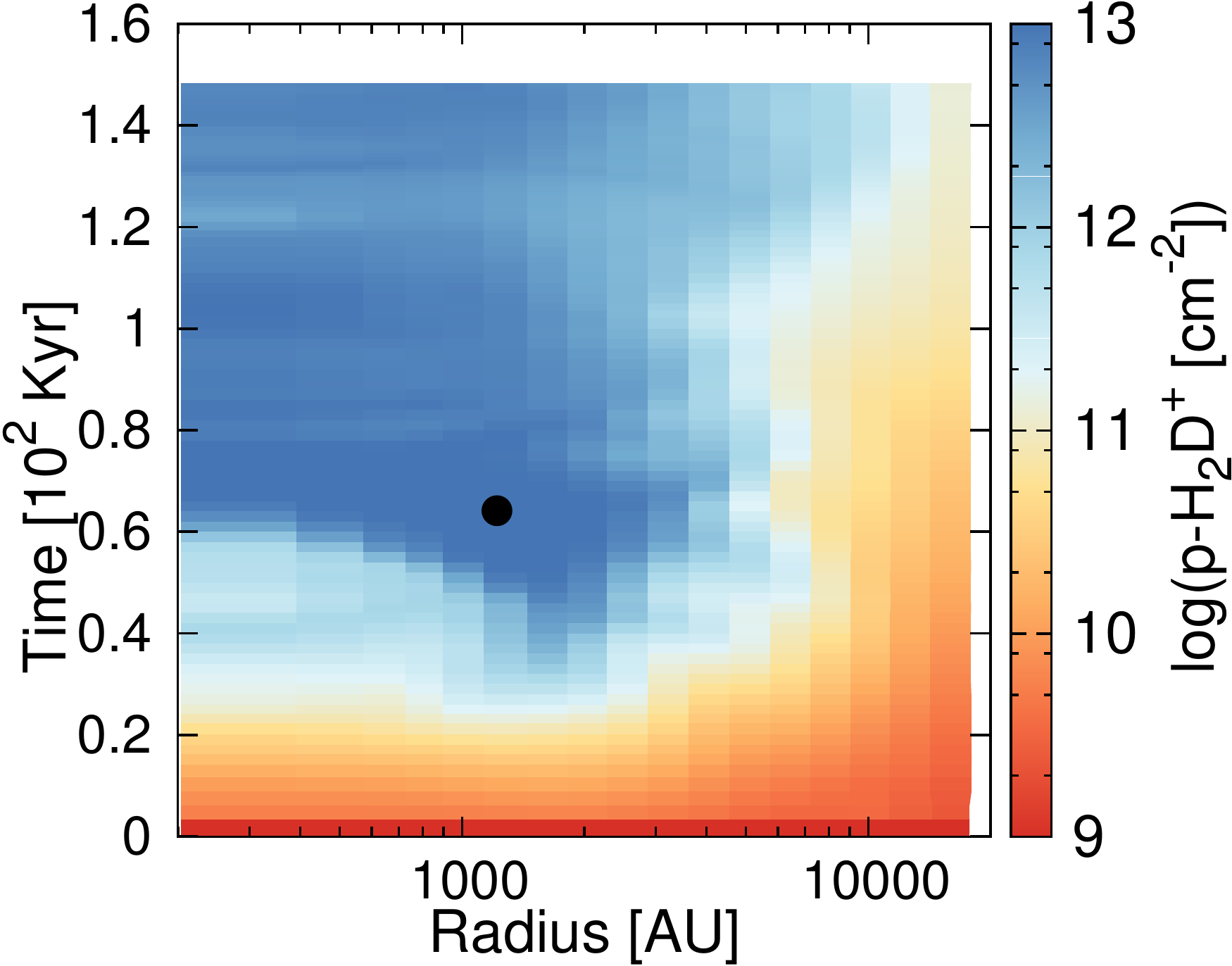} &\includegraphics[width=0.22\textwidth,angle=-90]{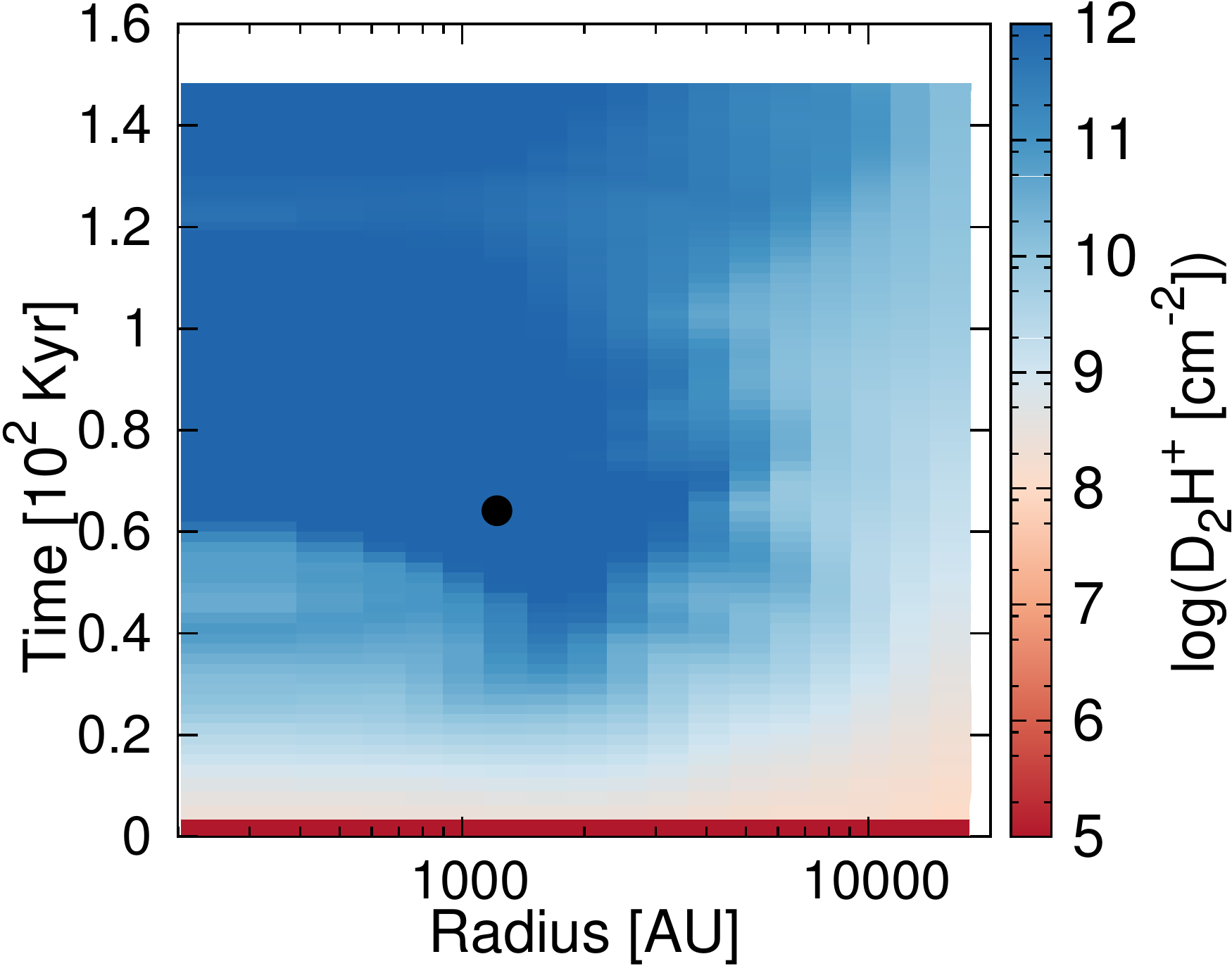} \\
		\rotatebox[origin=r]{90}{Lmu2.5M0.5\qquad\qquad} &\includegraphics[width=0.22\textwidth,angle=-90]{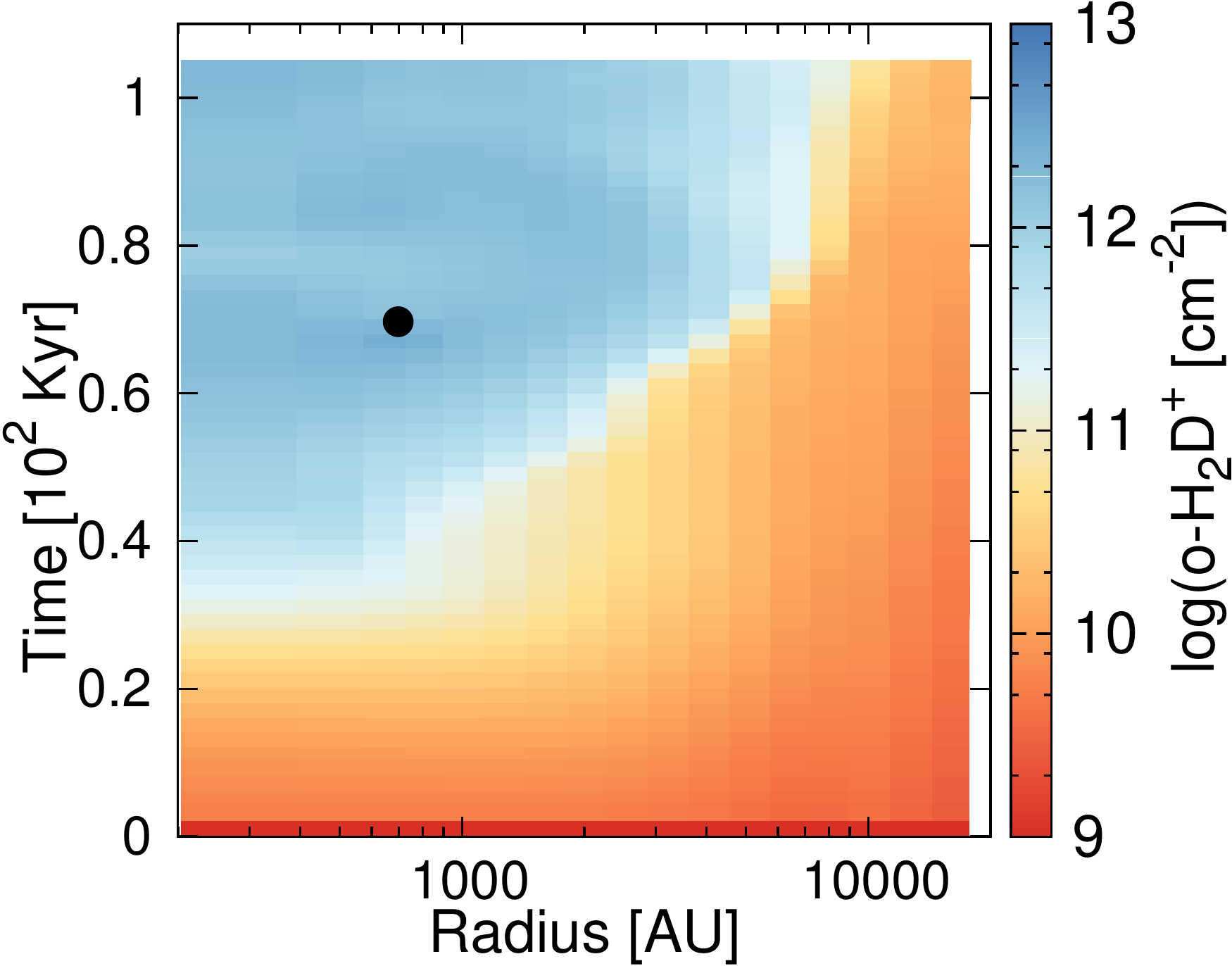} &\includegraphics[width=0.22\textwidth,angle=-90]{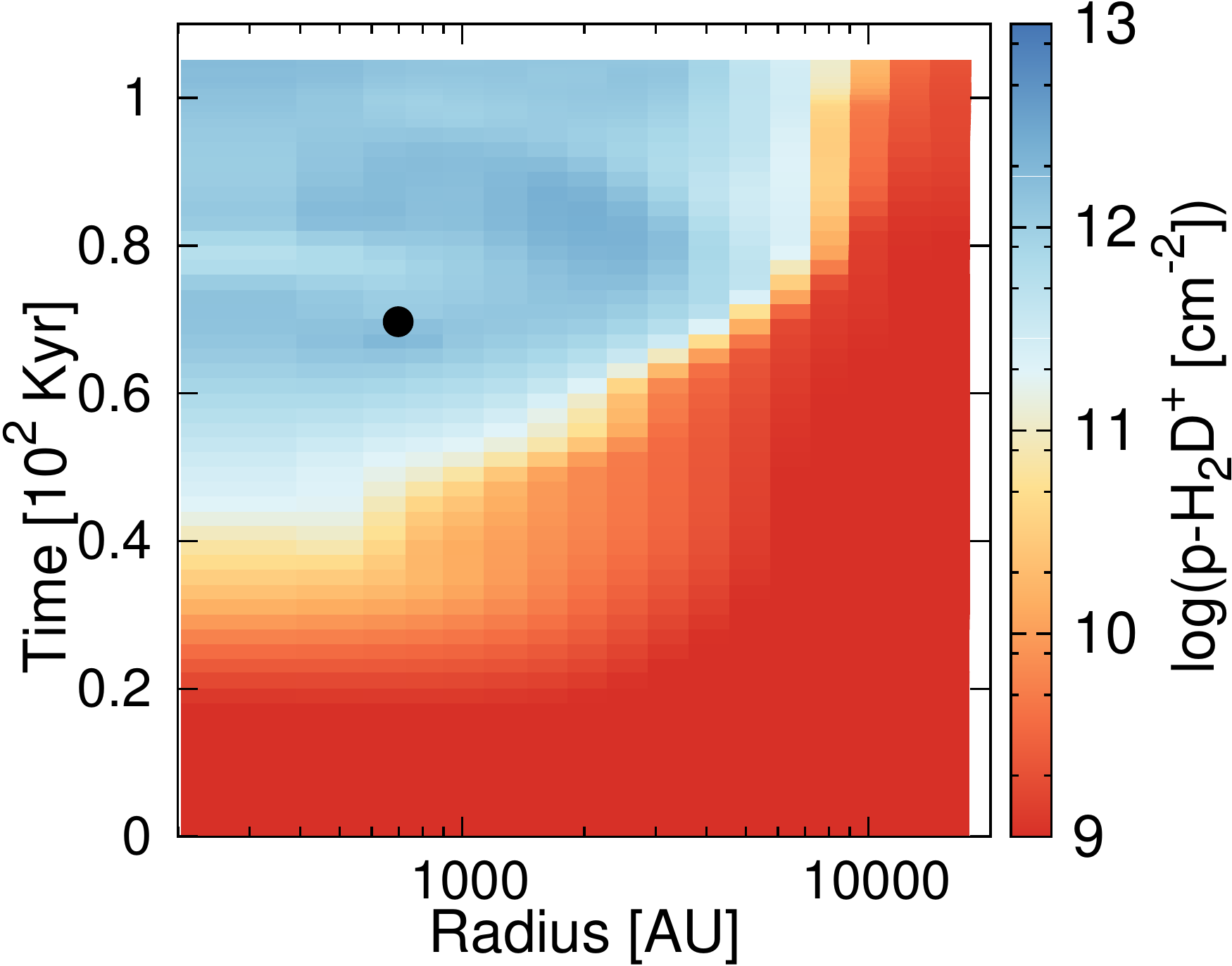} &\includegraphics[width=0.22\textwidth,angle=-90]{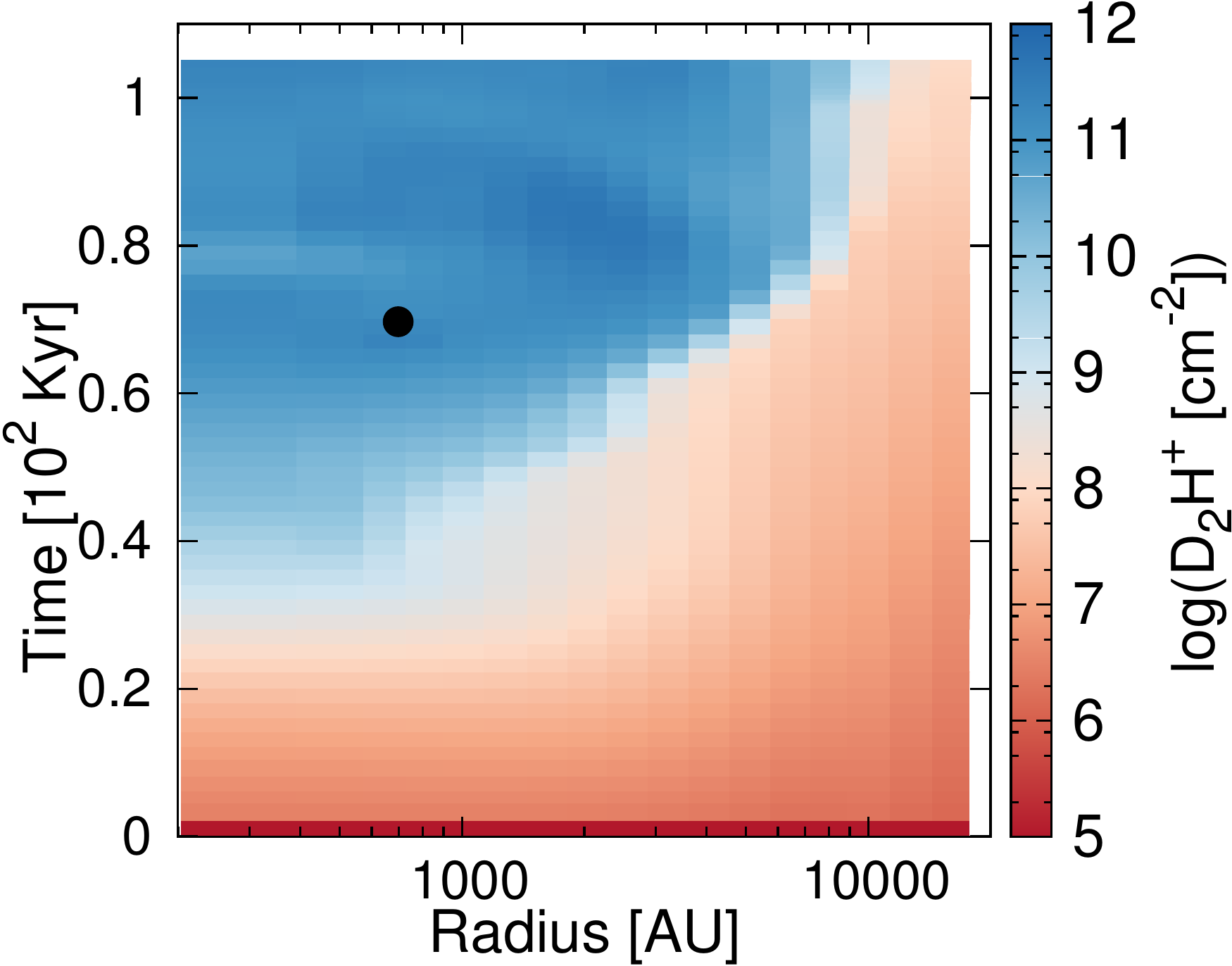} \\
				\rotatebox[origin=r]{90}{Hmu10M2\qquad\qquad} &\includegraphics[width=0.22\textwidth,angle=-90]{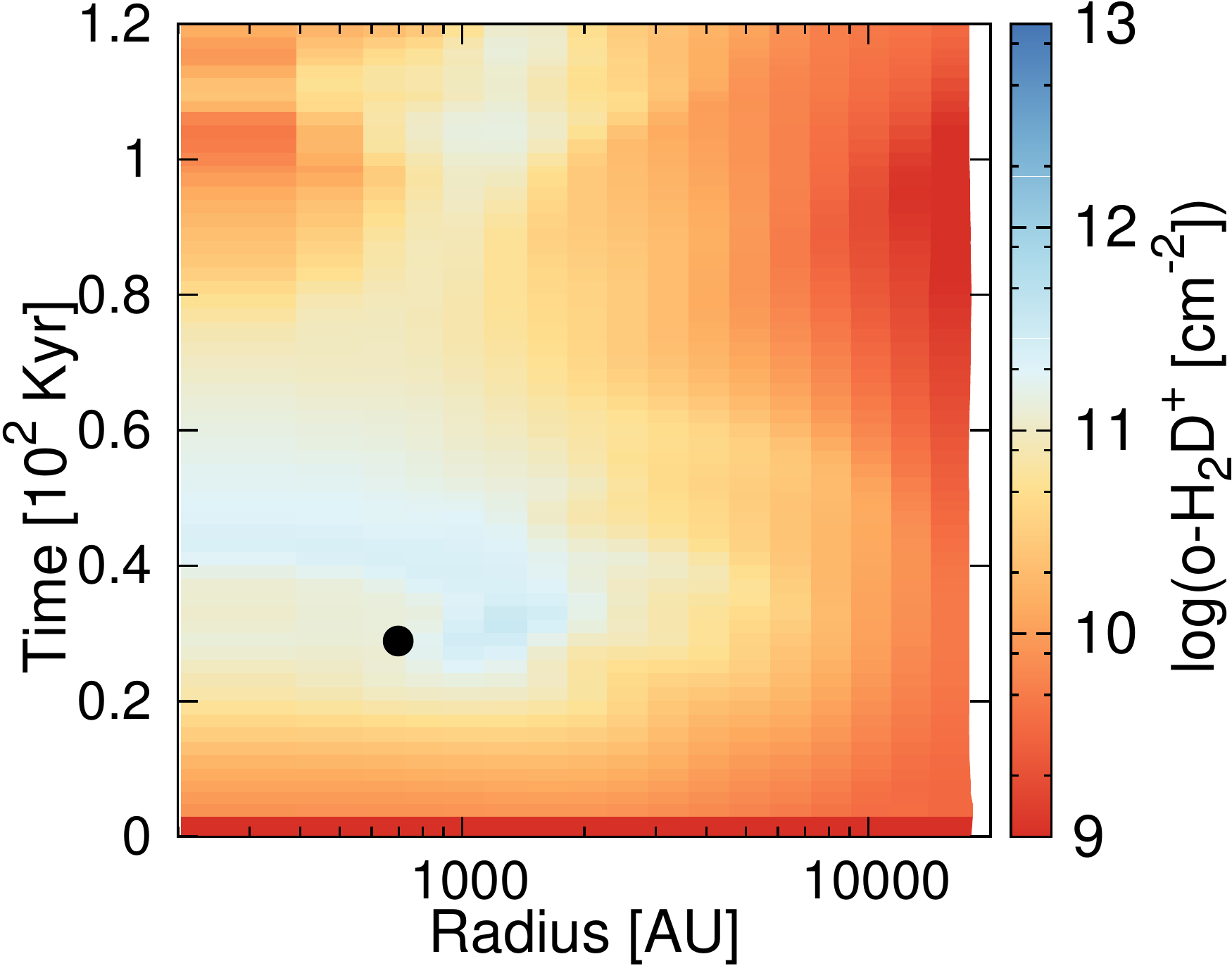} &\includegraphics[width=0.22\textwidth,angle=-90]{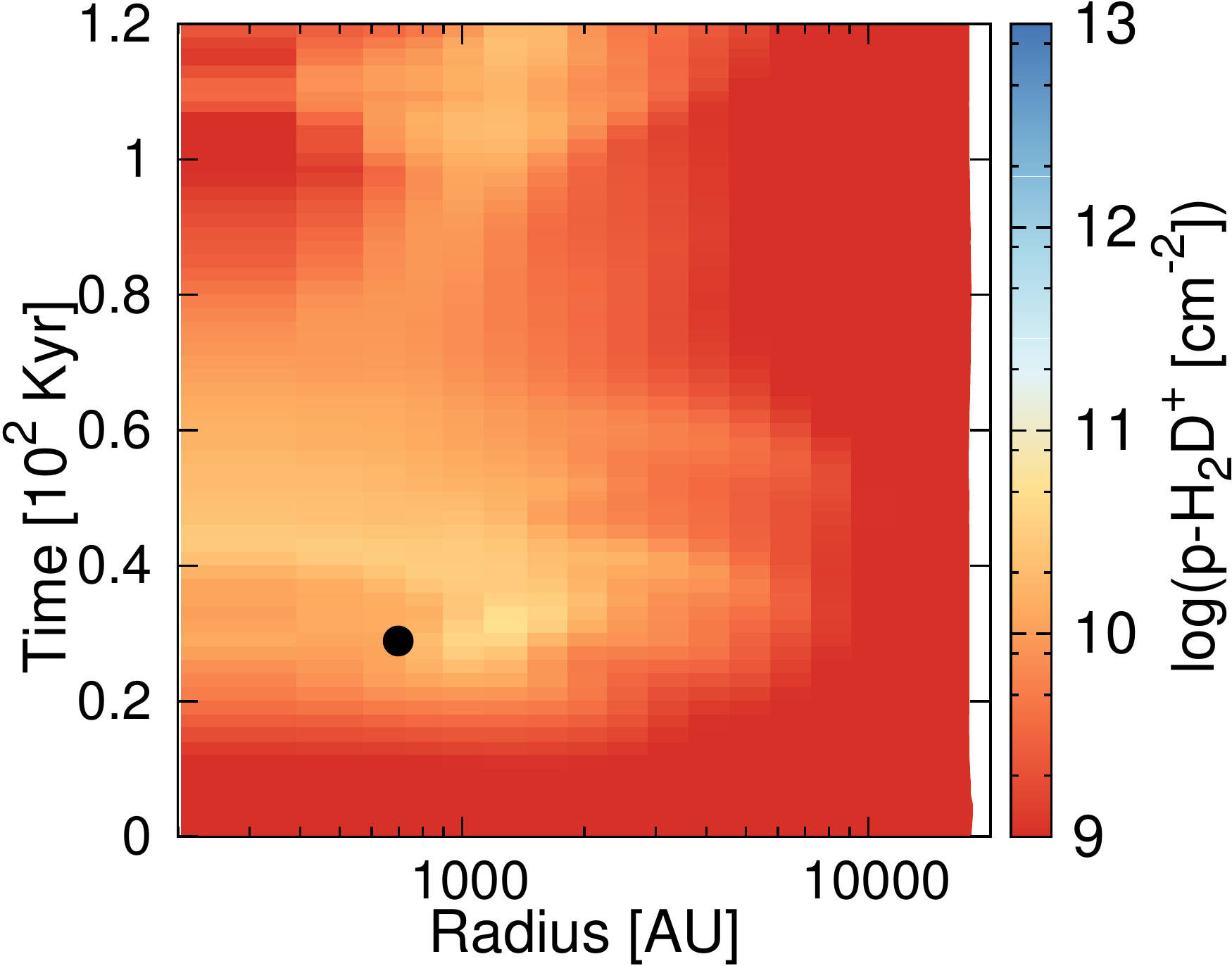} &\includegraphics[width=0.22\textwidth,angle=-90]{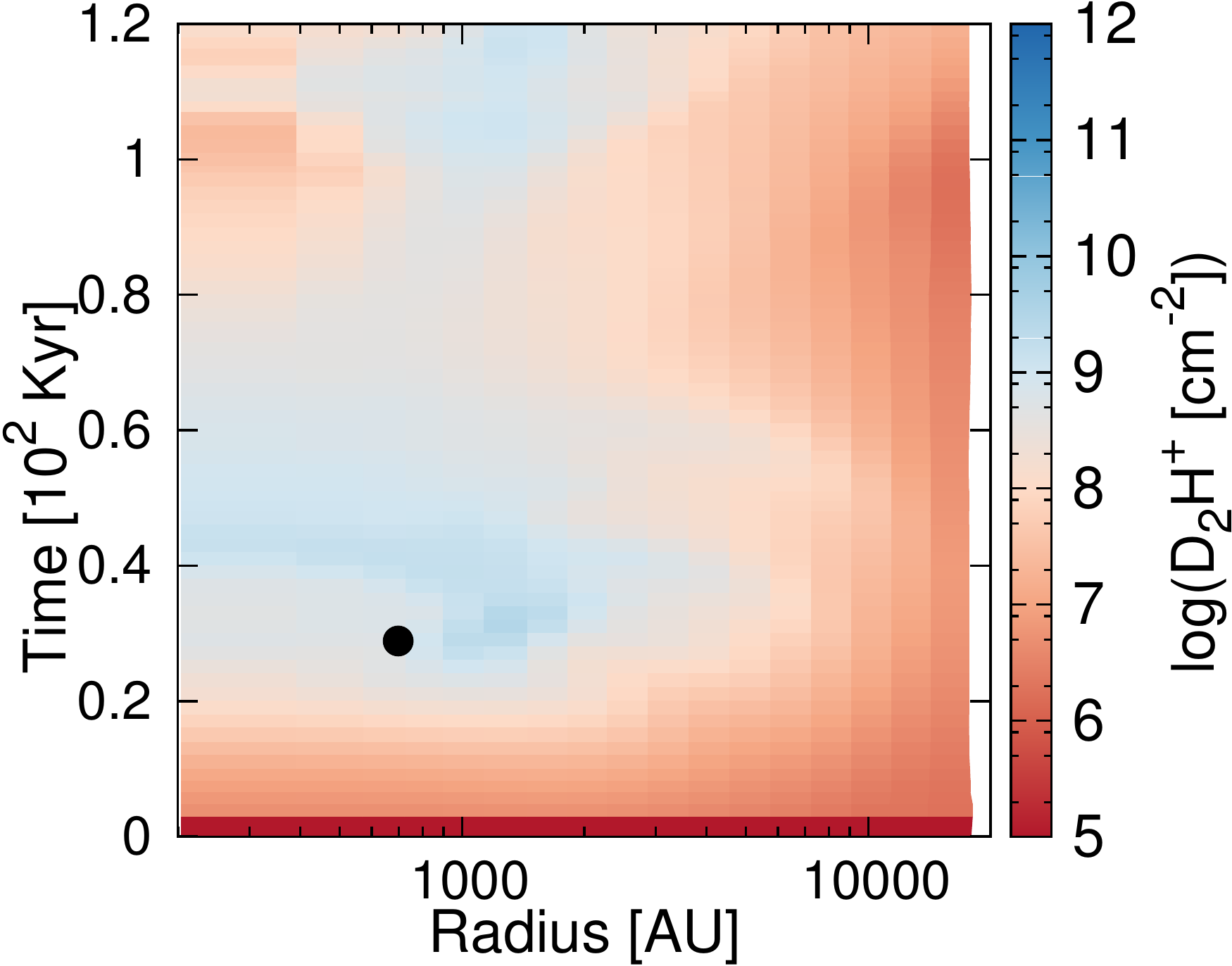} \\

		\end{tabular}
	\end{center}
	\caption{Radial profiles of the spin states of H$_2$D$^+$ and D$_2$H$^+$ for the selected runs.}
	\label{figBasti5}
\end{figure*}

\begin{figure*}
	\begin{center}
		\begin{tabular}{clr}
		\rotatebox[origin=r]{90}{\Large{Lmu10M2}\qquad\qquad} &\includegraphics[width=0.33\textwidth,angle=-90]{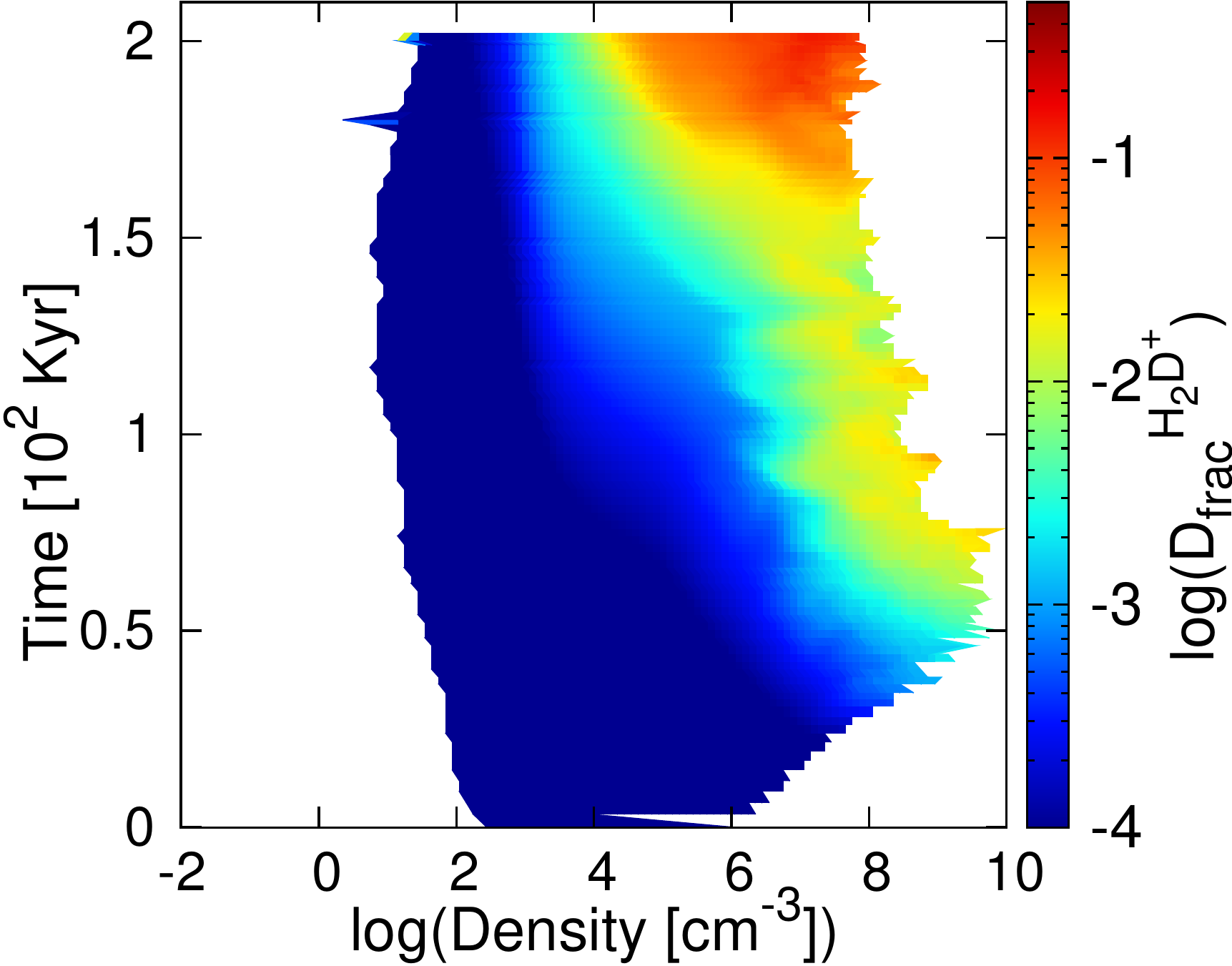} &\includegraphics[width=0.33\textwidth,angle=-90]{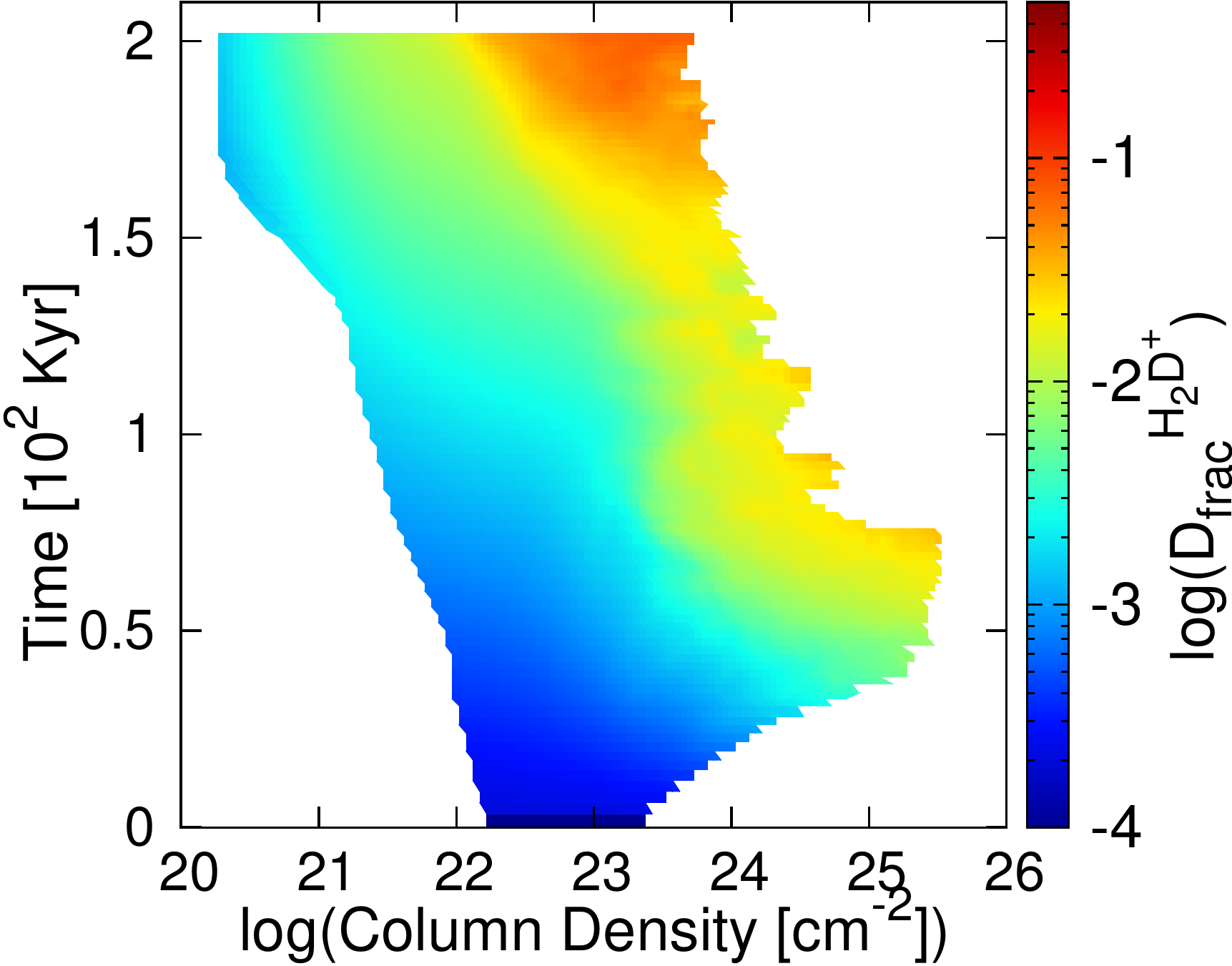} \\
		\rotatebox[origin=r]{90}{\Large{Lmu10M2S2}\qquad\qquad} &\includegraphics[width=0.33\textwidth,angle=-90]{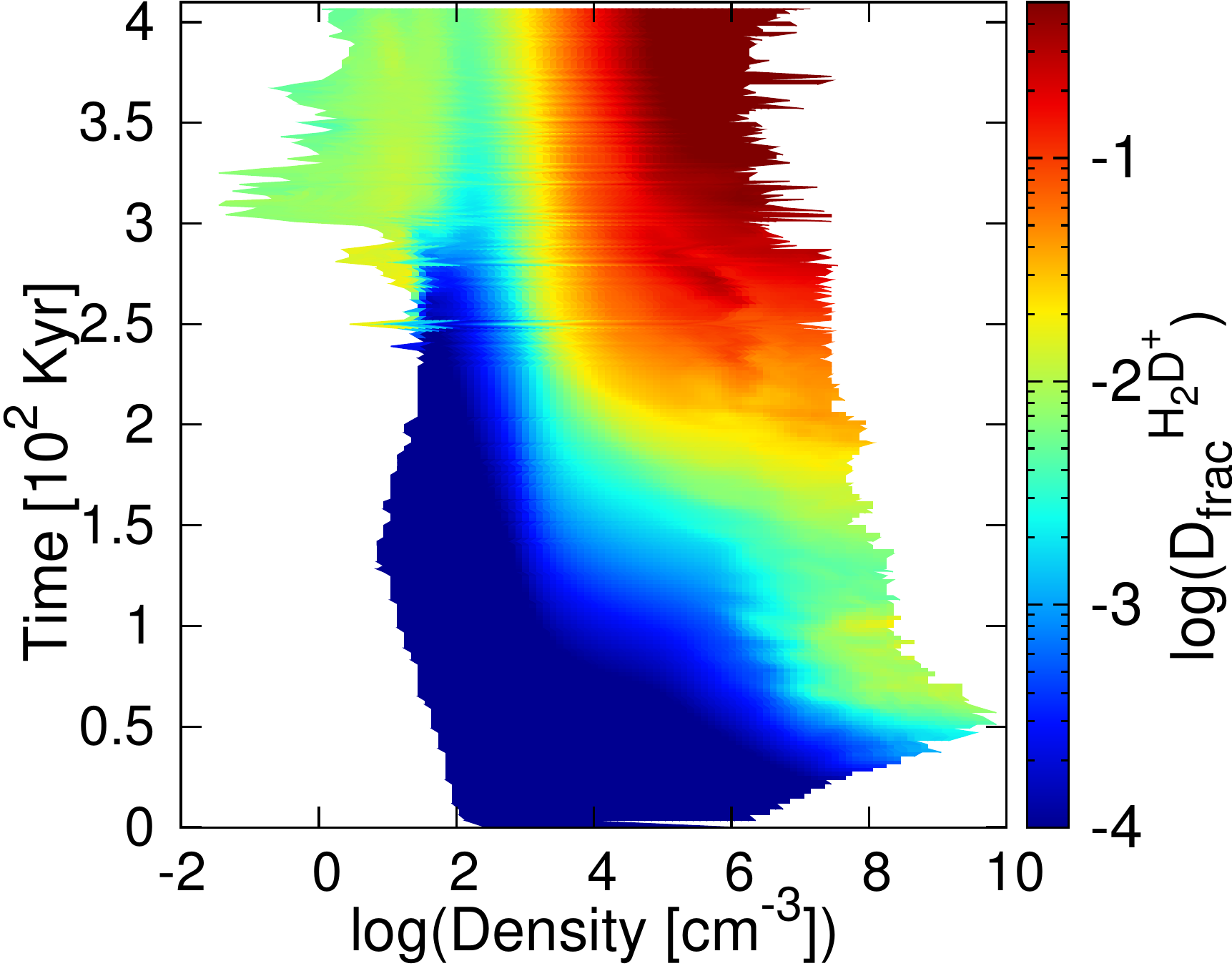} &\includegraphics[width=0.33\textwidth,angle=-90]{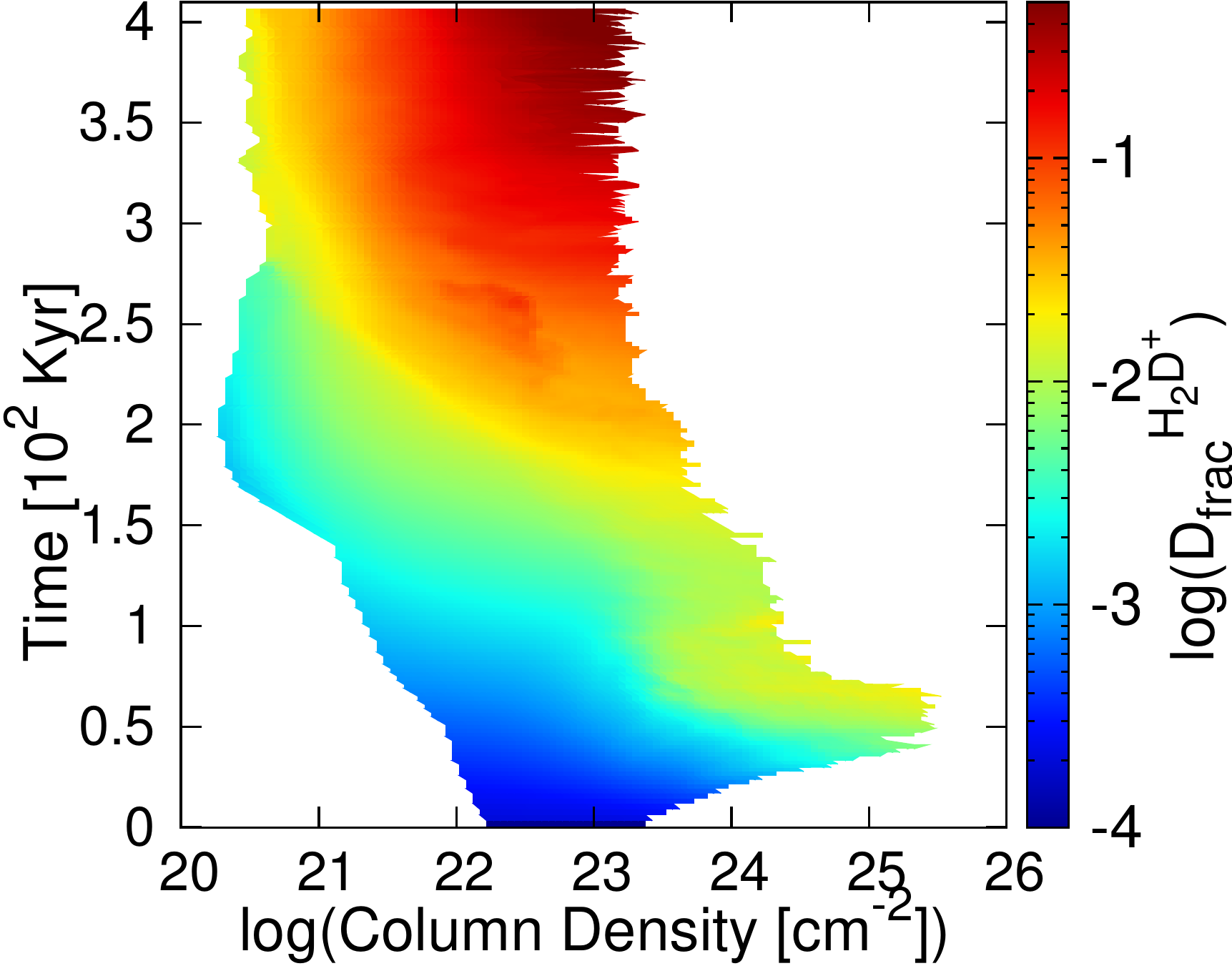} \\
		\rotatebox[origin=r]{90}{\Large{Lmu10M2OPR0.1}\qquad\qquad}  &\includegraphics[width=0.33\textwidth,angle=-90]{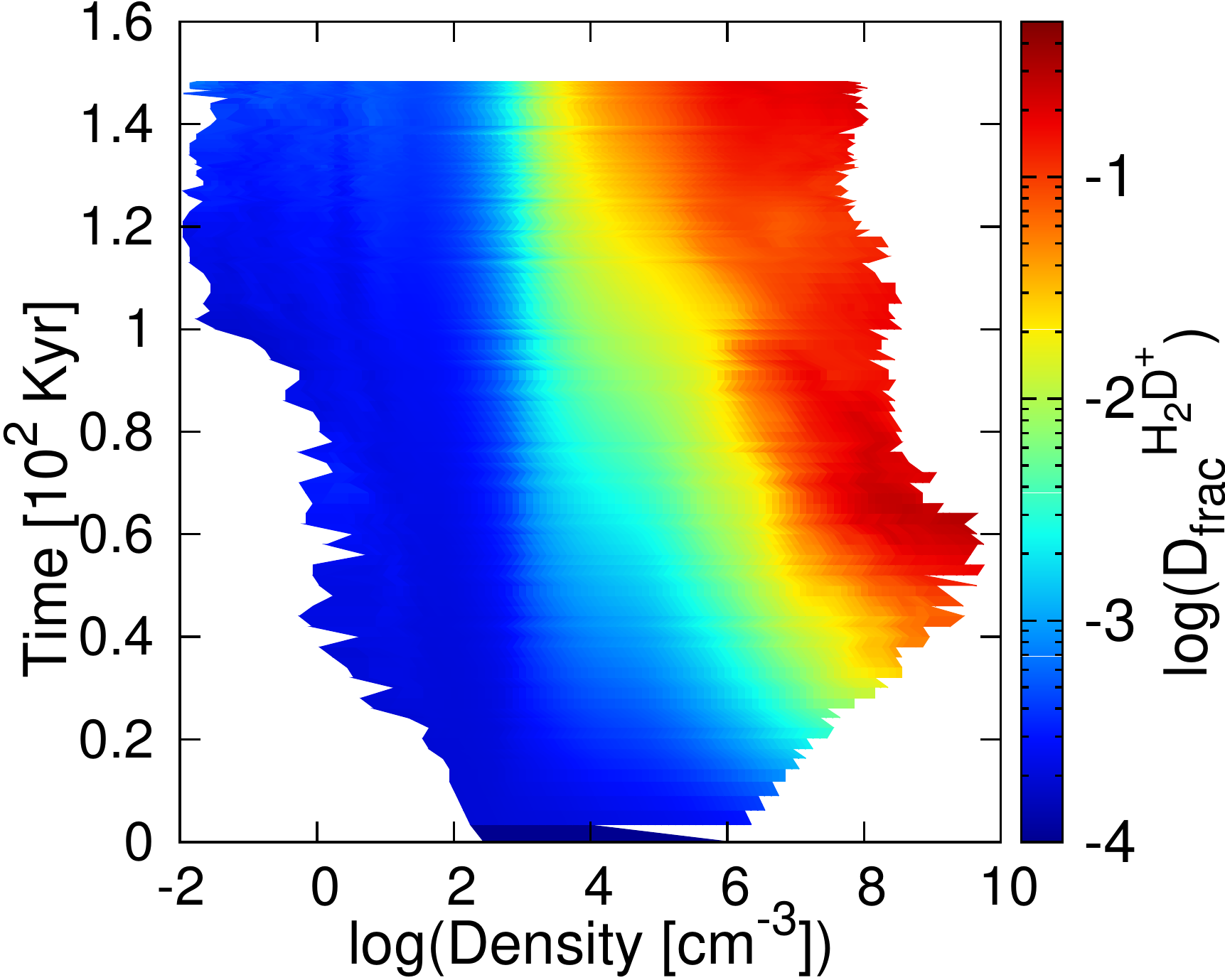} &\includegraphics[width=0.33\textwidth,angle=-90]{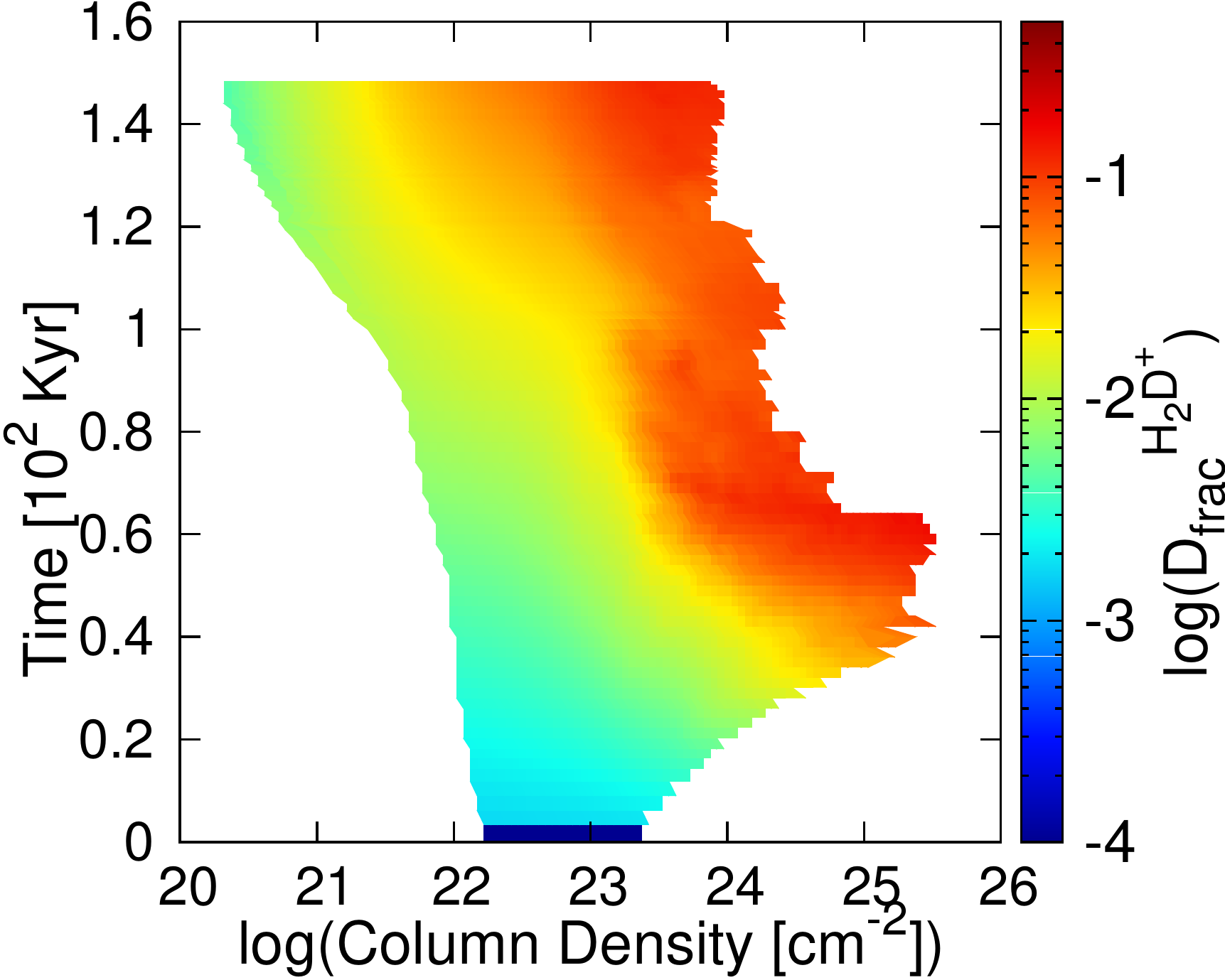} \\
		\end{tabular}
	\end{center}
	\caption{\ita{Left:} The mass--weighted deuterium fraction as function of time and volume density for selected runs. \ita{Right:} Column--density weighted deuterium fraction as function of time and total gas column--density for the same runs. Note that values below $n\sim3\times10^4\,\mathrm{cm}^{-3}$ are biased 
	as in this regime the assumption of complete depletion fails.}
	\label{figBasti9}
\end{figure*}

\begin{figure*}
	\begin{center}
		\begin{tabular}{clr}
		\rotatebox[origin=r]{90}{\Large{Lmu2.5M0.5}\qquad\qquad}  &\includegraphics[width=0.33\textwidth,angle=-90]{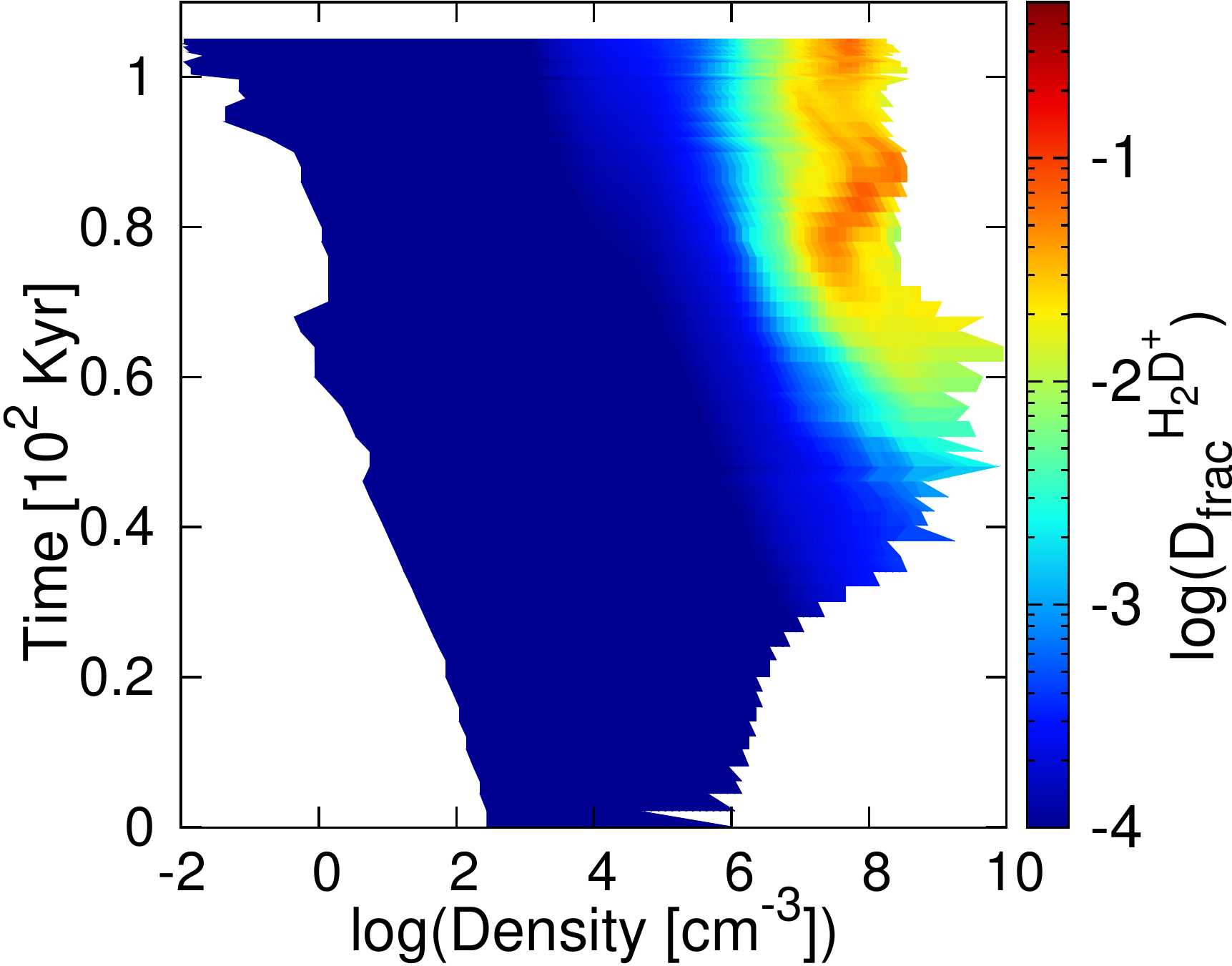} &\includegraphics[width=0.33\textwidth,angle=-90]{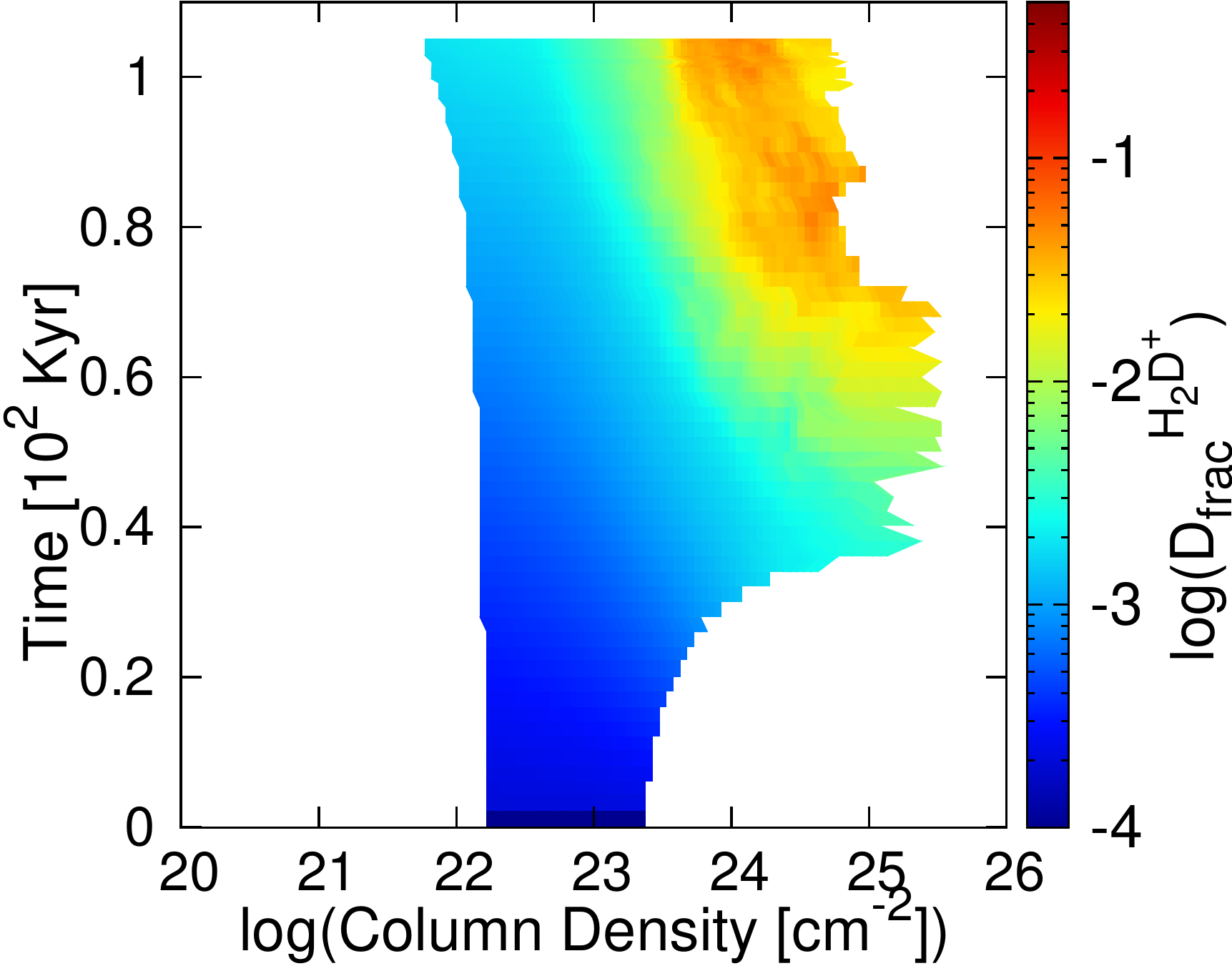}\\
		\rotatebox[origin=r]{90}{\Large{Hmu10M2}\qquad\qquad}  &\includegraphics[width=0.33\textwidth,angle=-90]{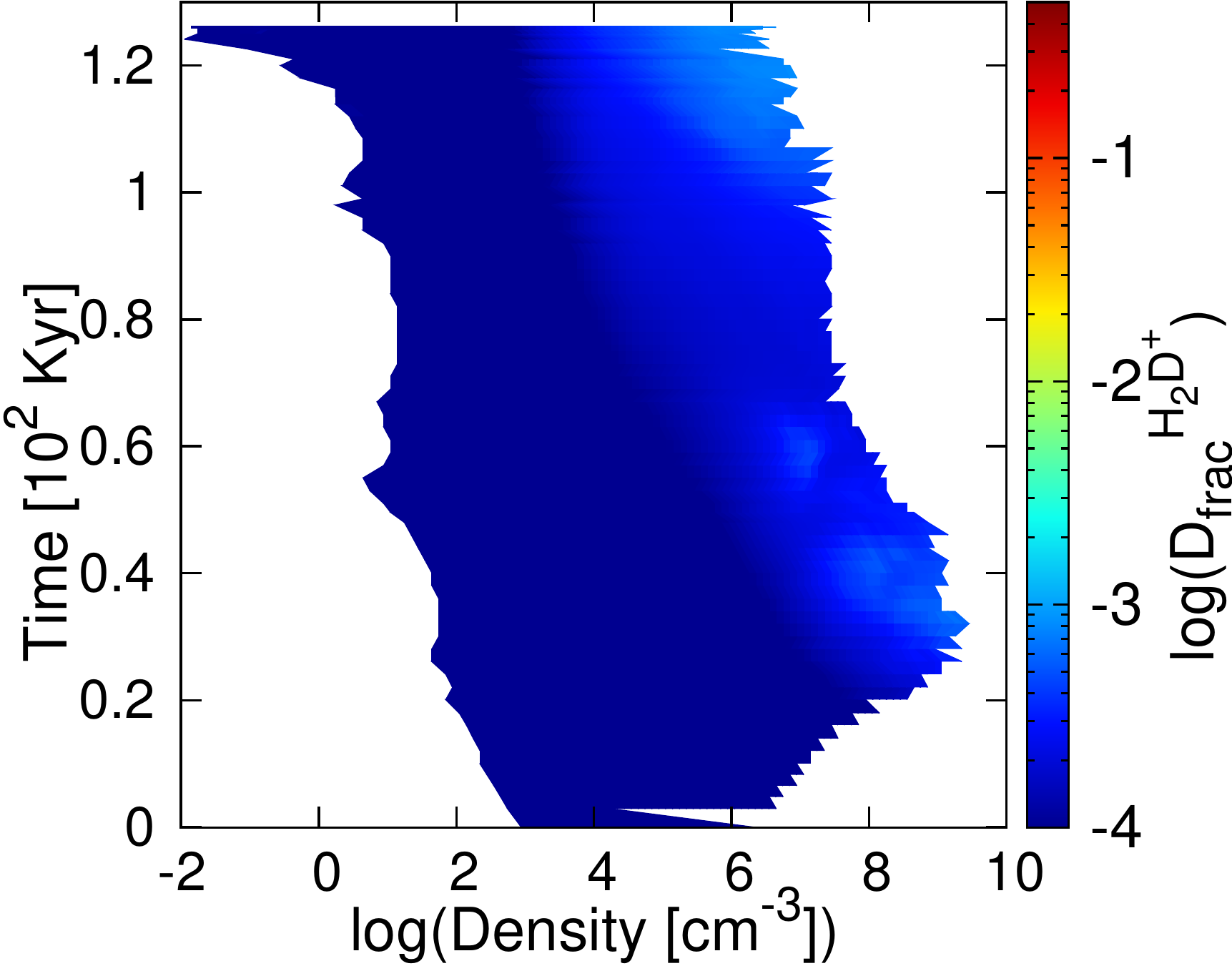} &\includegraphics[width=0.33\textwidth,angle=-90]{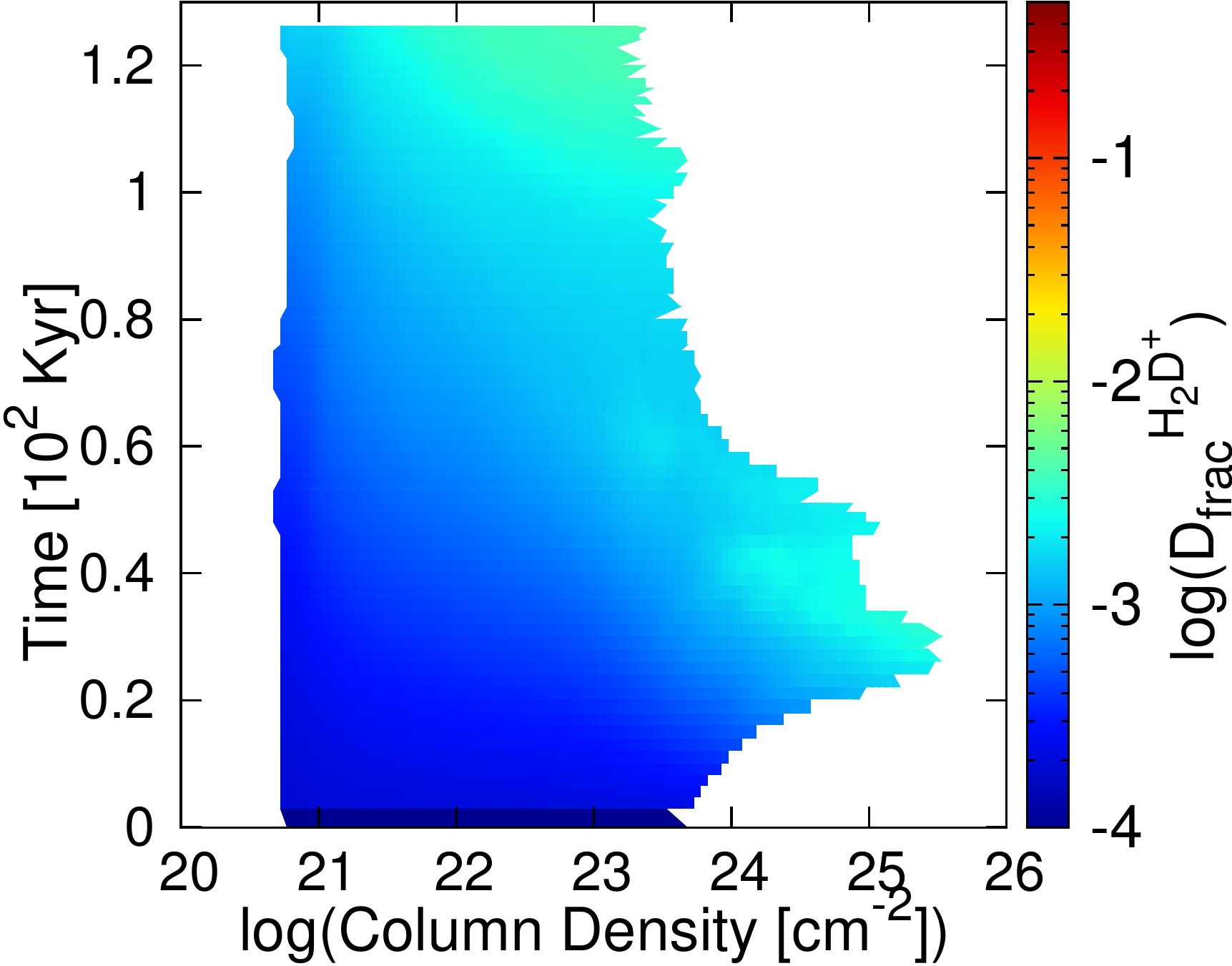} \\
		\end{tabular}
	\end{center}
	\caption{Same as Fig. \ref{figBasti9} for runs Lmu2.5M0.5 and Hmu10M2.}
	\label{figBasti10}
\end{figure*}

For the same runs, we now discuss the column density of \mbox{o--H$_2$D$^+$}, \mbox{p--H$_2$D$^+$} and \mbox{D$_2$H$^+$} as a function of radius and time, as illustrated in Fig.~\ref{figBasti5}. As for the deuterium fraction in general, we see here again that the abundances are correlated both with the local density, as well as with time, where the second is more important on the long term, in the sense that also in the lower--density components, high deuterium fractions and the related column densities are eventually reached. The column densities in particular are even less sensitive to the accretion of mass onto the sink particle, as their main contribution apparently is not due to the dense component, which is most rapidly accreted, but relevant contributions are present from the lower and intermediate--density material. For Lmu10M2S2, where the density strongly decreases after sink formation, there is some decrease in these quantities at late times, while Lmu10M2, Lmu10M2OPR0.1, Lmu2.5M0.5 and Hmu10M2 show a rather smooth behavior, and otherwise reflect the features already discussed. In particular, for the case of Lmu10M2OPR0.1, we note again that high column densities of \mbox{o--H$_2$D$^+$}, \mbox{p--H$_2$D$^+$} and \mbox{D$_2$H$^+$}  are reached early on, after about 20,000~yrs and that these are in agreement with typical observed values.\\
To provide a complementary view, the same runs are analyzed showing the average deuterium fraction both as a function of density and time in the left panel of Figs.~\ref{figBasti9} and \ref{figBasti10}, and as a function of column density and time in the right panel. The figures show clearly that the deuterium fraction increases most rapidly at high densities, and then the region with high deuterium fraction moves outward towards the lower densities as a function of time. In case of Lmu10M2OPR0.1, it is clearly visible that higher deuterium fractions are reached for lower gas densities, confirming the overall enhancement of the deuteration process. Lmu10M2 and Lmu10M2S2 are rather similar, except that the second simulation has not been evolved as long, while Hmu10M2 has a lower deuterium fraction, due to the more efficient accretion onto the sink, which is reducing the high--density gas where deuteration would be efficient. We however note that also this run would eventually reach high deuterium fractions if evolved for longer. The plots showing deuterium fraction as a function of column density and time show roughly similar trends, thus indicating at least an approximate correlation between gas density and column density.



\subsection{Comparison at the time of sink particle formation}

While we have so far compared the simulations as a function of physical time, we also want to explore the chemical properties when comparing them in different simulations at the same evolutionary stage. A relatively clean way to do that in our simulation is when considering the time of sink particle formation. For this purpose, we have calculated the radial profiles of the OPR of H$_2$ and H$_2$D$^+$, the deuterium fraction, the column density of \mbox{D$_2$H$^+$} as well as the column density of the spin states of H$_2$ and H$_2$D$^+$, for the simulations Lmu10M2, Lmu5M2, Mmu10M2, Hmu10M2 and Mmu5M2. This is to predominantly explore the dependence on dynamical parameters, while before we also analyzed the dependence on chemical conditions (the dependence on which is trivial).
In all cases, the OPR of H$_2$ is high on large scales, while it drops to lower values within the central 1,000~au in those simulations where the sink particle forms later. 
A very similar behavior is visible also for the OPR of H$_2$D$^+$. The deuterium fraction, on the other hand, is increasing towards the center, and shows the highest value in those simulations where the sink particle forms later. This is reflected also in the D$_2$H$^+$ column densities.
Regarding the column density of the spin states of H$_2$, it is clearly visible that the column density of the para state is basically independent of radius, while the ortho component is decreasing with decreasing radius. The latter thus naturally explains the decrease in its OPR. The decrease again is stronger for simulations where more time has passed. For the spin states of H$_2$D$^+$, the para column density is increasing towards the center, and more strongly in simulations where more time has passed before sink formation. The ortho column density shows a slightly decreasing trend with decreasing radius, but at least on large scales, the column density is somewhat higher in cases where the sink particle formed later. Overall, the para column density changes however more strongly, which gives rise to the dominant effect for the OPR. \\
Overall, one could be tempted to infer some trends, in particular that the deuterium fraction is higher if it took longer for the sink particle to form, i.e. if the collapse has been delayed for instance by turbulence or the magnetic field. 
Under the chemical parameters assumed in this study and motivated by some observational results, we could infer that a high level of $D_\mathrm{frac}$ may be reached well within one dynamical time despite the presence of additional support from turbulence or magnetic fields. The latter can, on the other hand, slow down the collapse and provide more time to build up the deuterium fraction. One has to be cautious when investigating such trends though, because the deuterium fraction depends strongly also on the initial OPR of H$_2$, as well as additional parameters not included in this study,  such as the cosmic ray ionization rate or the depletion factor of the heavy elements. It is thus in principle unclear whether deuteration can be used as a chemical clock.
However, while in particular a comparison of different star--forming regions might be difficult, one could hope that the conditions are relatively similar within one star--forming complex, implying similar or the same initial abundances, similar OPR and similar cosmic ray ionization rates. These assumptions should be tested in future studies, both observationally and with simulations, and if it is likely that the chemical initial conditions of the clumps are relatively uniform, one could tentatively employ deuteration as an approximate chemical clock, at least within one star--forming region. It is however clear that the latter still requires further investigation.

\begin{figure*}
	\begin{center}
		\begin{tabular}{ccc}
		\includegraphics[width=0.28\textwidth,angle=-90]{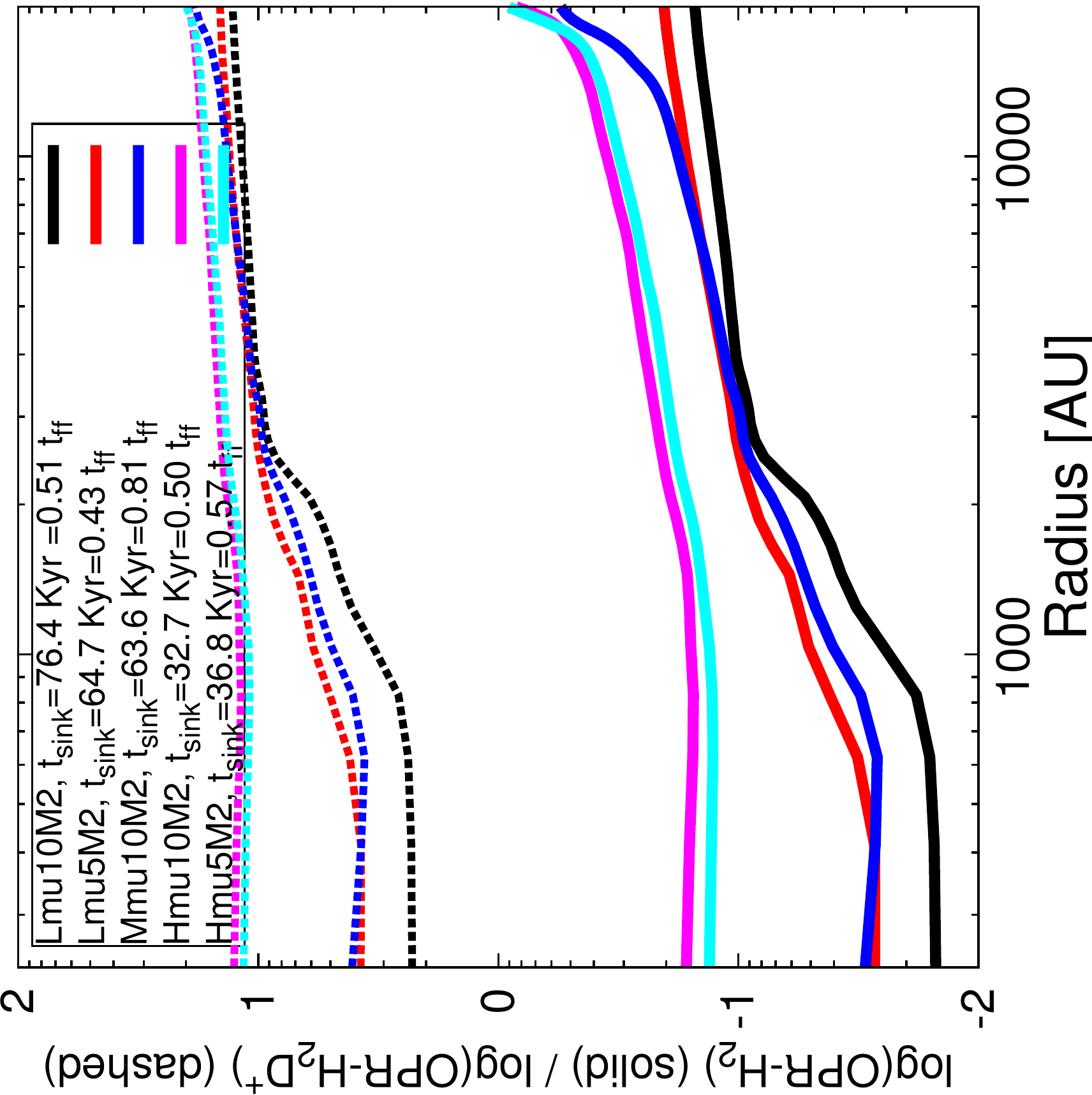} &\includegraphics[width=0.28\textwidth,angle=-90]{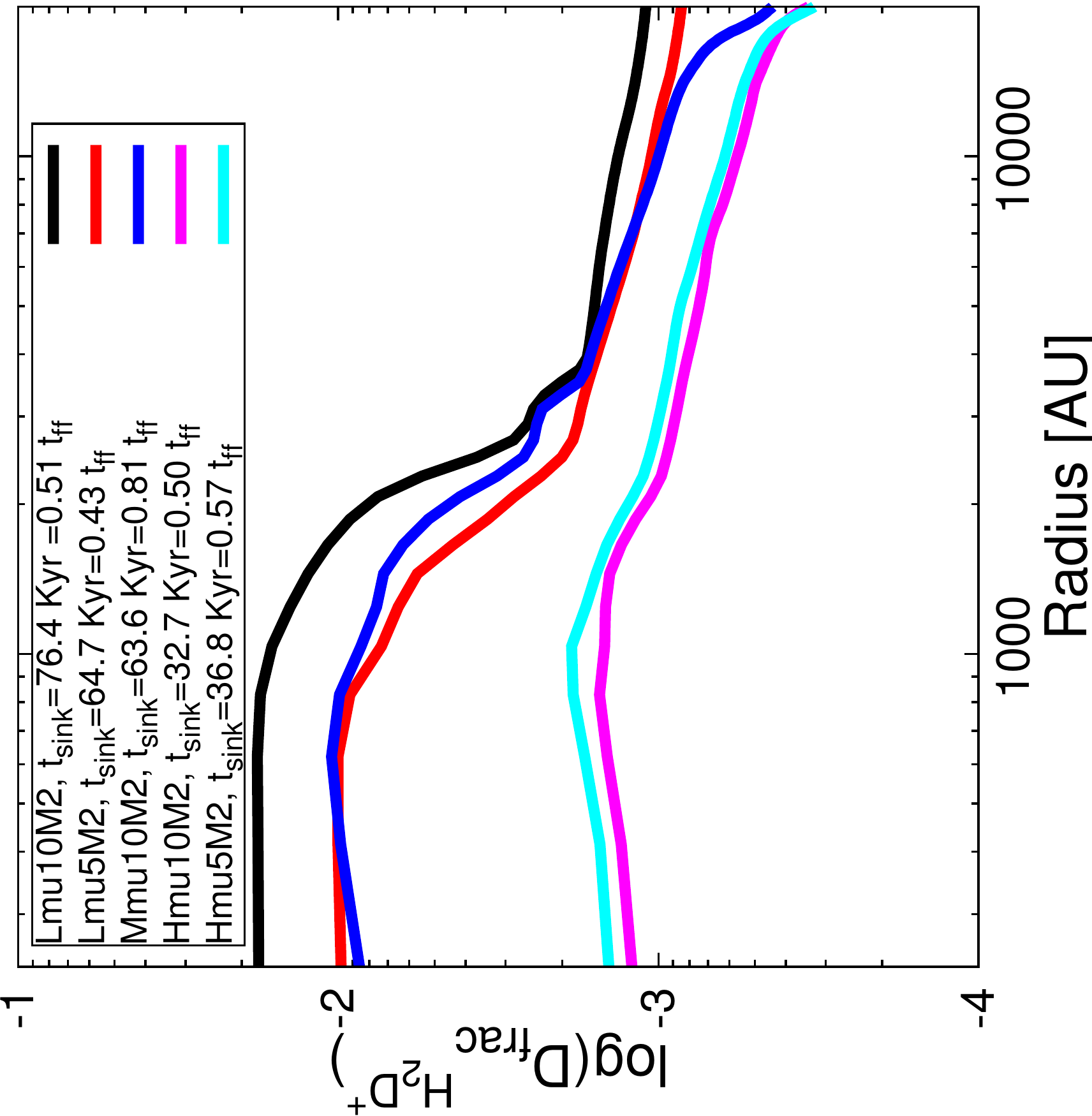} &\includegraphics[width=0.28\textwidth,angle=-90]{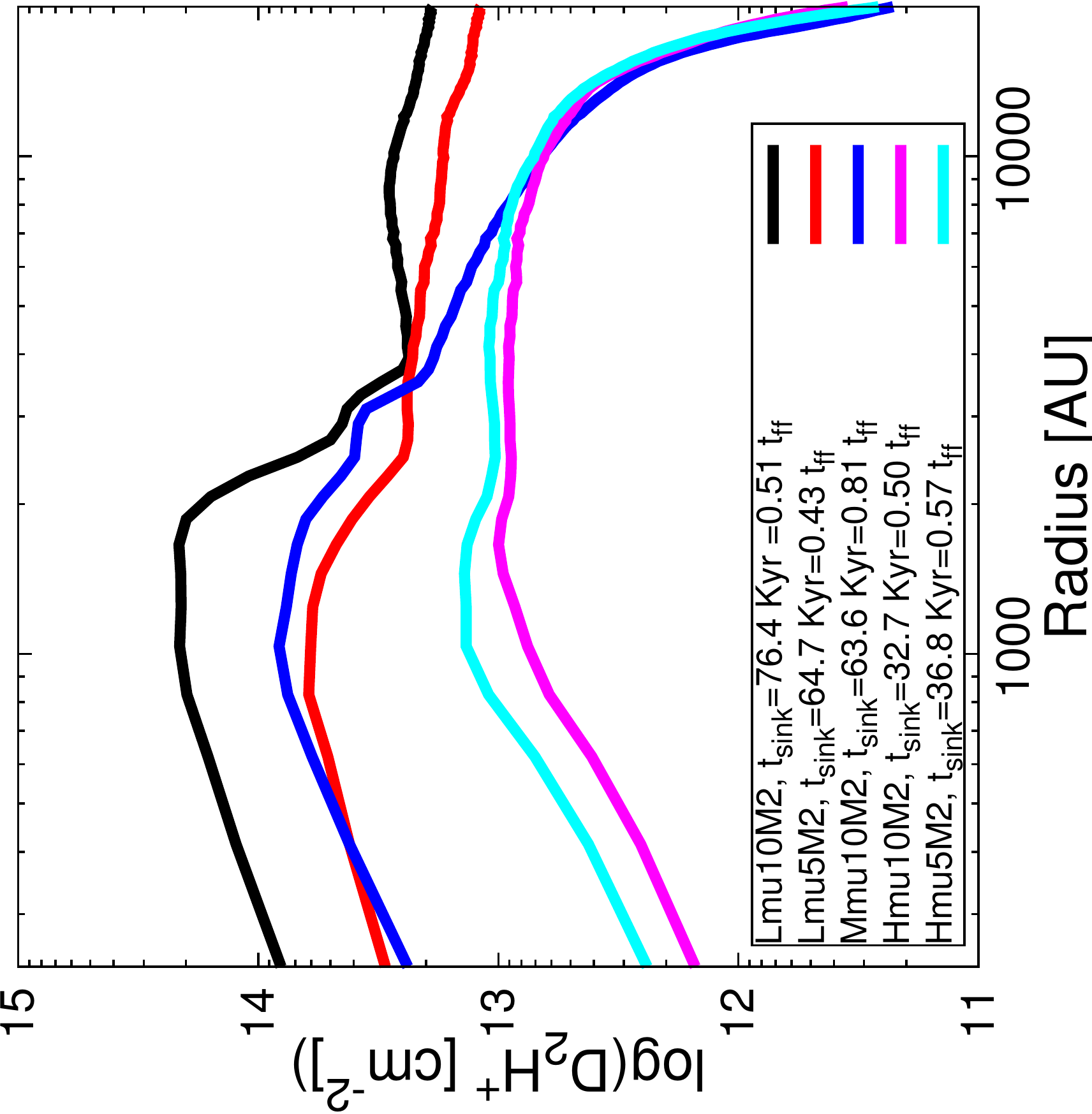} \\
		\includegraphics[width=0.28\textwidth,angle=-90]{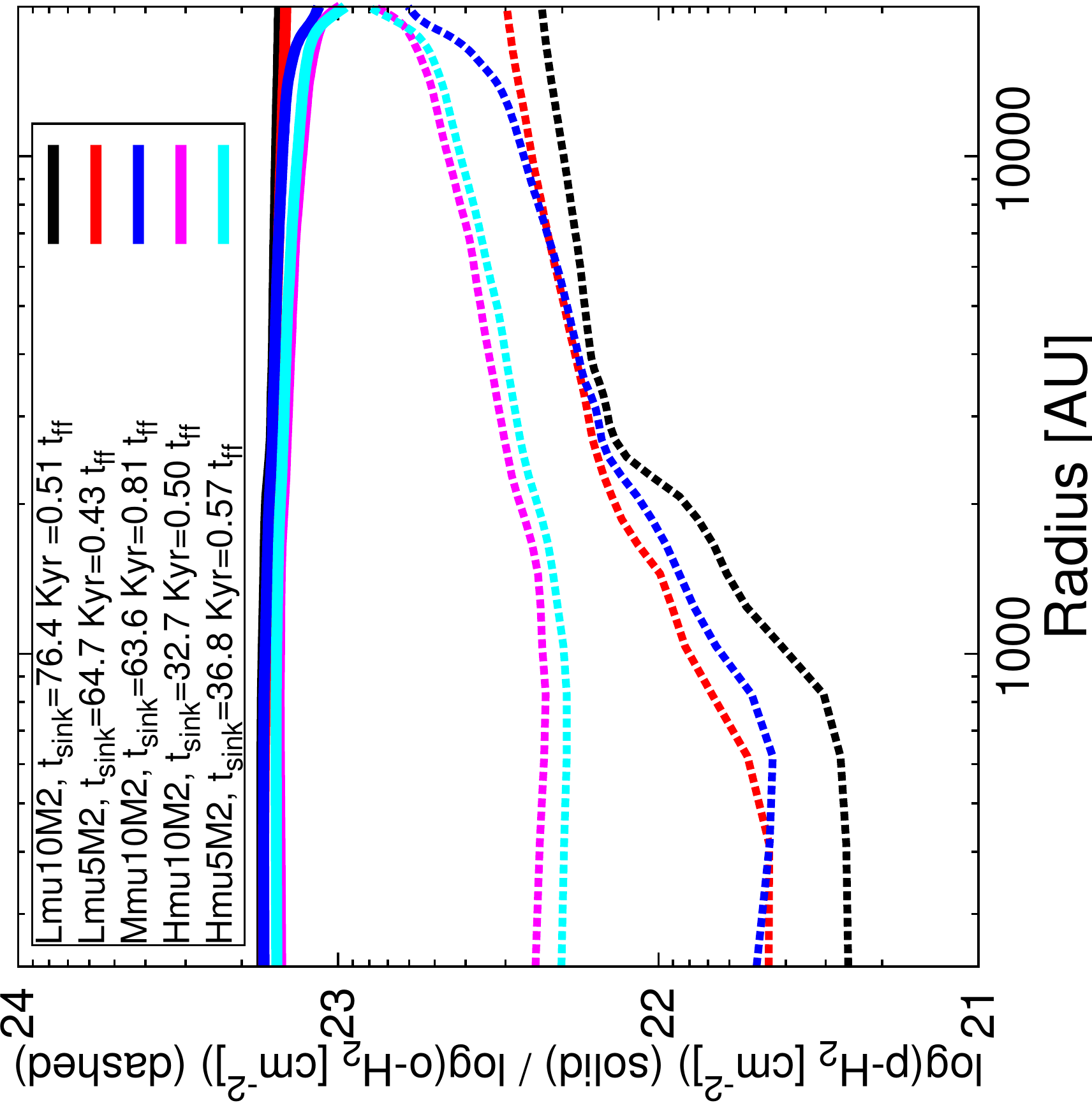} &&\includegraphics[width=0.28\textwidth,angle=-90]{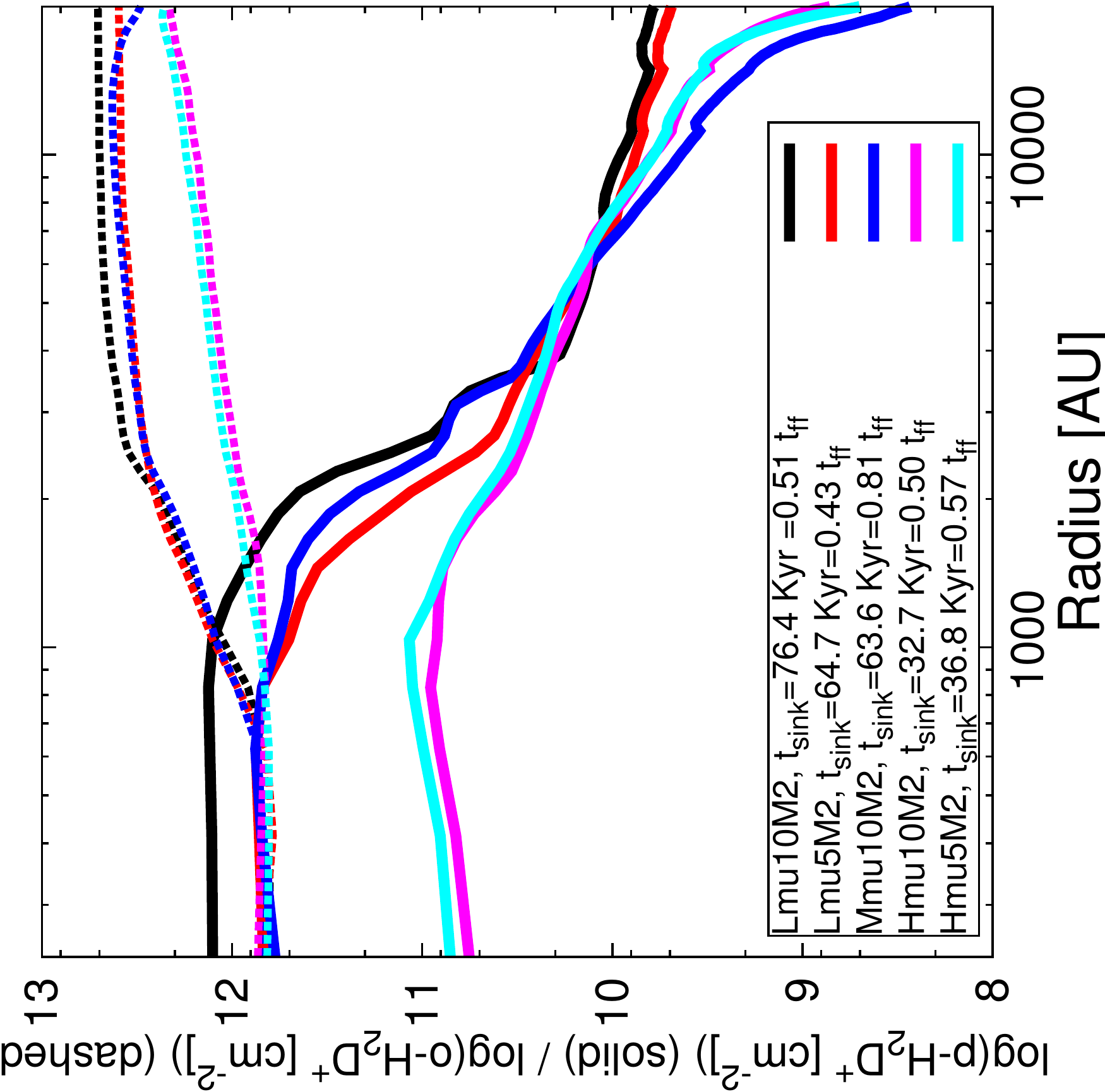} 
		\end{tabular}
	\end{center}
	\caption{Radial profiles of the OPR of H$_2$ and H$_2$D$^+$ (top left), the deuterium fraction (top middle), the column density of D$_2$H$^+$ as well as the column density of the spin states of H$_2$ (bottom left) and of H$_2$D$^+$ (bottom right). 
	Data is taken at the time when the sink particle has formed as the cores are then in a similar evolutionary state. The respective formation time is given in the figure legend in absolute and relative timescales. For simplicity we only show results from simulations having the same initial turbulent Mach number.}\label{evolution}
\end{figure*}

\subsection{Comparison of observed and modeled abundances}
\subsubsection{Observational constraints on the depletion in high--mass cores}
In the current setup, we are mainly concerned with the assumption that in the early stages of the star formation process the gas experiences complete depletion of heavy elements (depletion factor $f_D$ larger than 10).
Before comparing the model results with observations it is fundamental to understand the uncertainties  of the deuterium fractionation measurements and of the CO depletion factor, crucial for the $[\mathrm{D}/\mathrm{H}]$ enhancement. In addition to the calibration uncertainty and rms noise of the spectra, other sources can contribute to, and even dominate, the uncertainty budget.
In the following we list the sources of uncertainties that need to be considered in the derivation of molecular- and dust column densities, $N_\mathrm{mol}$ and $N_\mathrm{dust}$, respectively.

\paragraph*{Expected abundance:} The CO depletion factor is observationally determined as the ratio of the expected and calculated abundances.
Galactocentric elemental gradients imply a variation of the expected abundance of CO at a given galactocentric radius \citep{Fontani2006}. Propagating the uncertainties on these gradients gives a scatter of a factor of $\sim2$ on the expected CO abundance \citep{Wilson1992,Luck2011}.

\paragraph*{Optical depth/isotopic ratios:} Lines of abundant species can have a non--negligible optical depth, which implies that the line flux does not scale linearly with column density. If not properly taken into account, the derived $N_\mathrm{mol}$ will only be a lower limit. An optical depth of $0.5$ implies that the column density derived under the assumption of optically thin emission is underestimated by $30\%$.
CO lines have very high optical depths towards regions of high--mass star formation and this may also be relevant for the main isotopologue in the calculation of the deuterium fraction, using high--resolution data.
Optically thin transitions from rarer isotopologues are used to estimate the column density of carbon monoxide, and adjusted using the isotopic abundance ratios; for commonly used isotopologues the scatter of these ratios is $\sim30-50\%$ \citep[][]{Milam05,Giannetti2014}.

\paragraph*{Molecular excitation:} Assuming a typical excitation temperature or local thermodynamic equilibrium (LTE) may have a significant influence on the derived column density; the more transitions are observed the better the constraints on the excitation and thus $N_\mathrm{mol}$. An increased uncertainty is also introduced when using temperatures derived with different tracers.
The impact of non--LTE effects or erroneous excitation estimates on the deuterium fraction and on $f_\mathrm{D}$ can reach a factor of $5$ in the very cold gas \citep[e.g.][]{Kong2016ApJ}.

\paragraph*{Contamination from the envelope:} Material along the line--of--sight and in the beam contributes to the observed line flux.
When deriving $[\mathrm{D}/\mathrm{H}]$ it is usually assumed that the two isotopologues are cospatial, i.e. their emission originates in the same gas, which is an oversimplification \citep[e.g.][]{Miettinen2011,Kong2016ApJ}. The main isotopologue has a larger contribution from material unrelated to the core, reducing the measured deuterium fraction. Attempts to correct for this effect have been performed, introducing an uncertainty of a factor of $\sim2$ \citep[e.g.][]{Kong2016ApJ}.
Similarly, the contamination from the lower--density external material to the molecular line emission implies that the measured depletion factors are lower limits. As an example, \citet{Giannetti2014} show that, in high--mass clumps with typical depletion factors in the range $3-15$, it is possible to reproduce the observed lines of CO isotopologues assuming that all carbon monoxide is depleted in regions with densities exceeding $\mathrm{few}\times10^{4}\,\mathrm{cm}^{-3}$. In this case the observed depletion factor is only a measure of the extent of the depleted and undepleted regions \citep{Pagani2011}.

\paragraph*{Isothermal gas and dust:} The assumption of a single temperature in the clump or core affects both the derivation of $N_\mathrm{mol}$ and $N_\mathrm{dust}$. This has a larger impact on sources in later evolutionary stages, where luminous young stellar objects (YSOs) are heating the material from the inside and generating strong temperature gradients. 

\paragraph*{Dust opacity:} To derive the dust column density from sub--mm continuum emission the dust opacity at the wavelength used is needed, which can differ by a factor of $2$ depending on the dust model used \citep{Hildebrand1983,Ossenkopf1994}. On the other hand, the ice coating has a lower impact on the opacity, $\sim30\%$ \citep[cf.][]{Ossenkopf1994}.

\begin{figure}
	\begin{center}
		\includegraphics[scale=0.4,angle=-90]{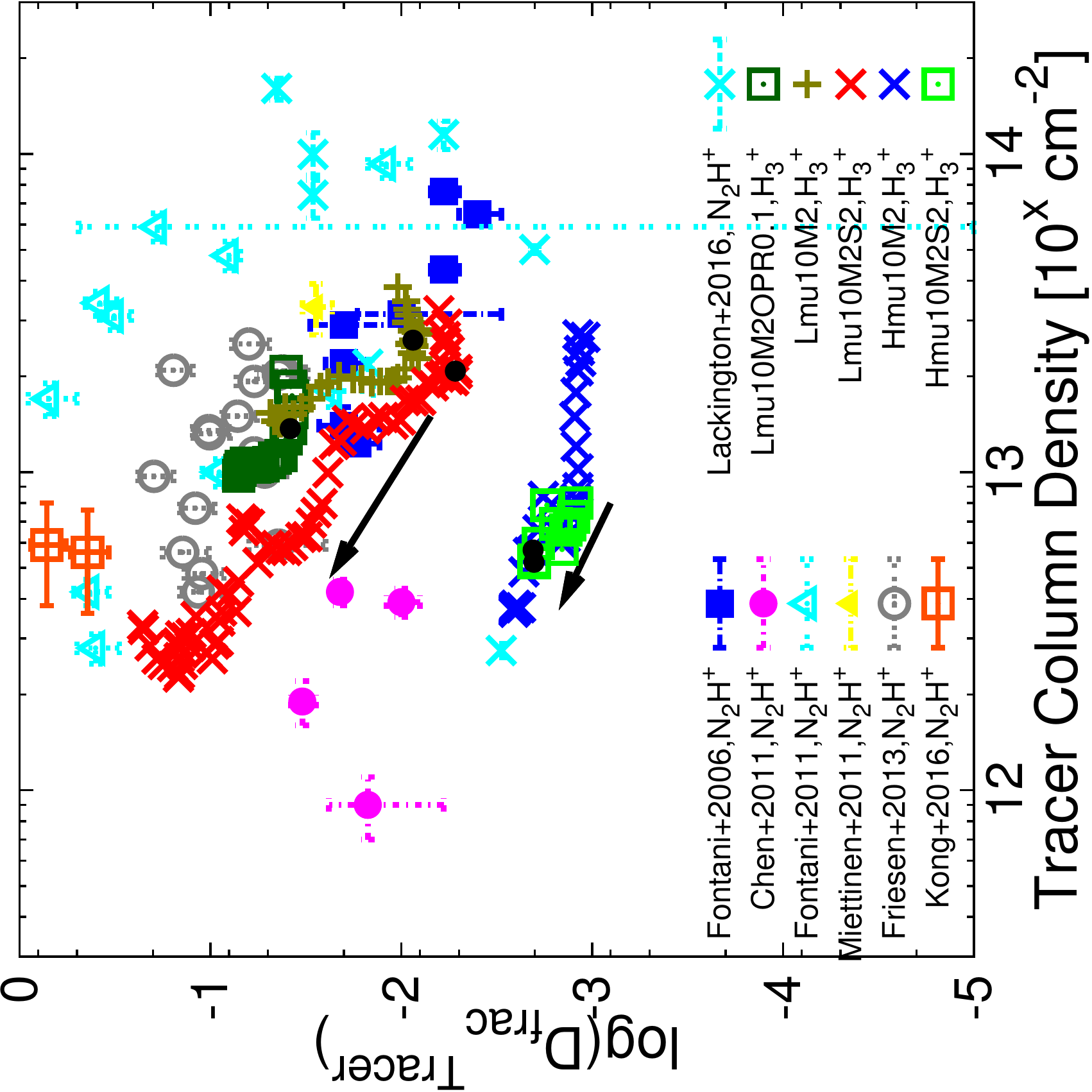}
		\caption{Comparison of selected simulations with observations of the deuterium fraction in high--mass and low--mass starless cores. Simulation data is only shown for times $t\geq 0.5\,t_\mathrm{ff}$. Observations are best matched by our simulations with initially low surface density of the core. Note that different observations are fitted by the same core at different evolutionary stages. The black arrows indicate the evolution in time. Black dots mark the column density and the value of 
		D$_\mathrm{frac}^{H_2D^+}$ at $t=113\,\mathrm{kyr}$, where run Hmu10M2S2 is terminated. Please note that we only show the data points by \citet[][]{Friesen13} that are detected both in N$_2$D$^+\left(3-2\right)$ and N$_2$D$^+\left(2-1\right)$. For a color version see 
		online manuscript.}
		\label{obs}
	\end{center}
	
\end{figure}

\paragraph*{Gas--to--dust ratio:} The dust column density is converted to $N_{\mathrm{H}_2}$ via the gas--to--dust ratio, that has a scatter of a factor of $2$ \citep[][]{Sodroski1994} around the local value of $\sim 100$.\\
Considering the sources of uncertainty described, we estimate that the overall uncertainties on the beam-- and line--of--sight averaged deuterium fraction and $f_\mathrm{D}$ are typically a factor of $\sim3-4$, up to an order of magnitude if the excitation is not well constrained.\\
While in our simulations the impact of freeze--out is not yet taken into account, our chemical network is well suited for high density cores ($n > 10^4\,\mathrm{cm}^{-3}$). We can then pursue a preliminary qualitative comparison of our results with observational data. For this purpose, we have compiled the available data from single--dish observations and interferometers in Table~\ref{obstab} for HMSCs and low--mass cores, and we plot them in Fig.~\ref{obs} along with the results from some representative simulations. These are in particular the models Lmu10M2, Lmu10M2S2, Lmu10M2OPR0.1, Hmu10M2 and Hmu10M2S2. The simulations are not plotted at a single time, but rather we calculated the average column density and the average deuterium fraction within the core at different times, and plot this sequence of points within the figure. The direction of time evolution is indicated by an arrow, showing that increasing time corresponds to decreasing column densities (due to collapse of the gas in the very central region onto the sink particle) and increasing deuterium fractions. We emphasise that the former effect 
may be due to insufficient accretion of material through the boundaries of the simulation volume and should thus be taken with caution.
We do however not suggest here that this interpretation can be generally applied, as the column density in some cores may have been higher or lower to begin with
-- though our low surface density runs do indeed agree with observational findings of high--mass quiescent clumps by \citet{Giannetti2013} --, and also the chemical evolution, as we saw before, depends strongly also on the OPR of H$_2$. Nevertheless, it will be useful to fill this diagram with observational data points as well as with results from more sophisticated numerical simulations in the future, to understand which parameter space is populated in reality and whether the same parameter space can be populated by simulations, and under what assumptions and conditions. 
Another point to caution on is the deuterium fraction itself, which we define here via H$_2$D$^+$, while many observational studies employ N$_2$D$^+$ as the main tracer (mainly due to the difficulties in observing H$_3^+$). It will be important to assess the differences between these tracers in future simulations employing more detailed networks.

	\begin{table*}
	\caption{Measurements of the deuterium fraction in regions of high--mass star formation. We include sources in the early stage of evolution where the estimates make use of N$_{2}$H$^{+}$ and its deuterated isotopologue, because their ratio is the most sensitive to the evolution according to \citet{Fontani2015}. We point 
	out that in light of beam and line--of--sight contamination the high values of the depletion factor in this table suggest that on a smaller scale CO is even less abundant, which is confirmed by higher--resolution observations \citep[][and references therein]{Zhang2015}.}
		\begin{tabular}{llllcc}
			\hline
			Ref. & $D_\mathrm{frac}$ & Objects  & $M_\mathrm{core} (M_\odot$) & $f_D$ & Instrument\\
			\hline
			Single-Dish     & & & & \\
			\hline
			\citet{Fontani2006}    & 0.006-0.02      & IRAS sources & 30-290 & 0.7-35.8 & IRAM\\		
			\citet{Fontani2011}    & 0.012-0.7       & HMSC in IRDCs & - & - & IRAM\\					%
			\citet{Miettinen2011}  & 0.01-0.028      & IRDCs  & 63-183 & 0.8-2.7 & APEX\\				%
			\citet{Fontani2012}    & 0.003           & IRDCs & 55-3280 & 5-78 & APEX\\						%
			\citet{Pineda2013}     & 0.004-0.5       & NGC 2264-D & - & - &APEX\\						%
			\citet{Gerner2015}     & 0.03-0.15       & IRDC18151, IRDC18223 & - & - &  SMT\\			%
			\citet{Kong2016MNRAS}     & 0.02-0.09       & IRDC G035.39-00.33 & 70 & 1.2-2.4 & IRAM\\		
			\citet{Lackington2016} & 0.003-0.14      & IRDCs & - & - & APEX\\							%
			\hline
			Interferometer  & & & & \\
			\hline
			\citet{Chen2010}       & 0.051           & IRDC G28.34 & 36 & - & SMA/SMT \\				%
			\citet{Chen2011}       & 0.02-0.11       & IRDCs & - & 6-16 & SMA/SMT\\					%
			\citet{Kong2016ApJ}      & 0.081-0.16-0.32 & IRDC G028.37+00.07  & 16 &  & ALMA\\			%
			\citet{Kong2016ApJ}      & 0.075-0.15-0.30 & IRDC G028.37+00.07 &62 &  & ALMA\\				%
			\hline
\label{obstab}
		\end{tabular}
\end{table*}

\section{Discussion and conclusions}\label{summary}
In the following, we will discuss the main results found in this study, discuss their limitations and the underlying assumptions, and provide a preliminary assessment whether deuterium fractions could serve as chemical clocks, indicating potential directions for future improvements. We further discuss how a better understanding of the deuteration processes could be achieved on the long term.

\subsection{Main results}
The main goal of our study was to assess how deuteration occurs in massive cores and to explore how this process is affected by the dynamical evolution, including the dependence on key parameters such as the turbulent Mach number, the gas column density, the mass--to--flux ratio as well as the distribution of the magnetic field within the core. For this purpose, we have performed 3D magneto--hydrodynamical simulations fully coupled for the first time with a chemical network \citep{Walmsley2004} that describes deuteration chemistry under the assumption of complete depletion.\\
In all simulations, we do see that deuteration increases with time, even though this may be temporarily affected by the dynamical evolution, as the dynamics of the collapse bring substantial amounts of gas into the central region of the core, thus reducing the density in the ambient medium and initially also decreasing the deuterium fraction. However, the deuteration process does catch up again with time, and large deuterium fractions are reached within one dynamical time in all of our models within the central 10,000~AU. This process is of course faster when the OPR is low, i.e. about 0.1, while the majority of our simulations were performed with a conservative ratio of 3.\\
Considering column--density weighted radial profiles, we find that the turbulent Mach number introduces a characteristic peak in the deuterium fraction, as the turbulence leads to the formation of a central turbulent core during the collapse with roughly homogeneous density. The highest density peak is typically offset from the center of the cloud, which we define through the peak in the column density, and also the deuterium fraction is typically highest around that point. For larger turbulent Mach numbers, these peaks of the deuterium fractions tend to occur on larger scales, due to the increasing size of the inner turbulent core.\\
We further investigated the effect of the gas surface density on the deuterium fraction, finding that it can lead to some differences in the time evolution, as cores with different column densities typically have different virial ratios and different free--fall times. Cores with shorter free--fall times typically form more compact central structures, and the peak in the deuterium fraction is thus close to the interior. The impact on the mean deuterium fraction is however rather small.\\
In addition, we investigated the impact of the mass--to--flux ratio, exploring values for $\mu/\mu_{\rm crit}$ of 10, 5 and 2.5 (going from several $\mu$G to 0.3 mG). While initially we find very similar results in all cases, i.e. hardly an effect on deuteration, the effect becomes more prominent at late times when the magnetic field has had more time to affect the dynamical evolution. In particular for cores where the virial parameter is high, a relevant mass--to--flux ratio can substantially delay the collapse, therefore delaying the build--up of higher--density structures and delay deuteration within the core. In the simulations pursued here, which were  supercritical to begin with, this effect amounts to about a factor of 2 and makes thus no strong difference. We also found that the distribution of the magnetic field makes no strong difference, as long as the mass--to--flux ratio is the same. In case of sub--critical cores, which we have not simulated here, even more time could be available to build up high deuterium fractions, so this might be further enhanced. This effect might be relevant in cores that are not fully depleted, and where deuteration could thus occur on a longer timescale. It is worth noting here that the evolution of highly magnetized cores might be regulated by ambipolar diffusion at high densities rather than free--fall, the former being usually longer. In cases of ambipolar diffusion regulated collapse, any conclusive statement about the 
age of the cores should then be drawn in terms of the ambipolar diffusion timescale.\\
A potentially important aspect that became clear in our analysis is that, while the deuterium fraction is not a monothonic function of time, for instance due to the dynamics of the collapse and the accretion of dense gas onto the sink particle, the gas column densities of o--H$_2$D$^+$ and p--H$_2$D$^+$ are much more robust in this respect, as also the lower and intermediate density gas has significant contributions, and they are less sensitive to dynamical changes like the accretion of gas onto the sink particles. It could then be an important point of future investigation whether these might be more reliable tracers of the time evolution.\\
What we further explored was the dependence of the deuterium fraction on density and time, as well as column density and time, finding (not unexpectedly) that deuteration occurs faster for the higher--density gas. While this may appear a trivial result, one may consider to improve the usage of deuteration measurements as chemical clocks by taking the local density or column density into account, thus allowing a more accurate interpretation of the results.\\
When comparing the deuterium fraction in our simulations at a fixed evolutionary stage, i.e. when the sink particle forms, we find a clear trend that the deuterium fraction correlates with the amount of time until sink particle formation. This is not unexpected, as the chemical initial conditions in these simulations were the same. Taking these into account may lead to a larger scatter and it is thus important to further explore which degree of uniformity in the chemical initial conditions can be expected.\\
Finally, as a preliminary comparison of our results with real observations, we have plotted deuterium fraction vs gas column density for a sample of observational data, as well as the average values at different times from a representative number of our simulations. Despite the dependence on the initial conditions of the cores, in particular their column density, as well as chemical initial conditions like the OPR or the depletion factor, it is nevertheless important to populate such a diagram to define the parameter space covered by observed astrophysical cores, and to explore whether the same parameter space is consistent with numerical simulations. At least within the presently available data and the assumptions made in our simulations, they currently appear broadly consistent. 

\subsection{Limitations and uncertainties}
After describing the main results, we now describe the main limitations and assumptions within our model.

\subsubsection{Chemical model: caveats}
Our main goal in this work is to study the effect of dynamical quantities, such as magnetic fields and turbulence, on the evolution of H$_2$D$^+$ and deuterium fractionation. For this reason we decided to employ a reduced but widely used network from the literature, reported by \citet{Walmsley2004} and presented in the previous sections.
The latter has been developed to follow the evolution of deuterated species in the absence of heavy elements. Over the years it has been employed in many studies as a basic compilation of the most relevant reactions for such studies \citep[e.g.][]{Vastel2006,Sipila2010,Pagani2013,Kong2015}. In the following we discuss some of the uncertainties of the model. 
The model by \citet{Walmsley2004} does not include the nuclear spin modifications of D$_2$, D$_2$H$^+$, and D$_3^+$ which has been extended by \citet{Flower2004} in a subsequent work.  
The most relevant uncertainties are connected to two important reaction rates: 
\mbox{H$_3^+$ + H$_2$}, recently updated by \citet{Hugo2009} and the H$_3^+$ dissociative recombination rate coefficient \citep{Pagani2009}. \citet{Sipila2010} presented a quantitative study of the uncertainties produced by employing different rates for these two relevant processes. They reported an increase in the steady--state deuteration of H$_3^+$ when using the new rate by \citet{Hugo2009} instead of the rate reported by
 \citet{Walmsley2004} and \citet{Flower2004} and adopted from the experiment by \citet{Gerlich2002}. The former experimental results differ by a factor of $\sim$4 compared to the new theoretical/experimental work by \citet{Hugo2009}; this discrepancy is mainly produced by the backward reaction \mbox{H$_2$D$^+$ + H$_2$} which forms H$_3^+$. Overall they reported a lower steady--state H$_2$D$^+$ OPR, and a slightly higher D$_2$H$^+$ para--to--ortho ratio than predicted by using the coefficients adopted in \citet{Flower2004}.
Additional improvements also aimed at exploring uncertainties related to reactions involving metal--species have been reported by \citet{Vastel2012}, \citet{Sipila2013}, \citet{Sipila2015}, and \citet{Kong2015}, but those detailed networks are far from the scope of this study.\\
To conclude, despite the above uncertainties, and considering that in our simulations we are far from the equilibrium, our network can still be considered accurate enough to study the effect of dynamics on the deuteration and allow us to explore in detail the formation of H$_2$D$^+$. 

\subsubsection{Resolution and sink particles}
Clearly, the coarse resolution of our numerical model prevents us from studying the fragmentation of the collapsing core in more detail. However, as already 
stated above, the choice of the resolution was a compromise between the amount of detail and the desire to simulate a much longer time, as a finer resolution will also imply a lower timestep.  Furthermore, to ensure that we always fully resolve the Jeans length in our simulation and to avoid numerical artifacts like artificial fragmentation, we introduced sink particles to our model. These sink particles represent the no--longer resolved gas which is in a state of collapse, and accrete further high--density gas from their surrounding. As the accreted gas is no longer on the grid, the latter affects the average deuterium fractions and the derived column densities in the region close to the sink particle, and in that respect, the values we give here could thus tentatively be considered as a lower limit. Especially for the column densities of the different species, we note that the effect is likely less pronounced, as their main contributions are due to the lower--density and intermediate--density gas.

{\subsubsection{Heavy element depletion}
Our chemical model is based on the assumption of full depletion. Our modeled cores have average densities of \mbox{$\left<n\right>_\mathrm{core}=1.3\times10^4-3.0\times10^5\,\mathrm{cm}^{-3}$}. Following \citet[][]{Caselli1999}, the CO depletion timescale can be expressed as a function of the H$_2$  number density \mbox{$t_\mathrm{dep}\sim10^9/\left(Sn\left(\mathrm{H}_2\right)\right)\,\mathrm{yr}$}, where $S$ is the sticking coefficient, which we assume to be unity. For the core densities in this study, this estimate yields $t_\mathrm{dep}\sim100-10\,\mathrm{kyrs}$, which is less than the turbulent crossing time in the cores. Hence, depletion is fast compared to possible processes which act to dilute the core. Even if compared to the free--fall time of the core, the depletion proceeds faster, further arguing in favor of our assumption. In addition, we have argued that observed CO depletion factors are only a lower 
limit due to the large uncertainty in the observations. Simulations by \citet{Hocuk14} for instance find CO depletion factors of $\sim100-1000$ at gas densities of $n\sim10^4-10^5\,\mathrm{cm}^{-3}$ that are consistent with our assumptions.\\
Nevertheless, models with incomplete depletion will be important to assess the deuteration at lower densities, where a somewhat enhanced deuterium fraction may already build up, and where the OPR may start to decrease during the chemical evolution. In that sense, the full depletion assumption effectively poses a constraint on our initial conditions, which we hope to relax in future investigations.

\subsubsection{Explored parameter space}
While we have been able to explore a set of parameters, including the core mass, the turbulent Mach number, the mass--to--flux ratio, the distribution of the magnetic field and the OPR, the sample we could explore was certainly not exhaustive, as each individual simulation requires a significant amount of computing time. In future investigations, it is important to further explore the dependence on dynamical parameters, including a larger range of core masses from high--mass to the low--mass cores, but also to explore more extreme cases of the mass--to--flux ratio, with values close to unity. Such simulations are numerically challenging due to the lower timestep implied by the strong magnetic field, and will require to take additional processes like ambipolar diffusion into account. Nevertheless, they should be part of a complete exploration. 
In addition, also the dependence on chemical initial conditions requires further investigation. While we explored here the impact of the OPR, additional important parameters are the cosmic--ray ionization degree and the depletion factor of the gas. Only with such simulations, we can for instance assess the impact of incomplete depletion and address how much chemical evolution occurs already in the lower--density material. This is important, as it could contribute to more uniform chemical initial conditions, which will be important for the interpretation of the deuterium fraction. On the long term, it will be relevant to explore the dynamical formation of cores within a filament, leading to a natural prediction for the chemical initial conditions, which may give us further clues on this point.

\subsection{Deuterium fractionation - A chemical clock?}
It is often discussed in the literature whether deuteration processes can serve as a chemical clock, due to their strong dependence on time, which we also find in our simulations. While we believe that definite conclusions cannot be drawn at this point, we can state the following based on our simulation results:\\
In case that the chemical initial conditions are the same, our results indeed suggest that deuteration may act as a chemical clock, as the influence of the dynamical parameters like gas surface density, turbulent Mach number, mass--to--flux ratio and magnetic field distribution -- at least within the limits explored here -- have no strong impact on the increase of the deuterium fraction with time. The deuterium fraction is not a monothonic function of time, as in particular the dynamics of the collapse can strongly reduce the amount of gas on intermediate scales and bring a lot of gas to the center, thus temporarily decreasing the average fractions. We however do find that this does not strongly affect the observed column densities of the deuterated species, as their main contributions are due to lower and intermediate density material.\\ 
At the same time, based on our simulations with different OPRs as well as the work from other groups \citep[see e.g.][]{Goodson2016}, it is clear that the chemical initial conditions are highly relevant and may strongly influence how rapidly deuteration occurs. It is at least conceivable that large degeneracies exist when interpreting a certain deuterium fraction with respect to time and chemical initial condition. On the long term, one may try to relax this constraint by better understanding the chemical initial conditions, or one may try to minimize such variations by comparing cores from the same complex, with hopefully similar chemical properties.
In addition, based on our simulation results, we also suggest to explore more refined measures to determine chemical ages. 
Furthermore, cases with strong upper limits on the deuterium fraction may be particularly interesting, as that translates into an upper limit for the age of the core, in which chemical uncertainties can be taken into account. Such upper limits of course will be stronger in the presence of additional information.
On the contrary, if high deuterium fractions are measured, this may imply rather a lower limit on the respective age. Instead to consider deuteration as a strict chemical clock, we suggest to employ the tracer to infer upper and lower limits, and to employ additional data whenever possible to obtain more accurate results.\\
Two additional relevant questions have to be discussed in this respect: In these simulations, we employed H$_2$D$^+$ as the main tracer of deuterium fractions, while observational studies tend to prefer N$_2$D$^+$ for technical reasons. It is however important to compare both tracers observationally and on the basis of numerical simulations with more detailed networks, to understand potential differences in their evolution and their sensitivity to different conditions. Potentially, such combined information may then also result in stronger constraints.\\
A last, but rather fundamental point concerns the definition of the age of the core, which is trivial in a one--zone model with a constant density, but much more complex in a dynamical environment. Our tentative definition here was the time passed after starting the simulations from the initial conditions, but of course even before that time chemical evolution may have occured.  It is very likely that this question can only be settled once it is understood how much the lower--density gas contributes to setting the chemical initial conditions and the initial deuterium fractions, in order to provide a more well--defined starting point for the dynamical evolution.

\section*{Acknowledgement}
We thank Silvia Leurini for helpful discussion on the observational uncertainties and their presentation.
BK, SB, DRGS, and RB thank for funding through the DFG priority program `The Physics of the Interstellar Medium' (projects BO 4113/1-2, SCHL 1964/1-2, and BA 3706/3-2). Furthermore RB acknowledges additional funding from the DFG for this project via the grants BA 3706/4-1 and BA 3706/15-1. SB is grateful to Paola Caselli and Jonathan Tan for fruitful discussions on this topic during the KITP program 'The Cold Universe' in Santa Barbara. We are indebted to Tommaso Grassi for having helped with the modelling
and to Olli Sipil\"a for his contribution on benchmarking the Walmsley model. SB also thanks for the kind hospitality of the Department of Astronomy of Concepci\'on, UdeC, Chile, where part of this work has been completed. DRGS thanks for funding through Fondecyt regular (project code 1161247), the ''Concurso Proyectos Internacionales de Investigaci\'on, Convocatoria 2015'' (project code PII20150171), ALMA Conicyt (project 3116000001) and the BASAL Centro de Astrof\'isica y Tecnolog\'ias Afines (CATA) PFB-06/2007. The software used in this work was developed in part by the DOE NNSA
 ASC- and DOE Office of Science ASCR-supported Flash Center for Computational
 Science at the University of Chicago. The simulations were performed on the local GPU cluster \ita{HUMMEL} at the University of Hamburg as well as on HLRN--III (www.hlrn.de) under project--ID hhp00022. BK further gratefully acknowledges the Gauss Centre for Supercomputing e.V. (www.gauss-centre.eu) for funding this project (project--ID pr92pu) by providing computing time on the GCS Supercomputer SuperMUC at Leibniz Supercomputing Centre (LRZ, www.lrz.de).

\begin{appendix}
\section{Resolution study}
Figure \ref{figA1} shows the mass--weighted average deuterium ratio D$_\mathrm{frac}^{H_2D^+}$ as function of time for run Lmu10M2S2 and three resolutions, ranging from $\Delta x=940\,\mathrm{AU}$ to $\Delta x=59\,\mathrm{AU}$. Due to the formation 
of a sink particle we are only able to compare the results up to this point. However, as can be inferred from this figure, the evolution is very similar for all runs, showing that our results are numerically converged.
\begin{figure}
	\begin{center}
	\includegraphics[width=0.4\textwidth,angle=-90]{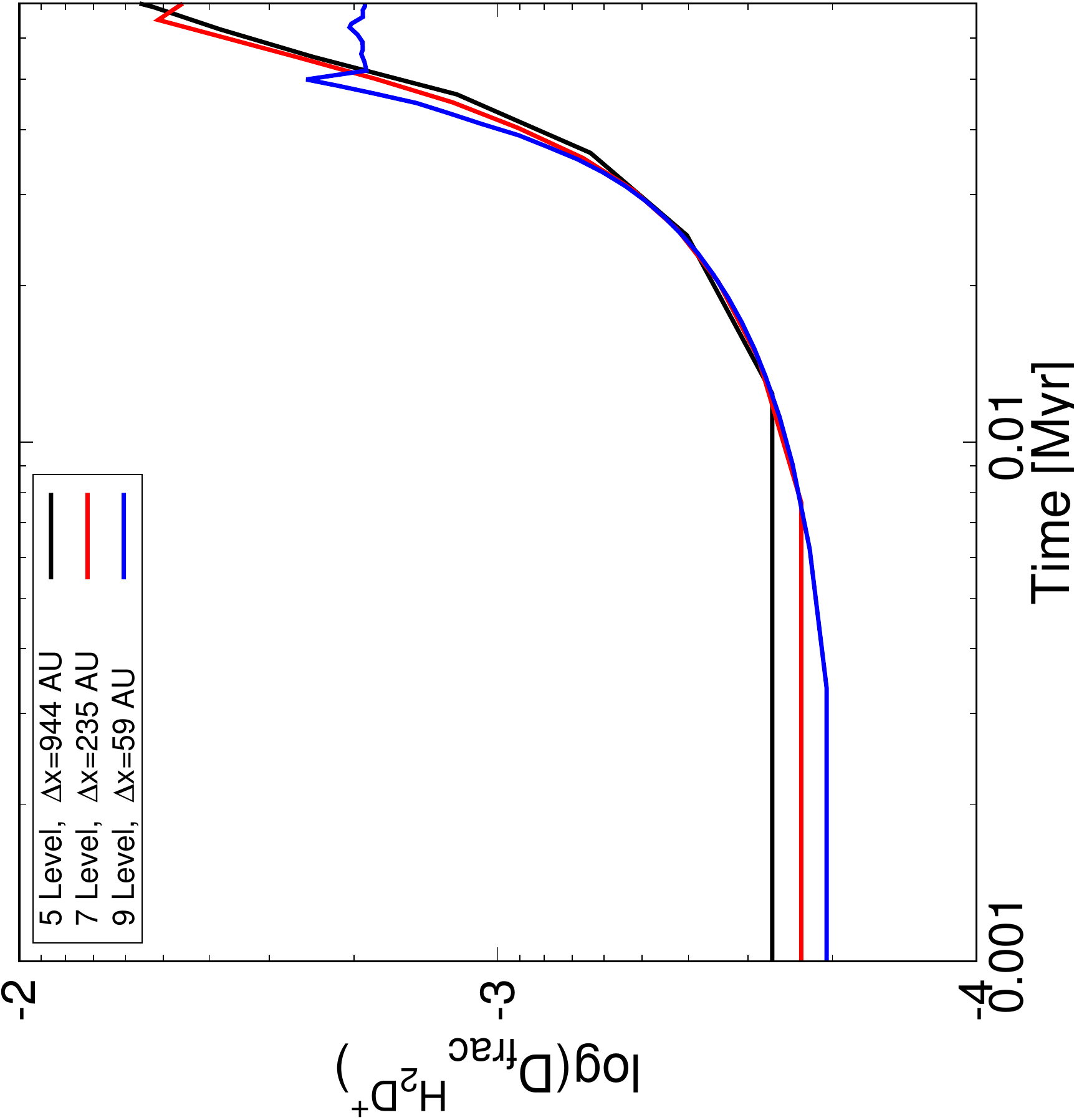}
	\caption{Convergence study of run Lmu10M2S2 showing good agreement between runs with different numerical resolution. The difference of the deuterium ratio at times $t<8\,\mathrm{kyr}$ is due to initial timestep being larger than the period of 
	dumping out data.}
	\label{figA1}
	\end{center}
\end{figure}
\end{appendix}

\bibliography{mybib_D} 

\bibliographystyle{mn2e}
\end{document}